# Electronic transport properties of few-layer graphene

## A Thesis

### Submitted to the
### Tata Institute of Fundamental Research, Mumbai
### for the degree of Doctor of Philosophy
### in Physics

### by

### Biswajit Datta

### School of Natural Sciences
### Tata Institute of Fundamental Research
### Mumbai
### July, 2019
### Final Version Submitted in December, 2019

# DECLARATION

This thesis is a presentation of my original research work. Wherever contributions of others are involved, every effort has been made to indicate this clearly, with due reference to the literature, and acknowledgement of collaborative research and discussions.

The work was done under the guidance of Professor Mandar M. Deshmukh, at the Tata Institute of Fundamental Research, Mumbai.

**Biswajit Datta**

In my capacity as supervisor of the candidate's thesis, I certify that the above statements are true to the best of my knowledge.

**Prof. Mandar M. Deshmukh**

Date:

*Dedicated to my family*

# Acknowledgements

First and foremost, I am privileged to thank my PhD supervisor Prof. Mandar M. Deshmukh for his abled guidance, continuous support, motivation and immense knowledge on the subject. I am thankful to him for allowing me to work in the Nanoelectronics laboratory. Apart from science, I truly value the discipline he has established in the group. I have benefitted by the practice of communicating in English among peers and conducting group meetings every week. It has significantly helped to improve my communication skill and confidence. His outreaching capacity helped me to learn and get connected to the scientists worldwide. I owe my utmost gratitude to him to show me the gateways to participate in several national and international conferences during my PhD tenure. His continuous encouragement and supervision enkindled flourishing my imagination that prepared me to take ownership of my work. I learnt how not to let go any opportunity, to pursue the expectations of the reviewers of the journals, from his determinacy and positive attitude, which rewarded me with good publications. I owe my sincerest gratitude and thankfulness to my supervisor to set deadlines that helped me to complete my thesis in time.

During my PhD, I had the fortune to collaborate with Professor Rajdeep Sensarma from the Department of Theoretical Physics of TIFR in my first two projects. I earnestly thank him for sharing his vast knowledge, potent inputs in my papers and the valuable time to engage in fruitful discussions. I also thank his students and my friends Abhisek and Santanu for helping me with the calculations. I owe my sincerest thanks to Prof Mandar Deshmukh, Prof. Rajdeep Sensarma and Prof. Rajamani Vijayaraghavan for monitoring my PhD work throughout as TMC (Three Member Committee) members.

I am grateful to the Infosys foundation to invite and provide financial support to the great scientists of the world. Our recent fruitful collaboration with Prof. Justin C. W. Song started when he visited TIFR to deliver an Infosys Foundation Lecture. We discussed one of our recent data and he was quick to recognize the importance of our work. This started a collaboration with him and his postdoc scholar Li-kun Shi which eventually resulted in a very interesting work being published in Science Advances. I am thankful to them for their help in making it successful. I also learnt from his excellent quality of scientific writing. He was even very kind to help me find out a postdoc position – his knowledge about the community was extremely useful to prepare an initial list of places I applied.

I was an awardee of the prestigious Prime Minister's Fellowship Scheme for Doctoral Research (2017-2019). I am grateful to my PhD supervisor Prof. Mandar M. Deshmukh, Dr. Shyam Kumar Choudhary, and Dr. Sumitesh Das from TATA Steel for their support in the Prime Minister's fellowship. I am also grateful to Confederation of Indian Industry (CII), especially Neha Gupta and Ravi Hira for their help during the Prime Minister's fellowship tenure.

I thank Digambar and Santosh for always helping me with any technical/instrumental problems and for keeping the clean room running. I am grateful to all the staff of TIFR Low Temperature Facility for providing us with a continuous supply of liquid Helium and Nitrogen. This allowed us to keep our flow cryostats cold all the time where I have done all the experiments. I am thankful to Atul and the technicians of our Central Workshop for making many spare parts of instruments . I thank Shilpa, Rudheer, and Purandare of the Nanomaterials Lab for generously helping us with the SEM and TEM imaging.

I am grateful to my awesome senior PhD students John Mathew and Sameer Grover who helped me to learn about low temperature measurements. I thank John for his general advice in and outside the lab. Sameer Grover was always the go to person whenever I needed help with Labview or any programming. His level of understanding and interest in instruments/programming always fascinated me. I thank Rajashree (project student) for helping me learn the basics of the fabrication process in the beginning. In the initial days, it was great to share the lab space with many wonderful colleague friends such as Abhinandan, Apoorv, Raj, Rajib and Anupama. Many times including the time of my postdoc search, I have benefitted from Apoorv's vast knowledge about the scientific community. A special thanks to Apoorv and Abhinandan for hosting me in New York after an APS March meeting.

I thank Hitesh Agarwal, another project student of our lab for helping me with experiments. He later moved to the Institute of Photonic Sciences, Barcelona (Spain) as a PhD student, I wish him all the best for his PhD. I also thank Vishakha, Ravi, Anirudh, and Shraddha who were involved in the same project as mine at different points of time. All of them have joined as PhD students in different institutions – I wish all the best for their career. Most recently, I worked with Ipsita and Jay in two different projects. Ipsita has moved to the Institute of Photonic Sciences and Jay is going to join in the University of Virginia as a PhD student. I admire them for their hard work and wish all the best for their PhD. I thank Aum, a very energetic project student in our lab, for teaching me calisthenics on what he was an expert. It was fun and refreshing to work with him.

I also express my heartfelt gratitude to Vibhor, Shamashis, and Sajal – the PhD alumni of our lab who are now themselves faculty in different institutes. Despite their busy schedule, they kindly reviewed our papers at different point of time and gave us valuable suggestions to improve our manuscripts. I thank Allan MacDonald and his student Fengcheng Wu, Jainendra Jain, Jim Eisenstein, Chandni U. and Sreejith GJ for engaging in scientific discussions with us and comments on our different manuscripts.

I am grateful to have the nicest set of people like my current colleagues. Special thanks to Pratap and Subhajit, the PhD students, for taking the lead to change many things for good in lab. I immensely thank Pratap for all the fruitful discussions we had and for his significant contribution on our recent work. I thank our current postdocs Subhamoy and Varma for bringing their expertise in our lab for which we have benefitted significantly. A special thanks to Sangani Varma and his wife Divyajyoti for inviting me for a delicious lunch, otherwise I would have not known about his excellent cooking skill apart from his extraordinary research skills. I thank all the current lab members Pratap, Supriya, Subhajit, Sanat, Lucky, Jay, Surya, Subhamoy and Sangani Varma for their company. and creating such a lively positive environment in the lab. I wish them all the best for their future.

I am grateful to my TIFR friends Krishnendu, Abhisek, Anirban, Emroj, Banaj, Kamalesh, Ahana, Suman and others for their company especially during Durga puja pandal hopping and a few short trips in Mumbai. A special thanks to all the people who accompanied during several monsoon and winter trekking in the Western Ghat Mountain Range – something I thoroughly enjoyed. Thanks to Krishnendu, Megha, Chanchal, Manaj, Suryanarayan, Tanay, Anirban, Sawani to name a few. I deeply thank my supervisor Prof. Mandar Deshmukh for arranging a few excellent monsoon trekkings.

I take this opportunity to thank a friend of mine who has made an enormous contribution to shape my life both academically and personally. I want to thank Animesh who has remained my best friend from school days. I owe him a lot – I would not have achieved many things academically without a true friend like him. Over the last fourteen years, he has become a part of my family.

Another person I am blessed to have in my life is Pratyusha – a truly multitalented person. Besides learning from her all the time, she has been my strength in all ups and downs for the last ten years. Her constant motivation, and support kept me ambitious and confident during my hard times. It feels amazing to see her happier than me in my success. I can't thank her enough for all that she has done for me.

I am very fortunate to meet a few teachers who have literally changed my life. I want to thank Mr. Partha Dasgupta – my science teacher in the tenth standard. He has a large contribution in laying the foundation of science in me. I am grateful to have Mr. Goutam Dasgupta, my Higher Secondary Physics teacher. Apart from being an excellent teacher his dedication for students made him very close to my heart. He really grew the interest of physics in me and I still miss his guidance. I also want to thank Prof. Alok Shukla and Prof. Ravinder Puri with whom I worked briefly during my masters in IIT Bombay. I feel fortunate to attend classes by such extraordinary teachers and thank them for their support.

In the end, I would like to acknowledge my family for their support in everything. I thank my parents for giving me such freedom and love. I wish they become even happier in the days to come. I thank my sweet elder sister Susmita who is really a friend of mine. It is great to have a sister like her who I can talk anything about. I also thank my eldest sister Suchitra for her love and support. I wish both of them a great success ahead.

# Contents

# List of Tables

# List of Figures

# Abstract

Starting from discovery of graphene [1] by Novoselov, K. S. *et al.* in 2004 the research on 2D materials is ever expanding. The quest for new functionalities has lead to the birth of innumerable 2D materials with electronic properties all the way from insulating, metallic to superconducting [2]. Different phenomena including charge density waves [3, 4], ferromagnetism [5, 6, 7], superconductivity [8, 9] have been observed in these materials surprisingly even down to monolayer thickness. The unique advantage of exfoliating thin layers from the bulk crystals and assembling layers of different crystals has kept this field at the forefront of condensed matter research over the last decade. Stacking different 2D materials with dissimilar properties one can achieve new functionalities absent in the individual layers. A prime example of this is emergent superconductivity in twisted bilayer graphene [10] where two graphene layers are stacked with a relative small twist angle. This is remarkable because graphene, purely made of carbon atoms has seemingly no connection with superconductivity. Another ingredient in few-layer graphene is layer degrees of freedom – the dispersion of bands can be tuned with number and stacking of layers. For $N$ layer ABC-stacked graphene the dispersion of the bands is $E \sim p^N$ [11]. Higher the exponents $N$, the flatter the band is at low momentum. For flat bands electronic interaction becomes important. Flat band is also the reason of unconventional superconductivity in twisted bilayer graphene. Moreover, electric field can also change the bandwidth hence it can tune the effect of electronic interaction.

In this thesis we will focus on a particular variant of few-layer graphene – ABA-stacked trilayer graphene. Bernal (ABA) stacked trilayer graphene (TLG) is a multiband sys-tem consisting of a pair of Dirac-like massless linear bands and a pair of massive quadratic bands [12, 13, 14]. We have studied the electronic properties of this system in detail and unfolded many interesting physics problems. The whole thesis is structured in the following way. In chapter 2 and chapter 3 we discuss the theoretical ingredients we will need to understand the experimental data presented in this thesis. Chapter 2 focuses on the band structure of graphene and few-layer graphene. In chapter 3 we focus on quantum Hall effect of graphene and few-layer graphene. In chapter 4 we discuss the device fabrication and characterization. Chapter 5, Chapter 6 and Chapter 7 present the original works which are published as papers.

In chapter 5 we describe the Berry's phase in ABA-trilayer graphene and discuss the intricacy of calculating Berry's phase in a multiband system in general. We show that Shubnikov-de Haas (SdH) oscillations of the quadratic band of ABA-TLG are shifted by a phase that sharply departs from the expected $2\pi$ Berry's phase. Our analysis reveals that, surprisingly, the anomalous phase shift is non-trivial and is inherited from the non-trivial Berry's phase of the linear Dirac band due to strong filling-enforced constraints between the linear and quadratic band Fermi surfaces. Given that many topological materials contain multiple bands, our work indicates how additional bands, which are thought to obscure the analysis, can actually be exploited to tease out the subtle effects of Berry's phase.

The content of chapter 6 can be divided into two parts. In the first part, we study the Landau level (LL) crossings at zero electric field arising from LLs of different band origin and degeneracy. We note that the LLs from the massless and massive bands disperse differently with the magnetic field which causes them to intersect meaning they become degenerate for some values of magnetic field. In particular, the valley degeneracy lifted $N_{\mathrm{M}}$=0 monolayer graphene (MLG)-like LL, independent of magnetic field, produces an intricate LL crossing pattern with a few ring-like structures. In a single band system, the Landau levels disperse linearly in the density-magnetic field phase space where the slope of the line is given by the filling factor ($\nu$). However, in a multiband system, as seen in our experiment, LL filling sequence (hence the filling factor of a given band) changes after a LL crossing point due to which a given LL shows a sudden change of slope. The ring-like structures, as we will see, is a consequence of crossing $N_{\mathrm{M}}$=0 MLG-like LL with $N_{\mathrm{B}}$=2 bilayer graphene (BLG)-like LL with different degeneracies.

In the second part, we study electronic interaction driven quantum Hall ferromagnetic states in ABA-trilayer graphene. In very clean systems and at sufficiently high magnetic field, exchange interaction of electrons can drive the system in a ferromagnetic state known as quantum Hall ferromagnet. One consequence of this is the enhancement of the LL gaps beyond the single particle picture which we confirm experimentally – this shows that Coulomb interaction is important in our devices. Moreover, we observe hysteresis as a function of filling factor and spikes in the longitudinal resistance ($R_{xx}$) which, together, signal the formation of quantum Hall ferromagnetic states at relatively low magnetic field.

In chapter 7 also there are two distinct aspects we focus on. In the first part, we study the effect of the trigonal warping – a three fold distortion of the Fermi surface in ABA-trilayer graphene. Trigonal warping, despite being small in magnitude, has important effects on the low energy physics of few-layer graphene. It stretches the Fermi circle along three directions enforcing discrete threefold rotational symmetry. In the presence of a magnetic field, it leads to a selection rule for the coupling between different Landau levels. In particular, we observe anticrossings between some LLs differing by 3 in LL index, which result from the breaking of the continuous rotational symmetry to $C_3$ symmetry by the trigonal warping. Our experiment provides smoking-gun evidence for the trigonal warping of the low energy bands.

In the second part, we discuss the effect of the non-uniform charge distribution in the ABA-stacked TLG. It is theoretically predicted that the charge distribution on the three layers of the ABA-stacked TLG can be non-uniform and this has an important effect on the substructure of the lowest Landau level. In particular, the sequence of the zeroth Landau levels between filling factors -6 to 6 in ABA-stacked trilayer graphene is unknown because it depends sensitively on this non-uniform charge distribution. Using the sensitivity of quantum Hall data on the electric field and magnetic field, in an ultraclean ABA-stacked TLG sample, we quantitatively estimate the non-uniformity of the electric field and determine the sequence of the zeroth LLs.

To summarize, in this thesis we have studied

- Intricacy of calculating Berry's phase in a multiband system [15]

- LL crossing physics and interaction driven quantum Hall ferromagnetic states in ABA-trilayer graphene [16]

- Finding the LL energy spectrum and the effect of broken continuous rotational symmetry in ABA-trilayer graphene [17]

**Chapter 2**

# Overview of graphene and few-layer graphene

In this chapter, we give an overview of monolayer and few-layer graphene. In particular, we discuss the band structure, the role of different symmetries, like mirror and inversion symmetry, and how the band structure gets affected when these symmetries are controllably broken in the experiment. We also discuss new concepts of pseudospin, chirality, winding number which are not relevant in normal 2DEG. Finally, we introduce the concept of Berry's phase and its connection to the pseudospin rotation in monolayer and few-layer graphene.

## 2.1 Monolayer graphene

### 2.1.1 Tight binding band structure

Graphene is a single atomic layer of carbon atoms arranged in a honeycomb lattice which is not a Bravais lattice. However, the structure can be thought of like a triangular lattice with a two atomic basis. This results in two atoms per unit cell. Fig. 2.1a shows a schematic of the crystal structure of graphene. The two lattice vectors are given by

$$\mathbf{a}_1 = \frac{a}{2}(3, \sqrt{3}) \,, \quad \mathbf{a}_2 = \frac{a}{2}(3, -\sqrt{3}) \,, \tag{2.1}$$

where a=1.42 $\mathring{A}$ is the carbon-carbon bond length which is related to the lattice con-

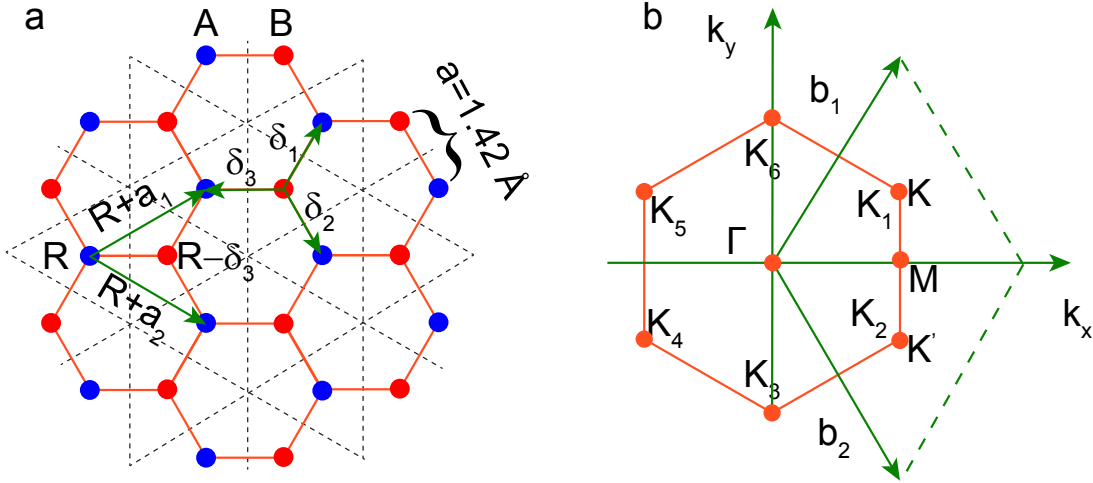

Figure 2.1: **Monolayer graphene lattice and Brillouin zone.** (a) Schematic of the crystal structure of graphene. The blue sites and red sites denote the A and B sublattice respectively. The two lattice vectors $\mathbf{a}_1$, $\mathbf{a}_2$ and three vectors joining the nearest neighbours $\boldsymbol{\delta}_1$, $\boldsymbol{\delta}_2$, $\boldsymbol{\delta}_3$ are also shown. The dashed grid shows the triangular graphene lattice. The whole crystal can be constructed by shifting the unit cell with two basis atoms by the lattice vectors $\mathbf{a}_1$ and $\mathbf{a}_2$. (b) First Brillouin zone of graphene. The reciprocal vectors are denoted by $\mathbf{b}_1$ and $\mathbf{b}_2$. The high symmetry points are marked. All the corners of the first Brillouin zone are high symmetry points but only two successive points ($\mathbf{K}$ and $\mathbf{K}'$) are independent. Rest of the corner points can be attained translating those two points by the reciprocal vectors.

stant by $|\mathbf{a}_1|=|\mathbf{a}_2|=\sqrt{3}a$. The reciprocal vectors are given by

$$\mathbf{b}_1 = \frac{2\pi}{3a}(1, \sqrt{3}), \quad \mathbf{b}_2 = \frac{2\pi}{3a}(1, -\sqrt{3}).  \tag{2.2}$$

Fig. 2.1b shows the first Brillouin zone which is a hexagon. The coordinates of the corners of the Brillouin zone are given by

$$\mathbf{K}_1 = \frac{\pi}{a}\left(\frac{2}{3}, \frac{2}{3\sqrt{3}}\right), \quad \mathbf{K}_2 = \frac{\pi}{a}\left(\frac{2}{3}, -\frac{2}{3\sqrt{3}}\right), \quad \mathbf{K}_3 = \frac{\pi}{a}\left(0, -\frac{4}{3\sqrt{3}}\right),$$

$$\mathbf{K}_4 = \frac{\pi}{a}\left(-\frac{2}{3}, -\frac{2}{3\sqrt{3}}\right), \quad \mathbf{K}_5 = \frac{\pi}{a}\left(-\frac{2}{3}, \frac{2}{3\sqrt{3}}\right), \quad \mathbf{K}_6 = \frac{\pi}{a}\left(0, \frac{4}{3\sqrt{3}}\right).  \tag{2.3}$$

All the corner points are not independent: $\mathbf{K}_1$, $\mathbf{K}_3$, $\mathbf{K}_5$ are related by the reciprocal lattice vector and are denoted by $\mathbf{K}$. Similarly, $\mathbf{K}_2$, $\mathbf{K}_4$, $\mathbf{K}_6$ are also related by the reciprocal lattice vector and are denoted by $\mathbf{K}'$.

The relative coordinate of the three nearest neighbours with respect to an atom can be written as

$$\boldsymbol{\delta}_1 = \frac{a}{2}(1, \sqrt{3})\,, \quad \boldsymbol{\delta}_2 = \frac{a}{2}(1, -\sqrt{3})\,, \quad \boldsymbol{\delta}_3 = -a(1, 0)\,. \tag{2.4}$$

Under the nearest neighbour approximation the tight binding Hamiltonian can be written as [18]

$$H = -\gamma_0 \sum_{<i,j>,\sigma} (a^{\dagger}_{\sigma,i} b_{\sigma,i} + \text{h.c.})\,, \tag{2.5}$$

where $< i, j >$ means that the summation is up to the nearest neighbours and $\sigma$ denotes the spin of the electrons. The coordinates of the three nearest neighbours of a B site (at $\boldsymbol{R} - \boldsymbol{\delta}_3$) are three A sites located at $\boldsymbol{R}$, $\boldsymbol{R} + \boldsymbol{a}_1$ and $\boldsymbol{R} + a_2$ (see Fig. 2.1a). Incorporating these coordinates, equation 2.5 can be written as

$$H = -\gamma_0 \sum_{\boldsymbol{R},\sigma} \left( a^{\dagger}_{\sigma}(\boldsymbol{R}) b_{\sigma}(\boldsymbol{R} - \boldsymbol{\delta}_3) + a^{\dagger}_{\sigma}(\boldsymbol{R} + \boldsymbol{a}_1) b_{\sigma}(\boldsymbol{R} - \boldsymbol{\delta}_3) + a^{\dagger}_{\sigma}(\boldsymbol{R} + \boldsymbol{a}_2) b_{\sigma}(\boldsymbol{R} - \boldsymbol{\delta}_3) + \text{h.c.} \right)\,. \tag{2.6}$$

This Hamiltonian can be diagonalized easily by changing the basis from the real space to the momentum space. The annihilation operator can be written in terms of its Fourier components $a_{\sigma}(\boldsymbol{R}) = \frac{1}{\sqrt{N}} \sum_n e^{-i\boldsymbol{k}_n.\boldsymbol{R}} a_{k_n,\sigma}$. Incorporating it to the equation 2.6 one gets

$$H_{MLG} = -\gamma_0 \left( s(k) a^{\dagger}_{k\sigma} b_{k\sigma} + \text{h.c.} \right)\,, \tag{2.7}$$

where $s(k) = 1 + e^{i\boldsymbol{k}.\boldsymbol{a}_1} + e^{i\boldsymbol{k}.\boldsymbol{a}_2}$. So, in the $(a_{k,\sigma}, b_{k,\sigma})$ basis the Hamiltonian can be written as

$$H_{MLG} = -\gamma_0 \begin{pmatrix} 0 & s^*(k) \\ s(k) & 0 \end{pmatrix}\,. \tag{2.8}$$

The energy eigenvalues give rise to the conduction and the valence bands shown in Fig. 2.2a.

$$E_{MLG} = \pm\gamma_0 |s(k)| \tag{2.9}$$

$$= \pm\gamma_0 \sqrt{1 + 4\cos\left(\frac{3}{2}k_x a\right)\cos\left(\frac{\sqrt{3}}{2}k_y a\right) + 4\cos^2\left(\frac{\sqrt{3}}{2}k_y a\right)} \tag{2.10}$$

$H_{MLG}$ can be expanded around $K$ and $K'$ valleys to get the effective low energy Hamiltonian:

$$\mathscr{H}_{MLG} = \frac{3\gamma_0 a}{2\hbar}\begin{pmatrix} 0 & p_x + i\xi p_y \\ p_x - i\xi p_y & 0 \end{pmatrix} = v_F \boldsymbol{\sigma}.\boldsymbol{p}\,, \tag{2.11}$$

where $\xi = \pm 1$ for the $K$ and $K'$ valleys respectively.

Energy eigenvalues and eigenvectors of $\mathscr{H}_{MLG}$ are

$$\epsilon_\pm = \pm v_F p\,, \quad |\psi_\pm\rangle = \frac{1}{\sqrt{2}}\begin{pmatrix} 1 \\ \pm\xi e^{i\xi\phi} \end{pmatrix}\,, \tag{2.12}$$

where $\pm$ refer to the conduction or valence bands, the Fermi velocity ($v_F$) is defined as $v_F = \frac{1}{\hbar}\frac{\partial\epsilon}{\partial k} = \frac{3\gamma_0 a}{2\hbar} \approx 10^6 \ \text{ms}^{-1}$, $p = \hbar k = \sqrt{p_x^2 + p_y^2}$ and $\phi$ is the polar momentum angle $\phi = \tan^{-1}(p_y/p_x)$. Energy dispersion of the eigenvalues of $\mathscr{H}_{MLG}$ is linear which looks exactly like that of massless Dirac fermion's with the velocity of light ($c$) replaced by $v_F$. Fig. 2.2b shows the contour plot of the conduction band. Fig. 2.2c shows the zoomed-in band structure showing the conduction and the valence bands touch at the six corners of the Brillouin zone. Fig. 2.2d shows a single cone in the $K$ valley.

The absence of a bandgap in graphene can be argued just from the symmetry consideration. We note that to generate a bandgap one needs to add a mass term proportional to $\sigma_z$ in the Hamiltonian (equation 2.11). The presence of a mass term sets the two sublattices on-site potential different. However, the two sublattices are related by the inversion symmetry. Since pristine graphene respects inversion symmetry, the different on-site potential on two sublattices is not allowed. That means a mass term is forbidden in monolayer graphene by inversion symmetry.

Moreover, there are several differences between a linear band compared to a usual quadratic band. For the usual quadratic dispersion $\epsilon = \frac{p^2}{2m}$, the velocity $v = \sqrt{2\epsilon/m}$ depends on the energy. The most striking difference for the massless Dirac dispersion is that the velocity of an electron is independent of its energy. On the other hand, the effective mass of a Dirac electron increases with an energy unlike the constant mass for an electron with quadratic dispersion, as we will see in the next section.

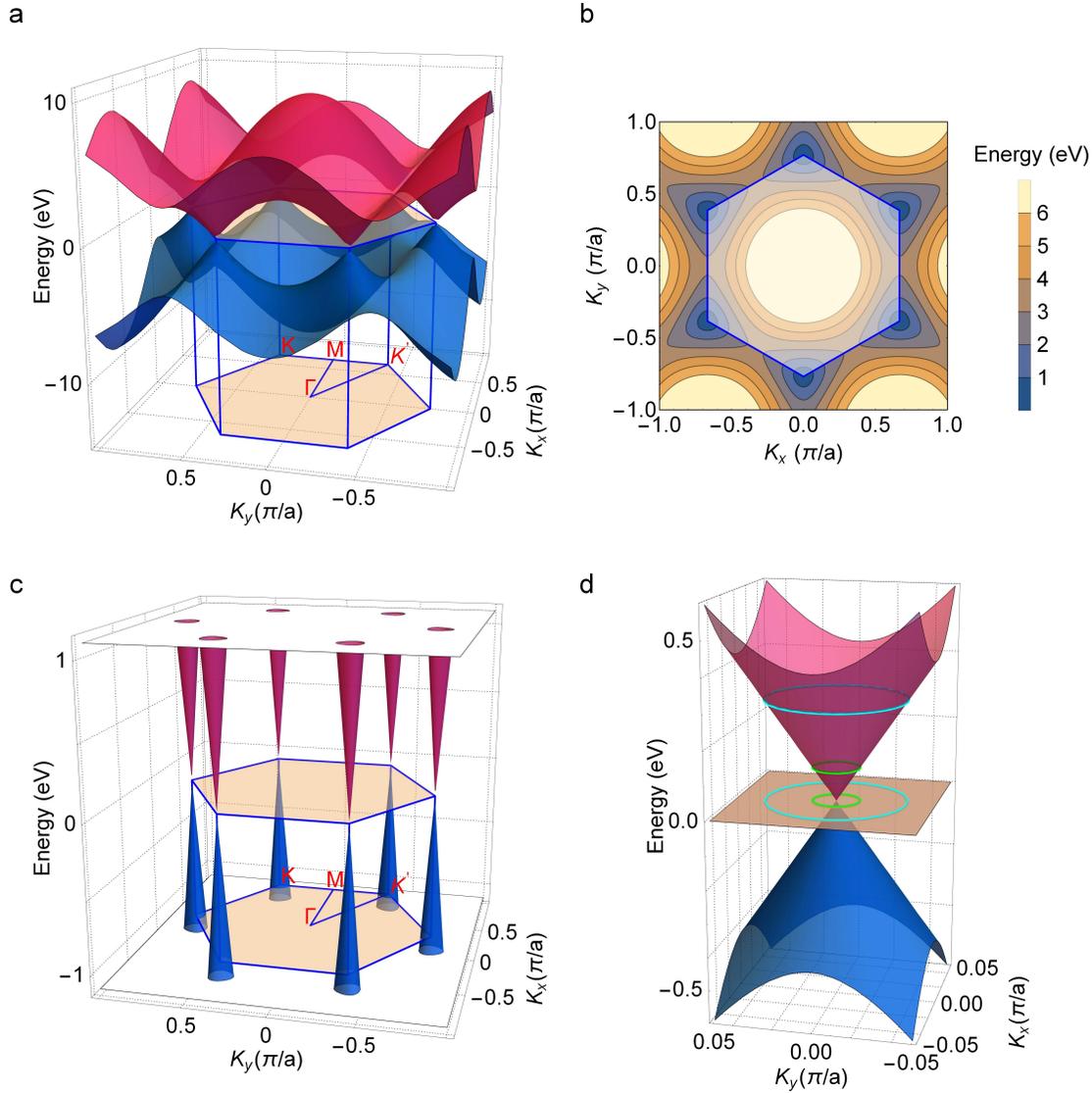

Figure 2.2: **Tight binding band structure of monolayer graphene.** (a) 3D plots of the conduction and the valence band. We take $\gamma_0$=3.1 eV, $\gamma_1$=390 meV. The first Brillouin zone is also marked by the blue hexagon. (b) Contour plot of the conduction band with the first Brillouin zone superimposed on it. It shows that the Fermi contours are trigonally warped at high energy. (c) The zoomed-in band structure. We can clearly see that the two cones touch at the six corner points of the Brillouin zone. (d) Further zoomed-in view of a single cone around the **K** point; the momentum coordinates are plotted with respect to the **K** point. Two Fermi contours and their projections on the $\mathbf{k}_x$-$\mathbf{k}_y$ plane are shown. It shows that the Fermi contours are circles at low energy.

## 2.1.2 Cyclotron mass

The cyclotron mass of electrons which determines the frequency of the cyclotron motion in the presence of a magnetic field can be written as

$$m^* = \frac{\hbar^2}{2\pi} \left[ \frac{\partial A(E)}{\partial E} \right]_{E=E_F} , \tag{2.13}$$

where $A(E)$ is the $\mathbf{k}$ space area enclosed by the electron orbit on the Fermi surface. Since at low energy the graphene band is conical, the cross-sectional area of the band is

$$A(E) = \pi k^2(E) = \frac{\pi E^2}{\hbar^2 v_F^2} . \tag{2.14}$$

Invoking equation 2.14 in equation 2.13 one gets

$$m^* = \frac{E_F}{v_F^2} = \frac{\hbar k_F}{v_F} . \tag{2.15}$$

Taking into account the spin and the valley degeneracies in graphene, the Fermi wave vector can be related to the density by $k_F = \sqrt{\pi n}$. Incorporating this we can calculate the effective mass as a function of density as shown in Fig. 2.3.

$$m^* = \frac{\hbar}{v_F} \sqrt{\pi n} . \tag{2.16}$$

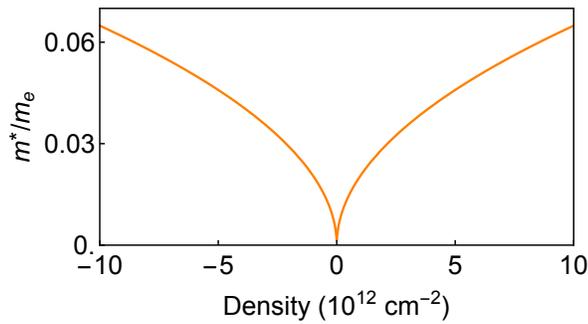

Figure 2.3: Cyclotron mass of electrons in graphene (normalized with free electron mass) as a function of density.

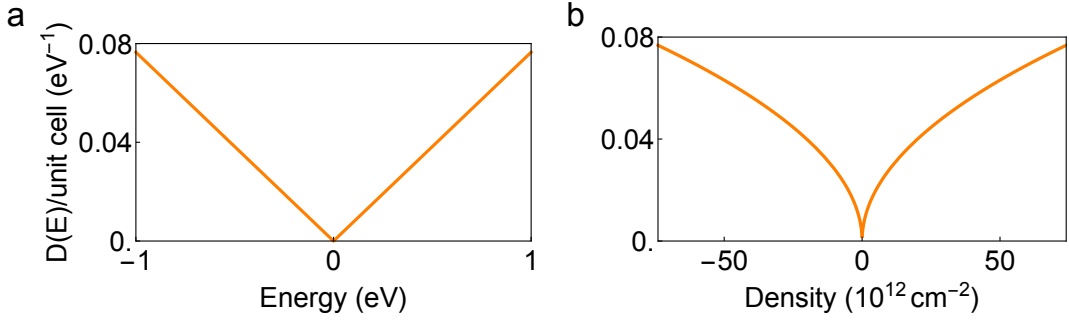

Figure 2.4: **Density of states of graphene.** (a) DOS(E) per unit cell as a function of energy. (b) DOS(E) per unit cell as a function of density.

### 2.1.3   Density of states

The density of states (DOS) at low energy can be easily found out using the dispersion relation in equation 2.12. $DOS(E)$ per unit area (incorporating the spin and valley degeneracies) can be defined as

$$DOS(E) = 4\frac{2\pi k}{(2\pi)^2}\frac{dk}{dE} = \frac{2}{\pi}\frac{E}{(\hbar v_F)^2} = \frac{8}{9\pi}\frac{E}{\gamma_0^2 a^2} \,. \qquad (2.17)$$

The DOS per unit cell can be written as

$$DOS(E) = \frac{8A_c}{9\pi}\frac{E}{\gamma_0^2 a^2} = \frac{4\sqrt{3}}{3\pi}\frac{E}{\gamma_0^2} \,, \qquad (2.18)$$

where the unit cell area is $A_c = 3\sqrt{3}a^2/2$. Fig. 2.4a and Fig. 2.4b show the DOS per unit cell as a function of energy and density respectively. The DOS is vanishingly small near the Fermi energy (of the pristine graphene) which makes graphene a semimetal.

### 2.1.4   Pseudospin, chirality and winding number

We note that the eigenvectors of graphene have two components which dictate the weight of the wavefunction on the A and B sublattice. The weights on the two sublattices are equal (but can have a phase difference) since the two sublattices are related by the inversion symmetry and graphene respects inversion symmetry. The sublattice degrees of freedom is known as pseudospin because mathematically it can be treated as the two components of a real spin. Sublattice A can be denoted as pseudospin up state $|\Uparrow\rangle = (1,0)^{\mathrm{T}}$ and sublattice B can be denoted as pseudospin down state $|\Downarrow\rangle = (0,1)^{\mathrm{T}}$ where T means the matrix transpose operation. In terms of the pseudospin up and down states the graphene wave function (equation 2.12) can be written as [19]

$$|\psi_{\pm}\rangle = \pm \frac{1}{\sqrt{2}} \left( |\Uparrow\rangle \pm \xi e^{i\xi\phi} |\Downarrow\rangle \right) \ . \tag{2.19}$$

We note that the two pseudospin components of the wavefunction have a phase difference ($\phi$) – known as pseudospin angle which defines the angle of the pseudospin vector. Moreover, for graphene, this angle is also the polar momentum angle. So, for graphene electrons, the direction of the momentum and the pseudospin vector are always aligned (see Fig. 2.5). In other words, the electrons in graphene are chiral and they have a definite chirality (also known as helicity). This can be seen by defining a helicity operator which projects electrons spin in the direction of its momentum [18]:

$$\hat{h} = \frac{1}{2} \boldsymbol{\sigma}.\hat{\boldsymbol{p}} \ . \tag{2.20}$$

From the definition, we can see that the eigenvectors of $\mathscr{H}_{MLG}$ (equation 2.12) are also the eigenvectors of the helicity operator:

$$\hat{h} |\psi_{\pm}\rangle = \pm \frac{1}{2} \xi |\psi_{\pm}\rangle \ . \tag{2.21}$$

It means that near $K$ valley electrons have positive helicity and holes have negative helicity. Further, near $K'$ valley it is opposite i.e. electrons have negative helicity and holes have positive helicity.

The pseudospin vector in graphene rotates by a full circle when the momentum rotates by $2\pi$ since the momentum and pseudospin are aligned in graphene. The number of times the pseudospin vector rotates when the momentum rotates by $2\pi$ is known as the pseudospin winding number ($n_w$) which is 1 for monolayer graphene. This pseudospin winding number determines the Berry's phase in graphene and few-layer graphene [20], as we will see later. It is important to notice that pseudospin degrees of freedom exist because of the two atomic bases in graphene. The concept of pseudospin is not there in normal two-dimensional electron gas (2DEG).

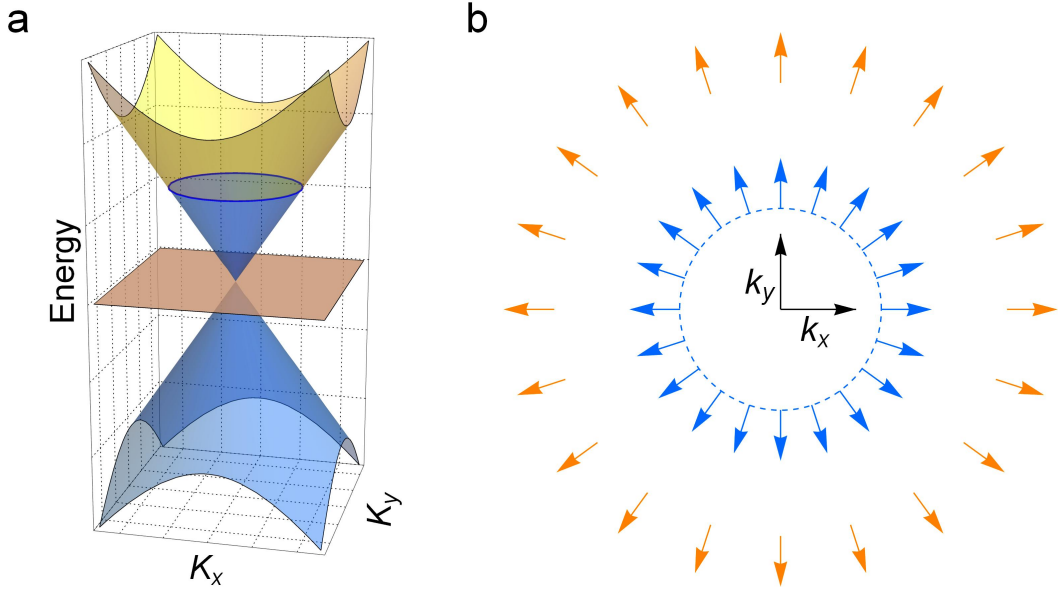

Figure 2.5: **Pseudospin and chirality in monolayer graphene.** (a) Schematic of the graphene cone showing the Fermi circle in blue. The orange plane denotes the chemical potential for the charge neutral graphene. (b) The blue dashed circle represents the Fermi circle. The blue and the orange arrows at each point show the direction of the momentum and the pseudospin vector respectively. As discussed in the text, they are aligned in monolayer graphene.

### 2.1.5 Berry's curvature and Berry's phase of gapless graphene

Phase plays a very important role in quantum mechanics. Even though the absolute value of the phase does not have a physical consequence, the phase difference is a measurable quantity in the experiment. According to the adiabatic theorem of quantum mechanics, when a parameter of a quantum mechanical system is changed adiabatically (slowly), the system remains in the same quantum state of the instantaneous Hamiltonian. When the parameter is changed over a closed loop, the wavefunction picks up a geometric phase apart from the usual dynamical phase because the system goes through all the state of the instantaneous Hamiltonian. This additional geometric phase is known as the Panchratnam Berry's (PB) phase, Pancharatnam's phase, or Berry's phase. This phenomena was discovered by S. Pancharatnam (1956) [21] and later generalized by M. Berry (1984) [22]. For electrons moving in a solid within a periodic lattice, this parameter can be crystal momentum. Berry phase is accrued by the electrons as they move in a closed loop in the Bloch crystal momentum **k** space. The Schrödinger equation for the electrons in a periodic lattice can be written in **k** space as

$$\hat{H} \left| u_{n,\boldsymbol{k}} \right\rangle = \left[ \frac{\hbar^2}{2m}(-i\nabla + \boldsymbol{k})^2 + V(\boldsymbol{r}) \right] \left| u_{n,\boldsymbol{k}} \right\rangle = \epsilon_n(\boldsymbol{k}) \left| u_{n,\boldsymbol{k}} \right\rangle \, , \qquad (2.22)$$

where $u_{n,\boldsymbol{k}}(\boldsymbol{r})$ is the periodic part of the Bloch wave function of n-th band. Berry's phase over a closed loop in the momentum space can be defined as [23]

$$\gamma_n = \oint_{\boldsymbol{k}} d\boldsymbol{k}. \, \boldsymbol{A}_n(\boldsymbol{k}) \, , \qquad (2.23)$$

where $\boldsymbol{A}_n(\boldsymbol{k})$ is known as Berry's connection or Berry's vector potential $\boldsymbol{A}_n(\boldsymbol{k}) = i \left\langle u_{n,\boldsymbol{k}} \middle| \frac{\partial}{\partial \boldsymbol{k}} u_{n,\boldsymbol{k}} \right\rangle$. This name is intentional and is called so because it acts like a real vector potential in momentum space.

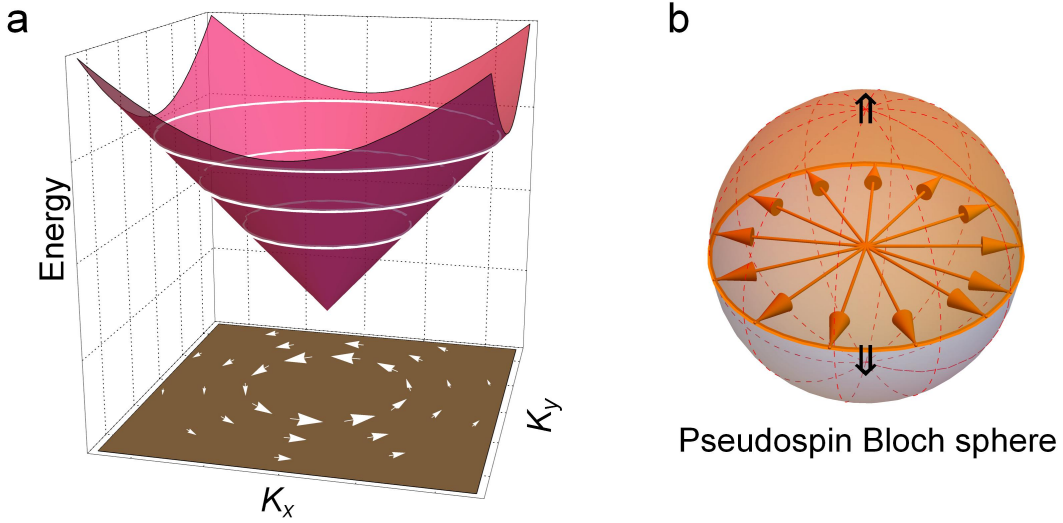

**Figure 2.6: Berry's connection and pseudospin rotation.** (a) Vector plot of Berry's connection. Three Fermi contours are overlaid on the conduction band of the gapless graphene. The corresponding vector potentials are plotted on the $\boldsymbol{k}_x - \boldsymbol{k}_y$ plane. The size of the arrows is proportional to the magnitude of the vector potential and they point to the direction of the vector potential at each point. (b) Pseudospin Bloch sphere on which the north pole and the south pole represent the pseudospin up and pseudospin down states respectively. The pseudospin vector lies on the equator of the Bloch sphere since the graphene wavefunction is an equal superposition of the two pseudospin components. When the momentum vector rotates on the Fermi surface (in presence of a magnetic field) the pseudospin rotates too by the same angle on the equator of the Bloch sphere. Berry's phase picked up by the wavefunction can be thought as half of the solid angle subtended by the pseudospin on the Bloch sphere.

Equation 2.23 can also be written as a surface integral:

$$\gamma_n = \int_{\boldsymbol{S}} d^2\boldsymbol{k}.\left(\boldsymbol{\nabla} \times \boldsymbol{A}_n(\boldsymbol{k})\right) = \int_{\boldsymbol{S}} d^2\boldsymbol{k}.\,\boldsymbol{\Omega}_n(\boldsymbol{k})\,, \tag{2.24}$$

where $\boldsymbol{\Omega}_n(\boldsymbol{k})$ is known as Berry's curvature $\boldsymbol{\Omega}_n(\boldsymbol{k}) = \boldsymbol{\nabla} \times \boldsymbol{A}_n(\boldsymbol{k})$. This acts like a pseudo magnetic field for the electrons in the momentum space. Equation 2.24 shows that Berry's phase is just the flux of Berry's curvature.

Using the relation

$$\langle n'|\nabla n\rangle = \frac{\langle n'|\,\nabla H\,|n\rangle}{E_n - E_n'}\,, \tag{2.25}$$

Berry's curvature can also be rewritten in terms of the Hamiltonian as [23]:

$$\boldsymbol{\Omega}^n(\boldsymbol{k}) = -\mathrm{Im}\sum_{n'\neq n}\frac{\langle n|\,\nabla H\,|n'\rangle \times \langle n'|\,\nabla H\,|n\rangle}{(\epsilon_n - \epsilon_n')^2}\,, \tag{2.26}$$

where the sum is over all the remaining bands and we have denoted $|u_{n,\boldsymbol{k}}\rangle = |n\rangle$ for simplicity. In 2D only the $\hat{z}$ component of the Berry's curvature is non-zero which can be written down explicitly as

$$\Omega_3^n(\boldsymbol{k}) = -\mathrm{Im}\sum_{n'\neq n}\frac{\langle n|\,\partial H/\partial\boldsymbol{k}^x\,|n'\rangle\,\langle n'|\,\partial H/\partial\boldsymbol{k}^y\,|n\rangle - \langle n|\,\partial H/\partial\boldsymbol{k}^y\,|n'\rangle\,\langle n'|\,\partial H/\partial\boldsymbol{k}^x\,|n\rangle}{(\epsilon_n - \epsilon_n')^2}\,. \tag{2.27}$$

For graphene, since there are only two bands, the summation has only one element. Using the wave functions in equation 2.12 one can easily see that the numerator of the equation 2.27 vanishes identically. This results in zero Berry's curvature in graphene with only one exception at the Dirac point. Since the two bands are degenerate at the Dirac point the denominator of the equation 2.27 also becomes zero and the Berry's curvature diverges. So, Berry's curvature of graphene is singular at the Dirac point and zero everywhere else.

The vanishing Berry's curvature can also be argued from symmetry consideration [24]. Berry's curvature behaves like a pseudo-magnetic field, so under time reversal, it is odd. So under time reversal $\boldsymbol{\Omega}(\boldsymbol{k}) \longrightarrow -\boldsymbol{\Omega}(-\boldsymbol{k})$ (**k** also goes to -**k** as momentum is odd under time reversal). If time reversal symmetry is protected then

$$\mathbf{\Omega}(\boldsymbol{k}) = -\mathbf{\Omega}(-\boldsymbol{k})\,. \tag{2.28}$$

Under spatial inversion Berry's curvature is even. So under spatial inversion $\mathbf{\Omega}(\boldsymbol{k}) \longrightarrow \mathbf{\Omega}(-\boldsymbol{k})$ (**k** goes to **-k** as momentum is odd under spatial inversion as well). If spatial inversion symmetry is protected then

$$\mathbf{\Omega}(\boldsymbol{k}) = \mathbf{\Omega}(-\boldsymbol{k})\,. \tag{2.29}$$

If both the symmetries are protected, equation 2.28 and equation 2.29 are satisfied simultaneously and as a result, Berry's curvature has to vanish. This is true for graphene as it respects both spatial inversion and time reversal symmetry.

Using the wave functions (in equation 2.12) one can easily calculate the Berry's vector potential analytically:

$$\boldsymbol{A}_n(\boldsymbol{k}) = \frac{1}{2|\boldsymbol{k}|}\hat{\phi}\,. \tag{2.30}$$

It shows Berry's vector potential diverges at the Dirac point and falls as $\frac{1}{|\boldsymbol{k}|}$ with increasing momentum. Fig. 2.6a shows the vector plot of Berry's vector potential for the conduction band. We see that it circulates in the $\boldsymbol{k}_x - \boldsymbol{k}_y$ plane around the Dirac point and hence its curl – Berry's curvature should be along $\hat{z}$ direction.

Berry's phase can be easily found out by integrating Berry's vector potential over a closed loop in $\boldsymbol{k}$ space

$$\gamma_\pm = \oint_{\boldsymbol{k}} d\boldsymbol{k}. \frac{1}{2|\boldsymbol{k}|}\hat{\phi} = \pm\pi\,, \tag{2.31}$$

which returns $\pm\pi$ for the conduction and valence band respectively. The direction of the integration is given by the cross product of magnetic field and group velocity vector $[\boldsymbol{B} \times \partial\epsilon/\partial\boldsymbol{k}]$. We note that even though Berry's vector potential depends on $\boldsymbol{k}$, Berry's phase is path independent.

The $\pm\pi$ Berry's phase can be seen physically in terms of the pseudospin rotation in graphene akin to the real spin in a magnetic field. In presence of a magnetic field, the momentum vector rotates on the Fermi surface. The pseudospin rotates by the same angle since the pseudospin vector points towards the momentum (see Fig. 2.6b). Following the well-known result for the real spin half particles, Berry's phase can be written as the half of the solid angle subtended by the pseudospin on the Bloch sphere:

$$\gamma_{\pm} = \pm \frac{1}{2} n_w \times \Omega_s \,, \tag{2.32}$$

where $\Omega_s$ is the solid angle (not to be confused with Berry's curvature denoted by $\Omega$) and $n_w$ is the winding number which is $n_w = 1$ for graphene. The solid angle subtended by the pseudospin is $\Omega_s = 2\pi$ when it rotates a full circle on the equator. So, Berry's phase becomes $\gamma_{\pm} = \pm \pi$ for the conduction band and valence band respectively.

### 2.1.6   Berry's curvature and Berry's phase of gapped graphene

We have seen in section 2.1.5 that one has to break either inversion symmetry or time reversal symmetry to generate a non-zero Berry's curvature. In graphene, inversion symmetry can be broken by setting different on-site potentials for the A and B sublattices. One way to achieve this is to make a heterostructure of graphene and hexagonal boron nitride (hBN) with a twist angle between their crystallographic axes. This creates a Moiré superlattice which breaks the inversion symmetry of graphene and introduces a gap (mass term) in the spectrum.

Introducing a mass term in the equation 2.11, it can be written as

$$\mathscr{H}_{gapped\ MLG} = \begin{pmatrix} \Delta & \hbar v_F(k_x + i\xi k_y) \\ \hbar v_F(k_x - i\xi k_y) & -\Delta \end{pmatrix} \tag{2.33}$$

$$= \hbar v_F(\sigma_x k_x + \xi \sigma_y k_y) + \Delta \sigma_z \tag{2.34}$$

$$= \boldsymbol{R.\sigma} \,, \tag{2.35}$$

where $\boldsymbol{R} = (\hbar v_F k_x, \xi \hbar v_F k_y, \Delta)$, $\xi = \pm 1$ for the $K$ and $K'$ valleys respectively.

Using the wave function of this Hamiltonian into the equation 2.27 the (the only non-zero) $\hat{z}$ component of the Berry's curvature of the gapped monolayer graphene can be calculated as

$$\Omega_3 = \pm \xi \frac{\hbar^2 v_F^2 \, \Delta}{2(\Delta^2 + \hbar^2 v_F^2 |\boldsymbol{k}|^2)^{3/2}} \tag{2.36}$$

$$= \pm \xi \frac{\hbar^2 v_F^2}{2} \frac{R_3}{|\boldsymbol{R}|^3} \, . \tag{2.37}$$

For a Dirac Hamiltonian in 3D, all the components of Berry's curvature are non-zero which leads to

$$\boldsymbol{\Omega}(\boldsymbol{R}) = \pm \xi \frac{\hbar^2 v_F^2}{2} \frac{\boldsymbol{R}}{|\boldsymbol{R}|^3} \, , \tag{2.38}$$

where $R_3 = \hbar v_F k_z$. In this case $k_z$ acts like a variable mass term in the $\boldsymbol{k}_x - \boldsymbol{k}_y$ plane. so, equation 2.38 is simplified to [24]

$$\boldsymbol{\Omega}(\boldsymbol{k}) = \pm \xi \frac{1}{2} \frac{\boldsymbol{k}}{|\boldsymbol{k}|^3} \, . \tag{2.39}$$

Equation 2.39 looks similar to the electric field distribution due to a static electric charge in the real space. So, mathematically Berry's curvature (pseudo-magnetic field) can be thought as if it is generated by a point source of magnitude $1/2$ at $\boldsymbol{k} = 0$. The Dirac point acts like a magnetic monopole i.e. the divergence of Berry's curvature is non-zero $\nabla_{\boldsymbol{k}}.\boldsymbol{\Omega}(\boldsymbol{k}) \neq 0$.

Using Berry's curvature (given in equation 2.36) Berry's phase of gapped graphene can be calculated as:

$$\gamma_{\pm} = \int_S d^2 \boldsymbol{k}. \, \boldsymbol{\Omega}(\boldsymbol{k}) \tag{2.40}$$

$$= \pm \pi \left[ 1 - \frac{\Delta}{\sqrt{\Delta^2 + \hbar^2 v_F^2 |\boldsymbol{k}|^2}} \right] \, . \tag{2.41}$$

Equation 2.36 shows that the smaller the bandgap the more concentrated the Berry's curvature is around the band origin. In the limit of a zero bandgap Berry's curvature diverges at the Dirac point and is zero elsewhere — as discussed for gapless graphene. Fig. 2.7a shows the 3D surface plot of Berry's curvature (red) and Berry's phase (Blue) as a function of $k_x$ and $k_y$ for a bandgap of 10 meV. It shows that Berry's curvature falls rapidly to zero as $|\boldsymbol{k}|$ departs from the band extremum. Since Berry's phase can be considered as the volume under the Berry's curvature surface, Berry's phase shows the opposite trend – it steadily increases with increasing $|\boldsymbol{k}|$, until the Berry's curvature itself decays to zero. Fig. 2.7b shows the 1d slice of the 3d plots.

The variation of Berry's phase can also be understood in terms of the pseudospin orientation. Right at the Dirac point ($\boldsymbol{k}$=0) the gapped graphene Hamiltonian (equation 2.33) is just proportional to $\sigma_z$ whose eigenvectors are $|\Uparrow\rangle = (1,0)^T$ and $|\Downarrow\rangle = (0,1)^T$. So, at the Dirac point, the pseudospin vector points towards the north pole and south pole for the conduction and valence band respectively. For large $|\boldsymbol{k}|$ the wavefunction consists of an almost equal combination of $|\Uparrow\rangle$ and $|\Downarrow\rangle$ for which the pseudospin vector lies on the equator of the Bloch sphere. If the pseudospin vector makes an angle $\theta$ with the z-axis, then rotation of the pseudospin vector about the z-axis subtends a solid angle $\Omega_s = 2\pi(1 - \cos\theta)$ at the origin. Incorporating this in equation 2.32 we can calculate Berry's phase as

$$\gamma_\pm = \pm\frac{1}{2}n_w \times 2\pi(1 - \cos\theta)\,, \tag{2.42}$$

where $n_w = 1$. Fig. 2.7c, Fig. 2.7d and Fig. 2.7e show the pseudospin direction for $\theta = 0°$, $60°$ and $85°$ for which the Berry's phase is 0, $\pi/2$ and 2.86 respectively.

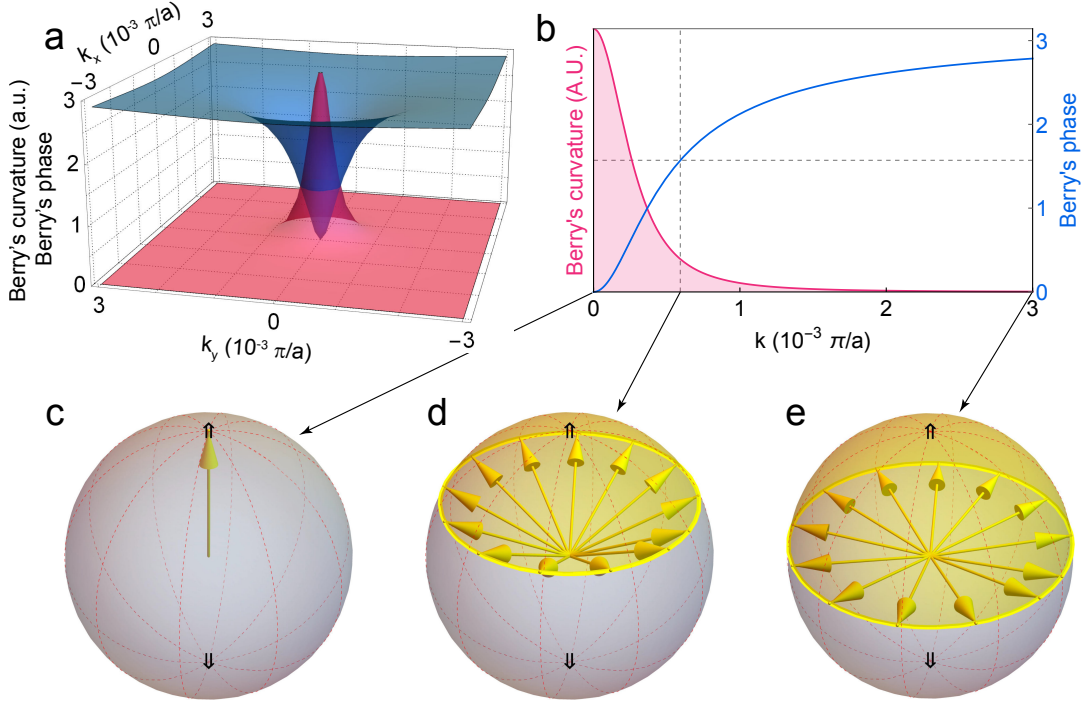

Figure 2.7: **Berry's curvature and Berry's phase of gapped graphene.** (a) 3D plot of ($\hat{z}$ component of the) Berry's curvature (red) and Berry's phase (Blue) in the momentum space. We note that Berry's curvature is a local manifestation of the geometric properties of the wavefunctions in the $\boldsymbol{k}$ space, so it is a function of momentum coordinates. On the other hand, Berry's phase is the integrated Berry's curvature on the Fermi surface – it is not a local geometric property in $\boldsymbol{k}$ space. However, when the inversion symmetry is broken, Berry's phase can be a function of Fermi energy which can be related to the scalar value of the Fermi wave vector. We have plotted this dependence of Berry's phase on the wave vector. The bandgap ($2\Delta$) is set to $2\Delta =10$ meV. Since Berry's phase is the flux (surface integration) of the Berry's curvature, Berry's phase at $|\boldsymbol{k}| = |\boldsymbol{k}_F|$ can be thought of as the volume under the Berry's curvature surface up to $|\boldsymbol{k}| = |\boldsymbol{k}_F|$. As we can see that Berry's curvature is maximum near the band extremum, the dominant contribution of the volume under the surface (Berry's phase) comes from a narrow region around the origin. Berry's phase asymptotically flattens out to the expected value of $\pi$ outside this region. (b) Line plot of the Berry's curvature (red) and Berry's phase (Blue) as a function of $\boldsymbol{k}$. The grid line position indicates the $\boldsymbol{k}$ value where Berry's phase falls to $\pi/2$. (c), (d) and (e) show the orientation of the pseudospin in the Bloch sphere for three values of Berry's phase respectively.

## 2.2 Bilayer graphene

### 2.2.1 Tight binding band structure

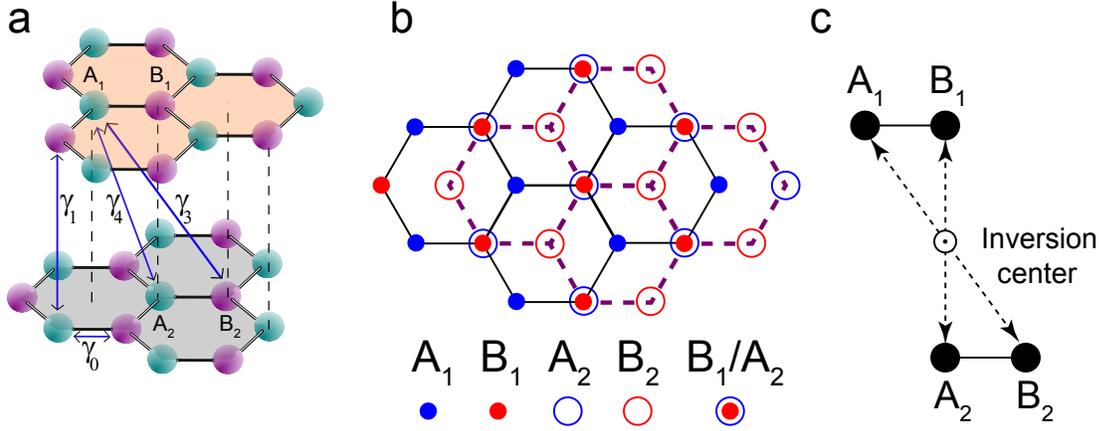

Figure 2.8: **Crystal structure of Bernal stacked bilayer graphene.** (a) 3D schematic of the crystal structure with all the hopping parameters. The blue sites and red sites denote the A and B sublattice respectively. (b) Top view of the lattice which shows that one graphene plane is shifted by one carbon-carbon atomic bond length with respect to the other. Circles of different radius are used to mark the top layer and bottom layer sites for the ease of visualization. (c) The unit cell of Bernal stacked bilayer graphene with the inversion center in the middle of the two layers. The dashed arrows join the pair of atoms which interchange their position under the inversion operation ($\boldsymbol{r} \longrightarrow -\boldsymbol{r}$) with respect to the inversion center. We can see that under inversion operation the unit cell remains unchanged which means inversion symmetry is protected in the pristine Bernal stacked bilayer graphene.

Bernal stacked bilayer graphene consists of two stacked graphene planes in which one plane is shifted by one carbon-carbon atomic bond length in the $\frac{\boldsymbol{\hat{a}_1}+\boldsymbol{\hat{a}_2}}{\sqrt{2}}$ direction with respect to the other plane (see Fig. 2.8a and Fig. 2.8b, lattice vectors $\boldsymbol{a}_1$ and $\boldsymbol{a}_2$ are shown in Fig. 2.1). In the tight biding formalism, we consider the 2p$_z$ orbitals of the four atoms in the unit cell labeled by $A_1$, $B_1$, $A_2$ and $B_2$ where the suffix is the layer index. Fig. 2.8b shows that $B_1$ and $A_2$ atoms vertically sit on top of each other. These two atoms are known as the dimer sites which have higher energy because of their close proximity with each other. Fig. 2.8c shows that the unit cell of the pristine bilayer graphene has inversion symmetry. The inversion symmetry can be broken by introducing a layer dependent potential. Considering the potential of the top layer and bottom layer to be $\Delta$ and $-\Delta$ respectively, the full tight-binding Hamiltonian of the bilayer graphene in the ($A_1$, $B_1$, $A_2$, $B_2$) basis can be written as [25]

$$H_{BLG} = \begin{pmatrix} \Delta & -\gamma_0 s(k) & \gamma_4 s(k) & -\gamma_3 s^*(k) \\ -\gamma_0 s^*(k) & \Delta & \gamma_1 & \gamma_4 s(k) \\ \gamma_4 s^*(k) & \gamma_1 & -\Delta & -\gamma_0 s(k) \\ -\gamma_3 s(k) & \gamma_4 s^*(k) & -\gamma_0 s^*(k) & -\Delta \end{pmatrix}. \qquad (2.43)$$

There are several hopping parameters which need to be considered in bilayer graphene to describe the low energy band structure accurately. Geometrically there are three kinds of hopping: (a) In-plane hopping between nearest neighbours

$$\gamma_0 : A_1 \leftrightarrow B_1, \quad A_2 \leftrightarrow B_2 \qquad (2.44)$$

(b) Vertical hopping between the dimer sites

$$\gamma_1 : A_2 \leftrightarrow B_1 \qquad (2.45)$$

(c) Skewed hopping between the non-dimer sites

$$\gamma_3 : A_1 \leftrightarrow B_2 \qquad (2.46)$$

and skewed hopping between the dimer and non-dimer sites

$$-\gamma_4 : A_1 \leftrightarrow A_2, \quad B_1 \leftrightarrow B_2. \qquad (2.47)$$

Skew hopping parameters have an in-plane component. All the hopping parameters with an in-plane component comes with the geometric factor (defined in section 2.1.1) $s(k) = 1 + e^{i\boldsymbol{k}.\boldsymbol{a_1}} + e^{i\boldsymbol{k}.\boldsymbol{a_2}}$. This takes into account the fact that in triangular lattice an electron from a site can hop into three sites around it. This is why $\gamma_0$, $\gamma_3$ and $\gamma_4$ appear with $s(k)$ in the Hamiltonian [26]. On the other hand, there is no such factor with $\gamma_1$ because it describes a strictly vertical hopping.

Since there are four atoms in the unit cell there are four bands plotted in Fig. 2.9a. There are two conduction bands and two valence bands. Fig. 2.9b shows the contour plot of the low energy conduction band. The contours are trigonally warped which reflects the triangular symmetry of the lattice. Fig. 2.9c shows the zoomed-in band structure revealing that the low energy conduction and valence bands touch each other at the six corners of the Brillouin zone. Fig. 2.9d shows the band structure around the $K$ valley. This shows, unlike in the monolayer graphene, the bands disperse quadratically around the band origin. The full Hamiltonian can be expanded to get an effective Hamiltonian which describes the bands near the two valleys [25].

By linearizing $\gamma_i s(k)$ around the two valleys we get

$$\gamma_i s(k) \approx -v_i(p_x - i\xi p_y) \,, \tag{2.48}$$

where $v_i = 3\gamma_i a/2\hbar$ and $v_0 = v_F$.

By defining a complex momentum $\pi = p_x + ip_y$ we get $\gamma_i s(k) \approx -v_i \pi^\dagger$. After linearizing the terms, the Hamiltonian in equation 2.43 can be simplified near the two valleys as

$$H_B = \begin{pmatrix} \Delta & v_F\pi^\dagger & -v_4\pi^\dagger & v_3\pi \\ v_F\pi & \Delta & \gamma_1 & -v_4\pi^\dagger \\ -v_4\pi & \gamma_1 & -\Delta & v_F\pi^\dagger \\ v_3\pi^\dagger & -v_4\pi & v_F\pi & -\Delta \end{pmatrix} \,. \tag{2.49}$$

We can get some insight by writing the approximate analytical solution for the low energy bands (keeping only $\gamma_0$ and $\gamma_1$). By setting $\Delta$, $v_3$ and $v_4$ to zero we get the two low energy bands as

$$\epsilon_\pm \approx \pm \frac{\gamma_1}{2} \left[ \sqrt{1 + 4v_F^2 p^2/\gamma_1^2} - 1 \right] \,. \tag{2.50}$$

This shows that the dispersion for small momentum is roughly quadratic and given by $\epsilon_\pm \approx \pm p^2/2m$ where $m = \gamma_1/2v_F^2$. However, the dispersion at high momentum is roughly linear and given by $\epsilon_\pm \approx \pm pv_F$. This crossover happens at $p = \gamma_1/2v_F$.

In the transport measurements typically density can be varied up to $10^{13}$ cm$^{-2}$ by using an electrostatic gate. To see the corresponding energy, we express the energy in terms of density:

$$\epsilon[eV] \approx \frac{p^2}{2me} \tag{2.51}$$

$$= \frac{\hbar^2 v_F^2 \times 2\pi n}{e^2 \gamma_1[eV]}. \tag{2.52}$$

This shows that the Fermi energy increases by $\sim 349$ meV when density is tuned by $\sim 10^{13}$ cm$^{-2}$. Fig. 2.9d shows that the higher energy band is above this energy. We discussed earlier that the dimer sites $B_1$ and $A_2$ have higher energy compared to the non-dimer sites $A_1$ and $B_2$. So, it is useful to have a low energy and small momentum ($p < \gamma_1/2v_F$ or $\epsilon < \gamma_1/4$) effective Hamiltonian which describes the two low energy bilayer graphene bands. The $2 \times 2$ Hamiltonian (neglecting $\gamma_4$) in ($A_1$, $B_2$) basis near the $K$ valley can be written as [26]:

$$H_B = -\frac{1}{2m} \begin{pmatrix} 0 & (\pi^\dagger)^2 \\ \pi^2 & 0 \end{pmatrix} + v_3 \begin{pmatrix} 0 & \pi \\ \pi^\dagger & 0 \end{pmatrix} + 2\Delta \left[ \frac{1}{2} \begin{pmatrix} 1 & 0 \\ 0 & -1 \end{pmatrix} - \frac{v_F^2}{\gamma_1^2} \begin{pmatrix} \pi^\dagger \pi & 0 \\ 0 & -\pi \pi^\dagger \end{pmatrix} \right], \tag{2.53}$$

where $\pi = p_x + ip_y$. The first term in the equation 2.53 produces a quadratic dispersion. The second term is linear in momentum and hence is relevant at very small momentum. For most of the energy range $\left(\frac{\gamma_1}{4}\left(\frac{v_3}{v}\right)^2 < |\epsilon| < \gamma_1/4\right)$ the first term is dominant. The third term takes into account the asymmetry between the two layers where $2\Delta$ is the potential difference between the two layers due to an external electric field.

The eigenvalues and eigenvectors of the first term of equation 2.53 are given by

$$\epsilon_\pm = \pm\frac{p^2}{2m}, \quad |\psi_\pm\rangle = \frac{1}{\sqrt{2}} \begin{pmatrix} 1 \\ \mp e^{i2\phi} \end{pmatrix}. \tag{2.54}$$

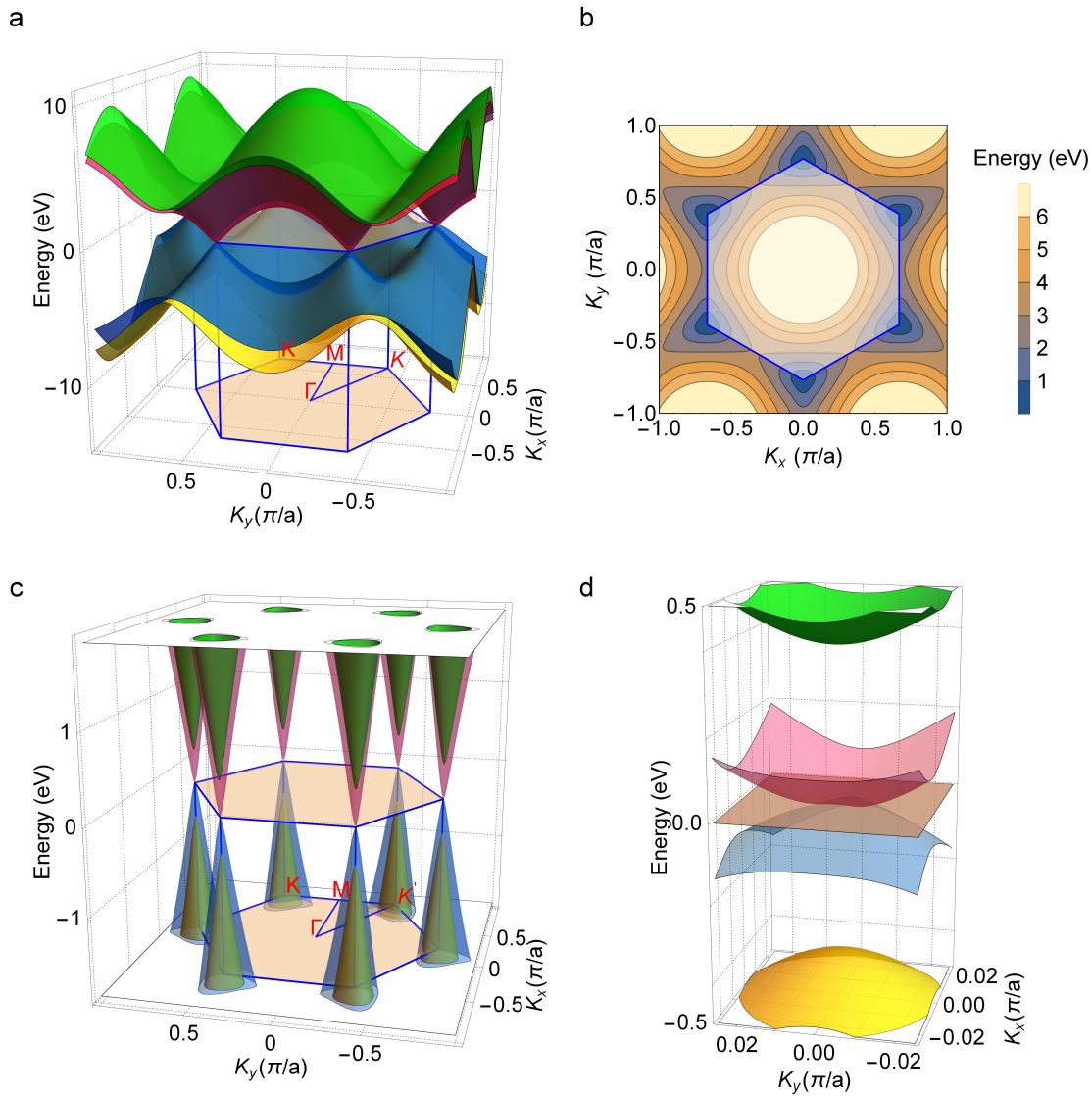

Figure 2.9: **Tight binding band structure of Bernal stacked bilayer graphene.**
(a) 3D surface plot of the four bands. We take $\gamma_0$=3.1 eV, $\gamma_1$=390 meV, $\gamma_3$=315 meV, $\gamma_4$=120 meV. The first Brillouin zone is also overlaid. (b) Contour plot of the low energy conduction band. (c) Zoomed-in 3D plot of the band structure showing that the low energy conduction and valence bands touch each other at the six corners of the Brillouin zone. (d) Further zoomed-in band structure around the $K_+$ valley where the momentum is measured relative to the $K_+$ point.

## 2.2.2 Inversion symmetry and band gap

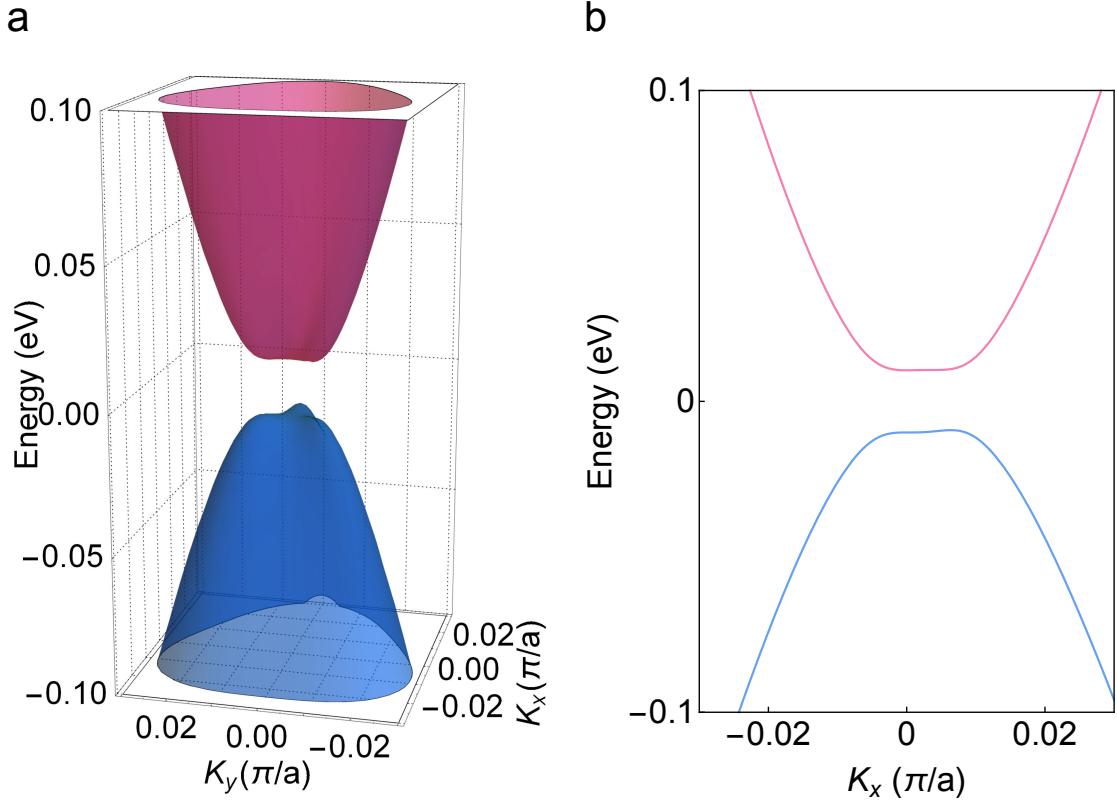

**a**　　　　　　　　　　　　　　　　**b**

Figure 2.10: **Breaking of inversion symmetry and opening of a band gap.** (a) 3D surface plot of the low energy two bands around the $K_+$ valley for a potential difference between the two layers $2\Delta$=20 meV. (b) Band structure plotted along $k_x$ direction showing a band gap $\sim 2\Delta$=20 meV.

We note that under inversion operation $A_1$ and $B_2$ atoms change their positions $A_1 \overset{r \to -r}{\longrightarrow} B_2$, see Fig. 2.8c. When a potential difference between the two layers ($2\Delta$) is introduced, it sets different on-site potentials for these two sites (see the third term in the equation 2.53). So, the presence of a finite $\Delta$ breaks the inversion symmetry. Since the matrix containing $\Delta$ is also proportional to $\sigma_z$, it introduces a bandgap in the spectrum. This is why breaking inversion symmetry and opening up a bandgap is intimately related in bilayer graphene. Fig. 2.10 shows that a bandgap $\sim 2\Delta$=20 meV opens up in the bilayer graphene for a potential difference $2\Delta$=20 meV between two layers. A similar connection between inversion symmetry and bandgap also exists in monolayer graphene.

### 2.2.3   Trigonal warping and the Lifshitz transition

Low energy Fermi surface of monolayer graphene is a circle. It becomes triangular at higher energies as the Fermi contour follows the symmetry of the lattice (embedded in the function $s(k)$); this becomes prominent at higher energy. In few-layer graphene, however, there are other sources of trigonal warping even at low energy. This can be seen explicitly if we take into account the second linear term of the Hamiltonian (equation 2.53) along with the first isotropic quadratic term. The resulting energy is given by [26]

$$\epsilon^2 = (pv_3)^2 - \frac{p^3 v_3}{m} \cos(3\phi) + \left(\frac{p^2}{2m}\right)^2 ,   (2.55)$$

where $\phi$ is the polar momentum angle. The second term of the equation 2.55 proportional to $v_3$ clearly breaks the isotropy of the Fermi surface. In Fig. 2.11 we show the effect of $v_3$ by calculating the band structure with $v_3$ (Fig. 2.11a and Fig. 2.11c) and without $v_3$ (Fig. 2.11b and Fig. 2.11d). Fig. 2.11a and Fig. 2.11c show that at very low energy the topology of the Fermi surface changes – it breaks into four discrete Fermi pockets. This is because the term proportional to $v_3$ is linear in momentum and hence dominates over the quadratic term at very low energy. This transition when the topology of the Fermi surface changes is known as the Lifshitz transition. This happens when the energy is less than threshold energy:

$$\epsilon < \frac{\gamma_1}{4} \left(\frac{v_3}{v}\right)^2 .   (2.56)$$

The Lifshitz transition is clearly demonstrated in Fig. 2.11c where we show two Fermi contours before (blue) and after (green) the Lifshitz transition respectively. Fig. 2.11b and Fig. 2.11d show that the trigonal warping (at low energy) goes away and the bands become rotationally symmetric for $v_3 = 0$.

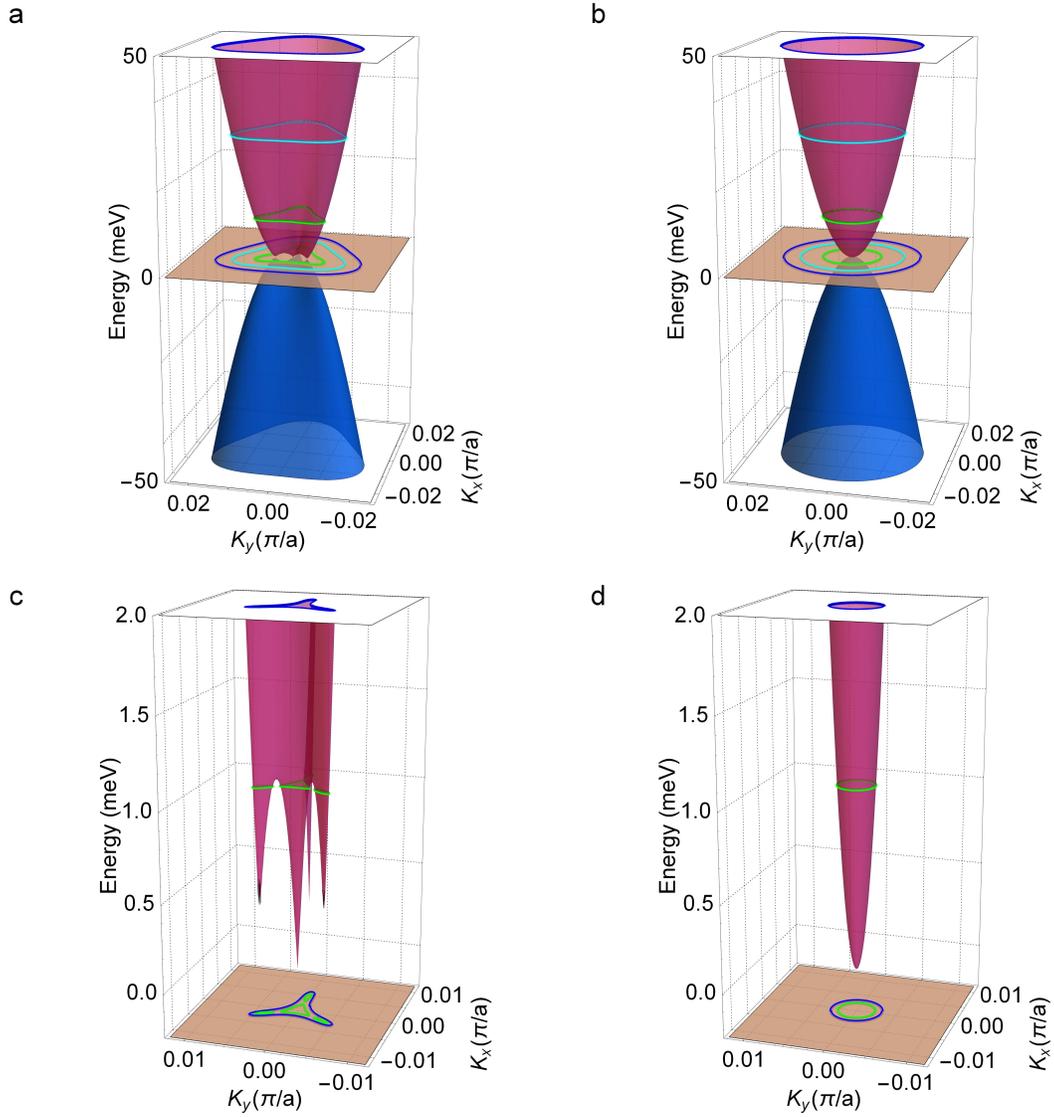

Figure 2.11: **Trigonal warping and the Lifshitz transition in bilayer graphene.** (a) Low energy bands around the $K_+$ valley which shows that the bands are strongly trigonally warped. Three Fermi contours are overlaid on the conduction band. The projection of the Fermi contours is also shown on the $k_x - k_y$ plane. (b) The band structure with $\gamma_3 = 0$ plotted in the same range showing no trigonal warping. This shows that the trigonal warping at low energy is imposed by $\gamma_3$. (c) Further zoomed-in conduction band showing that the Fermi surface breaks into four separate pockets close to the bottom of the band – this is known as Lifshitz transition. (d) The corresponding band structure with $\gamma_3 = 0$ plotted in the same range showing no trigonal warping and no Lifshitz transition.

### 2.2.4 Pseudospin, chirality and Berry's phase

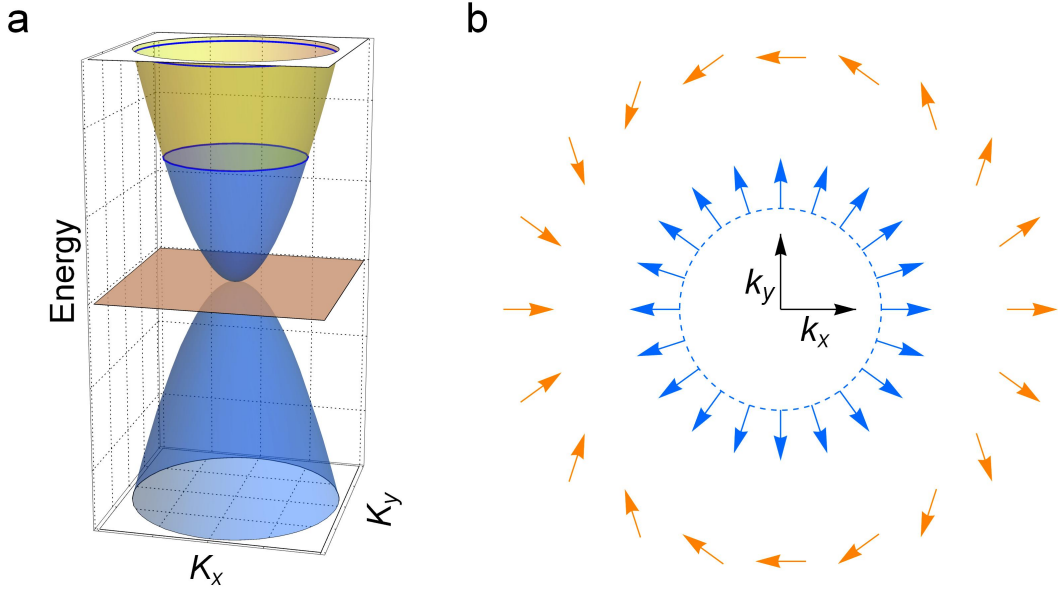

Figure 2.12: **Pseudospin and chirality in bilayer graphene.** (a) Schematic of the bilayer graphene band structure showing the Fermi circle in blue. (b) The blue dashed circle represents the Fermi circle. The blue and the orange arrows at each point show the direction of the momentum and the pseudospin respectively. As discussed in the text, in bilayer graphene the pseudospin vector rotates twice as fast as the momentum vector.

Writing the complex momenta in polar coordinates $\pi = pe^{i\phi}$ the first term in the equation 2.53 can be written as [12]

$$H_B^1 = -\frac{p^2}{2m}\begin{pmatrix} 0 & e^{-i2\phi} \\ e^{i2\phi} & 0 \end{pmatrix} \tag{2.57}$$

$$= -\frac{p^2}{2m}(\sigma_x \cos(2\phi) + \sigma_y \sin(2\phi)) \tag{2.58}$$

$$= -\frac{p^2}{2m}\boldsymbol{\sigma}.\hat{\boldsymbol{n}}\,, \tag{2.59}$$

where $\hat{\boldsymbol{n}} = (\cos(2\phi), \sin(2\phi))$. We note that the monolayer graphene Hamiltonian (equation 2.11) and the bilayer graphene Hamiltonian (equation 2.57) are a part of the family of Hamiltonian described by

$$H_J = f(|\boldsymbol{p}|)\boldsymbol{\sigma}.\hat{\boldsymbol{n}} \sim \begin{pmatrix} 0 & (\pi^\dagger)^J \\ \pi^J & 0 \end{pmatrix}\,, \tag{2.60}$$

where $f(|\boldsymbol{p}|)$ depends on the dispersion $f(|\boldsymbol{p}|) \sim |\boldsymbol{p}|^J$, $\hat{\boldsymbol{n}}$ is an unit vector $\hat{\boldsymbol{n}} = (\cos(J\phi), \sin(J\phi))$. This Hamiltonian is chiral with degree of chirality (or pseudospin winding number) $J$. This means that the direction of momentum and pseudospin are related, more precisely the pseudospin vector rotates $J$ times for a single rotation of the momentum vector. So, the bilayer graphene has pseudospin winding number $n_w = 2$.

Using the wavefunction given in equation 2.54 Berry's phase of bilayer graphene can be easily shown to be $2\pi$. This can also be understood in terms of the pseudospin winding number and the solid angle subtended by the pseudospin when the momentum rotates by $2\pi$

$$\gamma_\pm = \pm\frac{1}{2}n_w \times \Omega_s = \pm\frac{1}{2}2 \times 2\pi = \pm2\pi \,. \tag{2.61}$$

In fact, it turns out that for the generalized Hamiltonian in equation 2.60, the electrons pick up a Berry's phase of $J\pi$ on making a closed loop on the Fermi surface and bilayer graphene is a special case with $J = 2$.

## 2.3 Trilayer graphene

### 2.3.1 Tight binding band structure

It turns out that there are two stable stacking in trilayer graphene – Bernal (ABA) and rhombohedral (ABC). In ABA trilayer graphene the top and bottom layers are exactly aligned, whereas the middle layer is shifted by one carbon-carbon atomic bond. In ABC trilayer graphene all the successive layers are shifted by one carbon-carbon atomic bond with respect to each other. In our study, we will focus on the ABA trilayer graphene. The full tight binding Hamiltonian of ABA trilayer graphene can be written in the $(A_1, B_1, A_2, B_2, A_3, B_3)$ basis as [14]

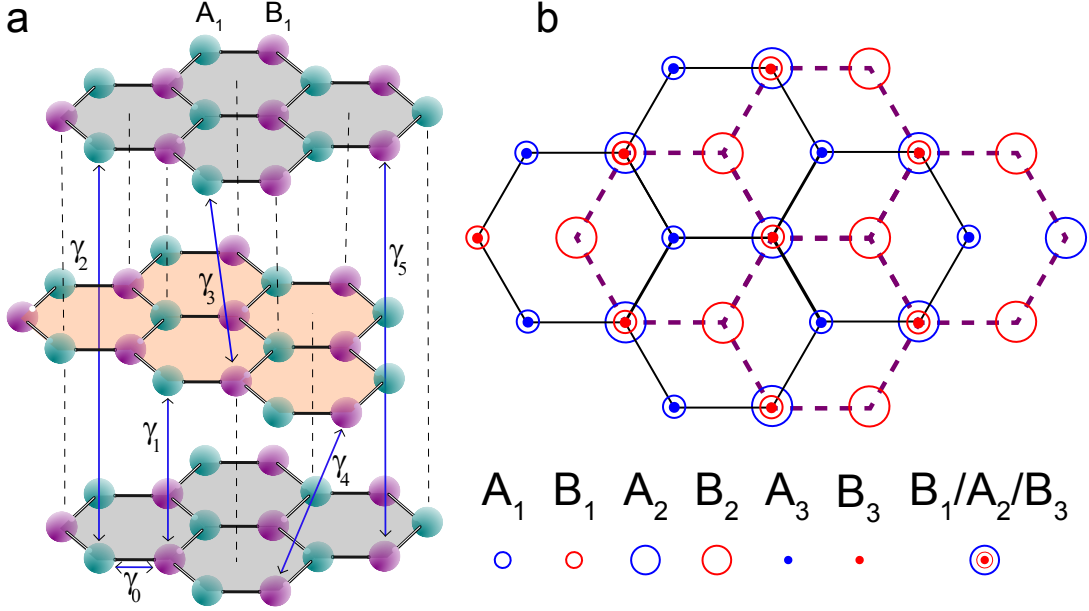

Figure 2.13: **Crystal structure of ABA-stacked trilayer graphene.** (a) 3D schematic of the crystal structure of ABA-TLG with all the hopping parameters marked in the diagram. (b) Top view of the crystal structure. Since the top and the bottom layers are exactly aligned, the hexagon grids for the two layers fall on top of each other. The middle layer represented by the dashed hexagon grid is shifted by one carbon-carbon atomic bond with respect to the other layers.

$$
H_{TLG} = \begin{pmatrix}
\Delta_1 + \Delta_2 & -\gamma_0 s(k) & \gamma_4 s(k) & -\gamma_3 s^*(k) & \gamma_2/2 & 0 \\
-\gamma_0 s^*(k) & \delta + \Delta_1 + \Delta_2 & \gamma_1 & \gamma_4 s(k) & 0 & \gamma_5/2 \\
\gamma_4 s^*(k) & \gamma_1 & \delta - 2\Delta_2 & -\gamma_0 s(k) & \gamma_4 s^*(k) & \gamma_1 \\
-\gamma_3 s(k) & \gamma_4 s^*(k) & -\gamma_0 s^*(k) & -2\Delta_2 & -\gamma_3 s(k) & \gamma_4 s^*(k) \\
\gamma_2/2 & 0 & \gamma_4 s(k) & -\gamma_3 s^*(k) & -\Delta_1 + \Delta_2 & -\gamma_0 s(k) \\
0 & \gamma_5/2 & \gamma_1 & \gamma_4 s(k) & -\gamma_0 s^*(k) & \delta - \Delta_1 + \Delta_2
\end{pmatrix}
$$

$$
\tag{2.62}
$$

$$
= H_0 + H_{\Delta_1} + H_{\Delta_2}, \tag{2.63}
$$

where geometrically there are three kinds of hopping (see Fig. 2.14a):

(a) In-plane hopping between nearest neighbours

$$
\gamma_0 : A_1 \leftrightarrow B_1, \quad A_2 \leftrightarrow B_2, \quad A_3 \leftrightarrow B_3 \tag{2.64}
$$

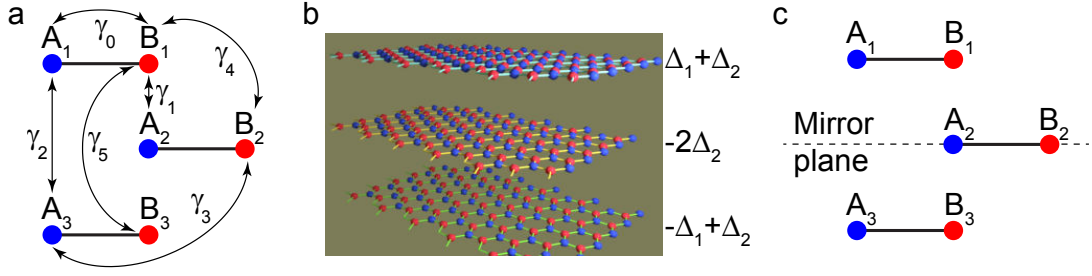

Figure 2.14: **Band parameters and symmetry of the lattice.** (a) Unit cell of the ABA trilayer graphene with all the hopping parameters. (b) Potential distribution across the three layers of ABA trilayer graphene. $\Delta_1$ which is the potential due to an external electric field breaks the mirror symmetry of the lattice. (c) Shows the plane in the middle layer with respect to which the unit cell has mirror symmetry.

(b) Vertical hopping between the dimer sites

$$\gamma_1 : A_2 \leftrightarrow B_{1,3}, \quad \gamma_5 : B_1 \leftrightarrow B_3 \tag{2.65}$$

Vertical hopping between the non-dimer sites

$$\gamma_2 : A_1 \leftrightarrow A_3 \tag{2.66}$$

(c) Skewed hopping between the non-dimer sites

$$\gamma_3 : A_{1,3} \leftrightarrow B_2 \tag{2.67}$$

and skewed hopping between the dimer and non-dimer sites

$$-\gamma_4 : A_{1,3} \leftrightarrow A_2, \quad B_{1,3} \leftrightarrow B_2 \,. \tag{2.68}$$

The three atoms on-site $B_1$, $A_2$ and $B_3$ sit directly on top of each other and is called a trimer. The onsite potential of the trimer atoms is higher: $\delta$ is the difference of the onsite potential of the trimer atoms and other atoms. $2\Delta_1$ is the potential drop between the top and bottom layers due to an external electric field. $\Delta_2$ describes the deviation of the middle layer potential from the mean potential (see Fig. 2.14b). $\Delta_1$ and $\Delta_2$ can be defined from the potential of the three layers $U_1$, $U_2$, $U_3$ as [27]

$$\Delta_1 = \frac{U_1 - U_3}{2} \tag{2.69}$$

$$\Delta_2 = \frac{U_1 + U_3 - 2U_2}{6} \tag{2.70}$$

$$U_1 + U_2 + U_3 = 0 \,. \tag{2.71}$$

From the definitions it is clear that $\Delta_1$ is proportional to the average linear electric field inside the trilayer graphene. $\Delta_2$ describes the non-linear part of the electric field inside the three layers. The last constraint just defines the average potential energy of the system as the zero energy level. Solving equations 2.69, 2.70 and 2.71 simultaneously we get $U_1 = \Delta_1 + \Delta_2$, $U_2 = -2\Delta_2$ and $U_3 = -\Delta_1 + \Delta_2$.

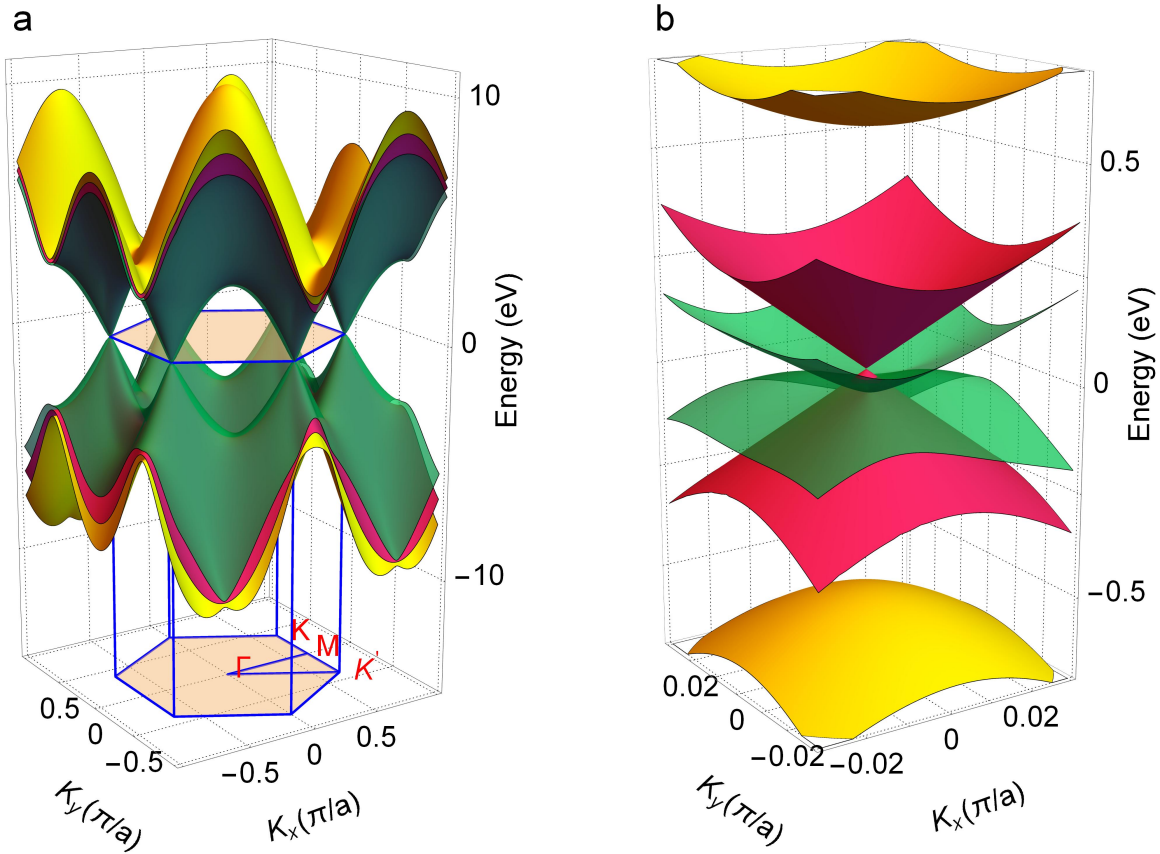

Figure 2.15: **Tight binding band structure of ABA trilayer graphene.** (a) 3D surface plot of the six bands. We take $\gamma_0$=3.1 eV, $\gamma_1$=390 meV, $\gamma_2$=-20 meV, $\gamma_3$=315 meV, $\gamma_4$=120 meV, $\gamma_5$=18 meV, $\delta$=20 meV. The first Brillouin zone is also overlaid. (b) Zoomed-in 3D plot of the band structure around the $K_+$ valley showing that there is no energy gap at the six corners of the Brillouin zone.

The full Hamiltonian (equation 2.62) consists of three parts [14]. $H_0$ is the Hamiltonian of the unbiased ABA-trilayer graphene. $H_{\Delta_1}$ and $H_{\Delta_2}$ take into account the effect of the linear and non-linear part of the electric field respectively.

$$H_0 = \begin{pmatrix} 0 & -\gamma_0 s(k) & \gamma_4 s(k) & -\gamma_3 s^*(k) & \gamma_2/2 & 0 \\ -\gamma_0 s^*(k) & \delta & \gamma_1 & \gamma_4 s(k) & 0 & \gamma_5/2 \\ \gamma_4 s^*(k) & \gamma_1 & \delta & -\gamma_0 s(k) & \gamma_4 s^*(k) & \gamma_1 \\ -\gamma_3 s(k) & \gamma_4 s^*(k) & -\gamma_0 s^*(k) & 0 & -\gamma_3 s(k) & \gamma_4 s^*(k) \\ \gamma_2/2 & 0 & \gamma_4 s(k) & -\gamma_3 s^*(k) & 0 & -\gamma_0 s(k) \\ 0 & \gamma_5/2 & \gamma_1 & \gamma_4 s(k) & -\gamma_0 s^*(k) & \delta \end{pmatrix} \quad (2.72)$$

$$H_{\Delta_1} = \begin{pmatrix} \Delta_1 & 0 & 0 & 0 & 0 & 0 \\ 0 & \Delta_1 & 0 & 0 & 0 & 0 \\ 0 & 0 & 0 & 0 & 0 & 0 \\ 0 & 0 & 0 & 0 & 0 & 0 \\ 0 & 0 & 0 & 0 & -\Delta_1 & 0 \\ 0 & 0 & 0 & 0 & 0 & -\Delta_1 \end{pmatrix} \quad (2.73)$$

$$H_{\Delta_2} = \begin{pmatrix} \Delta_2 & 0 & 0 & 0 & 0 & 0 \\ 0 & \Delta_2 & 0 & 0 & 0 & 0 \\ 0 & 0 & -2\Delta_2 & 0 & 0 & 0 \\ 0 & 0 & 0 & -2\Delta_2 & 0 & 0 \\ 0 & 0 & 0 & 0 & \Delta_2 & 0 \\ 0 & 0 & 0 & 0 & 0 & \Delta_2 \end{pmatrix} \quad (2.74)$$

The six eigenvalues of $H_0$ are plotted in Fig. 2.15a. The zoomed-in bands around $K_+$ valley is shown in Fig. 2.15b. The line cuts of the bands are plotted in Fig. 2.16. They show that the low energy band structure consists of a pair of linear and a pair of quadratic bands. There is another pair of quadratic bands which are gapped out at an energy scale $\sim 0.5$ eV irrelevant to transport studies.

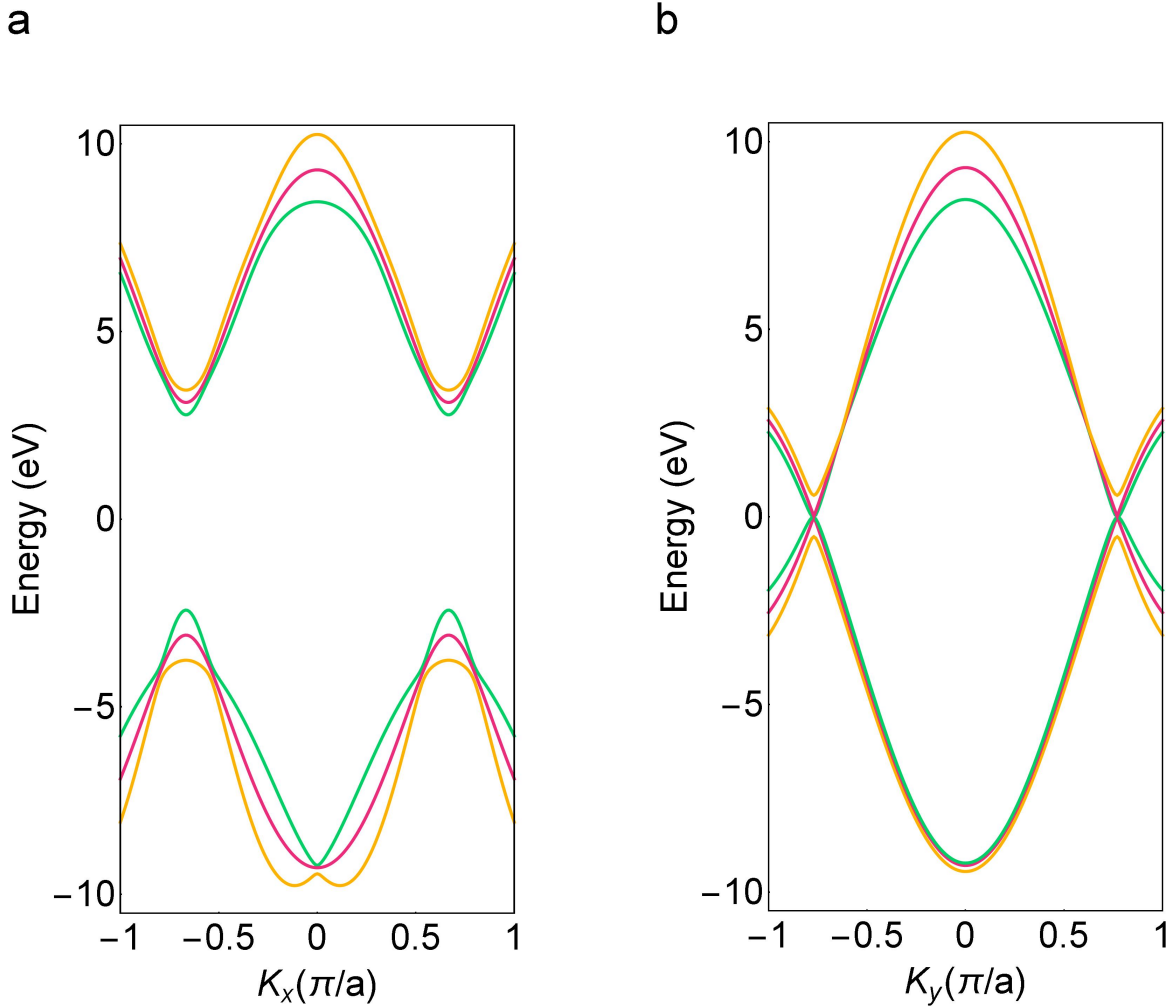

Figure 2.16: **Line cuts of the bands of ABA trilayer graphene.** (a) and (b) Band structure along $k_x$ and $k_y$.

## 2.3.2  Classification of the bands based on mirror symmetry

We note that ABA-trilayer graphene has a mirror symmetry [14, 27] with respect to its middle plane (see Fig. 2.14c). It is useful to exploit the symmetry of the Hamiltonian (equation 2.62). To do this we go to a new basis ($\frac{A_1 - A_3}{\sqrt{2}}$, $\frac{B_1 - B_3}{\sqrt{2}}$, $\frac{A_1 + A_3}{\sqrt{2}}$, $B_2$, $A_2$, $\frac{B_1 + B_3}{\sqrt{2}}$) where the new basis is anti-symmetric and symmetric linear combinations of the old basis ($A_1$, $B_1$, $A_2$, $B_2$, $A_3$, $B_3$). Mathematically, this amounts to do a transformation $H_{TLG} \to U H_{TLG} U^{-1}$ where

$$U = \begin{pmatrix} \frac{1}{\sqrt{2}} & 0 & 0 & 0 & -\frac{1}{\sqrt{2}} & 0 \\ 0 & \frac{1}{\sqrt{2}} & 0 & 0 & 0 & -\frac{1}{\sqrt{2}} \\ \frac{1}{\sqrt{2}} & 0 & 0 & 0 & \frac{1}{\sqrt{2}} & 0 \\ 0 & 0 & 0 & 1 & 0 & 0 \\ 0 & 0 & 1 & 0 & 0 & 0 \\ 0 & \frac{1}{\sqrt{2}} & 0 & 0 & 0 & \frac{1}{\sqrt{2}} \end{pmatrix}. \tag{2.75}$$

We first see how the unbiased ABA-trilayer graphene Hamiltonian ($H_0$) changes under the basis transformation

$$H_0' = U H_0 U^{-1} \tag{2.76}$$

$$= \begin{pmatrix} -\gamma_2/2 & -\gamma_0 s(k) & 0 & 0 & 0 & 0 \\ -\gamma_0 s^*(k) & \delta - \gamma_5/2 & 0 & 0 & 0 & 0 \\ 0 & 0 & \gamma_2/2 & -\sqrt{2}\gamma_3 s^*(k) & \sqrt{2}\gamma_4 s(k) & -\gamma_0 s(k) \\ 0 & 0 & -\sqrt{2}\gamma_3 s(k) & 0 & -\gamma_0 s^*(k) & \sqrt{2}\gamma_4 s^*(k) \\ 0 & 0 & \sqrt{2}\gamma_4 s^*(k) & -\gamma_0 s(k) & \delta & \sqrt{2}\gamma_1 \\ 0 & 0 & -\gamma_0 s^*(k) & \sqrt{2}\gamma_4 s(k) & \sqrt{2}\gamma_1 & \delta + \gamma_5/2 \end{pmatrix}. \tag{2.77}$$

We see that $H_0$ becomes block diagonal with a $2 \times 2$ anti-symmetric block and a $4 \times 4$ symmetric block. This is expected and it just reflects the mirror symmetry of the Hamiltonian. Now we investigate the symmetry properties of $H_{\Delta_1}$. We note that this term sets the potential of the top and bottom layers to $\Delta_1$ and $-\Delta_1$ respectively which gets flipped on a mirror symmetry transformation about the middle layer. In other words, $H_{\Delta_1}$ does not respect the mirror symmetry of $H_0$.

$$H'_{\Delta_1} = U H_{\Delta_1} U^{-1} \tag{2.78}$$

$$= \begin{pmatrix} 0 & 0 & \Delta_1 & 0 & 0 & 0 \\ 0 & 0 & 0 & 0 & 0 & \Delta_1 \\ \Delta_1 & 0 & 0 & 0 & 0 & 0 \\ 0 & 0 & 0 & 0 & 0 & 0 \\ 0 & 0 & 0 & 0 & 0 & 0 \\ 0 & \Delta_1 & 0 & 0 & 0 & 0 \end{pmatrix} \tag{2.79}$$

As expected, since $H_{\Delta_1}$ does not respect the mirror symmetry, under the basis transformation $H_{\Delta_1}$ does not become block-diagonal. It generates off-diagonal elements which couple the $2 \times 2$ anti-symmetric block with the $4 \times 4$ symmetric block.

For $H_{\Delta_2}$, we notice that this term sets the potential of the top, middle and bottom layers to $\Delta_2$, $-2\Delta_2$ and $\Delta_2$ respectively which remain unchanged on a mirror symmetry transformation about the middle layer. So, $H_{\Delta_2}$ respects the mirror symmetry of $H_0$.

$$H'_{\Delta_2} = U H_{\Delta_2} U^{-1} \tag{2.80}$$

$$= \begin{pmatrix} \Delta_2 & 0 & 0 & 0 & 0 & 0 \\ 0 & \Delta_2 & 0 & 0 & 0 & 0 \\ 0 & 0 & \Delta_2 & 0 & 0 & 0 \\ 0 & 0 & 0 & -2\Delta_2 & 0 & 0 \\ 0 & 0 & 0 & 0 & -2\Delta_2 & 0 \\ 0 & 0 & 0 & 0 & 0 & \Delta_2 \end{pmatrix} \tag{2.81}$$

As expected, since $H_{\Delta_2}$ respects the mirror symmetry, under the basis transformation $H_{\Delta_2}$ becomes diagonal. We can sum up the individual terms of $H_{TLG}$ in the new basis:

$$H'_{TLG} = H'_0 + H'_{\Delta_1} + H'_{\Delta_2} \tag{2.82}$$

$$= \begin{pmatrix} -\gamma_2/2 + \Delta_2 & -\gamma_0 s(k) & \Delta_1 & 0 & 0 & 0 \\ -\gamma_0 s^*(k) & \delta - \gamma_5/2 + \Delta_2 & 0 & 0 & 0 & \Delta_1 \\ \Delta_1 & 0 & \gamma_2/2 + \Delta_2 & -\sqrt{2}\gamma_3 s^*(k) & \sqrt{2}\gamma_4 s(k) & -\gamma_0 s(k) \\ 0 & 0 & -\sqrt{2}\gamma_3 s(k) & -2\Delta_2 & -\gamma_0 s^*(k) & \sqrt{2}\gamma_4 s^*(k) \\ 0 & 0 & \sqrt{2}\gamma_4 s^*(k) & -\gamma_0 s(k) & \delta - 2\Delta_2 & \sqrt{2}\gamma_1 \\ 0 & \Delta_1 & -\gamma_0 s^*(k) & \sqrt{2}\gamma_4 s(k) & \sqrt{2}\gamma_1 & \delta + \gamma_5/2 + \Delta_2 \end{pmatrix} \tag{2.83}$$

$H'_0$ and $H'_{\Delta_2}$ make the block diagonal parts of the full Hamiltonian. The anti-symmetric $2 \times 2$ block is similar to a monolayer graphene Hamiltonian with a small mass term in the diagonal. This is why it is labeled as $H_{MLG}$. Similarly, The symmetric $4 \times 4$ block is similar to a bilayer graphene Hamiltonian with a small mass term in the diagonal. This is why it is labeled as $H_{BLG}$.

$$H'_0 + H'_{\Delta_2} = \begin{pmatrix} [H_{MLG}]_{2\times2} & [0]_{2\times4} \\ [0]_{4\times2} & [H_{BLG}]_{4\times4} \end{pmatrix}, \tag{2.84}$$

where

$$H_{MLG} = \begin{pmatrix} -\gamma_2/2 + \Delta_2 & -\gamma_0 s(k) \\ -\gamma_0 s^*(k) & \delta - \gamma_5/2 + \Delta_2 \end{pmatrix} \tag{2.85}$$

$$H_{BLG} = \begin{pmatrix} \gamma_2/2 + \Delta_2 & -\sqrt{2}\gamma_3 s^*(k) & \sqrt{2}\gamma_4 s(k) & -\gamma_0 s(k) \\ -\sqrt{2}\gamma_3 s(k) & -2\Delta_2 & -\gamma_0 s^*(k) & \sqrt{2}\gamma_4 s^*(k) \\ \sqrt{2}\gamma_4 s^*(k) & -\gamma_0 s(k) & \delta - 2\Delta_2 & \sqrt{2}\gamma_1 \\ -\gamma_0 s^*(k) & \sqrt{2}\gamma_4 s(k) & \sqrt{2}\gamma_1 & \delta + \gamma_5/2 + \Delta_2 \end{pmatrix} \tag{2.86}$$

We can separate the off-diagonal terms of $H'_{TLG}$ containing $\Delta_1$ which hybridizes the bands described by $H_{MLG}$ and $H_{BLG}$:

$$H'_{TLG} = \begin{pmatrix} [H_{MLG}]_{2\times 2} & [H_{hyb}]_{2\times 4} \\ [H^T_{hyb}]_{4\times 2} & [H_{BLG}]_{4\times 4} \end{pmatrix}, \tag{2.87}$$

where $H_{hyb}$ is given by

$$H_{hyb} = \begin{pmatrix} \Delta_1 & 0 & 0 & 0 \\ 0 & 0 & 0 & \Delta_1 \end{pmatrix}. \tag{2.88}$$

Equation 2.83 is the full Hamiltonian in the new basis which describes the energy bands in the whole Brillouin zone. Often, we are interested in the low energy excitation near the two valleys. So, it is useful to expand $s(k)$ near the two valleys and make a small momentum - low energy Hamiltonian. As shown before, keeping only the linear term we get $\gamma_i s(k) \approx -v_i \pi^\dagger$ where we define a complex momentum $\pi = p_x + i\xi p_y$. After linearizing $H_{MLG}$ and $H_{BLG}$ we get

$$H_{MLG} = \begin{pmatrix} -\gamma_2/2 + \Delta_2 & v_F \pi^\dagger \\ v_F \pi & \delta - \gamma_5/2 + \Delta_2 \end{pmatrix} \tag{2.89}$$

$$H_{BLG} = \begin{pmatrix} \gamma_2/2 + \Delta_2 & \sqrt{2}v_3\pi & \sqrt{2}v_4\pi^\dagger & v_0\pi^\dagger \\ \sqrt{2}v_3\pi^\dagger & -2\Delta_2 & v_0\pi & -\sqrt{2}v_4\pi \\ -\sqrt{2}v_4\pi & v_0\pi^\dagger & \delta - 2\Delta_2 & \sqrt{2}\gamma_1 \\ v_0\pi & -\sqrt{2}v_4\pi^\dagger & \sqrt{2}\gamma_1 & \delta + \gamma_5/2 + \Delta_2 \end{pmatrix}. \tag{2.90}$$

Equation 2.89 is analogous to the monolayer graphene Hamiltonian (equation 2.11) and equation 2.90 is analogous to the bilayer graphene Hamiltonian (equation 2.49).

$H_{BLG}$ describes four bands. It is not surprising that the low energy bands reside primarily on the low energy non-dimer sites $A_1 + A_3$ and $B_2$. The higher energy bands are spanned by $B_1 + B_3$ and $A_2$ with a large band gap $\sim \sqrt{2}\gamma_1 \approx 0.55$ eV. High energy bands do not participate in the transport measurements since we cannot tune the density high enough (see equation 2.52). So, it is helpful to have a further low energy $2 \times 2$ effective Hamiltonian for the low energy bands. The $4 \times 4$ bilayer graphene-like block (equation 2.90) can be decomposed in a $2 \times 2$ effective Hamiltonian as (similar to equation 2.53):

$$H_{BLG} \approx -\frac{1}{2m} \begin{pmatrix} 0 & (\pi^\dagger)^2 \\ \pi^2 & 0 \end{pmatrix} + \sqrt{2}v_3 \begin{pmatrix} 0 & \pi \\ \pi^\dagger & 0 \end{pmatrix}$$

$$+ \begin{pmatrix} \gamma_2/2 + \Delta_2 & 0 \\ 0 & -2\Delta_2 \end{pmatrix} + \frac{v_F^2}{2\gamma_1} \begin{pmatrix} (\delta - 2\Delta_2)\pi^\dagger\pi & 0 \\ 0 & (\delta + \gamma_5/2 + \Delta_2)\pi\pi^\dagger \end{pmatrix},$$
$$(2.91)$$

where $m = \sqrt{2}\gamma_1/2v_F^2$. We note that the band mass of the BLG-like block of the trilayer graphene is $\sqrt{2}$ times heavier than the band mass of a stand alone pristine bilayer graphene. This happens because the role of $\gamma_1$, $\gamma_3$ and $\gamma_4$ in a pristine bilayer graphene (Equation 2.49) are played by $\sqrt{2}\gamma_1$, $\sqrt{2}\gamma_3$ and $\sqrt{2}\gamma_4$ respectively in the BLG-like block of the trilayer graphene (equation 2.90).

Fig. 2.17 shows the low energy bands of ABA trilayer graphene calculated from equation 2.89 and equation 2.90. A small gap between the pair of linear bands is evident. A similar bandgap also exists between the pair of quadratic bands (the BLG-like valence band is not shown to avoid clutter). We see from Fig. 2.17a that the BLG-like band is trigonally warped. Like in bilayer graphene the trigonal warping of ABA trilayer graphene at low energy is controlled by the hopping parameter $\gamma_3$ (the second term in equation 2.91). So, as expected, the BLG-like band becomes isotropic when $\gamma_3$ is set to zero (see Fig. 2.17b).

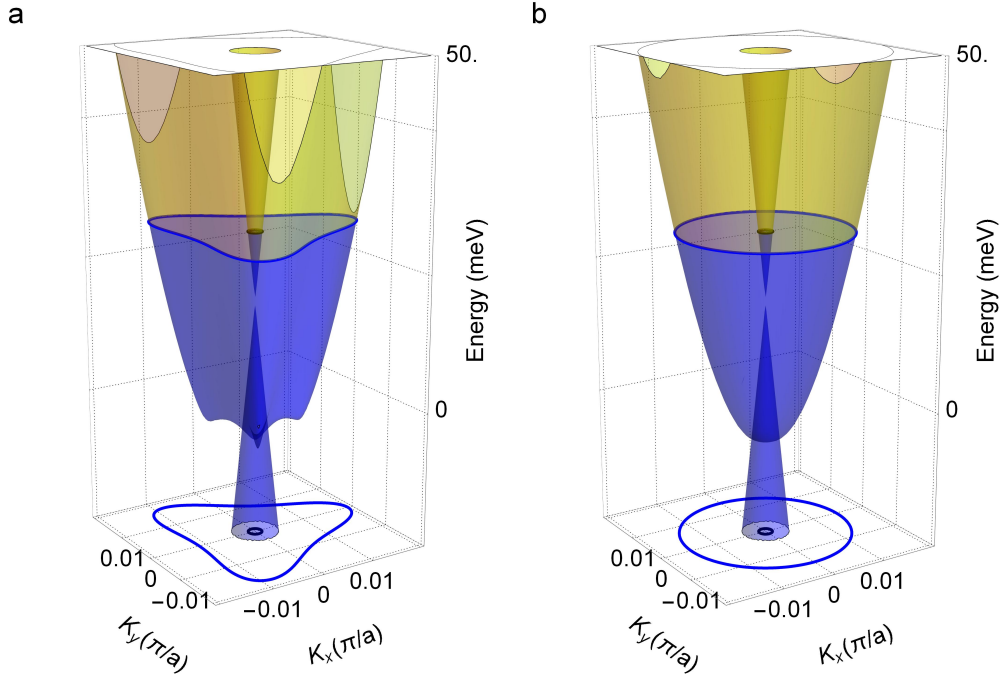

Figure 2.17: **The low energy bands of ABA trilayer graphene with and without $\gamma_3$.** (a) Low energy band structure of ABA-TLG around $K_+$ point showing one pair of linear (MLG-like) bands and one quadratic (BLG-like) conduction band. The BLG-like valence band is not shown to avoid clutter. The filled and empty parts of the bands are shown in blue and the yellow colour respectively. Fermi surface is shown as two thick blue contours – the inner one for the linear band and the outer one for the quadratic band (note that this is not the intrinsic Fermi level of the ABA-TLG which is at zero energy). We see that the Fermi surface of the linear band is isotropic. On the other hand, the Fermi surface of the quadratic band is trigonally warped. (b) The band structure with $\gamma_3=0$ plotted in the same range showing no trigonal warping. This shows that the trigonal warping of the quadratic band at low energy is imposed by $\gamma_3$.

### 2.3.3 Numerical calculation of Berry's phase

The Hamiltonian of ABA-TLG is a $6 \times 6$ dimensional matrix. Since fifth degree and onwards polynomials have no explicit analytical solutions, it is no longer possible to get the analytical wavefunctions of the trilayer graphene Hamiltonian which are required to evaluate the Berry's phase (equation 2.23). However, we can always solve for the eigenvectors at each $\boldsymbol{k}$ point numerically. To calculate the Berry's phase from the numerical wavefunctions, the gradient operator of equation 2.23 can be written on a one-dimensional path in momentum space [20] which gives

$$\gamma = -i \lim_{N\to\infty} \sum_{j=0}^{N-1} log \langle j|j+1\rangle \ , \tag{2.92}$$

where $|j\rangle$ is the wavefunction of the Hamiltonian at the j'th location in the momentum space. Fig. 2.18 shows a 1D path (blue circle) in momentum space for which the wavefunctions need to be calculated at a set of discrete grid points (red points). Fig. 2.19 shows the numerically calculated Berry's phase of the pair of linear bands and the pair of quadratic bands. We note that since the wavefunctions are numerically calculated at each $\boldsymbol{k}$ point, this approach of calculating Berry's phase can be readily extended to the case when the electric field is non-zero.

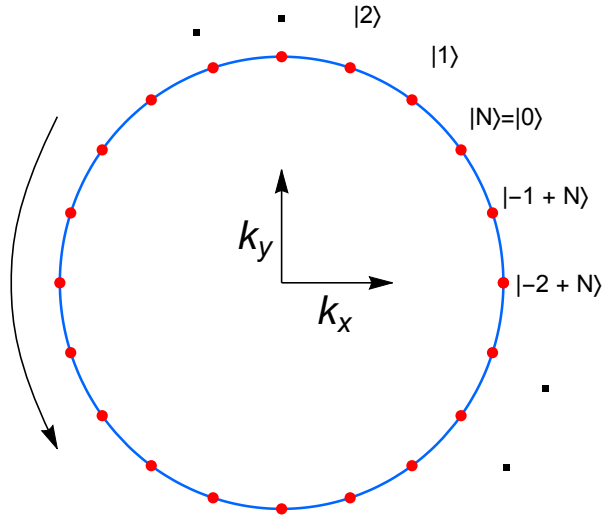

Figure 2.18: **Numerical calculation of Berry's phase.** Shows a snapshot of the instantaneous wavefunctions at some points of the loop for which Berry's phase needs to be calculated. $j = 0$ and $j = N$ denote the same point.

All the tight binding calculations in this chapter are reproduced by the candidate by following the previous theoretical works. The interacting part of the theoretical calculation (Chapter 6) and the numerical LL spectrum calculation in presence of the electric field (Chapter 7) were done in collaboration with Prof. Rajdeep Sensarma's group.

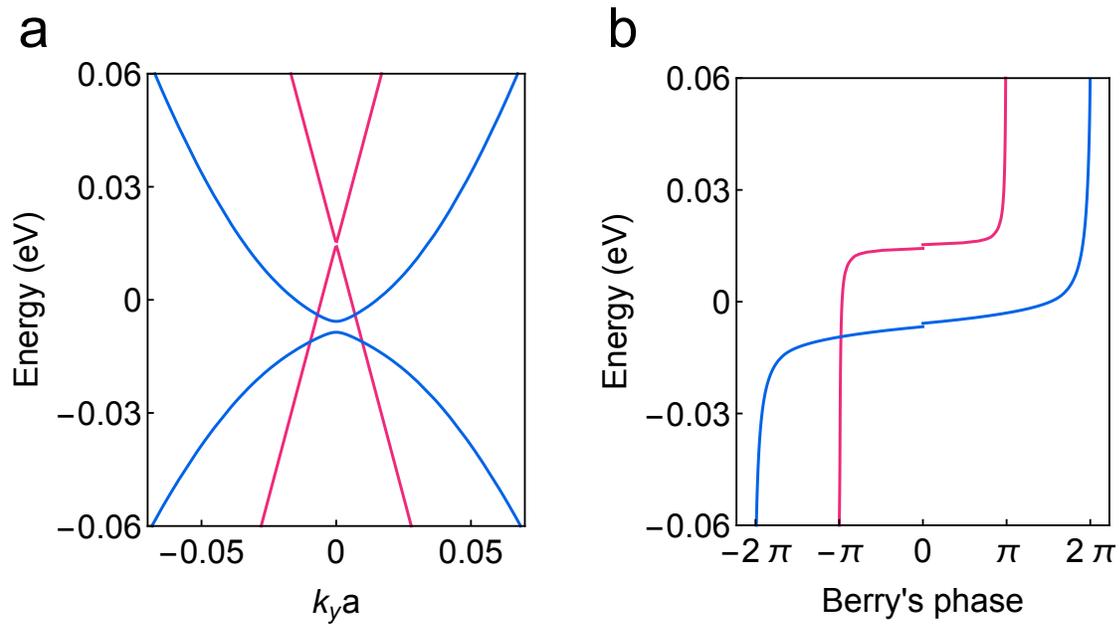

Figure 2.19: **Numerically calculated Berry's phase of ABA trilayer graphene at zero electric field.** (a) Band structure of ABA-trilayer graphene. (b) Berry's phase of the linear band (red) and quadratic band (blue) plotted in the same energy range as the band structure.

# Electrons in a magnetic field

In this chapter, we will discuss the electronic states in the presence of a magnetic field. When free electrons are subjected to a constant magnetic field they move in a circle due to the Lorentz force. This remains valid even in a crystal lattice. In the quantum regime, the radius of these cyclotron orbits and the energy of the electrons get quantized. These quantized energy levels are called Landau levels which give rise to quantum Hall effect in two dimensions. Here, we will briefly introduce the quantum Hall effect and then discuss how it can be understood in terms of the Landau levels. Then we will go into the details of Landau level spectrum in monolayer, bilayer and ABA-trilayer graphene. We will also discuss the quantization of electron orbits in $\boldsymbol{k}$ space known as Onsager rule and its connection to Berry's phase. In the end, we will describe how Berry's phase can be extracted studying the quantum oscillations in the experiment.

## 3.1  A brief overview of quantum Hall effect

When electrons are subjected to a magnetic field ($B$) perpendicular to the current direction in a conductor, it develops a voltage across the edges of the conductor both perpendicular to the current direction and magnetic field – known as classical Hall effect. Within the Drude model, the resistivity parallel and perpendicular to the current flow (in 2D) can be calculated as

$$\rho_{xx} = \frac{m}{ne^2\tau}, \quad \rho_{xy} = \frac{B}{ne}, \tag{3.1}$$

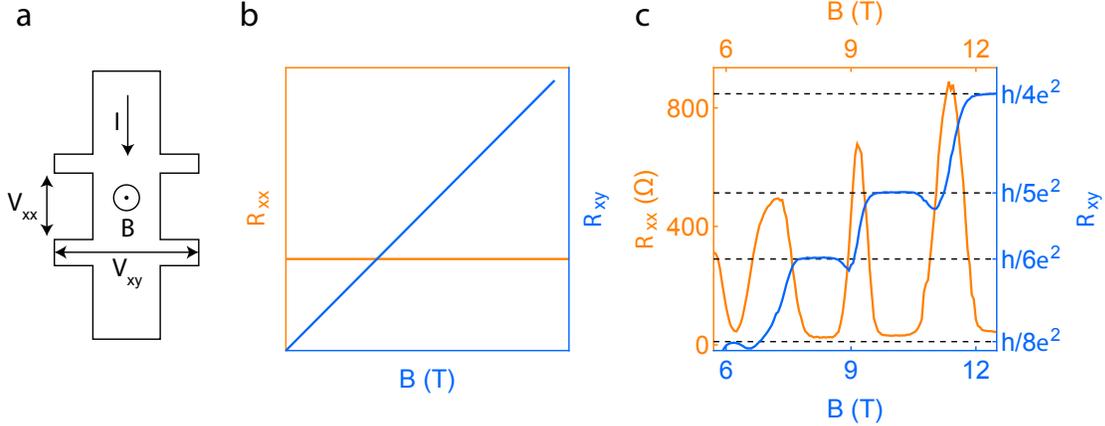

Figure 3.1: **Hall effect.** (a) Typical Hall bar geometry of a sample for Hall measurement. (b) Classical Hall effect where the Hall resistance is expected to be linear with the magnetic field and the longitudinal resistance is expected to be insensitive to the magnetic field. (c) Integer quantum Hall effect: the Hall resistance develops quantized plateaus at $R_{xy} = \frac{h}{Ne^2}$ where $N$ is an integer. Unlike in classical Hall effect the longitudinal resistance shows oscillations. The minima of the $R_{xx}$ oscillations coincide with the $R_{xy}$ plateaus. It is observed [28] that the oscillatory part of the longitudinal resistance ($\Delta R_{xx}$) can be expressed as a derivative of the Hall resistance with magnetic field $|\Delta R_{xx}| \approx B \times dR_{xy}/dB$.

where $m$, $n$, $e$, $\tau$ are mass, 2D density, charge and scattering time of electrons respectively. It shows that the longitudinal resistivity ($\rho_{xx}$) captures the information about the scattering. However, the Hall resistivity ($\rho_{xy}$) is independent of the scattering details – it captures something more intrinsic – the density of the system, see Fig. 3.1b. Moreover, in 2D the Hall resistance and Hall resistivity are identical which makes Hall resistance a fundamental quantity.

It was observed by Klaus von Klitzing in 1980 that when electrons are confined to move in a two-dimensional plane and subjected to a strong magnetic field at a low enough temperature, the Hall resistance develops plateaus for some values of the magnetic field instead of showing the linear field dependence expected from equation 3.1. The Hall resistance/resistivity at these plateaus are quantized at

$$R_H = \frac{h}{\nu e^2} \,, \tag{3.2}$$

where $h$ is Planck constant and $\nu$ is an integer. This is known as integer quantum Hall effect (IQHE) [29], see Fig. 3.1c. It is remarkable – for the first time, it was seen that a macroscopic quantity is quantized in units of the fundamental constants of nature. The ratio $h/e^2$ is known as the quantum of resistance and has a value $\sim$25.818 k$\Omega$.

In quantum Hall experiments we typically bias the sample with a constant current and measure the longitudinal and Hall voltage. This gives us longitudinal and Hall resistivity from which the longitudinal and Hall conductivity can be calculated by inverting the resistivity tensor as

$$\sigma_{xx} = \frac{\rho_{xx}}{\rho_{xx}^2 + \rho_{xy}^2}, \quad \sigma_{xy} = \frac{\rho_{xy}}{\rho_{xx}^2 + \rho_{xy}^2}. \tag{3.3}$$

We note that when $\rho_{xx} \to 0$, equation 3.3 shows that (for $\sigma_{xy} \neq 0$) $\sigma_{xx} \to 0$ as well. This looks contradictory because we typically associate $\rho_{xx} \to 0$ to the perfect conductor and $\sigma_{xx} \to 0$ to the perfect insulator. We will see that these two statements are not in contradiction but they have a meaning in presence of a magnetic field [30]. It happens because here the resistivity is a tensor (not a scalar). We note that from equation 3.1, $\rho_{xx} \to 0$ limit is achieved at $\tau \to \infty$ limit i.e. in absence of any scattering. In this limit magnetic field bends the direction of current ($\boldsymbol{J}$) perpendicular to the applied electric field ($\boldsymbol{E}$) direction. The Lorentz force acting on an electron is $\boldsymbol{F} = e(\boldsymbol{E} + \boldsymbol{v} \times \boldsymbol{B})$. So, the work done due to the electromagnetic field $= \boldsymbol{F}.d\boldsymbol{l} = e\boldsymbol{E}.\boldsymbol{v}dt \sim \boldsymbol{E}.\boldsymbol{J} = 0$ (since current is perpendicular to the applied electric field). No work done means no dissipation [30] and hence zero resistivity $\rho_{xx} = 0$. On the other hand, $\sigma_{xx} \to 0$ means no current is flowing parallel to the applied electric field [30].

## 3.2 Introduction to Landau levels

Now we go to the microscopics of the system to understand how quantum Hall effect occurs. Integer quantum Hall effect can be understood as the quantization of the kinetic energy of electrons due to an applied magnetic field. If the magnetic field is in the $\hat{z}$ direction the vector potential in the Landau gauge can be written as $\boldsymbol{A} = xB\hat{y}$. Replacing $\boldsymbol{p}$ by $\boldsymbol{p} + e\boldsymbol{A}$ we get the Hamiltonian for free charged particles in a magnetic field as [30]

$$H = \frac{1}{2m}\left(p_x^2 + (p_y + eBx)^2\right). \tag{3.4}$$

Since the Hamiltonian (equation 3.4) does not depend on $y$, we have translational invariance in the $\hat{y}$ direction. So, the eigenstates of this Hamiltonian will also be eigenstates of $p_y$. The eigenstates of $p_y$ are plane waves which motivates us to choose

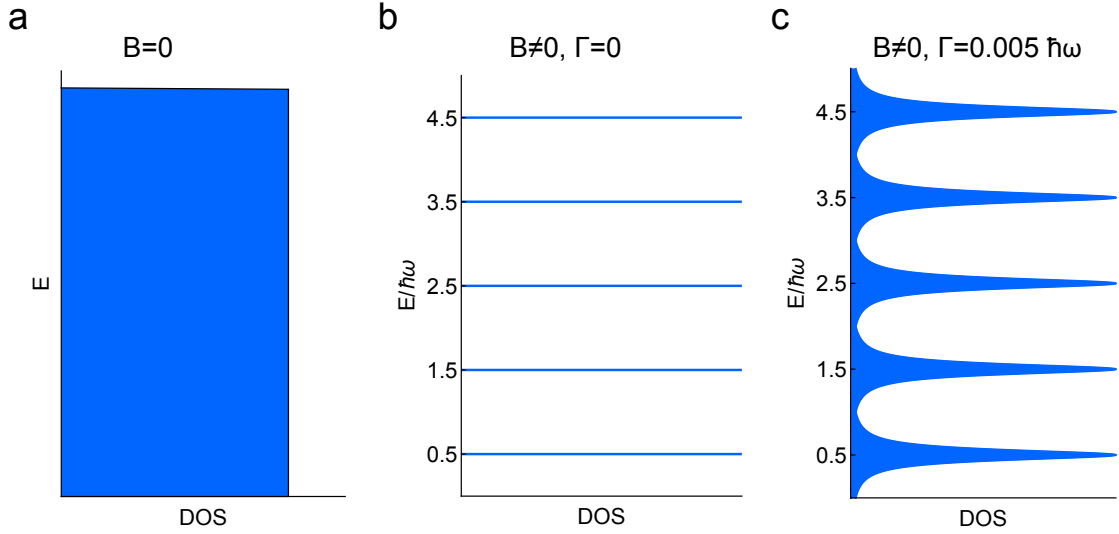

Figure 3.2: **Landau levels and the density of states in a normal 2DEG.** (a) Density of states in a metal which is almost constant in a narrow energy window. (b) Landau levels at a finite magnetic field. Ideally, the LLs have zero width with a huge $B/\Phi_0$ fold degeneracy per unit area. (c) In real systems, the LLs get broadened due to scattering which gives them a finite width. Here each LLs are broadened with a Lorentzian function which mimics the density of states. The FWHM of each Lorentzian is taken to be $2\Gamma = 0.01\hbar\omega_c$.

a solution of equation 3.4 of the form [30]

$$\psi_{k_y}(x, y) = e^{ik_y y} \phi_{k_y}(x).$$ (3.5)

The effect of the Hamiltonian (equation 3.4) acting on this wavefunction is just to replace $p_y$ operator with its eigenvalue $\hbar k_y$. So after doing this replacement, we get

$$H_{k_y} = \frac{1}{2m} \left( p_x^2 + (\hbar k_y + eBx)^2 \right)$$ (3.6)

$$= \frac{1}{2m} p_x^2 + \frac{1}{2} m \omega_c^2 \left( x + k_y l_B^2 \right)^2,$$ (3.7)

where $\omega_c = eB/m$ is the cyclotron frequency and $l_B$ is the magnetic length

$$l_B = \sqrt{\hbar/eB}.$$ (3.8)

At $B$=1 T magnetic field the magnetic length $l_B \sim 25$ nm. Magnetic length is the characteristic length scale for any quantum phenomena of electrons in the magnetic field.

Equation 3.7 is a Hamiltonian of a one-dimensional harmonic oscillator for which we know the eigenvalues

$$E_N = \hbar\omega_c(N + \frac{1}{2}),$$ (3.9)

where $\omega_c = eB/m$ is the cyclotron frequency and $N$ is called the LL index. These energy levels are known as Landau levels, see Fig. 3.2. We notice for an electron with quadratic dispersion Landau levels are equispaced (with separation $\hbar\omega_c$) and the energy of the LLs are proportional to the applied magnetic field. We also note that the energy of the LLs does not depend on $k_y$ i.e. the LLs are very flat and hence do not carry current.

As an aside, here we note the relation of the classical radius of the cyclotron orbit and magnetic length. The classical radius of the cyclotron orbit is given by

$$r_c = \frac{mv}{eB}$$ (3.10)

$$\Rightarrow r_c^2 = \frac{2mE}{e^2B^2}.$$ (3.11)

On the other hand, magnetic length has no classical analog – it is the width of the quantum mechanical wavefunction. Even though they have a different meaning, they are proportional to each other

$$\frac{r_c^2}{l_B^2} = \frac{2E}{\hbar\omega_c}$$ (3.12)

$$= 2(N + \frac{1}{2}).$$ (3.13)

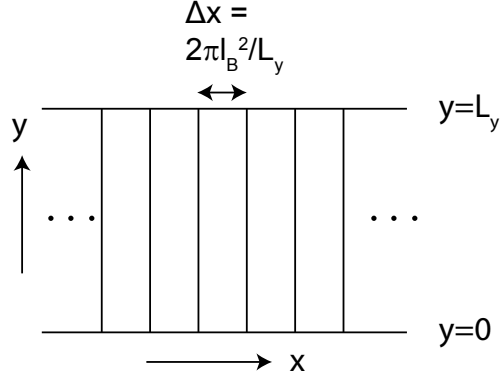

Figure 3.3: **Degeneracy of the LLs.** In the Landau gauge ($\boldsymbol{A} = xB\hat{y}$) location of the LLs depends on the $k_y$ momentum which can be mapped to the $x$ position. So, the states can be thought to lie on a regular grid in a strip. All the states with different $k_y$ in a single LL are degenerate. So, the number of grid points in the strip is the degeneracy of each LL.

## 3.2.1 Degeneracy of the LLs

We note that our initial problem of particles moving in a plane was two dimensional, however, the Hamiltonian is mapped into a problem of a one-dimensional Harmonic oscillator, see equation 3.7. To understand the origin of degeneracy we write down the (unnormalized) wavefunction – it is a multiplication of plane wave along $\hat{y}$ direction and the solution of a one-dimensional harmonic oscillator in the $\hat{x}$ direction [30]

$$\psi_{N,k_y}(x,y) \sim e^{ik_y y} H_N\left(x + k_y l_B^2\right) e^{-\left(x + k_y l_B^2\right)^2 / 2l_B^2} \,, \tag{3.14}$$

where $H_N$ is N'th order Hermite polynomial that appears in the solution of a one-dimensional harmonic oscillator. We note that the wavefunction of the LLs needs two quantum number $N$ and $k_y$, whereas the energy depends only on a single quantum number, N, the LL index. This indicates that there is a degeneracy associated with the other quantum number $k_y$.

We notice from equation 3.14 that $l_B$ is actually the spread of the quantum mechanical wave function around $x = -k_y l_B^2$ point around which the wavefunction is exponentially localized. We see that interestingly the momentum $k_y$ has become the position of the wavefunction along $\hat{x}$ direction. If the sample dimension is $L_x \times L_y$, then $k_y$ would be quantized in units of $2\pi/L_y$ i.e.

$$k_y = \frac{2\pi N_y}{L_y} \, . \tag{3.15}$$

So, the distance between two states in $\hat{x}$ direction would be

$$\Delta x = \Delta k_y l_B^2 \tag{3.16}$$

$$= \frac{2\pi}{L_y} l_B^2 \, . \tag{3.17}$$

We have seen earlier that the all the possible states with different $k_y$ in a single LL are degenerate and all these states form a grid along the $\hat{x}$ axis of the sample in real space (see Fig. 3.3). So, the degeneracy of a LL is the number of possible $k_y$ values which is alternatively the number of grid points along $\hat{x}$ axis [30, 31]

$$\mathcal{N} = \sum k_y \tag{3.18}$$

$$= \frac{L_x}{\Delta x} \tag{3.19}$$

$$= \frac{L_x L_y}{2\pi l_B^2} \, . \tag{3.20}$$

The degeneracy can also be found by counting all the states with different $N_y$ values in the equation 3.15. So the total number is the degeneracy of the each LL

$$\mathcal{N} = \sum N_y \tag{3.21}$$

$$= \sum \frac{L_y}{2\pi} k_y \tag{3.22}$$

$$= \frac{L_y}{2\pi} \frac{1}{l_B^2} \int_0^{L_x} dx \tag{3.23}$$

$$= \frac{L_x L_y}{2\pi l_B^2} \, . \tag{3.24}$$

Here we note a few important points about the LL degeneracy

- The degeneracy of each LL is not an intrinsic quantity – it scales with the sample dimension ($L_x \times L_y$).

- The degeneracy of each LL per unit area $= \frac{1}{2\pi l_B^2} = \frac{B}{\Phi_0}$ where $\Phi_0 = \frac{h}{e}$ is the quantum of flux. So, the degeneracy is proportional to the applied magnetic field.

- The degeneracy of each LL is huge – for a sample with area $10 \times 10 \ \mu\text{m}^2$ at 10 T it is $\sim 2.5 \times 10^5$!

### 3.2.2 Filling factor

The integer $\nu$ in equation 3.2 is known as the filling factor which means how many Landau levels (with distinct energy) are filled. We have seen the degeneracy of each LL (per unit area) is $B/\Phi_0$. If the number density of electrons is $n$ then it will fill $\frac{n}{B/\Phi_0}$ number of LLs. So the filling factor ($\nu$) is defined as

$$\nu = \frac{n}{B/\Phi_0} = \frac{nh}{eB} \,. \tag{3.25}$$

### 3.2.3 Calculating Hall conductance

Let's say we have applied an electric field in the $\hat{x}$ direction, then the potential energy of an electron would be $V = -eEx$. So, the equation 3.6 gets modified as [30]

$$H_{p_y} = \frac{1}{2m}\left(p_x^2 + (p_y + eBx)^2\right) - eEx \tag{3.26}$$

$$= \frac{1}{2m}\left[p_x^2 + p_y^2 + e^2B^2\left(x + \frac{p_y}{eB} - \frac{mE}{eB^2}\right)^2 - \left(p_y - \frac{mE}{B}\right)^2\right] \tag{3.27}$$

$$= \frac{1}{2m}\left[p_x^2 + \hbar^2k_y^2 + e^2B^2\left(x + k_yl_B^2 - \frac{eE}{m\omega_c^2}\right)^2 - \left(\hbar k_y - \frac{mE}{B}\right)^2\right] \tag{3.28}$$

$$= \frac{1}{2m}\left[p_x^2 + e^2B^2\left(x + k_yl_B^2 - \frac{eE}{m\omega_c^2}\right)^2\right] + \frac{1}{2m}\left[\hbar^2k_y^2 - \left(\hbar k_y - \frac{mE}{B}\right)^2\right] \tag{3.29}$$

$$= \frac{1}{2m}\left[p_x^2 + e^2B^2\left(x + k_yl_B^2 - \frac{eE}{m\omega_c^2}\right)^2\right] + eEk_yl_B^2 - \frac{mE^2}{2B^2} \,. \tag{3.30}$$

This shows that the mean position of $x$ is shifted by $-k_y l_B^2 + eE/m\omega_c^2$ from the origin, therefore the wavefunction becomes

$$\psi_{N,k_y}(x,y) \sim e^{ik_y y} H_N \left( x + k_y l_B^2 - \frac{eE}{m\omega_c^2} \right) e^{-\left( x + k_y l_B^2 - \frac{eE}{m\omega_c^2} \right)^2 / 2 l_B^2}. \qquad (3.31)$$

So the average position is given by

$$\langle x \rangle = -k_y l_B^2 + \frac{eE}{m\omega_c^2} \qquad (3.32)$$

$$= -\frac{\hbar k_y}{eB} + \frac{mE}{eB^2}. \qquad (3.33)$$

As evident, the energy eigenvalue of the equation 3.30 is given by [30]

$$E_{N,k_y} = \hbar\omega_c(N + \frac{1}{2}) + eEk_y l_B^2 - \frac{mE^2}{2B^2}. \qquad (3.34)$$

We see that in the presence of an electric field the LLs with different $k_y$ are not degenerate anymore. In other words, now the LLs are not completely flat – they disperse with $k_y$, see Fig. 3.4. So, the velocity is

$$v_y = \frac{1}{\hbar} \frac{\partial E_{N,k_y}}{\partial k_y} = \frac{1}{\hbar} eE l_B^2 = \frac{E}{B}. \qquad (3.35)$$

We know the current density can be written as

$$\boldsymbol{j} = -e\boldsymbol{v}. \qquad (3.36)$$

In presence of magnetic field velocity is

$$m\boldsymbol{v} = \boldsymbol{p} + e\boldsymbol{A}. \qquad (3.37)$$

So, the total current can be written as the summation of the current carried by all the filled states

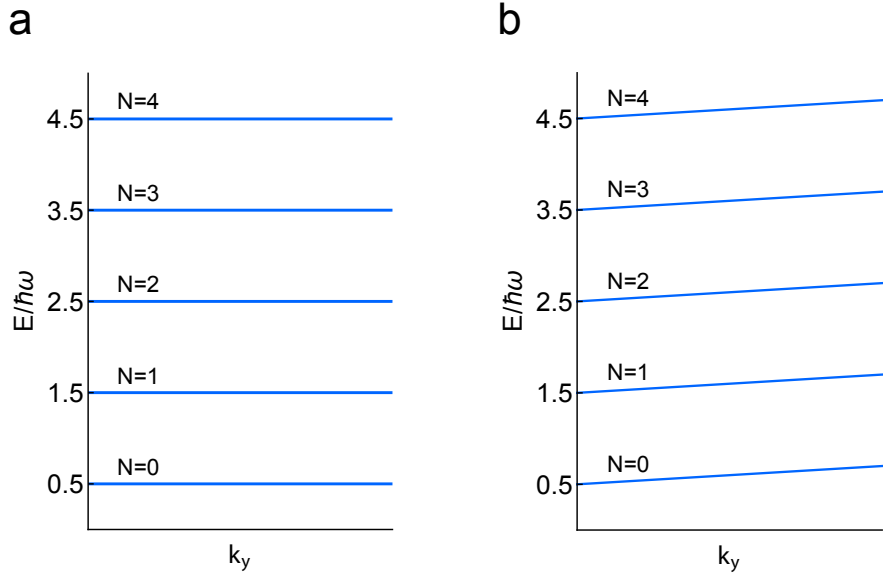

Figure 3.4:  **Dispersion of Landau levels.** (a) The LLs are completely flat at zero electric field.  All the $k_y$ states are strictly degenerate and hence the group velocity is zero.  (b) In presence of an applied electric field the LLs tilt which lifts the $k_y$ degeneracy.  As a result, the group velocity $\frac{1}{\hbar}\frac{\partial E_{N,k_y}}{\partial k_y}$ becomes non-zero.

$$\boldsymbol{j} = -\frac{e}{m} \sum_{i=1}^{\nu} \langle \psi | \, p_y + e\boldsymbol{A} \, | \psi \rangle \;, \tag{3.38}$$

where $\nu$ is number of filled LLs.

The current in the $\hat{y}$ direction is

$$j_y = -\frac{e}{m} \sum_{i=1}^{\nu} \sum_{k_y} \langle \psi_{N,k_y} | \, p_y + exB \, | \psi_{N,k_y} \rangle \;, \tag{3.39}$$

where $\boldsymbol{A}$ can be written as $\boldsymbol{A} = xB\hat{y}$ in the Landau gauge.  Incorporating $\langle x \rangle$ from equation 3.32 we get

$$j_y = -\frac{e}{m} \sum_{i=1}^{\nu} \sum_{k_y} \left( \hbar k_y - \hbar k_y + \frac{mE}{B} \right) \tag{3.40}$$

$$= -e\nu \sum_{k_y} \frac{E}{B} \;. \tag{3.41}$$

We have calculated the sum over $k_y$ before which is nothing but the degeneracy of the LLs at zero electric field i.e. $B/\Phi_0$ per unit area. So, the current density becomes

$$j_y = -e\nu\frac{E}{\Phi_0}. \tag{3.42}$$

In this calculation every state carries same current because equation 3.35 shows that every $k_y$ state has same group velocity (see Fig. 3.5 for LL in real space).

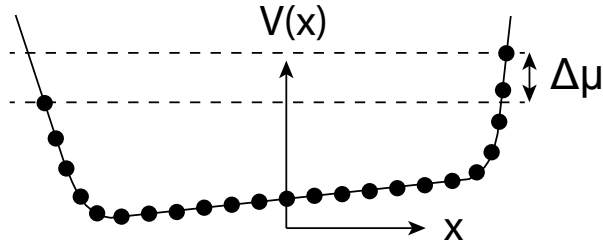

Figure 3.5: **LLs in real space in the presence of an electric field.** The LLs bend up (for electrons) or down (for holes) close to the edges of the sample due to the edge potential. An applied electric field in the $\hat{x}$ direction tilts the Landau level such that each $k_y$ states can carry a current. The solid circles denote the $k_y$ states as they can be thought as the position in the $\hat{x}$ direction. The chemical potential difference between the two edges creates the population imbalance between the left and the right moving electrons.

The current in the $\hat{x}$ direction is zero as we know the wave functions (in $x$) are harmonic oscillator states for which the average velocity is zero.

$$j_x = -\frac{e}{m}\sum_{i=1}^{\nu}\sum_{k_y}\left\langle\psi_{N,k_y}\right|p_x\left|\psi_{N,k_y}\right\rangle = 0 \tag{3.43}$$

From the current density ($\boldsymbol{j}$) and electric field ($\boldsymbol{E}$) we can formulate the conductivity tensor:

$$\begin{pmatrix} j_x \\ j_y \end{pmatrix} = \begin{pmatrix} \sigma_{xx} & \sigma_{xy} \\ -\sigma_{xy} & \sigma_{xx} \end{pmatrix}\begin{pmatrix} E_x \\ E_y \end{pmatrix} \tag{3.44}$$

$$\Rightarrow \begin{pmatrix} 0 \\ j_y \end{pmatrix} = \begin{pmatrix} \sigma_{xx} & \sigma_{xy} \\ -\sigma_{xy} & \sigma_{xx} \end{pmatrix}\begin{pmatrix} E_x \\ 0 \end{pmatrix} \tag{3.45}$$

Therefore, the longitudinal conductivity and the Hall conductivity (incorporating equation 3.42) can be calculated as

$$\sigma_{xx} = 0, \quad \sigma_{xy} = -\frac{j_y}{E_x} \tag{3.46}$$

$$= \frac{e\nu}{\Phi_0} = \frac{e^2}{h}\nu \,. \tag{3.47}$$

Using equation 3.3 the longitudinal resistivity and the Hall resistivity can be calculated as

$$\rho_{xx} = 0 \,, \quad \rho_{xy} = \frac{h}{\nu e^2} \,. \tag{3.48}$$

### 3.2.4   Edge states

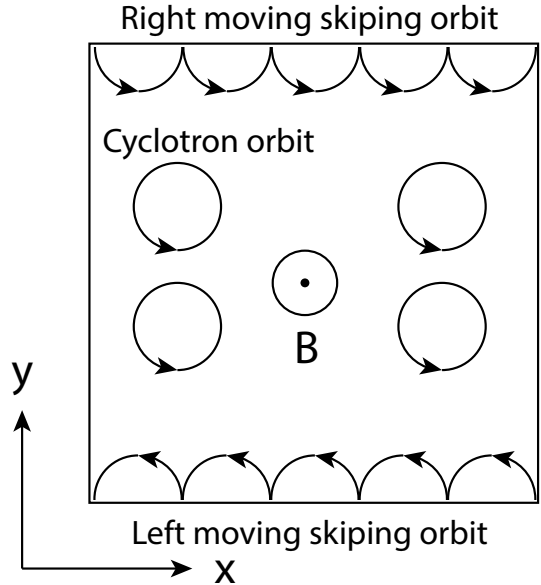

Figure 3.6:   **Classical electron skipping orbit and edge states.** Classically the edge states can be thought to originate from the semiclassical electron skipping orbits. Electrons at the edges reflect after they collide the sample boundary and travel in the opposite direction at the opposite edges. Bulk electrons are localized around potential hills, so they do not participate in the conduction.

In quantum Hall regime edge of the sample plays an important role – it hosts some modes of the system which are localized at the sample edges. The existence of edge modes can be motivated even classically. In presence of the cross electric field and magnetic field, the right moving and the left moving electrons go to opposite edges of the sample. Depending on the initial velocity direction of the electrons they rotate either clockwise or anticlockwise. However, at the edge of the sample, the electrons can not rotate a full circle – they reflect from the edges. This results in a semiclassical skipping orbit, see Fig. 3.6.

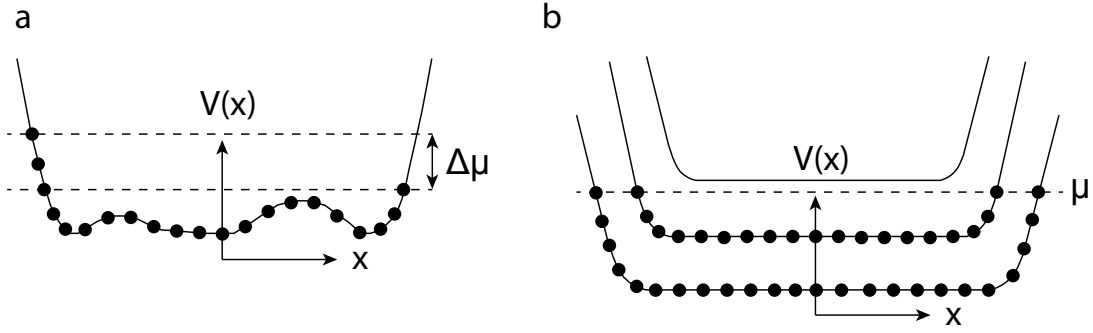

Figure 3.7: **Determination of Hall conductance only in terms of edge states.** (a) Schematic of a single LL in real space. It turns out that the Hall conductance of a single LL is $e^2/h$ which can be calculated only in terms of the applied chemical potential difference between the left movers and the right movers. The exact potential variation does not matter as long as the potential hills do not rise above the chemical potential. (b) Schematic of multiple LLs. Hall conductance due to N filled LL is $Ne^2/h$.

Quantum mechanically the edge states arise because the potential rises sharply at the physical edge of the sample. Let's say the sample is finite in $\hat{x}$ direction and the potential at the edge is $V(x)$. So, the Hamiltonian in the Landau gauge becomes [30]

$$H = \frac{1}{2m}\left(p_x^2 + (p_y + eBx)^2\right) + eV(x)\,.  \tag{3.49}$$

Keeping up to the linear term in the Taylor's expansion of $V(x)$ we get $V(x) \approx V(X) + (x - X)\partial V/\partial x + \dots$. We can shift the origin to neglect the constant term $V(x)$. We note that $\partial V/\partial x$ acts like an electric field in the $\hat{x}$ direction whose effect is to generate a $\hat{y}$ component of the velocity, see equation 3.35.

$$v_y = \frac{1}{\hbar} \frac{\partial E_{N,k_y}}{\partial k_y} \tag{3.50}$$

$$= \frac{e}{\hbar} \frac{\partial V}{\partial k_y} \tag{3.51}$$

From equation 3.33 we know that $k_y$ can be related to $x$: $\partial x = -\frac{\hbar}{eB} \partial k_y$. Therefore

$$v_y = -\frac{1}{B} \frac{\partial V}{\partial x} \,. \tag{3.52}$$

Since the slope of the potential $\left( \frac{\partial V}{\partial x} \right)$ is opposite at the two edges (see Fig. 3.5), $v_y$ has an opposite sign at the two edges of the sample. That means carriers at the two edges have an opposite velocity which agrees with the previous classical picture of the skipping orbits. Since these edge states carry current in one direction, they are called chiral. Chiral states are special – these electrons can not backscatter because the states with opposite velocity exist only at the other edge of the sample. So, to backscatter (reverse the velocity), an electron needs to traverse all the way through the bulk of the sample which is a long way for an electron and hence the probability of such scattering is very small.

The current carried by a single LL in the $\hat{y}$ direction can be calculated as

$$I_y = -e \int v_y \frac{dk_y}{2\pi} \tag{3.53}$$

$$= \frac{e}{2\pi} \int \frac{1}{B} \frac{\partial V}{\partial x} \frac{dx}{l_B^2} \tag{3.54}$$

$$= \frac{e}{\hbar} \Delta\mu \,. \tag{3.55}$$

where $\Delta\mu$ is the applied chemical potential difference which is also the Hall voltage $\Delta\mu = eV_H$. So, the Hall conductivity

$$\sigma_{xy} = \frac{I_y}{V_H} = \frac{e^2}{h}\,. \tag{3.56}$$

This is the quantized Hall conductivity of a single LL. If $N$ LLs are filled then the Hall conductivity becomes

$$\sigma_{xy} = N\frac{e^2}{h}\,. \tag{3.57}$$

We note that when there is no external electric field applied $v_y = 0$ in the bulk because the LLs are flat. So, all the currents in equation 3.53 is carried by the edge states.

## 3.3 Landau levels of monolayer graphene

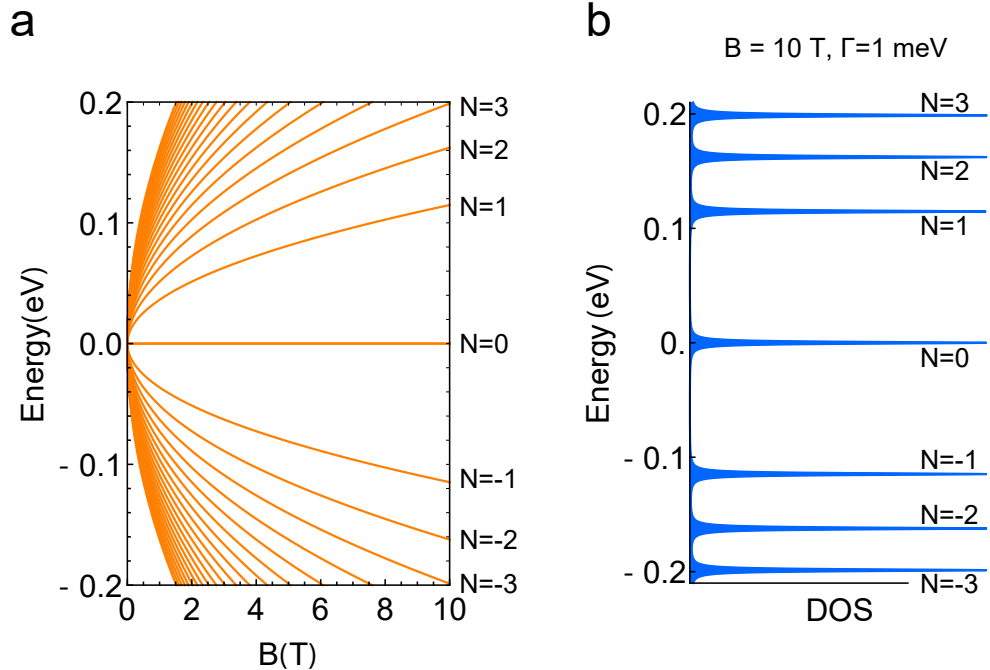

Figure 3.8: **Landau level diagram and the density of states of monolayer graphene.** (a) Landau level diagram with the marked LL indices. $N = 0$ LL is half electron-like and half hole-like. $\pm$ sign for LLs $N \geq 1$ denotes the electron-like and half hole-like LLs respectively. We see that the LLs at a given magnetic field are not equispaced unlike in normal 2DEG. (b) The density of states as a function of energy. Here each LLs are broadened with a Lorentzian function which mimics the density of states. The FWHM of each Lorentzian is $2\Gamma = 2$ meV.

We remember that the Hamiltonian of the monolayer graphene at zero magnetic field is given by

$$H_M = v_F \begin{pmatrix} 0 & p_x + i\xi p_y \\ p_x - i\xi p_y & 0 \end{pmatrix}, \tag{3.58}$$

where $\xi = \pm 1$ for the $K$ and $K'$ valleys respectively. In presence of magnetic field $\boldsymbol{p}$ is replaced by $\boldsymbol{\pi} = \boldsymbol{p} + e\boldsymbol{A}$. From $\boldsymbol{\pi}$ operator the creation and annihilation operators can be defined as

$$\pi_+ = \pi_x + i\pi_y = \frac{\sqrt{2}\hbar}{l_B} a^\dagger, \quad \text{and} \quad \pi_- = \pi_x - i\pi_y = \frac{\sqrt{2}\hbar}{l_B} a, \tag{3.59}$$

where $l_B = \sqrt{\hbar/eB}$ is the magnetic length. When $a$ and $a^\dagger$ act on the LL wavefunctions they lower and raise the LL index of the wavefunction respectively: $a\psi_N = \sqrt{N}\psi_{N-1}$ and $a^\dagger\psi_N = \sqrt{N+1}\psi_{N+1}$. So, at $K$ valley the Hamiltonian becomes

$$H_K = v_F \frac{\sqrt{2}\hbar}{l_B} \begin{pmatrix} 0 & a^\dagger \\ a & 0 \end{pmatrix} \tag{3.60}$$

$$= \hbar\omega_c \begin{pmatrix} 0 & a^\dagger \\ a & 0 \end{pmatrix}, \tag{3.61}$$

$$\tag{3.62}$$

where $\omega_c$ is defined as $\omega_c = v_F\sqrt{2}/l_B \sim \sqrt{B}$. Equation 3.60 can be written in $(|N\rangle, |N-1\rangle)$ basis to find its eigenvalues and eigenfunctions:

$$H_K = \hbar\omega_c \begin{pmatrix} 0 & \langle N| \sqrt{N} |N\rangle \\ \langle N-1| \sqrt{N} |N-1\rangle & 0 \end{pmatrix} \tag{3.63}$$

$$= \hbar\omega_c \begin{pmatrix} 0 & \sqrt{N} \\ \sqrt{N} & 0 \end{pmatrix}. \tag{3.64}$$

The eigenvalues of equation 3.64 are shown in Fig. 3.8 and are given by

$$E_N = \hbar\omega_c sgn(N)\sqrt{N} \quad \text{where } N = 0, \pm 1, \pm 2... \tag{3.65}$$

Eigen vectors of equation 3.64 in $(|N\rangle, |N-1\rangle)$ basis are given by $\begin{pmatrix} 1 & \pm 1 \end{pmatrix}^T$. Including the basis vectors the LL wavefunction becomes

$$|\psi_0\rangle = \begin{pmatrix} |0\rangle \\ 0 \end{pmatrix}, \quad \text{and } |\psi_N\rangle = \begin{pmatrix} |N\rangle \\ \pm |N-1\rangle \end{pmatrix} \quad \text{for } N \geq 1, \tag{3.66}$$

where $\pm$ refers to the electron and hole-like LLs respectively. Since each LL contributes $e^2/h$ in the Hall conductance, the Hall conductance when the Fermi energy is in between $N$ and $(N+1)$ LL is given by

$$\sigma_{xy} = g_s g_v (N + \frac{1}{2})\frac{e^2}{h} = \nu\frac{e^2}{h}, \quad \text{where } N = 0, \pm 1, \pm 2... \tag{3.67}$$

Here $g_s$ and $g_v$ are spin and valley degeneracy respectively. At low magnetic field when spin and valley degeneracy are not lifted $g_s = 2$ and $g_v = 2$. The factor $1/2$ comes because the $N = 0$ LL is half-filled when the system is charge neutral. Therefore, at the low magnetic field, the Hall conductivity of graphene shows plateaus at $\nu = \pm 2, \pm 6, \pm 10.....$

## 3.4 Landau levels of bilayer graphene

The predominant term of the bilayer graphene Hamiltonian at low energy is given by

$$H_B = \frac{1}{2m} \begin{pmatrix} 0 & (p_x + i\xi p_y)^2 \\ (p_x - i\xi p_y)^2 & 0 \end{pmatrix}, \tag{3.68}$$

where $m = \gamma_1/2v_F^2$. Using the creation and annihilation operator (see equation 3.59) the Hamiltonian at the $K$ valley can be rewritten as

$$H_B = \frac{1}{2m} \frac{2\hbar^2}{l_B^2} \begin{pmatrix} 0 & (a^\dagger)^2 \\ a^2 & 0 \end{pmatrix} \tag{3.69}$$

$$= \hbar\omega_c \begin{pmatrix} 0 & (a^\dagger)^2 \\ a^2 & 0 \end{pmatrix}, \tag{3.70}$$

where $\omega_c = eB/m$. Equation 3.70 can be written in $(|N\rangle, |N-2\rangle)$ basis to find its eigenvalues and eigenfunctions:

$$H_K = \hbar\omega_c \begin{pmatrix} 0 & \langle N| \sqrt{N(N-1)} |N\rangle \\ \langle N-2| \sqrt{N(N-1)} |N-2\rangle & 0 \end{pmatrix} \quad (3.71)$$

$$= \hbar\omega_c \begin{pmatrix} 0 & \sqrt{N(N-1)} \\ \sqrt{N(N-1)} & 0 \end{pmatrix} . \quad (3.72)$$

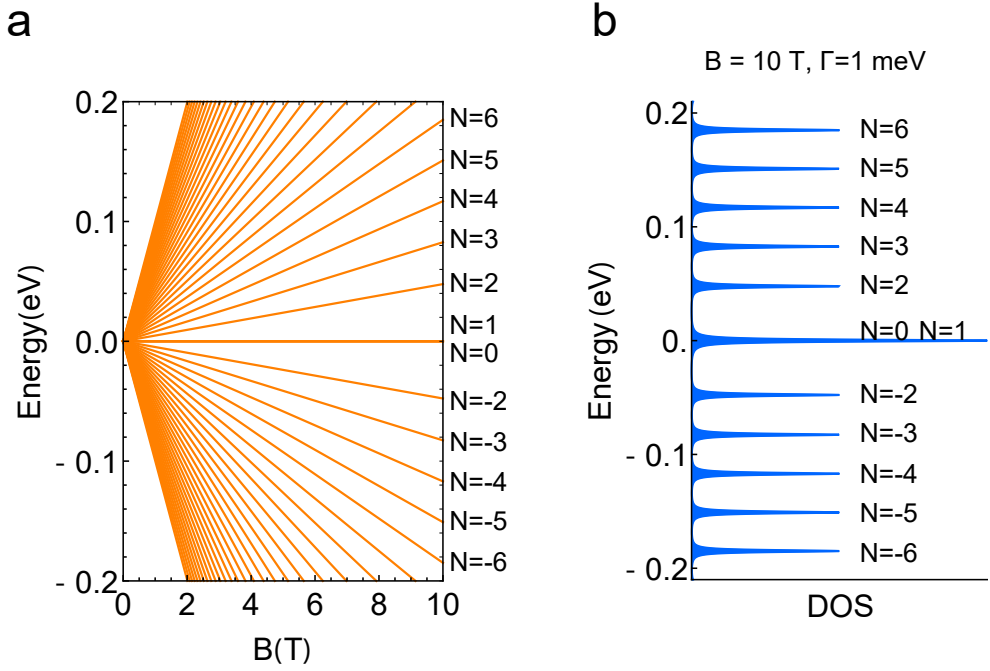

**Figure 3.9: Landau level diagram and the density of states of bilayer graphene.**
(a) Landau level diagram with the marked LL indices. $N = 0$ and $N = 1$ LLs of bilayer graphene are degenerate and they are half electron-like and half hole-like. $\pm$ sign for LLs $N \geq 2$ denotes the electron-like and half hole-like LLs respectively. (b) The density of states as a function of energy. DOS at zero energy is higher compared to other LLs because at zero energy there are two ($N = 0$ and $N = 1$) degenerate LLs. Here each LLs are broadened with a Lorentzian function which mimics the density of states. The FWHM of each Lorentzian is $2\Gamma = 2$ meV.

The eigenvalues of equation 3.72 are shown in Fig. 3.9 and are given by

$$E_N = \hbar\omega_c \, sgn(N)\sqrt{N(N-1)}, \quad \text{where } N = 0, 1, \pm 2, \pm 3... \quad (3.73)$$

We note that $N = 0$ and $N = 1$ LLs of bilayer graphene have the same energy. Hence the first LL of bilayer graphene has an additional 2 fold orbital degeneracy apart from the usual spin and valley degeneracy making the zero energy LL 8 fold degenerate in total.

Eigenvectors of equation 3.72 are

$$|\psi_0\rangle = \begin{pmatrix} |0\rangle \\ 0 \end{pmatrix}, \quad |\psi_1\rangle = \begin{pmatrix} |1\rangle \\ 0 \end{pmatrix}, \quad \text{and } |\psi_N\rangle = \begin{pmatrix} |N\rangle \\ \pm |N-2\rangle \end{pmatrix} \quad \text{for } N \geq 2, \quad (3.74)$$

where $\pm$ refers to the electron and hole-like LLs respectively. The Hall conductance when the Fermi energy is in between $N$ and $(N+1)$ LL is given by

$$\sigma_{xy} = g_s g_v N \frac{e^2}{h} = \nu \frac{e^2}{h}, \quad \text{where } N = \pm 1, \pm 2...(\neq 0). \quad (3.75)$$

Here $N = 0$ is not allowed to accommodate the 8 fold degeneracy of the zeroth LL. Therefore, at the low magnetic field, the Hall conductivity of graphene shows plateaus at $\nu = \pm 4, \pm 8, \pm 12.....$ We note that the sequence of the quantum Hall plateaus of bilayer graphene is different from that of the monolayer graphene. One more difference is that in monolayer graphene all plateaus are equispaced (separated by $4e^2/h$), but in bilayer graphene when the Fermi level crosses the charge neutrality point the Hall conductivity jumps by $8e^2/h$ – a direct consequence of the 8 fold zeroth energy LL.

## 3.5   Landau levels of ABA-trilayer graphene

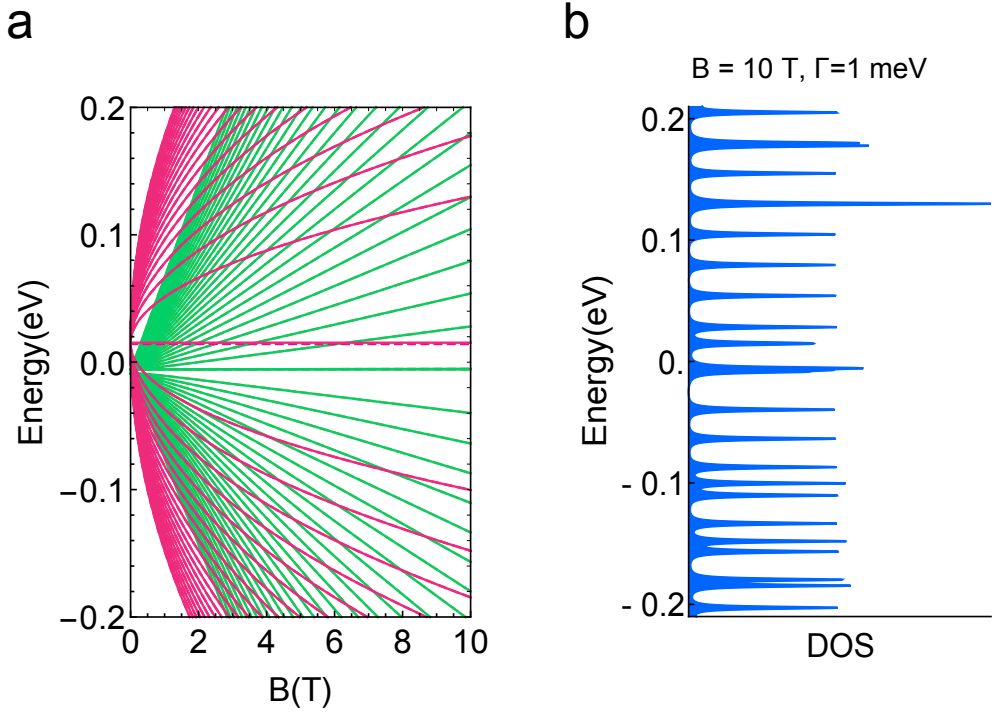

Figure 3.10:   **Landau level diagram and the density of states of ABA-trilayer graphene.** (a) Landau level diagram where the red and green colors denote the LLs coming from the linear band and the quadratic band respectively. (b) The total density of states due to LLs from both the linear and quadratic bands. Here each LLs are broadened with a Lorentzian function which mimics the density of states. The FWHM of each Lorentzian is $2\Gamma = 2$  meV.

We discussed in Chapter 2 that in zero electric field and at low energy, neglecting the trigonal warping, the ABA-trilayer graphene Hamiltonian can be thought as a superposition of a monolayer graphene-like and a bilayer graphene-like Hamiltonian [14]

$$H_{MLG} = \begin{pmatrix} -\gamma_2/2 + \Delta_2 & v_F \pi^\dagger \\ v_F \pi & \delta - \gamma_5/2 + \Delta_2 \end{pmatrix} \tag{3.76}$$

$$H_{BLG} \approx -\frac{1}{2m} \begin{pmatrix} 0 & (\pi^\dagger)^2 \\ \pi^2 & 0 \end{pmatrix} + \begin{pmatrix} \gamma_2/2 + \Delta_2 & 0 \\ 0 & -2\Delta_2 \end{pmatrix}$$
$$+ \frac{v_F^2}{2\gamma_1} \begin{pmatrix} (\delta - 2\Delta_2)\pi^\dagger \pi & 0 \\ 0 & (\delta + \gamma_5/2 + \Delta_2)\pi \pi^\dagger \end{pmatrix}, \tag{3.77}$$

where $m = \gamma_1/\sqrt{2}v_F^2$ and $\pi = \xi\pi_x + i\pi_y$ with $\xi = \pm 1$ for $K$ and $K'$ valleys respectively. This description remains valid also in quantum Hall regime. In presence of magnetic field, $\pi$ and $\pi^\dagger$ can be defined in the $K$ valley as [14]

$$\pi = \frac{i\sqrt{2}\hbar}{l_B}a^\dagger , \quad \text{and} \quad \pi^\dagger = -\frac{i\sqrt{2}\hbar}{l_B}a \quad (3.78)$$

and in the $K'$ valley as

$$\pi = \frac{i\sqrt{2}\hbar}{l_B}a , \quad \text{and} \quad \pi^\dagger = -\frac{i\sqrt{2}\hbar}{l_B}a^\dagger . \quad (3.79)$$

Eigenvalues of equation 3.76 and equation 3.77 are plotted together in Fig. 3.10a. The red and green LLs are from the monolayer graphene-like and the bilayer graphene-like Hamiltonian respectively.

## 3.6   Onsager rule and Berry's phase

We know that the available electronic states in the presence of a magnetic field can be found by quantizing the cyclotron orbits in momentum space. This can be achieved by applying the semi-classical Bohr-Sommerfeld quantization rule [32]

$$\oint \boldsymbol{p}.d\boldsymbol{r} = (N + \gamma)h , \quad (3.80)$$

where $\gamma$ is a purely quantum mechanical correction related to the Berry's phase ($\Phi_B$) as $\gamma = \frac{1}{2} - \frac{\Phi_B}{2\pi}$. In presence of magnetic field $\boldsymbol{p}$ is given by

$$\boldsymbol{p} = \hbar\boldsymbol{k} - e\boldsymbol{A} . \quad (3.81)$$

Incorporating this into the equation 3.80 we get

$$\oint (\hbar \boldsymbol{k} - e\boldsymbol{A}).d\boldsymbol{r} = (N + \gamma)h \,. \tag{3.82}$$

The momentum can be found from the semi-classical equation of motion

$$\hbar \frac{d\boldsymbol{k}}{dt} = -e(\boldsymbol{v} \times \boldsymbol{B}) \,, \tag{3.83}$$

where the group velocity ($\boldsymbol{v}$) is given by $\boldsymbol{v} = \frac{1}{\hbar} \frac{\partial E}{\partial \boldsymbol{k}}$.

From equation 3.83, $\boldsymbol{k}$ can be found as [32]

$$\boldsymbol{k}(t) - \boldsymbol{k}(0) = -\frac{eB}{\hbar}(\boldsymbol{r}(t) - \boldsymbol{r}(0)) \times \hat{B} \tag{3.84}$$

$$= -\frac{1}{l_B^2}(\boldsymbol{r}(t) - \boldsymbol{r}(0)) \times \hat{B} \,. \tag{3.85}$$

This equation tells us that position of an electron in real space is related to its position in the momentum space. Since the position vector of an electron in real space rotates in a loop – so does the momentum with the radius of orbit scaled by $1/l_B^2$. The cross product $(\boldsymbol{r}(t) - \boldsymbol{r}(0)) \times \hat{B}$ just tells us that the orbit in $\boldsymbol{k}$ space is perpendicular to both the orbit in real space and magnetic field.

Incorporating the solution for $\boldsymbol{k}$ in equation 3.82 we get

$$\oint (-eB\boldsymbol{r} \times \hat{B} - e\boldsymbol{A}).d\boldsymbol{r} = (N + \gamma)h \tag{3.86}$$

$$\Rightarrow \boldsymbol{B}. \oint (d\boldsymbol{r} \times \boldsymbol{r}) - \int_S \boldsymbol{B}.d\boldsymbol{S} = (N + \gamma)\Phi_0 \,, \tag{3.87}$$

where $\Phi_0 = h/e$ is the quantum of flux. We note that $\frac{1}{2} \oint (d\boldsymbol{r} \times \boldsymbol{r})$ is the area of the cyclotron orbit, so the first term of equation 3.87 becomes $2\Phi$ where $\Phi$ is the magnetic flux through the cyclotron orbit. The second term of equation 3.87 is clearly the magnetic flux -$\Phi$.

So, equation 3.87 becomes

$$\Phi = (N + \gamma)\Phi_0 \,.$$  (3.88)

This is the quantization of magnetic flux through the cyclotron orbit in real space. The corresponding quantization condition in $\boldsymbol{k}$ space can be found by noting that the real space orbit $(A)$ and $\boldsymbol{k}$ space orbit area $(a_k(\epsilon))$ are related by $l_B^4$ (equation 3.85)

$$A = l_B^4 a_k(\epsilon) \,.$$  (3.89)

So, equation 3.88 becomes

$$a_k(\epsilon_N) = \frac{2\pi}{l_B^2}(N + \gamma) \,.$$  (3.90)

This is known as Onsager rule [32]. This tells us that in presence of a magnetic field only certain $\boldsymbol{k}$ states are allowed for which the Fermi surface area satisfies equation 3.90. The $\boldsymbol{k}$ space area depends on the value of the magnetic field, Berry's phase of the orbit and a quantum number $(N)$. Fig. 3.11 shows the allowed electron trajectory in ABA-trilayer graphene for three different values of the magnetic field.

For massive fermions with quadratic dispersion $\gamma = 1/2$ and the energy is given by

$$\epsilon_N = \frac{\hbar^2 k^2}{2m}$$  (3.91)

$$= \frac{\hbar^2}{2m\pi}a_k$$  (3.92)

$$= \frac{\hbar^2}{2m\pi}\frac{2\pi}{l_B^2}(N + \frac{1}{2})$$  (3.93)

$$= \hbar\omega_c(N + \frac{1}{2}) \,,$$  (3.94)

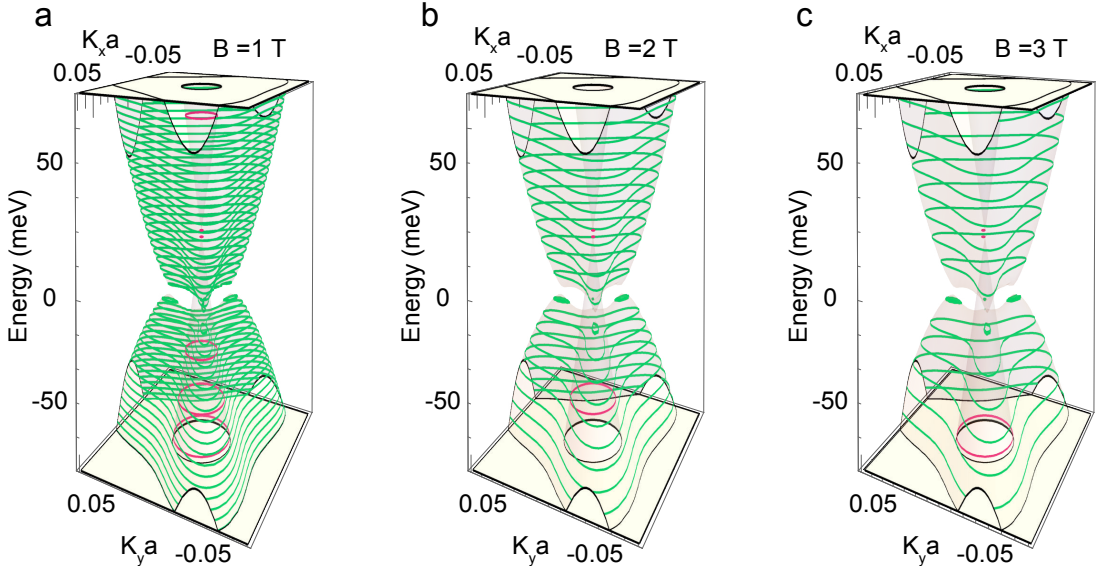

Figure 3.11: **Allowed momentum space orbits of ABA-trilayer graphene in presence of magnetic field.** (a), (b) and (c) show the $\boldsymbol{k}$ space orbits allowed by the Onsager rule for B= 1 T, 2 T and 3 T respectively. The red and green contours denote the electron orbits for the linear and quadratic band respectively.

where $\omega_c = eB/m$ is the cyclotron frequency. We note that this is precisely the same as the LL energies of massive fermions with quadratic dispersion we derived before (equation 3.9).

For Dirac Fermions with linear dispersion $\gamma = 0$ and the energy is given by

$$\epsilon_N = \hbar v_F k \tag{3.95}$$

$$= \hbar v_F \frac{\sqrt{2}}{l_B} \sqrt{N} \, . \tag{3.96}$$

Again this is the same as the LL energies of Dirac fermions derived before in equation 3.65.

# 3.7 Determining Berry's phase from quantum oscillations

We have seen in section 3.6 that the Landau level energies of a system depend on Berry's phase. For example, we have seen that for $\gamma = 0$ there exists a zero energy LL (example: graphene). On the other hand, for $\gamma = 1/2$ either there is no LL at zero energy (example: normal 2DEG ) or the zero energy LL has an additional 2 fold degeneracy (example: bilayer graphene). Consequently, Berry's phase affects the density of states as a function of energy.

It is easy to see the difference between graphene and normal 2DEG. Graphene has a DOS peak at zero energy and 2DEG has a DOS dip at zero energy. Since, the conductance scales with DOS, graphene shows a conductance peak at zero energy (see Fig. 3.8b) and the normal 2DEG shows a conductance dip at zero energy (see Fig. 3.2c). As a function of the inverse magnetic field, the conductance oscillates which can be expressed as a cos function. We notice that the phase of the cosine function of these two systems is different – the phase is zero for graphene because it starts with a maximum and $\pi$ for normal 2DEG because it starts with a minimum. Since this phase is related to the Berry's phase, by analyzing the SdH oscillation one can determine the Berry's phase.

At a fixed density, the conductance oscillations (SdH) can be written as [32]

$$\Delta G_{\mathrm{xx}} = G \cos[2\pi(\frac{B_{\mathrm{F}}}{B} + \gamma)] \,, \tag{3.97}$$

where $G$ is the oscillation magnitude, $B_{\mathrm{F}} = \frac{nh}{ge}$ is the SdH oscillation frequency in $1/B$ parameter space and the phase shift $\gamma = \frac{1}{2} - \frac{\Phi_{\mathrm{B}}}{2\pi}$. Here, $n$ is the charge density, $g$ is the LL degeneracy which is 4 for graphene and 2 for 2DEG, $\Phi_{\mathrm{B}}$ is the Berry's phase.

At the minima of the cosine function, the argument is an odd multiple of $\pi$ i.e.

$$2\pi(\frac{B_F}{B_N} + \gamma) = (2N+1)\pi \tag{3.98}$$

$$\Rightarrow \frac{B_F}{B_N} - \frac{\Phi_B}{2\pi} = N \,, \tag{3.99}$$

where N is the LL index corresponding to $N^{th}$ the minima in longitudinal conductance. Equation 3.98 shows that by fitting a line to the LL index (N) vs. $\frac{1}{B_N}$ plot and examining the intercept at $\frac{1}{B} = 0$ one gets the Berry's phase in units of $2\pi$. The slope ($B_F$) of the line is related to the band density as $n = \frac{4e}{h} \times B_F$ and Fermi surface area as $S_F = \frac{2\pi e}{h} \times B_F$.

The LL index at a given $R_{xx}$ dip can be found out from the simultaneously measured Hall conductance. But, in the cases where Hall data is not available the assignment of the LL index can be done in the following way. Let's take the example of graphene to illustrate this, see Fig. 3.12b. We note that $\frac{B_F}{B}$ in equation 3.97 can be written as

$$\frac{B_F}{B} = \frac{nh}{4eB} \tag{3.100}$$

$$= \frac{\nu}{4} \tag{3.101}$$

$$= N + \frac{1}{2} \,. \tag{3.102}$$

Equation 3.100 shows that at the $N^{th}$ minima, the integer part of the $\frac{B_F}{B}$ should be the correct LL index. As an example, we note that the $1/B$ frequency of the SdH oscillation in Fig. 3.12b is 10 T. So, at $B^{-1}$=0.05 $T^{-1}$, the value of $\frac{B_F}{B}$ is $\frac{1}{2}$ whose integer part is 0. So, the correct the LL index corresponding to the minima at $B^{-1}$=0.05 $T^{-1}$ is $N = 0$.

Fig. 3.12a and Fig. 3.12b show the schematic of two SdH oscillations and their corresponding LL index plots for Berry's phase zero and $\pi$ respectively. As explained the ($N$ axis) intercept of the LL index plot gives Berry's phase in units of $2\pi$.

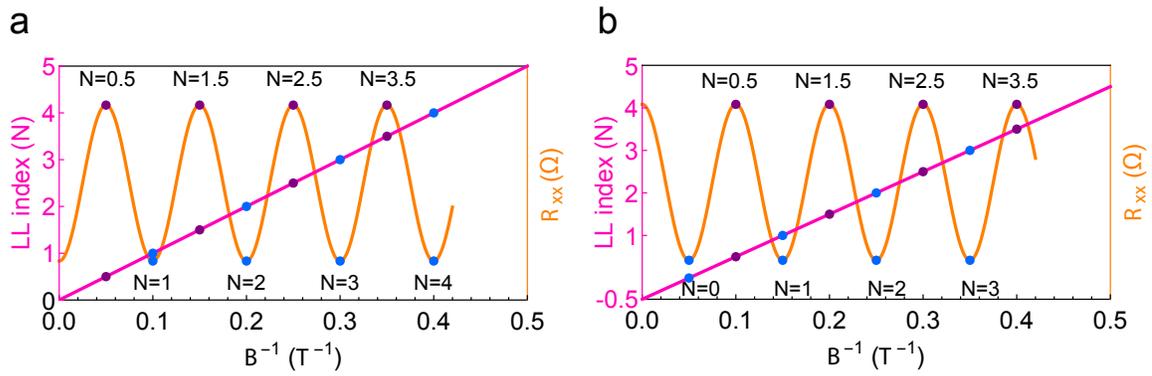

Figure 3.12: **Determining Berry's phase from quantum oscillations.** (a) Schematic of SdH oscillation and the LL index plot for Berry's phase zero. (b) Schematic of SdH oscillation and the LL index plot for Berry's phase $\pi$.

# Device fabrication and characterization

We study the transport properties of different systems in nanometer and micrometer length scale. This requires us to make electrical contacts in this length scale. In this chapter, we will describe how we thin down bulk materials to monolayer limit, stack different materials at our will and using the standard semiconductor fabrication method deposit contacts on those materials.

## 4.1 Exfoliation and finding thin layers of graphite

Graphite is a layered material in which different atomic layers are held together by Van der Waals force. Individual layers of graphite can be easily cleaved since there is no chemical bond in the out of the plane direction and Van der Waals force is relatively weak. The individual layers of graphite were isolated first by Novoselov, K. S. *et al.* [1]. This is remarkable because it was believed that due to thermal fluctuation no long-range order in 1D and 2D is possible hence an atomically thin material would be unstable [33].

Following graphene, a wide variety of other layered materials have been discovered [34, 35, 36]. Hexagonal boron nitride (hBN) – a layered insulator is another example. The process of extracting thin layers from a layered bulk crystal is known as exfoliation. There are a host of exfoliation techniques developed over the last decade [37]. The most common method is a (Scotch) tape-based mechanical exfoliation technique in which we put a bulk crystal on an adhesive tape and while peeling off the tape it can extract thin layers of that material. After peeling off the tape we put it on a substrate (typically 300 nm Silicon di-oxide coated on doped silicon) and gently rub the top surface of the tape with the blunt end of a tweezer (or round end of a dropper) for a few minutes. Keeping the substrate at an elevated temperature ($\sim 100$°C) at this stage increases the yield of graphene transfer to the substrate, but it is not a mandatory step. After the substrate cools down to the room temperature we gently remove the tape from the substrate. The rubbing and heating actions transfer a few thin layers of the material on the substrate. For even larger yield, we (Oxygen) plasma clean the substrate just before the exfoliation. The same exfoliation technique can be used to exfoliate other materials like hBN.

After the exfoliation, we look for thin flakes of graphite under an optical microscope. Multiple reflections and their interference of the light within the different layers (graphene, $SiO_2$ and Si) enhances the contrast of graphene high enough that an atomically thin material is visible with just a simple optical microscope. The reflected light intensity changes as a function of the graphene thickness. In thin-film limit, the reflected light intensity is proportional to the number of layers which can be calibrated and later used to find out the number of layers in a thin graphite sample. Atomic force microscopy (AFM) can also be used to find the thickness of multilayer graphene, but often in ambient due to different adsorbates on the surface, the thickness may not be accurate in the atomic scale.

## 4.2   Raman spectroscopy

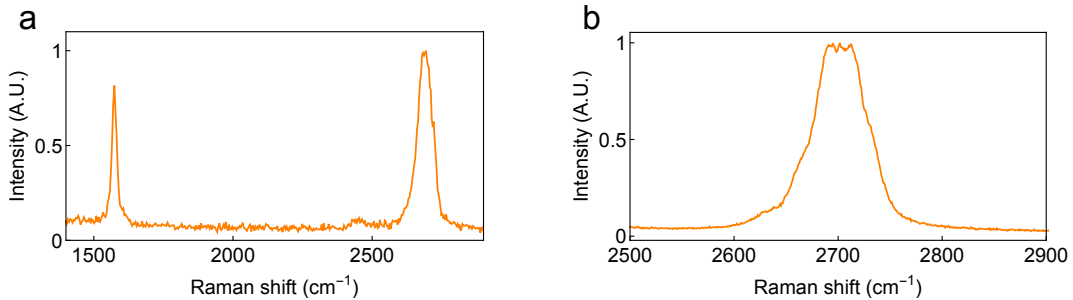

Figure 4.1: **Raman spectra of ABA-trilayer graphene.** (a) G peak and 2D peak – showing the ratio of the 2D peak intensity over G peak intensity is $\sim 1.1$ in trilayer graphene. The spectrum was recorded with a grating of 600 grooves/mm. Excitation wavelength is 532 nm for all the measurements. (b) High resolution spectrum of 2D peak. The spectrum was recorded with a grating of 1800 grooves/mm.

Raman spectroscopy is an important part of our device fabrication. For example,

it can be used to identify the number of layers in a few-layer graphene sample [38]. Graphene has two characteristic Raman peak: G peak ($\sim$1580 cm$^{-1}$) and 2D peak ($\sim$2690 cm$^{-1}$). The line shape (or width) and the position of the 2D peak are good metrics to characterize the stacking and the number of layers in thin graphite films. The ratio of the 2D peak intensity to G peak intensity ($I_{2D}/I_G$) is also an approximate measure of the thickness. For monolayer graphene, the 2D peak has a single well-defined peak and $I_{2D}/I_G$ $\sim$4. As the number of layers increases the 2D peak develops substructure and the width of the 2D peak increases as well. In bilayer and trilayer graphene $I_{2D}/I_G$ is $\sim$2 and $\sim$1.1 respectively. $I_{2D}/I_G$ $\sim$0.3 for bulk highly oriented pyrolytic graphite (HOPG). Fig. 4.1 shows the Raman spectra of an ABA-trilayer graphene sample.

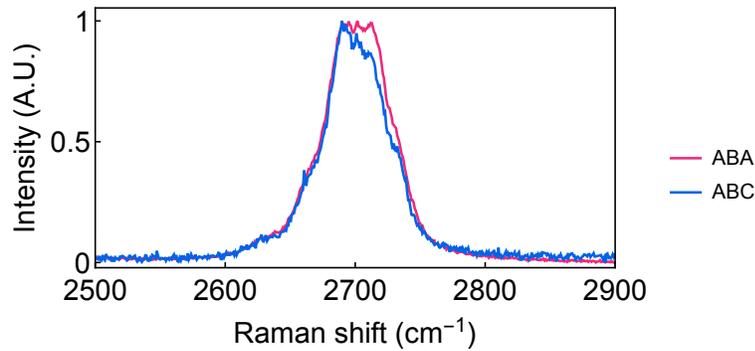

Figure 4.2: **Comparison between Raman spectra of ABA and ABC-trilayer graphene.** It shows that the 2D peak line shape of the ABA-trilayer graphene is symmetric with respect to the peak center. On the other hand, the 2D peak line shape of the ABC-trilayer graphene is asymmetric with respect to its peak center.

Although the difference is subtle, the Raman spectra can be used to differentiate between ABA and ABC-trilayer graphene [39]. Fig. 4.2 shows that the line shape of the 2D peak for ABA and ABC-trilayer graphene is different. The line shape is symmetric in ABA-trilayer graphene and asymmetric in ABC-trilayer graphene. This is why the FWHM of the 2D peak of the ABC-trilayer graphene is a little smaller than that of ABA-trilayer graphene. A more sophisticated and reliable way to distinguish between ABA and ABC stacking is the infrared microscopy [40] which was not used here because of the unavailability of the measurement setup. We cannot be completely eliminate the possibility of ABA and ABC mix stacking regions as we do not have a local probe. However, transport properties between all the electrical leads show distinct LL crossing signature of the ABA stacking.

Moreover, the Raman spectrum is also useful to spatially map the graphene in a hBN-

graphene-hBN stack when the graphene is not visible through the hBN in an optical microscope. This is useful for the subsequent steps of making electrical contacts to graphene. Since the characteristic Raman peak of graphene (G peak $\sim$1580 cm$^{-1}$ and 2D peak $\sim$2690 cm$^{-1}$) and hBN (1366 cm$^{-1}$) are well separated, they can be used to image the individual layers. Fig. 4.3 shows that using correct filters on the Raman spectra we can image both the graphene and hBN layers.

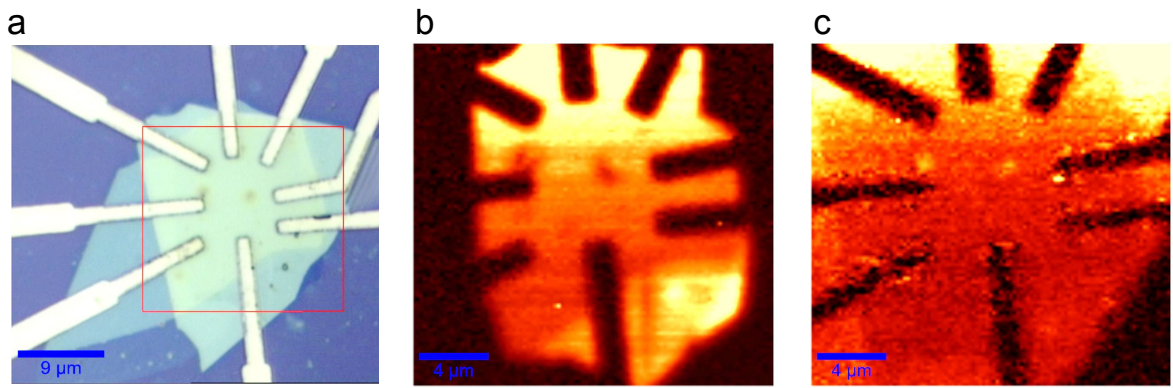

Figure 4.3: **Using Raman mapping to spatially map graphene.** (a) Optical image of a hBN sandwiched trilayer graphene device. Since the hBN are slightly thick $\sim$30 nm the graphene is not visible in the optical image. The red rectangle shows the area for which the Raman mapping is shown in the other panels. (b) Spatial mapping of the characteristic Raman peaks of graphene can be used to identify the graphene. Here we have plotted the summed intensity (area under the intensity curve) of the 2D peak (2690 cm$^{-1}$) between 2600 cm$^{-1}$ to 2800 cm$^{-1}$. (c) Similarly, the characteristic Raman peaks of hBN can be used to identify hBN. It shows the plotted summed intensity of the hBN peak (1366 cm$^{-1}$) between 1335 cm$^{-1}$ to 1392 cm$^{-1}$.

## 4.3 Device fabrication

Graphene and few-layer graphene are atomically thick materials with extremely high intrinsic mobility. However, due to the very high surface area, the electrons are exposed to the environment. The exposure to various chemicals during device fabrication leads to unintentional spatially non-uniform doping which can ruin the sample quality. To maintain the intrinsic quality of graphene high we use the following two different fabrication methods.

### 4.3.1 Fabrication of edge contact devices

This is the most frequent method of fabrication which leads to a one-dimensional electrical contact to a two dimensional material [41]. First the graphene is encapsulated within two hexagonal boron nitride (hBN) flakes [2, 42]. A dry transfer method is used to pick up all the successive layers. The pickup stamp consists of a piece of polydimethylsiloxane (PDMS) with a polypropylene carbonate (PPC) polymer layer on top. Graphene and hBN flakes are exfoliated on a silicon (Si) substrate. We first pick up a top hBN layer, followed by graphene and a bottom hBN layer. We deposit the whole stack on a fresh Si chip by raising the temperature to 70°C. After encapsulating the graphene we do electron beam lithography to define the electrodes on the resist. Then we plasma etch (Ar:O$_2$=1:1) the top hBN along with the graphene and evaporate metal (Cr (3 nm)/Pd (15 nm)/Au (30 nm)) to contact the graphene. A lift-off is done to remove the metal from all places on the chip other than the defined electrodes by lithography.

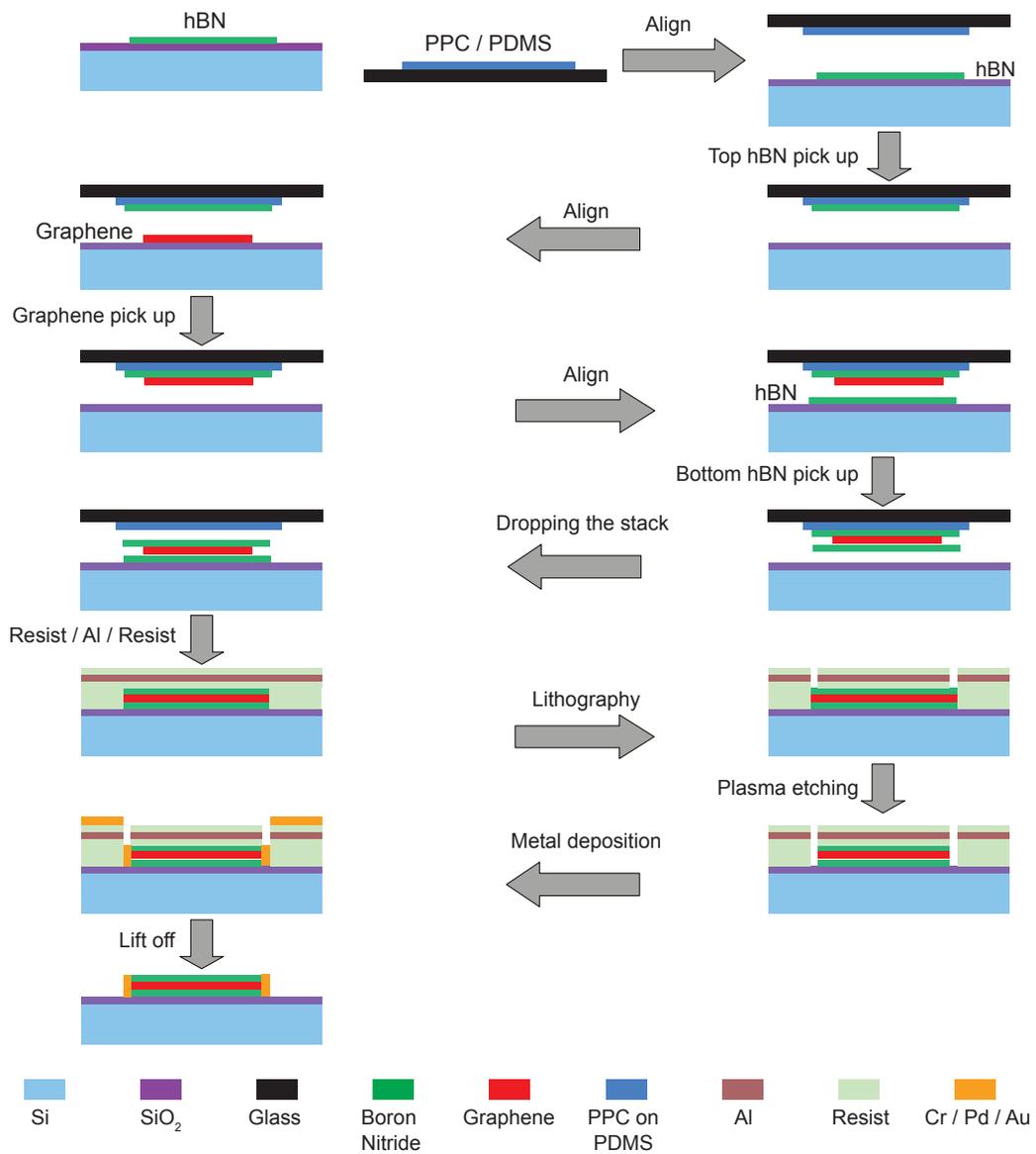

Figure 4.4: **Schematic of fabrication steps to make 1D contact on a hBN-graphene-hBN stack made by PPC method.**

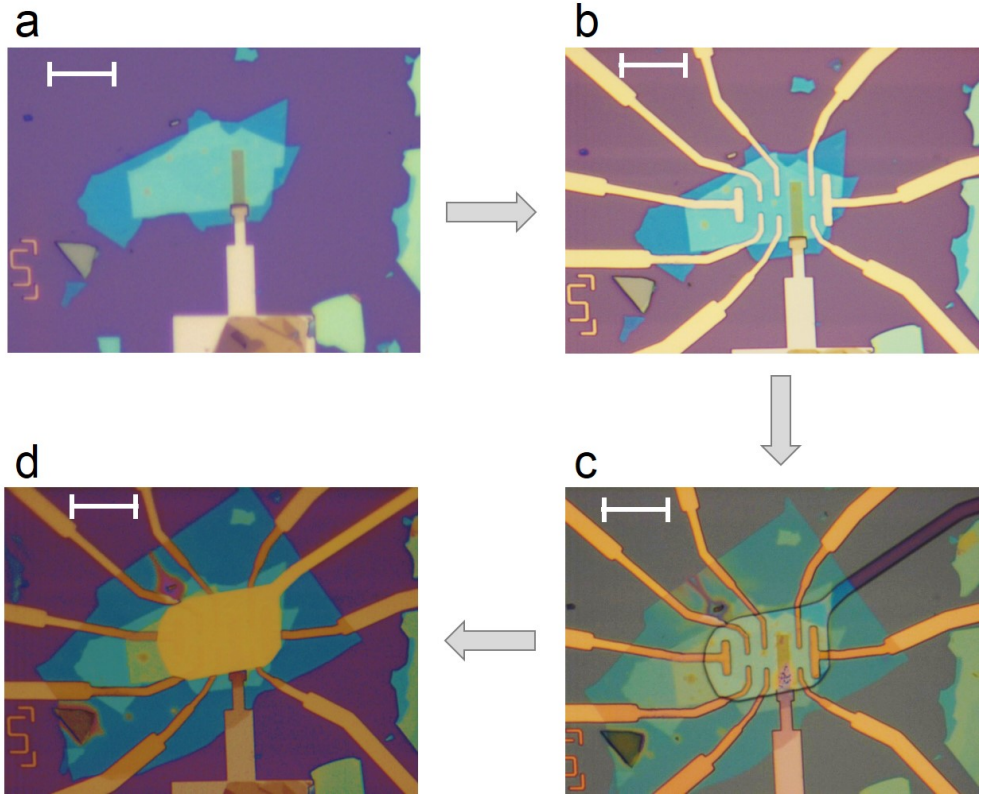

Figure 4.5: **Optical images of different steps of making 1D electrical contact.** (a) A hBN-graphene-hBN stack transferred on a SiO$_2$/Si$^{++}$ substrate with a pre-patterned local gate. (b) After etching the top hBN and deposition of metal contacts. (c) After making the electrical contacts a layer of hBN was transferred on top of the device which acts as an insulator between the contacts and gate electrode. E-beam lithography was done and the resist was developed to define the top gate electrode. (d) After depositing the metal to make the top gate. Scale bar: 10 µm.

Since the graphene is also etched during the top hBN etching, the resultant electrodes contact the graphene only along the edges, hence it is called a 1D contact or edge contact [41]. The flow chart of the whole fabrication process is shown in Fig. 4.4. Fig. 4.5 shows optical images of different steps of making electrical contact and local top gate in a real device.

### 4.3.2 Fabrication of via contact devices

We follow via contact [43] technique in cases where etching is detrimental to the channel or edge contact does not make good electrical connection to the channel. In this process, we pre-pattern the top hBN with an array of holes (of size few µm$^2$) which later serve as the contact regions. We use this pre-patterned top hBN to make the full hBN-graphene-hBN stack. No etching of the top hBN is required after making the stack since the top hBN has already holes on it. E-beam lithography is done to define electrodes which preciously merge to the holes on the top hBN. Since the graphene beneath the holes is already exposed we can directly metalize the graphene. We note that the metal electrodes overlap with the channel through the holes which have few μm² area, making this a 2D contact in nature, unlike the edge contact. This process of making via contact is useful for the materials such as TMDC for which edge contact shows poor performance. Fig. 4.6 shows optical images of different steps of making electrical contact by via method.

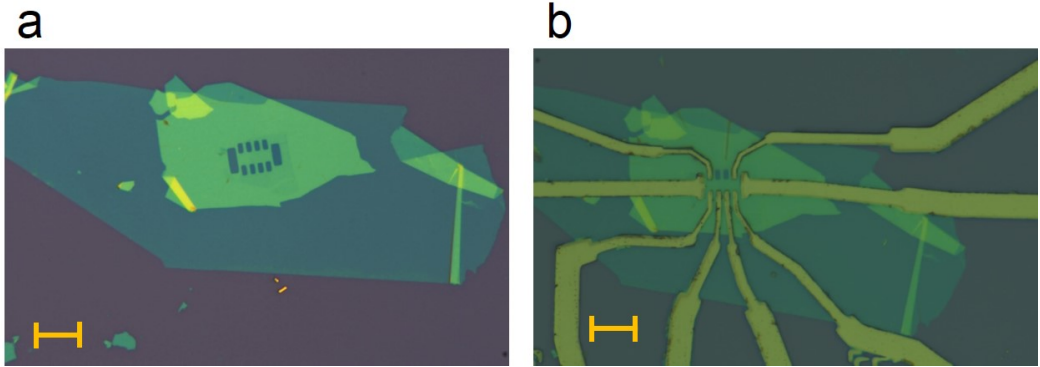

Figure 4.6: **Fabrication steps of making contacts by via method.** (a) Shows a complete stack for which top hBN was pre-patterned with rectangular etched holes. (b) After depositing the metal to the holes on the top hBN. No additional etching of the top hBN after making the stack is required as the graphene beneath the holes is already exposed and is ready to be metalized. Scale bar: 10 μm.

## 4.4 Measurement scheme

All the low-temperature measurements presented in this thesis are done in a liquid He flow cryostat at base temperature T=1.5 K. Some of the measurements are done at 300 mK using a He-3 insert. Standard low-frequency lock-in technique was used to do all current biased four-probe resistance measurements. Excitation current was 100 nA for most of the measurements but sometimes increased to a higher value to measure low resistances accurately at low magnetic fields. It is important to make sure that the Joule heating due to the dissipation in the device is not sufficient to raise the temperature beyond the base temperature of the cryostat. For example, if we flow 400 nA current through a device of resistance 1 kΩ, the power dissipation is 1.6 pW. The cooling power of a typical cryogen free dilution refrigerator at the base temperature (10-100 mK) is around 10-500 µW. In this case, the cooling power of the cryostat is much higher than the power dissipated in the device. However, the power dissipation can be significant if the device resistance is very high. Helium flow cryostats typically have the cooling power several order of magnitude higher (it can be close to a Watt at ∼2 K) than the cryogen free dilution refrigerators. So, cooling power is often not a limiting factor in Helium flow cryostats.

It is also important to keep the electron temperature lower than the temperature of measurement. For example, if we apply a 10 µV bias across a device then the effective electron temperature in the device becomes 10 µV/$k_B$=0.1 K where $k_B$ is the Boltzmann constant.

Reducing noise is very important to resolve small features in the data. For most of our metallic low resistance (up to ∼ 100 kΩ) samples we do AC measurements using the low frequency lock-in technique. Higher values of resistances are usually measured with DC current to eliminate capacitive contribution in the measurement. However, since in both AC and DC measurements we typically have several instruments such as voltage source, current source, voltmeter, current-meter, filters connected it is possible to generate ground loops which give noise in the measurement. We take care to eliminate such ground loops.

Most of our measurements are automated where we vary a few parameters (let's say gate voltage and magnetic field in a quantum Hall measurement) and record the resistance. In the end to visualize we colour plot the resistance as a function of these parameters. So, it is important to record the resistance of the device in the same sweep direction of a parameter (let's say gate voltage). Otherwise, the hysteresis between the up and down gate sweeps may lead to artefacts in the data which will complicate the analysis.

# Berry's phase in ABA-trilayer graphene

Quantum oscillations provide a striking visualization of the Fermi surface of metals including associated geometrical phases, such as Berry's phase, that play a central role in topological quantum materials. In this chapter we discuss the existence of a new quantum oscillation phase shift in a multiband system. In particular, we study ABA-trilayer graphene, the band structure of which is comprised of a weakly gapped linear Dirac band, nested within a quadratic band. We observe that Shubnikov-de Haas (SdH) oscillations of the quadratic band are shifted by a phase that sharply departs from the expected $2\pi$ Berry's phase and is inherited from the non-trivial Berry's phase of the linear band. Our analysis reveals that this happens due to the simultaneous filling of both the bands – required by the uniform Fermi energy. Moreover, we measure a continuous gate tuning of the extracted phase from $\pi$ to $-\pi$ across the weakly gapped linear Dirac band. Given that many topological materials contain multiple bands, our work indicates how additional bands, which are thought to obscure the analysis, can actually be exploited to tease out the effect of often subtle quantum mechanical geometric phases.

## 5.1 Introduction

The accumulation of a non-trivial geometric phase in quantum oscillations of a band is often a tell-tale sign of a rich underlying internal structure [23, 44, 45, 46]. These can arise from diverse settings including in strong spin-orbit coupled systems that possess real-space [47, 48] or momentum-space spin-texture [49], periodic driving by strong electromagnetic fields [50], and multi-orbital/site structure within a unit cell [51]. Even though such phases are often encoded in the subtle twisting of electronic wavefunctions, their impact on material response can be profound, being responsible for a wealth of unusual quantum behaviors that include unconventional magneto-electric coupling [52], an emergent electro-magnetic field for electrons [48], and protected edge modes [53] amongst others.

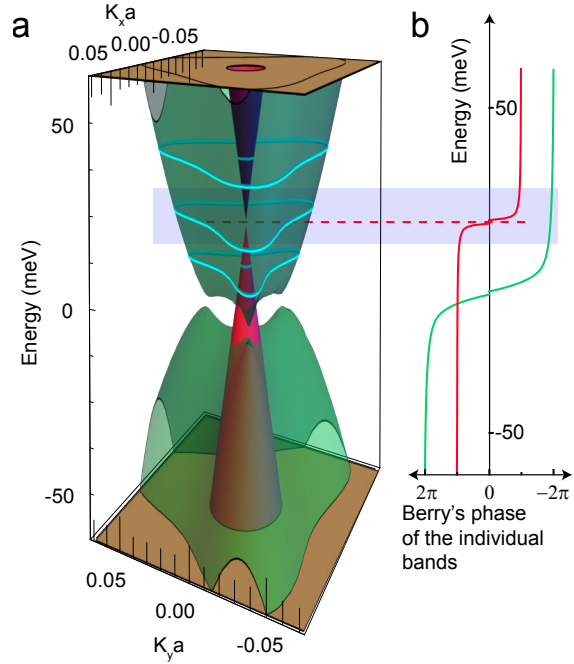

Figure 5.1: **Band diagram and Berry's phase of ABA-trilayer graphene.** (a) Band diagram of ABA-stacked trilayer graphene showing one pair of the conical band (colored red) and another pair of the quadratic band (colored green). Fermi surface at three different energies is overlaid for which Fermi energy lies in the valence band, band gap and in the conduction band of the MLG-like band. There is no contour from the MLG-like band when Fermi energy is in the band gap. (b) Calculated Berry's phase plot with same color codes for both the bands. Since the bands are gapped, Berry's phase of the individual bands goes to zero at the respective band edges. The shaded blue rectangle shows the range of Fermi energies of our interest around the Dirac band's gap.

A prominent example is the Berry's phase [21, 22, 54]. In anomalous Hall metals, the Berry's phase on the Fermi surface determines the (un-quantized part of the) anomalous Hall conductivity [55, 56]; non-trivial $\pi$ Berry's phase enforces the absence of back-scattering in topological materials [57]. Indeed, the value of the Berry's phase of electrons as they encircle a single, closed Fermi surface can be used as a litmus-test for topological bands — $\pi$ indicates a non-trivial band [58, 59, 60, 61, 62], whereas $2\pi$ indicates a massive quadratic band [20, 63, 64]. In the presence of a magnetic field ($B$), the (quantized) size of closed cyclotron orbits depends on both the magnetic flux threading the orbits as well as the Berry's phase of electrons. As a result, quantum oscillations of a closed Fermi surface can acquire phase shifts – a direct result of the Berry's phase of electrons [45]. This is visible in oscillations of both resistance and thermodynamic quantities like magnetization. Tracking such quantum oscillations phase shifts have emerged as a powerful probe for topological materials [65, 66, 66, 67, 68, 69].

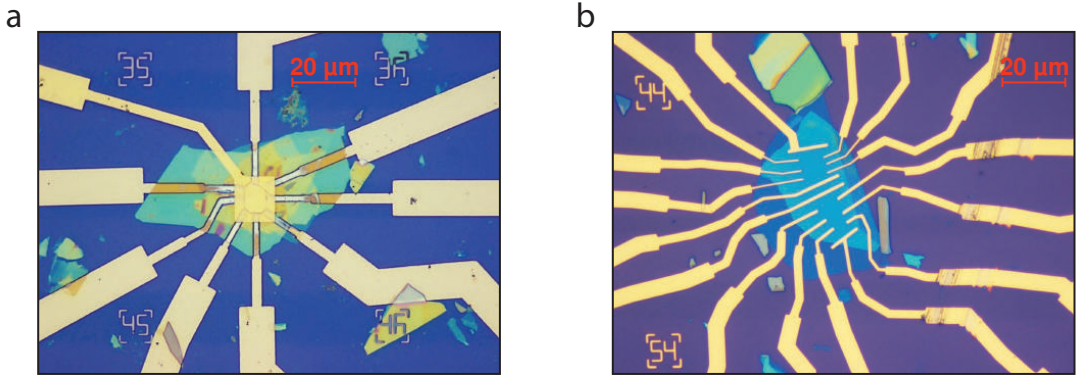

Figure 5.2: (a) Optical micrograph of an ABA-trilayer graphene device (device - A). The graphene is encapsulated by two layers of hBN. An additional hBN (for top gate insulator) was transferred on the completed device to make a uniform top gate. (b) Optical micrograph of another ABA-trilayer graphene device (device - B).

ABA-trilayer graphene is a very interesting system because it is the simplest system supporting the simultaneous existence of monolayer graphene (MLG)-like linear and bilayer graphene (BLG)-like quadratic band in experimentally accessible Fermi energy (see Fig. 5.1a) [14, 16, 70, 71, 72, 73]. Structure of the ABA-trilayer graphene lattice breaks the inversion symmetry intrinsically even at the zero electric field; this generates small mass terms in the Hamiltonian [14, 73]. As a result, both the pairs of bands are individually gapped as seen in Fig. 5.1a. Fig. 5.1a shows that when both these bands are filled, the Fermi surface of the ABA-trilayer graphene consists of two Fermi contours– the inner contour comes from the MLG-like band and the outer contour comes from the BLG-like band. Fig. 5.1b shows that the MLG-like Dirac cone has a robust $\pi$ Berry's phase which only reduces to zero in the vicinity of the MLG-like band edge. However, since the Dirac band gap is very small ∼1 meV [17, 74], it was not possible to resolve the Dirac band gap and tune the Fermi level through the gap controllably in most previous studies [70, 74]. Since the LL broadening in our device is small we can resolve the Dirac band gap and study the phase of the BLG-like SdH oscillations as the Fermi level is tuned through the MLG-like band gap. We note that the BLG-like conduction band has more-or-less a constant trivial Berry's phase $2\pi$ in the region of interest (around the Dirac cone gap). In our experiment, we probe a narrow energy window near the MLG-like band gap. In the following, we use "band gap" to refer to the MLG-like Dirac band's gap.

We study high mobility hexagonal boron nitride (hBN) encapsulated ABA-stacked trilayer graphene devices. Fig. 5.2 shows the optical micrograph of two devices. Dimension of device - A is 9.5 µm × 7.5 µm and the mobility near the density of our interest is ∼500,000 cm$^2$V$^{-1}$s$^{-1}$, see Fig. 5.3. We note that in presence of magnetic field quantum mobility is more relevant than the transport mobility. The Landau level broadening depends on quantum mobility which is significantly lower than the transport mobility in our device – discussed in section 6.5 in detail. All the analysis are done using the data from device - A because of higher quality . Data from device - B is shown in section 5.8. A metal top gate and a highly doped silicon back gate ensure independent tunability of charge carrier density and electric field.

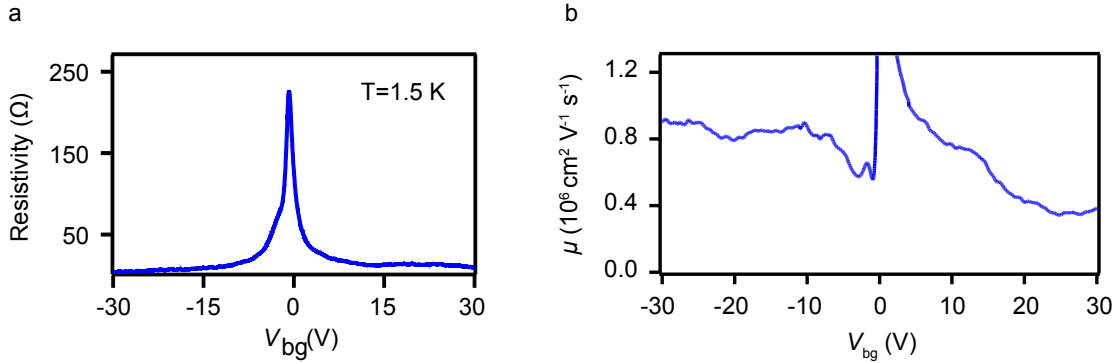

Figure 5.3: Drude mobility. a) Field effect gating of device - A at 1.5 K. b) Drude mobility calculated from four-terminal resistance of the device shown in panel (a).

## 5.2 Landau level, SdH oscillations and Berry's phase

In the presence of a magnetic field, the continuous band structure shown in Fig. 5.1a splits into Landau levels (LLs). The closed orbits **k** space area takes on quantized values that depend on Berry's phase (and magnetic flux). As the magnetic field is swept and the charge density is varied independently, LLs cross the Fermi surface giving rise to the density of states oscillations that result in longitudinal conductance ($G_{xx}$) oscillations [75]. At a fixed density, the conductance oscillations (SdH) can be written as $\Delta G_{xx} = G\cos[2\pi(\frac{B_F}{B}+\gamma)]$ where G is the oscillation magnitude, $B_F = \frac{n_s h}{ge}$ is the SdH oscillation frequency in $1/B$ parameter space and the phase shift $\gamma = \frac{\Phi_B}{2\pi} - \frac{1}{2}$. Here, $n_S$ is the density in S sub-band for a multiband system, $g$ is the LL degeneracy which is 4 for graphene, and $\Phi_B$ is the Berry's phase. Fig. 5.4a shows our measured SdH oscillation in $G_{xx}$ as a function of $B$ and $V_{BG}$. The corresponding band structure at zero magnetic field is shown in Fig. 5.4b. Theoretically calculated LL diagram (Fig. 5.4c) shows that the MLG-like and the BLG-like LLs disperse as $\sim \sqrt{B}$ and $\sim B$ respectively [14, 16, 73]. The distinct dispersion of the LLs along with the corresponding Hall conductance enables easy identification of the MLG-like and the BLG-like LLs [16, 17, 74, 76].

The central result of our study – that of an unusual phase shift in the quadratic BLG-like band – is vividly illustrated in Fig. 5.4d. It shows three slices of BLG-like SdH oscillations at different densities away from the crossing points which correspond to Fermi levels in the valence band, in the gap, and in the conduction band of the MLG-like Dirac cone respectively. We emphasize that for all these three densities, the Fermi level lies in the conduction band of the BLG-like band. The SdH oscillations above and below the gap clearly show a $\pi$ phase shift from the SdH oscillation at the gap. This is intriguing since the BLG-like band in this energy range has a constant trivial Berry's phase (see Fig. 5.1b). It is clear from the experiment that the Fermi level position in the Dirac band has a bearing on the phase of the BLG-like band. The experimental ability to "tune out" the role of the Dirac band using Fermi energy is crucial to the analysis. It serves as an inbuilt control in our experiment.

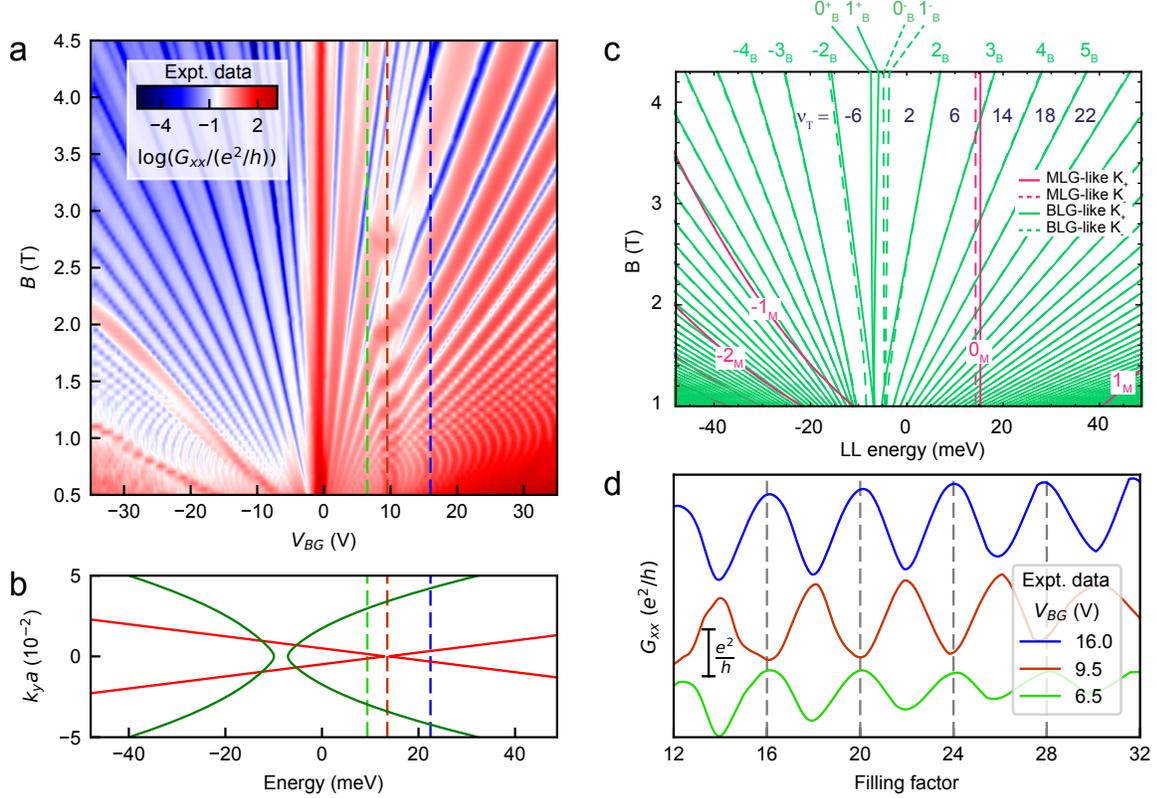

Figure 5.4: **Quantum oscillation of ABA-trilayer graphene.** (a) Colour scale plot of experimentally measured $G_{xx}$ as a function of back gate voltage and magnetic field. The vertical feature parallel to the magnetic field axis at $V_{BG} \sim 10\ V$ corresponds to the LL crossings of $N_M=0$ LL with other BLG-like LLs. This $V_{BG}$ also corresponds to the band gap of the MLG-like bands. (b) Calculated energy band diagram shown in the same energy range as the experimental fan diagram shown in panel (a). (c) Theoretically calculated LL energies of the spin degenerate Landau levels as a function of magnetic field. Red and green lines denote LLs originating from the Dirac and the quadratic bands respectively. Solid and dashed lines denote LLs from $K_+$ and $K_-$ valleys respectively. (d) Experimentally measured SdH oscillations ($G_{xx}$) as a function of filling factor below the band gap (green), in the band gap (red) and above the band gap (blue) which show that the phase of the SdH oscillation in the band gap is $\pi$ shifted compared to the other two. The curves are shifted in the vertical direction for clarity. Gate voltage and approximate energy locations of the three SdH oscillation slices are marked with dashed lines of the corresponding color in the fan diagram (a) and in the bandstructure (b) respectively.

## 5.3 Determination of the BLG-like LL index from the experimental Hall conductance

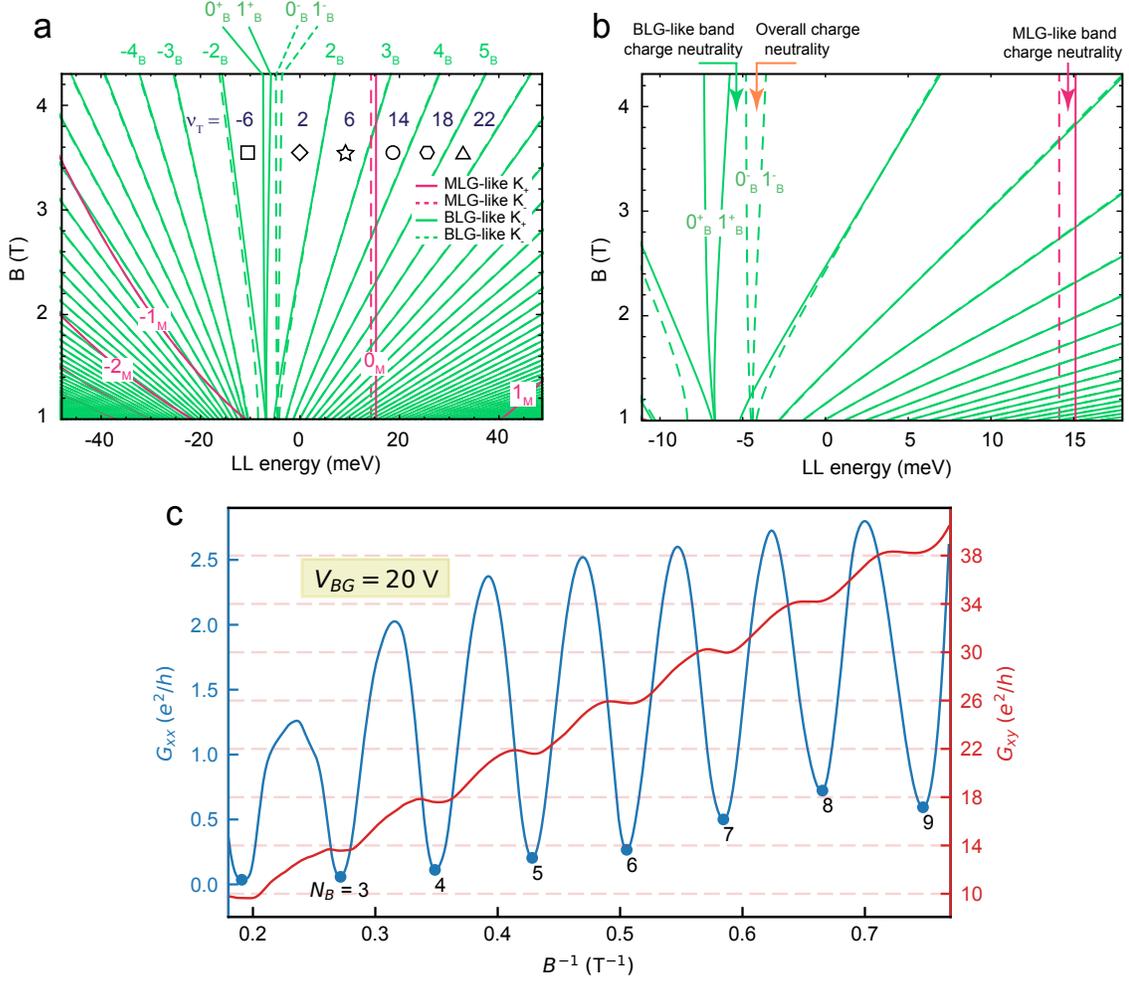

Figure 5.5: **Calculation of the BLG-like LL index from the total filling factor.** (a) Different regions in the LL diagram are marked for which we show the calculation of the LL index. (b) Zoomed-in LL diagram showing the charge neutrality points of the MLG-like bands, BLG-like bands and the overall charge neutrality point of the system. Band specific filling factors are counted from their band specific charge neutrality points whereas the total filling factor is counted from the overall charge neutrality point. (c) Experimentally measured $G_{xx}$ and $G_{xy}$ as a function of the magnetic field on the electron side. LL indices at all the minima are marked and are given by $N_B = \frac{1}{4}(\nu_T - 2)$.

Following the previous theoretical study [14], we numerically calculate the LL energy diagram shown in Fig. 5.5a. We consider the full tight binding Hamiltonian of ABA-trilayer graphene [14] with all the hopping parameters. We have used the following band parameters for all calculations: $\gamma_0$=3.1 eV, $\gamma_1$=390 meV, $\gamma_2$=-20 meV, $\gamma_3$=315 meV, $\gamma_4$=120 meV, $\gamma_5$=18 meV, $\delta$=20 meV, $\Delta_2$=4.3 meV which were calculated by matching the experimental LL crossing points with theory [17].

Fig. 5.5a shows that the MLG-like and BLG-like LL origins are shifted. We note that overall charge neutrality of the system is located in between $0_B$ and $1_B$ electron-like LLs (Fig. 5.5b) where Hall conductance goes to zero. Total filling factor (counted from the overall charge neutrality point) can be written as the sum of MLG-like filling factor ($\nu_M$) counted from the MLG-like band origin and BLG-like filling factor ($\nu_B$) counted from the BLG-like band origin: $\nu_T = \nu_M + \nu_B$. If $N_M$ is the LL index of the MLG-like LLs then filling factor of the MLG-like band above and below the band gap is given by $\nu_M = 4(N_M \pm 0.5)$. Similarly, if $N_B$ is the LL index of the BLG-like LLs then filling factor of the BLG-like band (for $N_B > 0$) is given by $\nu_B = 4N_B$. We find the total filling factor ($\nu_T$) from the experimentally measured quantized Hall conductance ($G_{xy}$) data. Fig. 5.5c shows a line slice of the $G_{xx}$ and $G_{xy}$ as a function of the magnetic field at $V_{BG}$=20 V on the electron side. Total filling factor is given by the integers where the quantum Hall $G_{xy}$ plateaus occur. Filling factor of the MLG-like band ($\nu_M$) can also be easily counted from the experimental fan diagram since the MLG-like LLs are very sparse and have a distinct parabolic dispersion. This allows us to calculate $N_B = \frac{1}{4}(\nu_T - \nu_M)$. Table 5.1 shows the calculated BLG-like LL indices at different filling factors marked in Fig. 5.5a.

Table 5.1: Extracted LL index at 4 T for different filling factors

| Below the MLG-like band gap | | | | | |
|---|---|---|---|---|---|
| Symbol | $\nu_T$ | $\nu_M = 4(N_M - 0.5)$ | $\nu_B = 4N_B$ | $N_M$ | $N_B$ |
| □ | -6 | -2 | -4 | 0 | -1 |
| ◇ | 2 | -2 | 4 | 0 | 1 |
| ☆ | 6 | -2 | 8 | 0 | 2 |
| Above the MLG-like band gap | | | | | |
| Symbol | $\nu_T$ | $\nu_M = 4(N_M + 0.5)$ | $\nu_B = 4N_B$ | $N_M$ | $N_B$ |
| ◯ | 14 | 2 | 12 | 0 | 3 |
| ◯ | 18 | 2 | 16 | 0 | 4 |
| △ | 22 | 2 | 20 | 0 | 5 |

## 5.4 Quantitative determination of SdH oscillations phase shift

We quantify the unusual phase shift via a detailed analysis of the SdH oscillations using the LL index plot [59]. Briefly, this involves fitting a line to the LL index (N) corresponding to a minimum in the $G_{xx}$ vs. the corresponding inverse magnetic field ($\frac{1}{B_N}$) plot and examining the intercept at $\frac{1}{B} = 0$. From the intercept in the LL index axis (Fig.5.6a), we see that the intercept is 0.5 (-0.5) in the valence (conduction) band and is zero in the middle of the band gap. The 0.5 (-0.5) value of the intercept corresponds to a $\pi$ (-$\pi$) phase shift of the SdH oscillations when the Fermi level lies away from the band gap even though the phase is extracted *only* from the BLG-like SdH oscillations. Fig.5.6b shows fits at several densities away from the gap (firmly in either conduction or valence band). While possessing different slopes, their intercepts assume only two quantized values: 0.5 or -0.5 depending on the Fermi energy inside the valence or conduction MLG-like Dirac cone. This reinforces the robustness of the unusual phase shift.

Strikingly, it is only when the Fermi energy is tuned through the MLG-like band's gap, that the intercept varies continuously from 0.5 to -0.5, see Fig.5.6c. We note the smooth gate-tuning through the bandgap is possible due to the gapless nature of the BLG-like conduction bands throughout the region of interest. Both the non-trivial values and the tunable nature of the unusual phase shift sharply departs from the traditional understanding of quantum oscillation being purely sensitive to the specific part of the Fermi surface it is sampling – BLG-like Fermi surface in the present case.

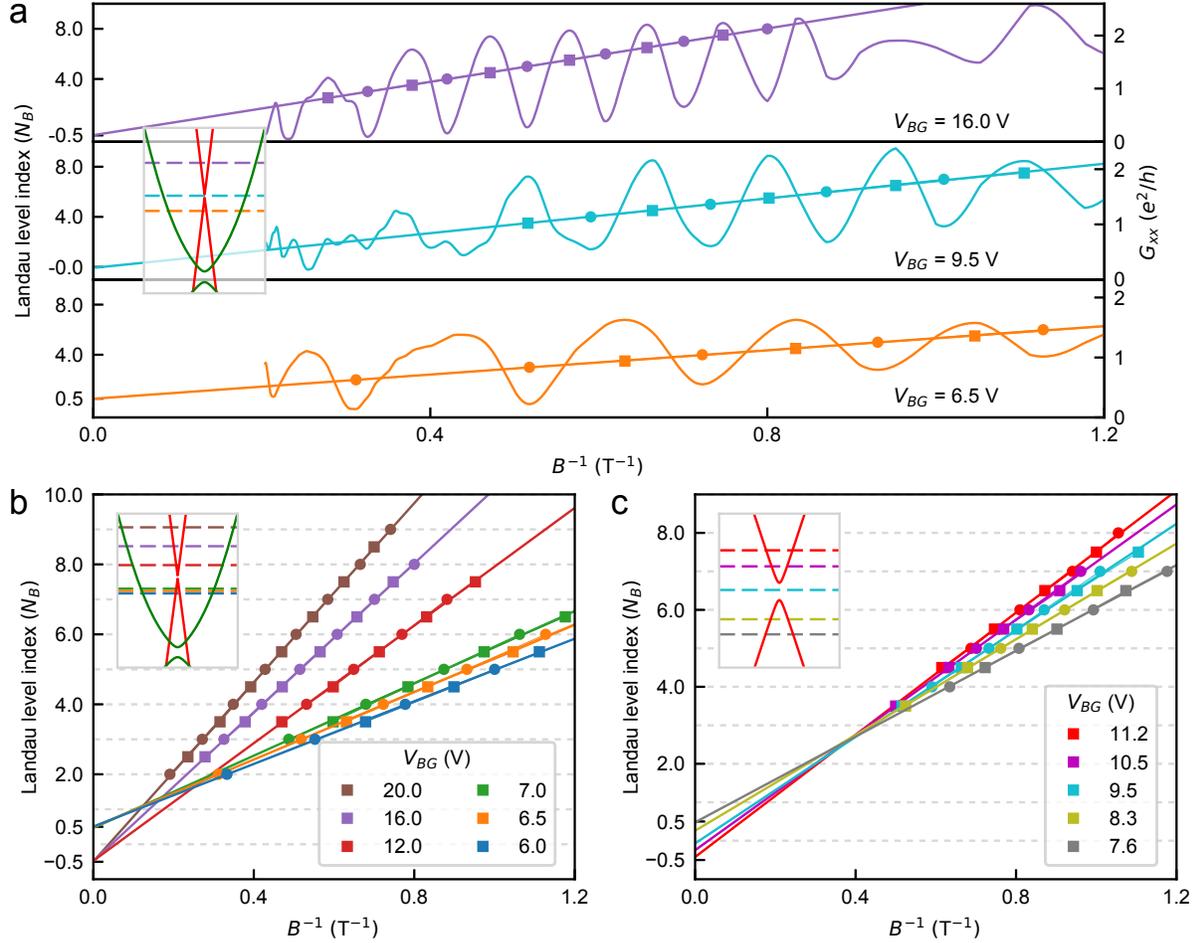

Figure 5.6: **Unusual SdH phase shift.** (a) SdH oscillations ($G_{xx}$) and the LL index vs inverse magnetic field fit below the band gap (orange), in the bandgap (cyan), and above the bandgap (purple). Circles and squares denote the SdH minima and maxima respectively. Inset of all the panels shows the band diagram and the Fermi energy locations for which the SdH fits are shown. (b) LL index vs inverse magnetic field fits at different densities away from the band gap. The linear fit produces $\pm\frac{1}{2}$ intercept when Fermi level lies in the MLG-like valence band and MLG-like conduction band respectively. (c) LL index vs inverse magnetic field fits at different densities close to the band gap. This shows that the intercept varies continuously from $1/2$ to $-1/2$ when the Fermi level goes from the valence to the conduction MLG-like band by tuning the density. The inset shows the zoomed-in band diagram very close to the band gap.

## 5.5 Origin of the unusual phase shift

We now focus on the origin of the unusual phase shift. In general, SdH oscillations depend on contributions from the Fermi surfaces of both the bands:

$$\Delta G_{xx} = G_M \cos\left[2\pi\left(\frac{B_{FM}}{B} + \gamma_M\right)\right] + G_B \cos\left[2\pi\left(\frac{B_{FB}}{B} + \gamma_B\right)\right], \qquad (5.1)$$

where $M$ and $B$ subscripts denote MLG-like and BLG-like bands respectively. As we explain below, the complex pattern of band fillings across multiple bands of distinct type (encoded in $(B_{FB}, B_{FM})$) control the SdH oscillations.

To unravel the pattern in ABA-trilayer graphene, there are two key effects to understand. First, MLG-like LLs possess large LL separation (first LL gap is ∼50 meV at 2 T) even at small magnetic fields. In contrast, the LL spacing of BLG-like LLs is far smaller (∼5 meV at 2 T). This means that multiple BLG-like LLs can be swept through (over large density and magnetic field windows) while keeping the filling factor of the MLG-like LLs constant in our experiment, see Fig.5.4c. This is most prominent between the $N_M = 0$ and $N_M = 1$ MLG-like LLs, where we were able to easily resolve and analyze ∼10 BLG-like LLs. Even though the filling factor of the BLG-like LLs steadily varies over this region, the filling factor of the MLG-like band remains *pinned* to 2 due to the particularly large first MLG-like LL energy spacing and the non-magnetic field dispersive nature of the $N_M = 0$ LL. As a result, in between MLG-like LLs [for e.g., that realized in the region $E(0_M) < E_F < E(1_M)$] MLG-like oscillations are frozen, and the SdH oscillations are dominated by the BLG-like band: $\Delta G_{xx} \approx G_B \cos[2\pi(\frac{B_{FB}}{B} + \gamma_B)]$.

Second, in SdH oscillation measurements, the total density is fixed (set by the gate voltage) while the magnetic field is varied. In ABA-trilayer graphene, the total density ($n_T = n_M + n_B$) is comprised of the individual band densities in each of the MLG-like ($n_M$) and the BLG-like ($n_B$) bands, which may reconfigure with the magnetic field while keeping $n_T$ constant. This constraint strongly influences the BLG-like SdH oscillations. To see this we express its oscillation frequency in terms of the total density via: $B_{FB} = \frac{n_B h}{4e} = \frac{(n_T - n_M)h}{4e} = B_{FT} - \frac{\nu_M B}{4}$, where $B_{FT} = \frac{n_T h}{4e}$ and $\nu_M = \frac{n_M h}{eB}$ is the filling factor of the MLG-like band. Crucially, for $E(0_M) < E_F < E(1_M)$ (above the MLG-like band gap) only $N_M$=0 electron like LL is filled, so the filling factor of the MLG-like band remains pinned to 2. This yields a BLG-like oscillation frequency as $\frac{B_{FB}}{B} = \frac{B_{FT}}{B} - 1/2$. Similarly, for $E(-1_M) < E_F < E(0_M)$ (below the MLG-like band gap) the filling factor of the MLG-like band remains pinned to -2 producing $\frac{B_{\text{FB}}}{B} = \frac{B_{\text{FT}}}{B} + 1/2$. Incorporating both cases into the BLG-like SdH oscillations, we obtain

$$\Delta G_{\text{xx}} \approx G_B \cos \left[ 2\pi \left( \frac{B_{\text{FT}}}{B} + \gamma_B \pm 1/2 \right) \right] \qquad (5.2)$$

that displays an unusual, non-trivial, and tunable phase shift, acquired due to the strong filling-enforced constraint above and below the bandgap. This yields an additional $\pi$ $(-\pi)$ phase shift in the BLG-like oscillations due to the fully-emptied (fully-filled) MLG-like lowest $N_M = 0$ LL. We note that when the Fermi energy is in the band gap, there is no additional phase shift in BLG-like band; this is consistent with the expectation that a completely filled (MLG-like valence) band does not influence the transport.

## 5.6   Determination of the phase of the BLG-like SdH oscillations when multiple MLG-like LLs are filled

In general, the filling enforced phase of the BLG-like SdH oscillations can be affected by the filling factor of the MLG-like band. Fig. 5.7 shows that the phase shift increases to $3\pi$, $5\pi$ ... when MLG-like higher LLs are filled one by one by changing Fermi energy – thus confirming the "filling-enforced" origin of the phase shift.

Since the first few MLG-like LLs have large gaps, it is possible that two successive MLG-like LLs contain several BLG-like LL oscillations. We fit such BLG-like LL oscillations contained between two successive MLG-like LLs away from the crossing regions. When the Fermi energy goes through the BLG-like LL oscillations in between $N_{\text{M}}$ and $(N+1)_{\text{M}}$ LLs, the MLG-like filling factor remains constant to $\nu_{\text{M}} = 4(N_{\text{M}} \pm 0.5)$ because of being in the LL gap of the MLG-like LLs. For this range of Fermi energy the equation 5.1 simplifies to

$$\Delta G_{\text{xx}} = G_{\text{B}} \cos \left[ 2\pi \left( \frac{B_{\text{FT}}}{B} + \gamma_B - \frac{\nu_{\text{M}}}{4} \right) \right] . \qquad (5.3)$$

If N and $B_N$ are the LL index of the BLG-like LLs and the corresponding magnetic field at an SdH oscillation minima, then the equation of the fitting line in the LL index plot is given by

$$N = \frac{B_{\text{FT}}}{B_{\text{N}}} + \frac{\Phi_{\text{B}}}{2\pi} - \frac{\nu_{\text{M}}}{4} \, . \tag{5.4}$$

Here the slope $B_{\text{FT}}$ relates to the total density ($n_{\text{T}} = \frac{4e}{h} \times B_{\text{FT}}$) and the total Fermi surface area ($S_{\text{FT}} = \frac{2\pi e}{\hbar} \times B_{\text{FT}}$). Now, only $N_{\text{M}} = 0$ and $N_{\text{M}} = -1$ hole like LLs are filled, when the Fermi energy lies below the MLG-like band gap between the $-1_{\text{M}}$ and $-2_{\text{M}}$ LLs i.e. $E(-2_{\text{M}}) < E_{\text{F}} < E(-1_{\text{M}})$. In this case the filling factor of the MLG-like band remains pinned to -6 making the equation of the fitting line $N = \frac{B_{\text{FT}}}{B_{\text{N}}} + \frac{\Phi_{\text{B}}}{2\pi} + \frac{3}{2}$. Since, Berry's phase $\Phi_{\text{B}} = 0$ for BLG-like LLs, this returns 1.5 intercept at $1/B = 0$ (see the orange line in Fig. 5.7). Similarly, $N_{\text{M}} = 0$, $N_{\text{M}} = -1$ and $N_{\text{M}} = -2$ hole like LLs are filled, when the Fermi energy lies below the MLG-like band gap between the $-2_{\text{M}}$ and $-3_{\text{M}}$ LLs i.e. $E(-3_{\text{M}}) < E_{\text{F}} < E(-2_{\text{M}})$. In this case the filling factor of the MLG-like band remains pinned to -10 making the equation of the fitting line $N = \frac{B_{\text{FT}}}{B_{\text{N}}} + \frac{\Phi_{\text{B}}}{2\pi} + \frac{5}{2}$. This results in 2.5 intercept at $1/B = 0$ (see the green line in Fig. 5.7).

We also fit the MLG-like LLs at the low field when the BLG-like LLs are not resolved. At very low magnetic field $B < 1$ T ($1 < 1/B < 3$ in Fig. 5.7), we resolve only MLG-like LLs since the LL spacing of the MLG-like bands are significantly larger than the BLG-like LLs. In this regime, the amplitude of the BLG-like SdH oscillations dies almost to zero, so, the SdH oscillation can be captured only in terms of the MLG-like LLs:

$$\Delta G_{\text{xx}} = G_{\text{M}} \cos \left[ 2\pi \left( \frac{B_{\text{FM}}}{B} + \gamma_M \right) \right] \, . \tag{5.5}$$

If N and $B_{\text{N}}$ are the LL index of the MLG-like LLs and the corresponding magnetic field, then the equation of the fitting line in the LL index plot is given by

$$N = \frac{B_{\text{FM}}}{B_{\text{N}}} + \frac{\Phi_{\text{M}}}{2\pi} \, . \tag{5.6}$$

Here the slope $B_{\text{FM}}$ relates to the MLG-like band density ($n_{\text{M}} = \frac{4e}{h} \times B_{\text{FM}}$) and the MLG-like Fermi surface area ($S_{\text{FM}} = \frac{2\pi e}{\hbar} \times B_{\text{FM}}$). Since, Berry's phase $\Phi_{\text{M}} = \pi$ for MLG-like LLs, this returns 0.5 intercept at $1/B = 0$. The red line in Fig. 5.7 shows that indeed the intercept of the MLG-like LLs is close to 0.5 confirming the nontrivial $\pi$ Berry's phase. This again confirms that the MLG-like band individually retains its $\pi$ Berry's phase and there is no hybridization between the bands. We note that the slope of the red line is almost an order of magnitude smaller than the orange and the green lines. This is because the Fermi surface area of the MLG-like band is roughly an order of magnitude smaller than the BLG-like Fermi surface area for this Fermi energy.

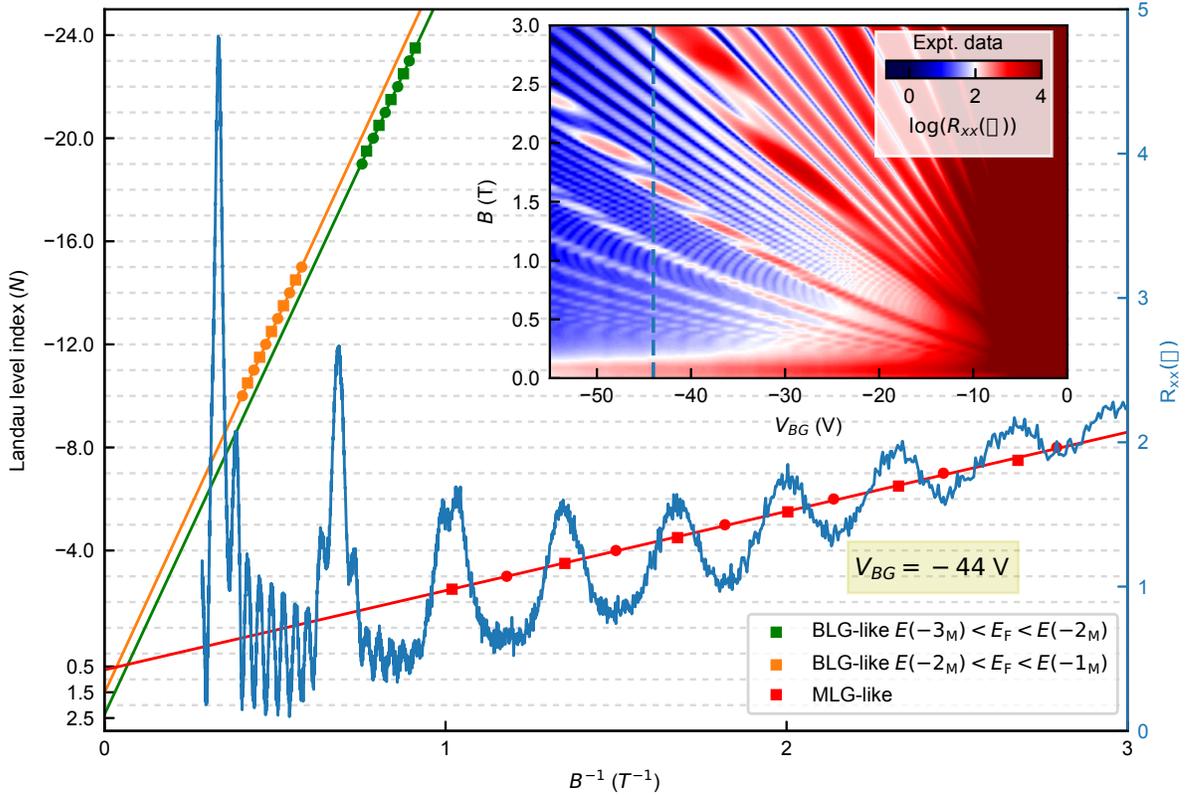

Figure 5.7: **SdH phase shift when multiple MLG-like LLs are filled.** An SdH oscillation line slice of the experimental data showing the beating pattern due to the two bands with different Fermi surface areas. Low and high-frequency oscillations come from the MLG-like and the BLG-like bands respectively. The LL index plots of BLG-like LLs are shown in orange and green color for which the Fermi energy satisfies E(-2M) < $E_F$ < $E(-1_M)$ and E(-3M) < $E_F$ < $E(-2_M)$ respectively. The phase of the BLG-like SdH oscillations in these two regions are shifted differently because of different fillings (of the MLG-like LLs) – this is evident in their different intercepts of 3/2 and 5/2 in the LL index plot. The red line is a fit of the MLG-like LLs when the BLG-like LLs are not resolved. This shows 1/2 intercept, as expected because of π Berry's phase of the MLG-like bands. The inset shows the color plot of the resistance on the hole side. The overlaid blue dashed line shows the gate voltage position ($V_{BG}$=-44 V) for which the SdH oscillation is plotted.

## 5.7 Shift in SdH oscillation frequency

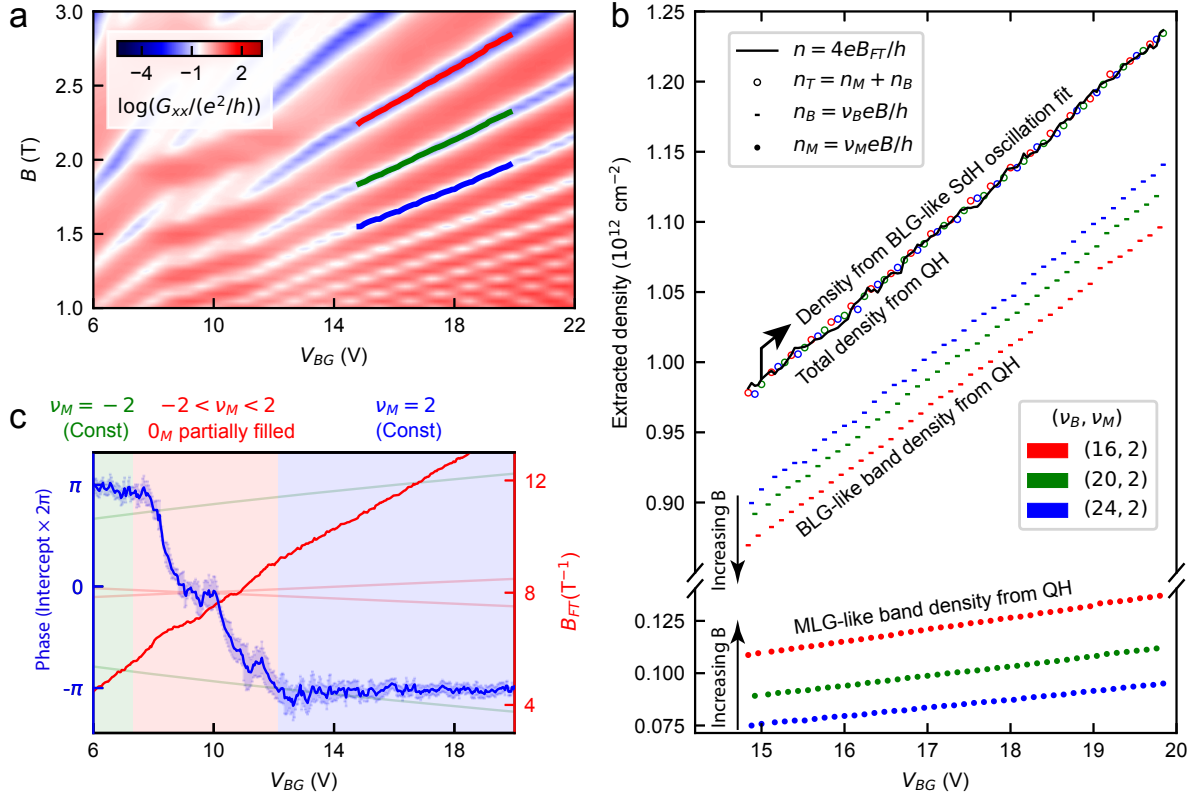

Figure 5.8: **Consequence of the filling enforced constraint on SdH frequency and phase.** (a) Zoomed in measured LL fan diagram showing three lines drawn for three filling factors along which we extract the density. (b) MLG-like band density ($n_M = \nu_M eB/h$) and BLG-like band density ($n_B = \nu_B eB/h$) calculated from the quantum Hall data are marked with filled circle and dash respectively for total filling factor $\nu_T = 18$ (red), 22 (green) and 26 (blue). The unfilled circles of corresponding colors show the total density $n_T = n_M + n_B$ for each total filling factors. At a constant gate voltage with the increasing magnetic field, the density of the MLG (BLG)-like band increases (decreases), keeping total density constant at all magnetic fields (filling factors). The black line shows the density calculated from the SdH frequency ($n_T = \frac{4e}{h} \times B_{FT}$) of the BLG-like SdH oscillations which is remarkably close to the combined density calculated from the quantum Hall data for all filling factors. (c) Intercept (blue) and slope (red) of the BLG-like SdH oscillation fit as a function of $V_{BG}$. Fitting errors are shown for the intercept. Corresponding band diagram is overlaid for visualization. Three regions are shaded with three different colors – green shade and blue shade indicate completely empty and completely filled $N_M=0$ LL respectively whereas the saffron shade indicates partially filled $N_M=0$ LL.

The filling-enforced constraint is further corroborated by the measured quantum oscillation frequency. In particular, Eq. 5.2 indicates that the BLG-like quantum oscillations have $1/B$ frequency that scales with the combined density of the MLG-like and the BLG-like bands; rather than the density of the BLG-like band only. To illustrate this, we focus on the solid lines in Fig. 5.8a, where the filling factor takes on precise quantized values $\nu_B = 16, 20, 24$ with $\nu_M = 2$, obtained directly from Hall conductance measurements (see section 5.3). Hence using $n_{M(B)} = \nu_{M(B)} eB/h$, as plotted in Fig. 5.8b, we can calculate the MLG-like band density (colored filled circles) and BLG-like band density (colored dash plots) on these lines. Total electron density $n_T = n_M + n_B$, which is independent of the filling factor at a given gate voltage, is shown by colored unfilled circles and matches exactly with the density obtained from the oscillation frequency $B_F$ (solid black line in Fig. 5.8b).

This unprecedented concordance, expected directly from Eq. (5.2), has a far-reaching consequence – it is assumed that quantum oscillations allow one to isolate a Fermi surface in a multiband system. So, using frequency to isolate the motion of electrons on Fermi surfaces is the de-facto method for mapping the Fermi surface. Our analysis shows that for a certain bandstructure this simple picture gets modified – in our case the SdH frequency of the BLG-like band not only depends on the BLG-like band Fermi surface area – but it also depends on the MLG-like band Fermi surface area. This, together with the unusual phase shift unequivocally display the strong effect of the filling-enforced-constraint present in a multiband system.

The unusual (non-trivial) zeroth Landau level filling enforced phase shifts, that we find in the BLG-like bands, can be attributed to Berry's phase of the MLG-like Dirac band, since the existence of the half-filled zeroth LL in a Dirac band is a direct consequence of its non-trivial Berry's phase. We extracted the phase shift (of the BLG-like quantum oscillations) over a fine grid as gate-voltage is tuned through the bandgap, see Fig.5.8c. This displays the smooth evolution of phase shift from $\pi \to 0 \to -\pi$ that closely tracks the smooth evolution of Berry's phase seen in Fig.5.1b expected for the gapped MLG-like band in inversion symmetry broken ABA-trilayer graphene.

## 5.8 Determination of the phase of the BLG-like SdH oscillations from another device

Here we show the quantum oscillation phase analysis from device - B. Fig. 5.9a shows the experimentally measured Landau level fan diagram. We have observed similar phase variation of the BLG-like SdH oscillations across the MLG-like Dirac band gap (Fig. 5.9b).

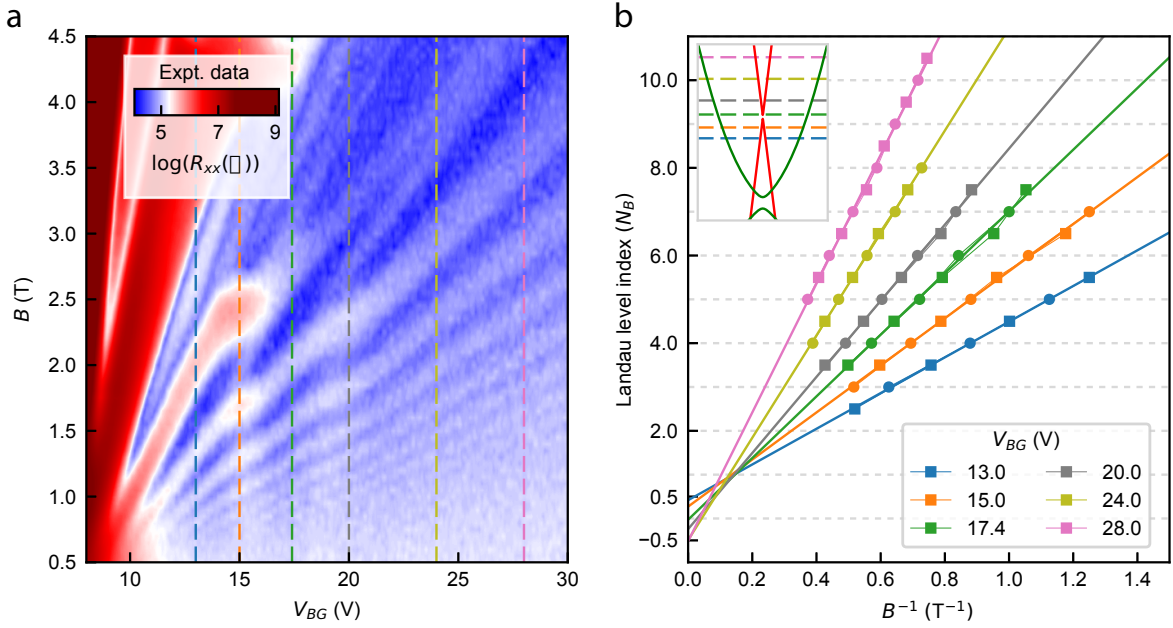

Figure 5.9:  (a) Experimentally measured Landau level fan diagram from device - B. (b) LL index plot for the BLG-like SdH oscillations at different densities across the Dirac band gap. Gate voltage for all the fittings are marked with dashed lines of the corresponding color in the fan diagram (a).

## 5.9   Determination of the phase from the simulated density of states (DOS)

### A. Fitting the DOS oscillation at a constant energy

We extract Berry's phase also by fitting the theoretical DOS oscillations [17]. Fig. 5.10a shows the DOS as a function of Fermi energy and magnetic field.  BLG-like DOS oscillation extrema are fitted at a line of constant energy.  Like in the experiment, integer (half-integer) LL indices are assigned at the minima (maxima) of the DOS oscillations.  We note that the DOS maxima positions correspond to the experimental $G_{xx}$ maxima.  All the fits (for the Fermi level below the MLG-like band gap, in the MLG-like band gap and above the MLG-like band gap) show zero intercepts – irrespective of the Fermi level position in the MLG-like band (see Fig. 5.10b).  This shows that the BLG-like band individually retains its $2\pi$ Berry's phase.

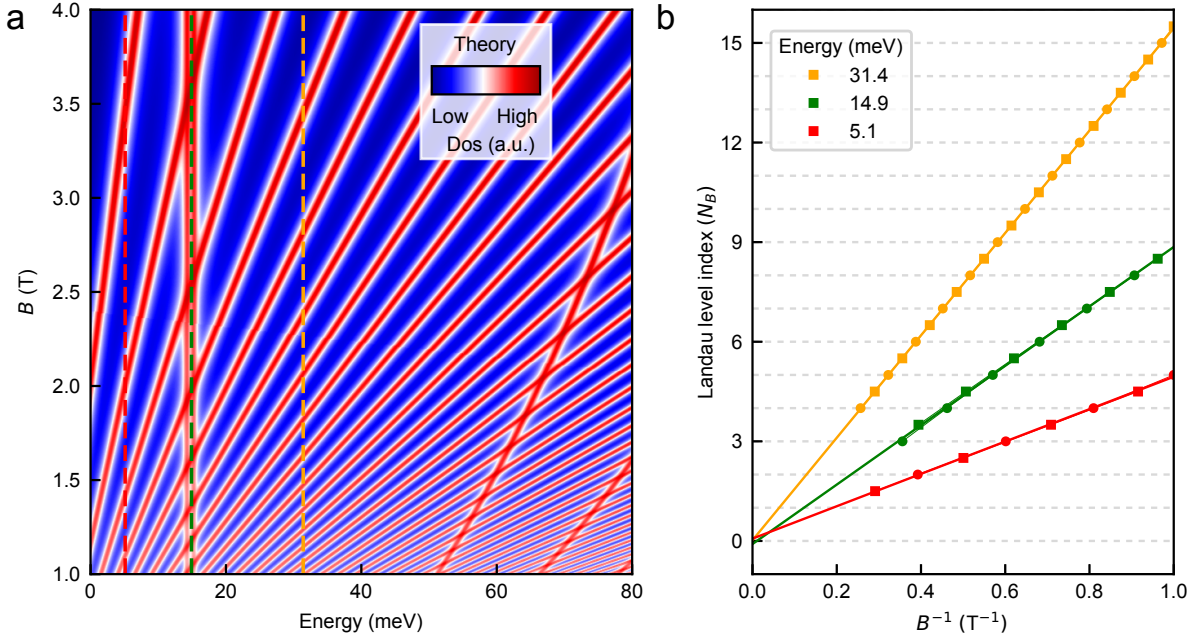

Figure 5.10: **Fitting using the DOS oscillations at a constant energy.** (a) Calculated DOS as a function of energy and magnetic field. (b) BLG-like LL Fits below the band gap (red), in the band gap (green) and above the band gap (yellow).

## B. Fitting the DOS oscillation at a constant density

We carry out similar LL fits as shown in Fig. 5.6 using the theoretically calculated DOS oscillations [17] (Fig. 5.11a) in the density magnetic field space. Like in the experiment, BLG-like DOS oscillation extrema are fitted at a line of constant density. We see similar anomalous phase: $\pi$ below the MLG-like band gap and $-\pi$ above the MLG-like band gap while it goes to zero in the MLG-like band gap (Fig. 5.11b). As we have explained in section 5.5, this additional phase picked up by the trivial BLG-like Fermi surface is because of the constraint on the total density in a multiband system. This constraint naturally occurs because experimentally the SdH oscillations are measured at a constant total density controlled by the gate voltage. The comparison between the fits done at a constant energy (Fig. 5.10b) and at a constant density (Fig. 5.11b) clearly shows the role of density constraint to determine the phase of the quantum oscillations in a multiband system.

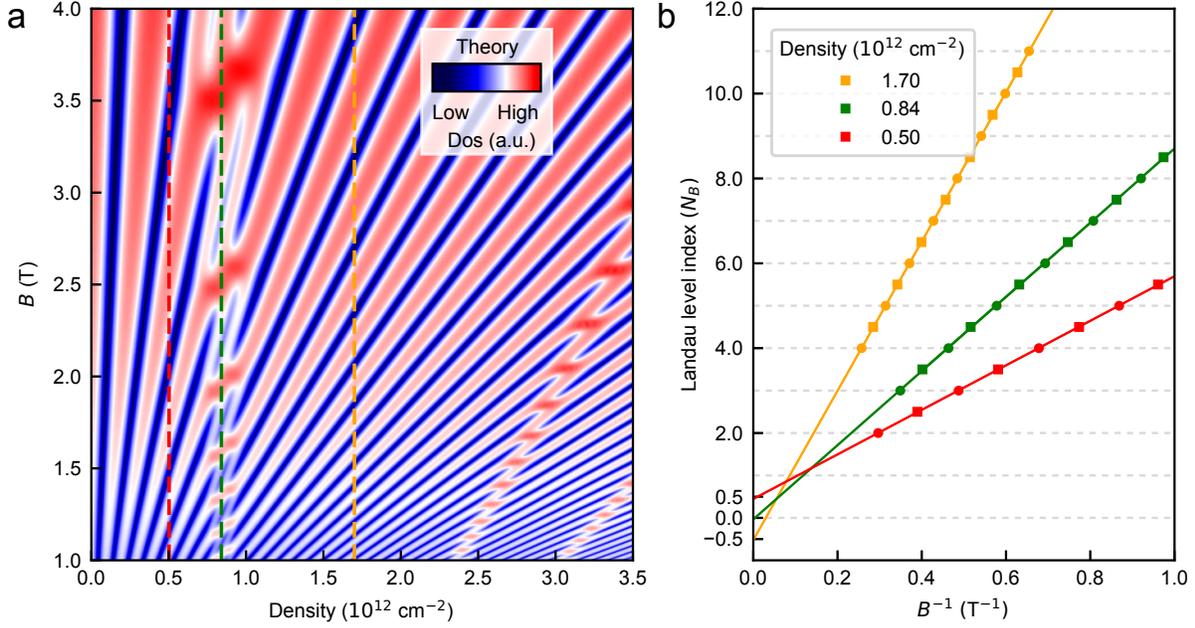

Figure 5.11: **Fitting using the DOS oscillations at a constant density.** (a) Calculated DOS as a function of density and magnetic field. (b) BLG-like LL Fits below the band gap (red), in the band gap (green) and above the band gap (yellow).

## 5.10 Summary

To summarize, here we unveil a new phase shift for quantum oscillations that appears in multi-Fermi-surface metals. In particular, we reveal how the quantum oscillations of a massive quadratic band (with a constant and trivial Berry's phase) can acquire non-trivial ($\pm\pi$) phase shifts that are gate-tunable. The phase shift of the quadratic band SdH oscillations depends on the position of the Fermi level in the coexisting Dirac band; the phase switches sharply from $\pi$ to $-\pi$ as the Fermi level is tuned from below the Dirac band gap to above it. Moreover, we show for the first time the continuous variation of Berry's phase induced quantum oscillation phase shift, as a function of gate voltage ($V_{\mathrm{BG}}$), in an inversion symmetry broken system close to the Dirac band edge. Here we stress the fact that Berry's phase of the two bands is not additive [23]. It is a routine exercise to map individual Fermi surfaces in a multiband system by isolating different frequencies in SdH oscillations [77] and one can, indeed, measure Berry's phase of individual bands [65]. Together with the fact that in ABA-trilayer graphene, a well-studied system, one can unambiguously map the band origin of the Landau levels using tight-binding calculation, our finding - the dependence of one band's quantum oscillation phase on the other is surprising. This method of phase detection of a non-trivial band using quantum oscillations from a coexisting band with more oscillations is novel; it can provide a surprising new facility to probe non-trivial quantum geometry. The existence of this filling-enforced phase is generically applicable to any multiband system – the unique band structure of ABA trilayer graphene, the gate tunability, and the high sample quality just enable this vivid visualization. While our study shows that additional care should be taken to extract the Berry's phase and the Fermi surface area in a multiband system, it could shed light also on other topological materials like Weyl semimetals [78] that host multiple bands.

# Landau level crossing physics and effect of electronic interaction in ABA-trilayer graphene

In this chapter, we study the Landau level (LL) crossings of mirror symmetry protected Bernal (ABA)-stacked trilayer graphene and the effect of electronic interaction. In a magnetic field, both the mass-less monolayer graphene (MLG)-like Dirac bands and massive bilayer graphene (BLG)-like bands develop their Landau levels. We see Shubnikov–de Haas (SdH) oscillations due to the LLs at a magnetic field as low as 0.3 T. These LLs cross (become degenerate) at different energies since the origin of the massless and massive bands are offset in energy and the fact that massless and massive LLs have different dispersion with the magnetic field. Of all the massless Landau levels $N_M = 0$ LL is special as it does not disperse with the magnetic field. This is why it crosses all other massive LLs and forms the most predominant set of the LL crossings. The $N_M = 0$ LL splits into four LLs when the spin and valley degeneracies are lifted at a relatively low magnetic field ($\sim 2.2$ T). Crossings of the four $N_M = 0$ LLs with two spin split $N_B = 2$ LLs (from the massive quadratic band) generate three "ring-like" structures on the electron side. On the other hand, the Landau level crossing pattern is very different on the hole side – a direct consequence of broken electron-hole symmetry in ABA trilayer graphene. Quantum Hall (QH) effect also provides a simple way to study the competition between single particle physics and electronic interaction. However, electronic interaction becomes important only in very clean graphene samples and so far the trilayer graphene experiments are understood within non-interacting electron picture. Here, we show the evidence of strong electronic interactions and quantum Hall ferromagnetism (QHF) seen in Bernal-stacked trilayer graphene. Due to high mobility ~500,000 cm$^2$V$^{-1}$s$^{-1}$ in our device compared to previous studies, we find all symmetry broken states and that Landau level gaps are enhanced by interactions; an aspect explained by our self-consistent Hartree-Fock calculations. Moreover, we observe hysteresis as a function of filling factor ($\nu$) and spikes in the longitudinal resistance ($R_{\mathrm{xx}}$) which, together, signal the formation of quantum Hall ferromagnetic states at a low magnetic field.

## 6.1   Introduction

Landau level crossings provide a novel opportunity to study the interaction between two LLs. This has been extensively studied in AlGaAs/GaAs systems and silicon quantum wells [79, 80, 81]. In a conventional single band system both the LL energy ($\hbar eB/m^*$) and the Zeeman energy ($g\mu_B B$) are linear in $B$ which does not result in any LL crossing, see Fig. 6.1a. We note that LL energy depends only on the perpendicular component of the magnetic field ($B_\perp$) but the Zeeman energy depends on the total magnetic field. Thus tilting the magnetic field with respect to the normal of the plane, two LLs (with successive LL index) of opposite spins can become degenerate (see Fig. 6.1b) when

$$\hbar\omega_c = g\mu_B B \tag{6.1}$$
$$\Rightarrow \hbar\frac{eB_\perp}{m^*} = g\mu_B\sqrt{B_\perp^2 + B_\parallel^2}, \tag{6.2}$$

where $\hbar$ is Planck's constant, $e$ is electronic charge, $m^*$ is the effective mass, $g$ is Lande' g-factor for electrons in a solid, $\omega_c$ is cyclotron frequency, $\mu_B$ is Bohr magneton, $B_\parallel$ and $B_\perp$ are the parallel and perpendicular component of the magnetic field respectively. At the crossing of two LLs, the existence of two different states with the same energy can kill the QH state. However, sometimes it gives rise to a new ground state (quantum Hall ferromagnet) near the crossing point.

In the systems with multiple bands, tilting the magnetic field is not essential to generate LL crossings. It is possible that a higher energy LL from the low energy band cross with a lower energy LL from the high energy band since the LL origins of the two bands are offset in energy (see Fig. 6.1c).

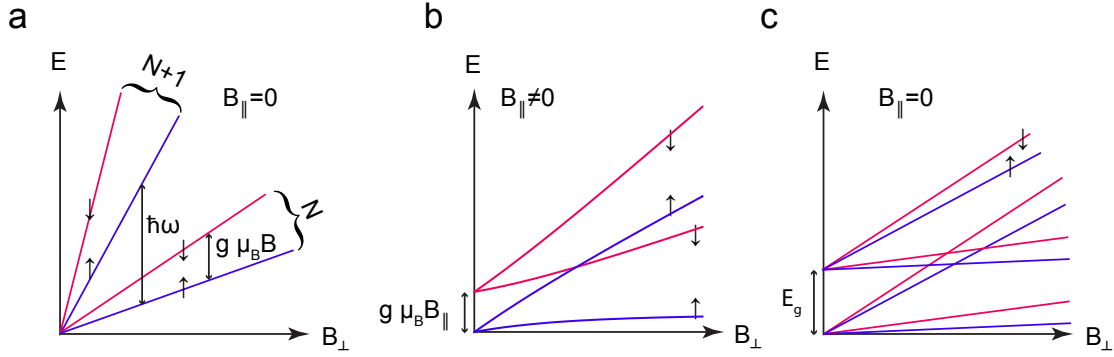

Figure 6.1: **Schematic of Landau level crossing.** a) The dispersion of LLs with magnetic field perpendicular to the two-dimensional electron gas (2DEG) plane in a single band system showing no LL crossing is possible. LLs with two successive LL indices are shown in the diagram. The red and the blue color denote the down and up spin respectively. b) The dispersion of LLs with $B_\perp$ in a single band system in presence of a constant non-zero in-plane magnetic field showing LL crossing is possible. c) The dispersion of LLs with $B_\perp$ in a two-band system showing LLs from the different bands can cross. $E_g$ is the band gap.

Band structure of ABA trilayer graphene, shown in Fig. 6.2a and Fig. 6.2b consist of both monolayer graphene-like linear and bilayer graphene-like quadratic bands [14, 82]. Fig. 6.2a shows the band structure setting $\gamma_2$, $\gamma_5$, $\delta$ and $\Delta_2$ to zero which results in the electron-hole symmetric bands. However, when all the hopping parameters are included we clearly see the electron-hole symmetry gets broken.

## 6.2 Landau level crossings at low magnetic field

Fig. 6.2c shows an optical image of the device where the ABA-TLG graphene is encapsulated between two hBN flakes [41]. Four probe resistivity ($\rho$) of the device is shown in Fig. 6.2d. The electron-hole asymmetry is not very evident in the zero magnetic field resistivity – although a little bump develops on the hole side (at around $V_{bg} \sim$ -3.5 V) in the resistivity at the lowest temperature 1.5 K. We will see later, that this electron-hole asymmetry is pronounced when the Landau levels form at a finite magnetic field. The low disorder in the device is reflected in high mobility $\sim$500,000 cm$^2$V$^{-1}$s$^{-1}$ on electron side and $\sim$800,000 cm$^2$V$^{-1}$s$^{-1}$ on hole side; this leads to carrier mean free path in excess of $\sim$7 μm.

We next consider the magnetotransport in ABA-TLG that reveals the presence of

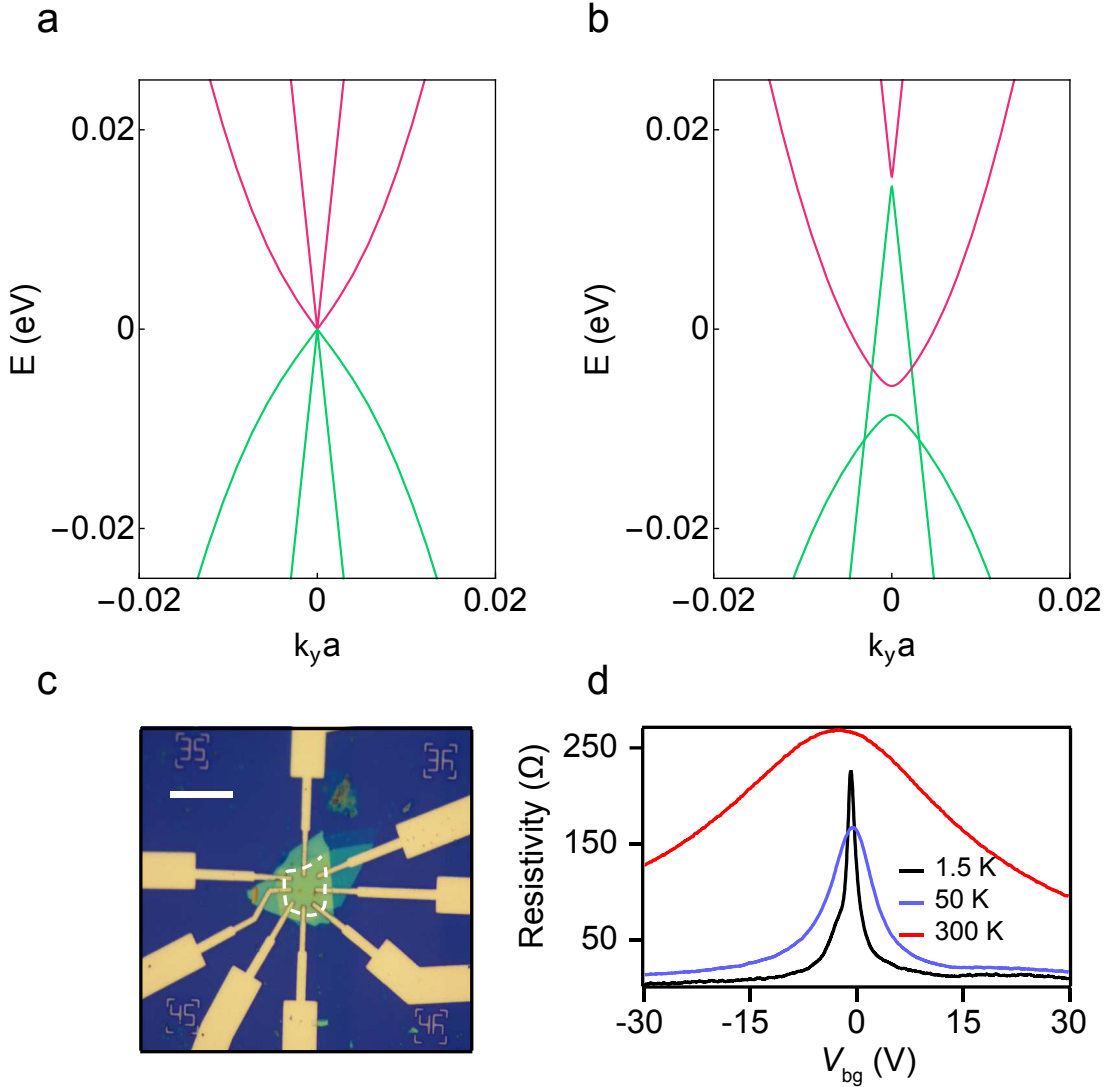

Figure 6.2: **Electron-hole asymmetry and electrical transport.** a) Low energy band structure of ABA-TLG around the $K_+$ point in the Brillouin zone for $\gamma_0 = 3.1$ eV, $\gamma_1 = 0.39$ eV, $\gamma_2 = 0$, $\gamma_3 = 0.315$ eV, $\gamma_4 = 0.12$ eV, $\gamma_5 = 0$, $\delta = 0$, $\Delta_1 = 0$, $\Delta_2 = 0$. We see that setting $\gamma_2$, $\gamma_5$, $\delta$ and $\Delta_2$ to zero results in an electron-hole symmetric band structure. Green and red lines denote the BLG-like bands the MLG-like bands along the $k_y$ direction. b) Band structure including all the band parameters $\gamma_0 = 3.1$ eV, $\gamma_1 = 0.39$ eV, $\gamma_2 = $ -0.02 eV, $\gamma_3 = 0.315$ eV, $\gamma_4 = 0.12$ eV, $\gamma_5 = 0.018$ eV, $\delta = 0.02$ eV, $\Delta_1 = 0$ , $\Delta_2 = 4.3$ meV. We clearly see that including the band parameters $\gamma_2$, $\gamma_5$, $\delta$ and $\Delta_2$ breaks the electron-hole symmetry. It also separately gaps out the MLG-like bands BLG-like bands. However, there is no band gap in total, semi-metallic nature of ABA-TLG is clear from the band overlap. c) Optical image of the hBN encapsulated trilayer graphene device; Scale bar, 20 μm. White dashed line indicates the boundary of the ABA-TLG. The graphene sample is a slightly distorted rectangle, but the electrodes are designed in a Hall bar geometry. Length and breadthwise distance between furthest electrodes are 9.3 μm and 7.8 μm respectively. This makes the aspect ratio to be 1.19. The substrate consists of 30 nm thick hBN and 300 nm thick Silicon dioxide (SiO$_2$) coated highly $p$ doped Silicon which also serves as a global back gate. d) Room temperature and low temperature four probe resistivity of the device as a function of $V_{bg}$.

LLs arising from both MLG-like and BLG-like bands. The LLs are characterized by
the following quantum numbers: $N_M(N_B)$ defines the LL index with M(B) indicating
monolayer (bilayer) graphene-like LLs, +(-) denotes the valley index of the LLs and
↑(↓) denotes the spin quantum number of the electrons. All the data are taken at
1.5 K. Accessing the monolayer graphene-like (MLG-like) LLs requires us to go to high
density and very low magnetic field when the degeneracy of LLs is low so that electrons
can occupy high LL index energy levels. Usually, the sample quality makes it harder to
resolve LLs at such low field (∼0.5 T - 3 T ), possibly it explains the reason why so far
crossings of MLG-like and BLG-like LLs were not observed at such a low field in the fan
diagram. Fig. 6.3a and Fig. 6.3d show the LL fan diagram at a low magnetic field on
hole and electron side respectively. Observation of LLs up to very high filling factor $\nu$
= 118 confirms the high quality of the device. Along with the usual straight lines in the
fan diagram, we find additional interesting parabolic lines which arise because of LL
crossing points between different MLG-like and BLG-like LLs. Fig. 6.3b and Fig. 6.3c
show the line slices of the hole side data along back-gate voltage ($V_{bg}$) axis at 1.45 T
and along **B** axis at $V_{bg}$ = -44 V respectively. Similarly, Fig. 6.3e and Fig. 6.3f show
the line slices of the electron side data along $V_{bg}$ axis at 1.22 T and along **B** axis at $V_{bg}$
= 50 V respectively. All the line slices show large $R_{xx}$ peak at the crossing points. It is
noticeable that only for **B** > 1 T magnetic field BLG-like LLs start to resolve whereas
the presence of MLG-like LLs can be seen in fan diagram (via crossings) at much lesser
magnetic field ∼0.25 T which is consistent with smaller cyclotron frequency of bilayer
graphene-like band compared to the monolayer graphene-like band.

It should be emphasized that there is no way to distinguish between the linear dis-
persing BLG-like LLs (with magnetic field) and square root dispersing MLG-like LLs
in the fan diagram as fan diagram shows only constant filling factor lines which are
always linear for a particular filling factor $\nu = \frac{nh}{Be}$ where $n$ is the number density of
carriers, given by $ne = C_{bg}V_{bg}$. That is why both in monolayer and bilayer graphene
constant filling factor regions in fan diagram are a set of straight lines but having dif-
ferent slope depending on different Hall quantization. It is the LL crossings in the fan
diagram which actually helps to realize the presence of two sets of differently dispersing
levels: every crossing gives rise to high DOS at the crossing points and it shows up as
longitudinal conductance maxima. Longitudinal conductance ($G_{xx}$) is given by $G_{xx} =$
$\frac{R_{xx}}{R_{xx}^2+R_{xy}^2}$ where $R_{xx}$ and $R_{xy}$ are the longitudinal and Hall resistance respectively. For
graphene $R_{xy} \gg R_{xx}$ at a Hall resistance plateau, so the formula reduces to $G_{xx} \approx \frac{R_{xx}}{R_{xy}^2}$

other than at the $\nu = 0$ plateau. Hence, for any quantum Hall plateau other than $\nu = 0$, conductance maxima also show up as resistance maxima. These crossings at low field appear at some discrete points in the fan diagram on a set of parabola as MLG-like LLs disperse as $E_M \sim \sqrt{B}$ and BLG-like LLs disperse as $E_B \sim B$. Experimental Landau level crossing points can be used to determine the hopping parameters. Here we have used the crossing points of monolayer graphene-like $N_M = 1$ Landau level (LL) with other bilayer graphene-like Landau levels from LL index $N_B = 17$ to $N_B = 26$ to calculate different hopping parameters. Since, $\gamma_0$ and $\gamma_1$ are known very precisely, we vary relatively smaller hopping parameters $\gamma_2, \gamma_5$ and $\delta$ a little over the known values for bulk graphite [83] to match the experimentally observed magnetic field values at the crossing points. We find $\gamma_0 = 3.1$ eV, $\gamma_1 = 0.39$ eV, $\gamma_2 = -0.028$ eV, $\gamma_5 = 0.01$ eV and $\delta = 0.021$ eV best describe our data.

We used the approximate $2 \times 2$ MLG-like and $2 \times 2$ BLG-like Hamiltonians [12, 13] (see equation 2.91 of Chapter 2) to calculate the low energy Landau level spectra (shown in Fig. 6.10b) which allows an analytical solution of the non-interacting Hamiltonian. Under this approximation higher order terms containing $\gamma_3$ and $\gamma_4$ are neglected. Experimentally Landau level orbital index of the Landau levels is determined by the help of filling factors determined from the experimental Hall conductance ($G_{xy}$) plot. The experimental crossing points of different monolayer graphene-like and bilayer graphene-like LLs used to determine the hopping parameters and their comparison with the theoretical ones (after obtaining the final band parameters) are shown in Table 6.1.

Table 6.1: Comparison of experimental and theoretical Landau level crossing points

| $N_M$ | $N_B$ | Experimental magnetic field (T) at crossing point | Theoretical magnetic field (T) at crossing point |
|---|---|---|---|
| 1 | 17 | 1.47 | 1.47 |
| 1 | 18 | 1.35 | 1.35 |
| 1 | 19 | 1.24 | 1.24 |
| 1 | 20 | 1.14 | 1.15 |
| 1 | 21 | 1.05 | 1.06 |
| 1 | 22 | 0.98 | 0.99 |
| 1 | 23 | 0.92 | 0.92 |
| 1 | 24 | 0.86 | 0.87 |
| 1 | 25 | 0.80 | 0.82 |
| 1 | 26 | 0.76 | 0.77 |

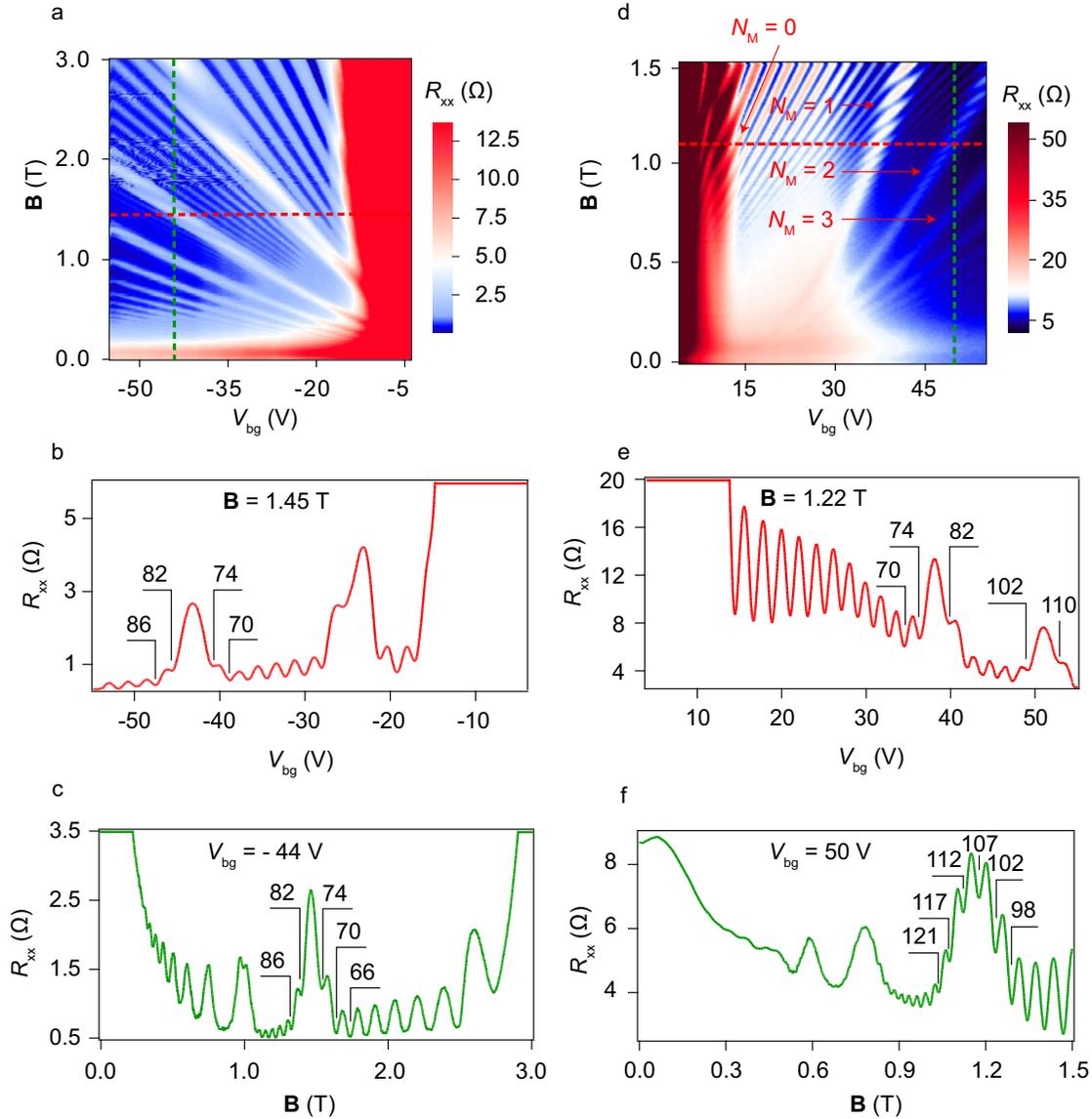

Figure 6.3: **Landau level crossings in ABA trilayer graphene at low magnetic field.**
a) $R_{xx}$ colour plot on hole side. The line slices of the data along the red and green lines are
shown in panel (b) and (c). b) Line slice of hole side data along $V_{bg}$ axis at **B** = 1.45 T. c)
Line slice of hole side data along **B** axis at $V_{bg}$ = -44 V. d) $R_{xx}$ colour plot on electron side.
The line slices of the data along the red and green lines are shown in panel (e) and (f). e)
Line slice of electron side data along $V_{bg}$ axis at **B** = 1.22 T. f) Line slice of electron side
data along **B** axis at $V_{bg}$ = 50 V.

## 6.3   Landau level crossings at high magnetic field

We now consider the LL fan diagram for a larger range of magnetic field. Fig. 6.4a
shows the calculated [12, 13, 14] energy dispersion of the spin degenerate LLs with $\mathbf{B}$.
Fig. 6.4b shows the main fan diagram where the measured longitudinal conductance
($G_{xx}$) is plotted as a function of $V_{bg}$ and $\mathbf{B}$. Due to lack of inversion symmetry, valley
degeneracy is not protected in ABA-TLG, it breaks up with increasing $\mathbf{B}$ and reveals all
the symmetry broken filling factors as seen in Fig. 6.4b. It is clear from the LL energy
diagram (Fig. 6.4a) that low energy regime is mostly populated by bilayer graphene-like
(BLG-like) LLs. So, at high field (roughly $\mathbf{B} > 5$ T) no monolayer graphene-like levels
exist (except $N_M = 0$) in our experimentally accessible density range $-4 \times 10^{12}$ cm$^{-2}$
to $4 \times 10^{12}$ cm$^{-2}$ which corresponds to -60 V to +60 V back gate voltage.

Fig. 6.4c shows measured $G_{xx}$ focusing on the $\nu = 0$ state, which shows a dip right at
the charge neutrality point, evident for $\mathbf{B} > 6$ T. Corresponding Hall conductance ($G_{xy}$)
shows a plateau at zero indicating the occurrence of the $\nu = 0$ state (see section 6.4 for
a longer discussion on the $\nu = 0$ state). While, the $\nu = 0$ plateau has been observed in
monolayer graphene [84] and in bilayer graphene [85] (for $\mathbf{B}$ more than ∼15 T - 25 T),
this is the first observation of $\nu = 0$ state in trilayer graphene at such low $\mathbf{B}$. A marked
reduction in disorder allows observation of the $\nu = 0$ state in our device.

Focusing on the electron side, Fig. 6.4d and Fig. 6.4e show the experimentally measured
LL fan diagram and labelled LLs, respectively. We see that the presence of $N_M = 0$
LL gives rise to a series of vertical crossings along the $\mathbf{B}$ axis as is expected from the
LL energy diagram (Fig. 6.4a). The highest crossing along the $\mathbf{B}$ axis appears when
$N_M = 0$ crosses with $N_B = 2$ LL at ∼5 T. From the complex fan diagram, seen in
Fig. 6.4d and Fig. 6.4e, we can see both above and below the topmost LL crossing
($V_{bg}$ ∼10 V and $\mathbf{B}$ ∼5 T), $N_M = 0$ LL is completely symmetry broken and $N_B = 2$ LL
quartet, on the other hand, becomes two-fold split at ∼3.5 T. The crossing between
$N_M = 0$ and $N_B = 2$ LLs gives rise to three ring-like structures. Calculated LL energy
spectra near the topmost crossing (Fig. 6.4e inset) shows that spin splitting is larger
than valley splitting for $N_B = 2$ LL but valley splitting dominates over spin splitting
for $N_M = 0$ LL. We note that valley splitting of $N_M = 0$ LL is very large compared to
other MLG-like LLs; which arises because MLG-like bands are gapped in ABA-TLG
unlike in monolayer graphene. As one follows the $N_M = 0$ LL down towards $\mathbf{B} = 0$ one
observes successive LL crossings of $N_M = 0$ with $N_B = 2, 3, 4 \ldots$. The sharp abrupt bends in the fan diagram occur due to the change of the order of filling up of LLs after crossings and the fact that the horizontal axis is charge density (proportional to $V_{\text{bg}}$ and not LL energy). When these crossings are extrapolated to $\mathbf{B} = 0$, we see that $N_{\text{M}}$ = 0 LL is valley split as expected from the LL energy diagram Fig. 6.4a.

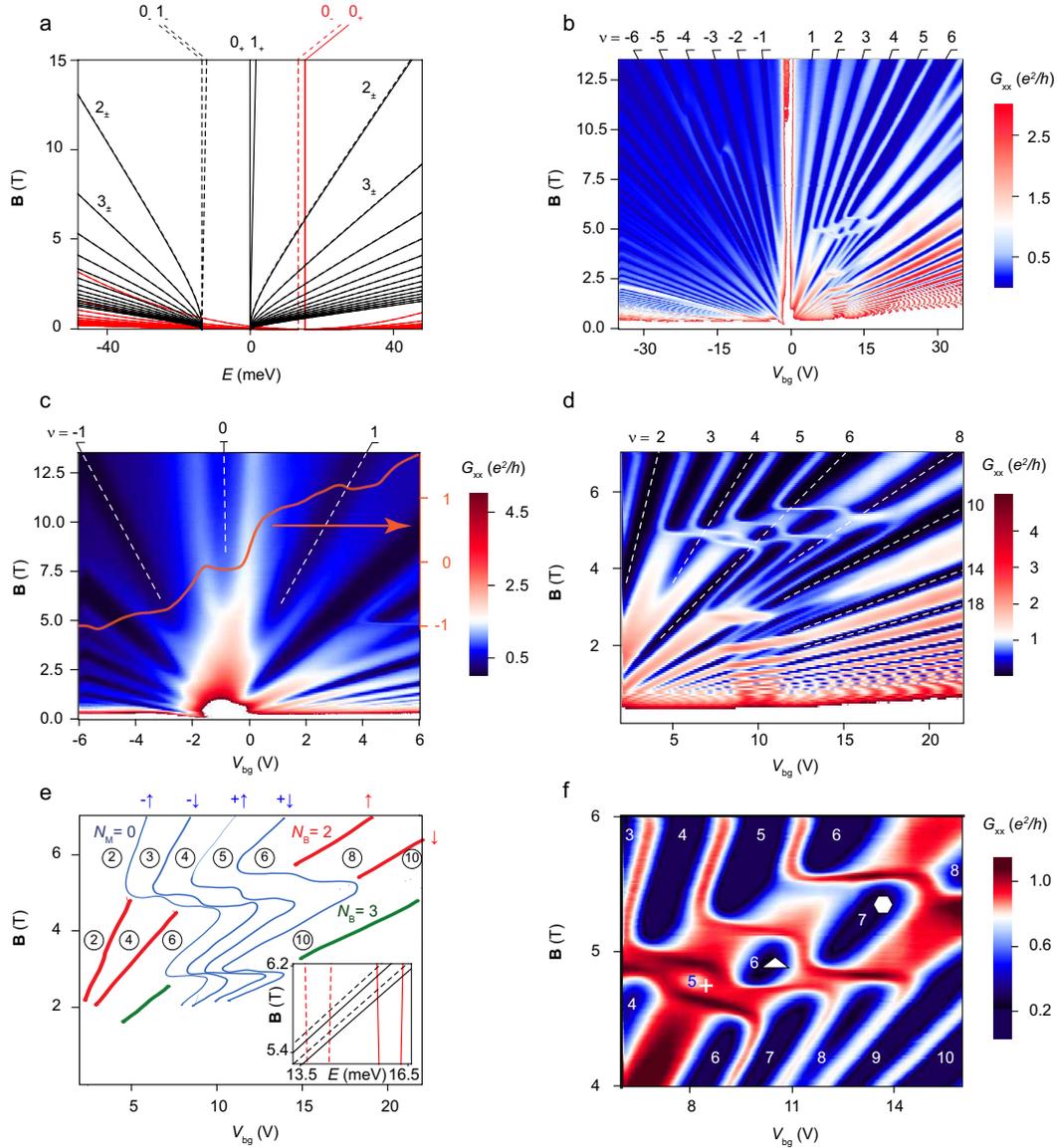

Figure 6.4: **Landau level crossings at high magnetic field.** a) Calculated low energy spectra using SWMcC parametrization of the tight binding model for ABA-TLG [12, 13, 14]. Red and black lines denote the MLG-like and BLG-like LLs respectively. Solid and dashed lines denote LLs coming from $K_+$ and $K_-$ valleys respectively. Labelled numbers represent the LL indices of the corresponding LLs. b) Colour scale plot of $G_{xx}$, showing the LL fan diagram. $\nu = 0$ feature is not seen in this colour scale as the lock-in sensitivity was set to a low value in this measurement to record the low resistance values accurately. The filling factors measured independently from the $G_{xy}$ are labelled in every plot. As a function of the **B**, one can observe several crossings on electron and hole side. c) Zoomed-in fan diagram around charge neutrality point showing the occurrence of $\nu = 0$ from ~6 T: $G_{xx}$ shows a dip and $G_{xy}$ shows a plateau at $\nu = 0$. The overlaid red line presents $G_{xy}$ at 13.5 T which shows the occurrence of $\nu = -1$, 0 and 1 plateaus. d) Zoomed-in recurrent crossings of $N_M = 0$ LL with different BLG-like LLs. e) The lines indicate the LLs seen in the data shown in Fig. 6.4d and their crossings. Circled numbers denote the filling factors. f) A further zoomed-in view of the parameter space showing LL crossing of fourfold symmetry broken $N_M = 0$ LL with spin split $N_B = 2$ LL.

## 6.4   $\nu = 0$ and other symmetry broken states

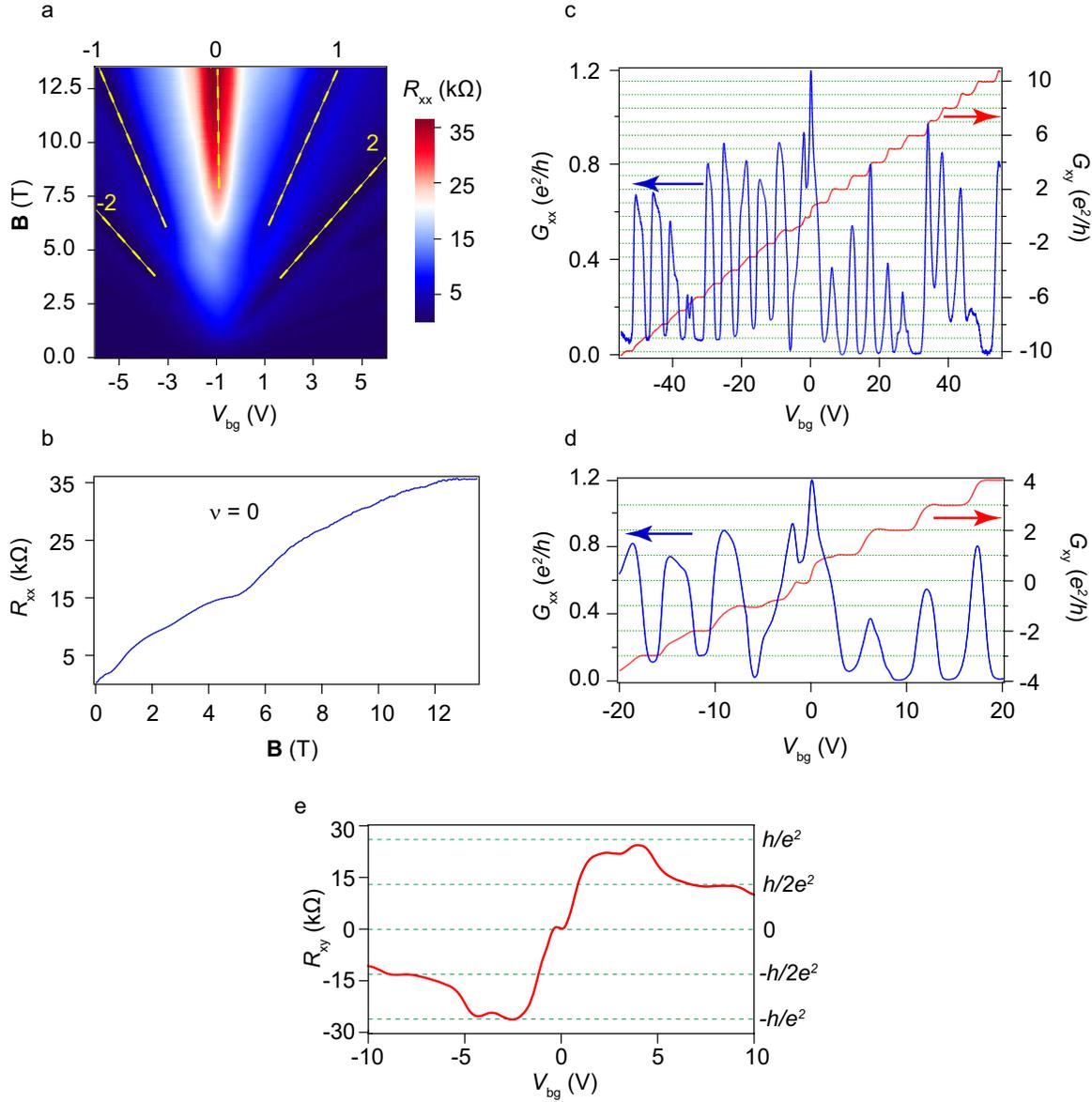

Figure 6.5:  $\nu = 0$ **and other symmetry broken states.** a) Colour plot of $R_{xx}$ in the vicinity of charge neutrality point. b) Shows the evolution of $R_{xx}$ with magnetic field at the charge neutrality point. c) $G_{xx}$ and all $G_{xy}$ plateaus from filling factors -10 to 10 at 13.5 T. d) Close up of panel (c) showing all filling factors from -3 to 4 at 13.5 T. e) $R_{xy}$ plateaus from $\nu$ = -2 to 2 including $\nu = 0$ at 13.5 T which shows a plateau at zero resistance.

In this section we study the nature of the $\nu = 0$ state.  Fig. 6.5a shows the $R_{xx}$ corresponding to the $G_{xx}$ shown in Fig. 6.4c.  Upon applying magnetic field the charge neutrality point resistance increases from $\sim 150\Omega$ to $\sim 35$ k$\Omega$ – showing $\sim 2\times10^4\%$ fold increase in the resistance (Fig. 6.5b).  The corresponding $R_{xy}$ shows a plateau at zero resistance, indicating the occurrence of the $\nu = 0$ state. Though $R_{xx}$ and $R_{xy}$ are the experimentally measured quantities, $G_{xx}$ and $G_{xy}$ are more fundamental in theory [86] and are related as $R_{xx} = \frac{G_{xx}}{G_{xx}^2 + G_{xy}^2}$. At $\nu = 0$ state both $G_{xx}$ and $G_{xy}$ go to zero, hence $R_{xx}$ starts showing very high value which increases with magnetic field. In our device, though $\nu = 0$ state starts developing from 6.5 T, it is still not fully developed till 13.5 T. Fig. 6.5c and Fig. 6.5d show the observed integer quantum Hall plateaus at 13.5 T. Fig. 6.5e shows a plateau at zero Hall resistance indicating the formation of $\nu = 0$ state.

## 6.5 Disorder strength and quantum scattering time calculation from Dingle plot

We estimate disorder strength $\Gamma$ from the magnetic field dependence of the SdH oscillations which is well studied in semiconductor two dimensional electron system (2DES) [87, 88] and recently in graphene [88, 89]. We calculate single particle quantum broadening ($\Gamma$) of the LLs which is related to the single particle scattering time or the quantum scattering time ($\tau_q$) by $\Gamma = \frac{\hbar}{2\tau_q}$. Magnetic field dependence of SdH oscillations is given by [87, 88] $\frac{\Delta R}{R_0} = 4X(T)\exp(-\pi/\omega_c\tau_q)$ where $\omega_c = \frac{eB}{m^*}$ is the cyclotron frequency. $\Delta R$ and $R_0$ are the oscillatory and non-oscillatory part of the resistance respectively. $X(T)$ is the temperature dependent amplitude, given by $X(T) = \frac{2\pi^2 k_B T/\hbar\omega_c}{sinh(2\pi^2 k_B T/\hbar\omega_c)}$. Fig. 6.6a shows the SdH oscillations at $V_{bg} = 23$ V after subtracting the non-oscillatory background resistance. The logarithm of the ratio of the oscillatory part ($\Delta R$) and non-oscillatory part ($R_0$), normalised by the temperature dependent amplitude ($X(T)$) is plotted against inverse magnetic field in Fig. 6.6b. This is known as Dingle plot.
The red line is the straight line fit where the slope is given by $-\pi m^*/e\tau_q$. We do the
fitting in the magnetic field range where only bilayer graphene-like LLs are present
and hence we use the BLG-like electron effective mass which is given by $m^* = \frac{\gamma_1}{\sqrt{2}v_F^2}$.
On electron side, we calculate quantum scattering time from the slope $\tau_q \approx 218$ fs
which yields $\Gamma \sim 1.5$ meV. The transport scattering time ($\tau_t = \frac{m^*\mu}{e}$) turns out to be
$\approx 10755$ fs using mobility $\mu = 3.9 \times 10^5$ cm$^2$V$^{-1}$s$^{-1}$ at the same density. The large
ratio of transport scattering time and quantum scattering time in our device ($\frac{\tau_t}{\tau_q} \approx$
49) is consistent with high mobility 2DES [87]. Transport scattering time is not very
sensitive to small angle scattering, it is the large angle scattering of the carriers which
decreases the mobility and hence reduces the transport scattering time. Large $\tau_t/\tau_q$
indicates that small angle long range Coulomb scattering is the dominant scattering
mechanism in our device [87, 88, 90].

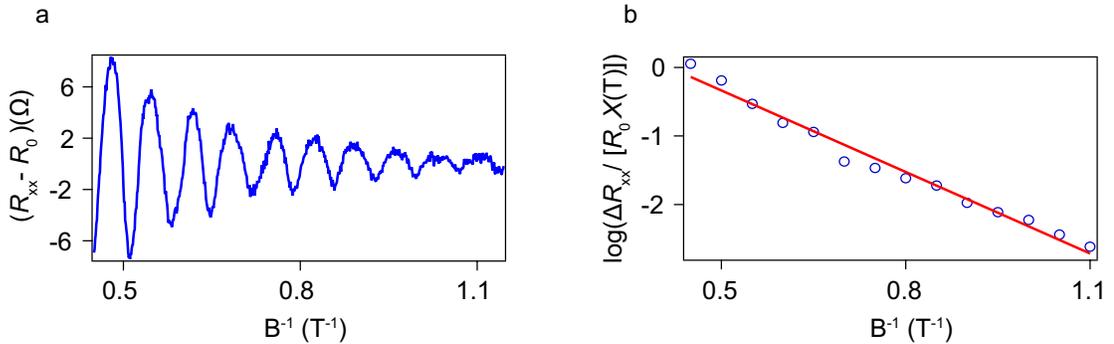

Figure 6.6: **Disorder strength calculation from Dingle plot on electron side at $V_{bg}$ = 23 V.** a) SdH oscillations after subtracting the non-oscillatory part ($R_0$). b) Straight line fit to the Dingle plot.

## 6.6  Role of electronic interactions

We next discuss experimental signatures that point towards the importance of elec-
tronic interaction. Observation of spin split $N_M = 0$ LL at **B** $\sim$2 T cannot be ex-
plained from the non-interacting Zeeman splitting for $\Gamma \sim 1.5$ meV on electron side,
estimated from the Dingle plot, see Fig. 6.6. Also, the large ratio of transport scat-
tering time ($\tau_t$) to quantum scattering time ($\tau_q$)($\frac{\tau_t}{\tau_q} \approx$ 49) indicates that small angle
scattering is dominant, a signature of the long-range nature of the Coulomb poten-
tial [87, 88, 90] (see section **??**). We also measure activation gap for the symmetry
broken states $\nu = 2, 3, 4, 5, 7$ at **B** = 13.5 T, and find significantly higher gaps than the
non-interacting spin-splitting. For $\nu = 3$ and 5, Fermi energy ($E_F$) lies in spin-polarized
gap of $N_M = 0$ LL in $K_-$ and $K_+$ valley respectively. Measured energy gap at $\nu =$
3 is $\sim$5.1 meV and at $\nu = 5$ is $\sim$2.8 meV whereas free electron Zeeman splitting is
$\sim$1.56 meV at **B** = 13.5 T (see Fig. 6.7 and Table 6.2). We note that typically the
transport gap tends to underestimate the real gap due to the LL broadening, so actual
single particle gap might be even larger. This shows the clear role of interactions even
with a conservative estimation of the Landau level gap.

Interaction results in symmetry broken states at low **B** that are QHF states. For the
data in Fig. 6.4d, $\nu = 2, 3, 4, 5$ are QHF states for **B** > 5.5 T. Similarly, $\nu = 7, 8, 9$
are also QHF states for 5.5 T > **B** > 4 T. In fact the LLs associated with $\nu = 3, 4, 5$

after crossing are the same MLG-like LLs which are responsible for $\nu = 7, 8, 9$ before crossing, see Fig. 6.4e. The crossings result in three ring-like structures marked by +, $\triangle$ and $\bigcirc$ in Fig. 6.4f.

The key role of interactions is also reflected in the hysteresis of $R_{xx}$ in the vicinity of the symmetry broken QHF states. Though QHF has been extensively studied in 2DES using semiconductors [80, 81] there are only a few reports of studying QHF in graphene [91, 92, 93, 94]. In our experiment, we vary filling factor by changing $V_{bg}$ at a fixed $\mathbf{B}$ (see Fig. 6.8a) and observe that the sweep up and down of $V_{bg}$ shows a hysteresis in $R_{xx}$ which can be attributed to the occurrence of pseudospin magnetic order at the symmetry broken filling factors [95]. Corresponding hysteresis is absent in simultaneously measured Hall resistance $R_{xy}$ (Fig. 6.8b). The pinning, that causes the hysteresis could be due to residual disorder within the system as the domains of the QHF evolve. Along a constant filling factor $\nu = 6$ line shown in Fig. 6.8c transport measurements show an appearance of $R_{xx}$ spikes around the crossing of $N_M = 0$ and $N_B = 2$ LLs, see Fig. 6.8d. One possible explanation for the spike in $R_{xx}$ [96] is the edge state transport along domain wall boundaries as studied earlier in semiconductors [81, 97]. The hysteresis in $R_{xx}$ comes due to the underlying QHF states [95]. We do not see any pronounced hysteresis for the filling factor $\nu > 5$ probably because those states are not polarized or the LL gaps are small so that the Coulomb interaction is not enough to stabilize the QHF states.

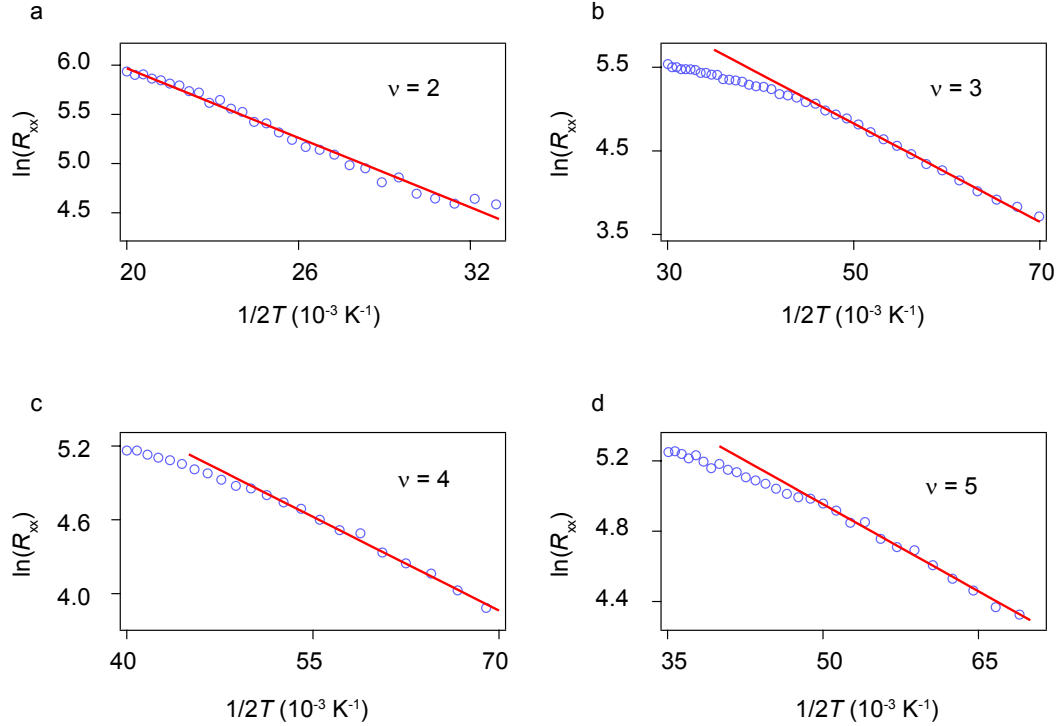

Figure 6.7: **Activation gap from the temperature dependence of the SdH oscillations.** a) For $\nu = 2$. b) For $\nu = 3$. c) For $\nu = 4$. d) For $\nu = 5$.

In Fig. 6.8a, hysteresis data is shown for a high field (13.5 T) which indicates the pseudospin ordering in the QHF states. Fig. 6.9a shows the hysteresis in $R_{xx}$ across the symmetry broken $N_M = 0$ LL at a lower magnetic field 3.8 T. Even at such a low magnetic field $N_M = 0$ LL is complete symmetry broken and gives rise to these QHF states. Fig. 6.9c shows the hysteresis at 13.5 T measured at 15 K. Beyond this temperature, LLs are not resolved properly. Fig. 6.9d shows resistance recorded at zero magnetic field that shows no hysteresis with gate sweep. It implies that hysteresis near the QHF states comes from the pseudospin order, and not from the charge traps in $SiO_2/Si^{++}$ substrate.

Table 6.2: Extracted Landau level energy gaps from Arrhenius plots.

| $\nu$ | $\Delta$ at 13.5 T (meV) |
|---|---|
| 2 | $10.14 \pm 0.02$ |
| 3 | $5.1 \pm 0.1$ |
| 4 | $4.37 \pm 0.04$ |
| 5 | $2.8 \pm 0.1$ |

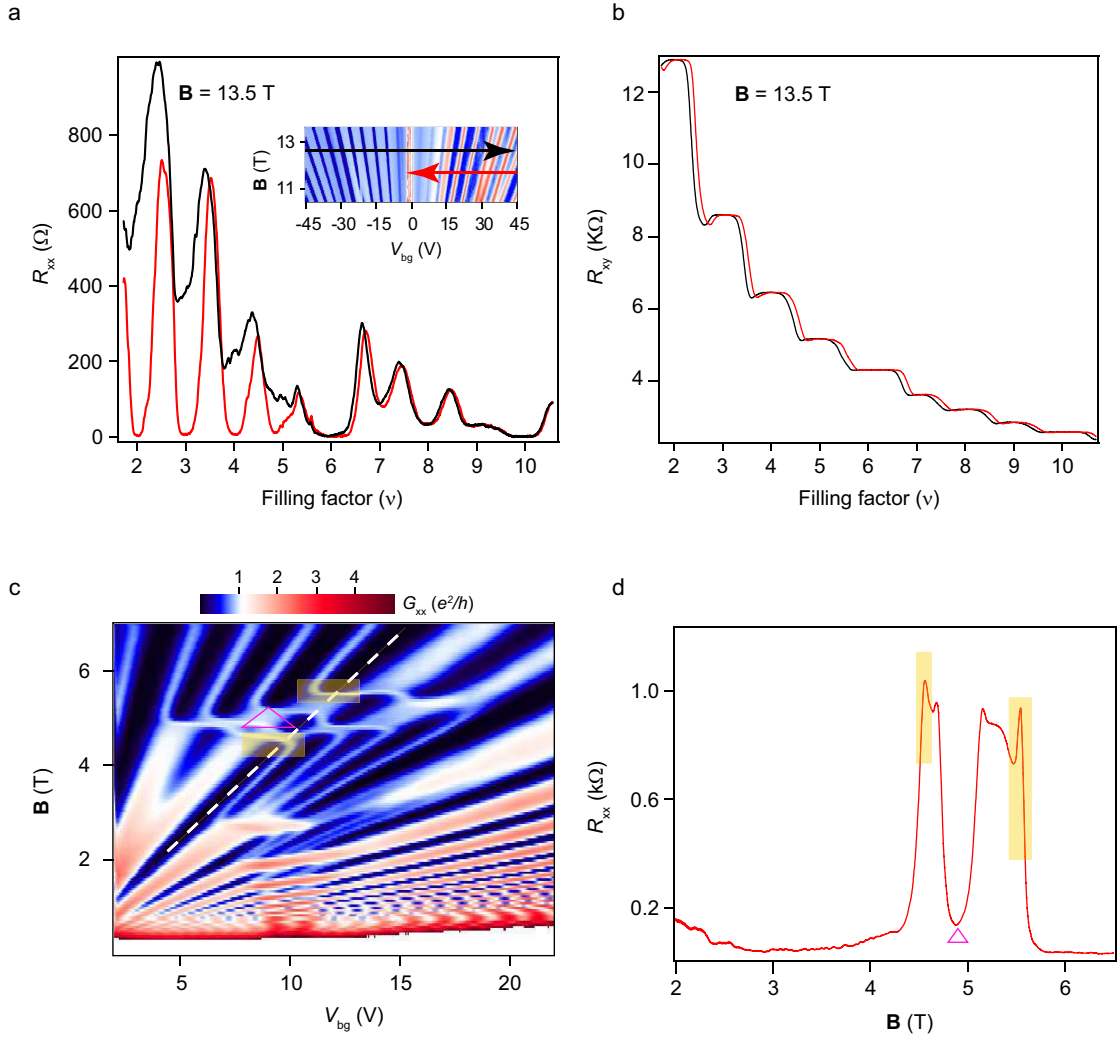

Figure 6.8: **Hysteresis in the longitudinal resistance as a function of the filling factor and observation of resistance spikes.** a) Measurement of $R_{xx}$ as a function of $V_{bg}$ at $\mathbf{B}$ = 13.5 T in the two directions as shown in (inset) the measurement parameter space. Largest hysteresis is seen for the spin and valley-polarized $N_M = 0$ LL. We have done gate sweep as slow as 3 mVs$^{-1}$ to check if the hysteresis goes away, however, it stays. Nature of hysteresis does not change. The sweep up and sweep down rates were the same. b) Simultaneously measured $R_{xy}$ that exhibits clear quantization plateaus in the two sweep directions. c) The overlaid white dashed line on the fan diagram shows the parameter space along which the $R_{xx}$ is plotted in the next panel. d) $R_{xx}$ plotted along the dashed line shown in the parameter space. Spikes in resistance, shaded in yellow, correspond to boundaries of the region marked △ in Fig.6.8c.

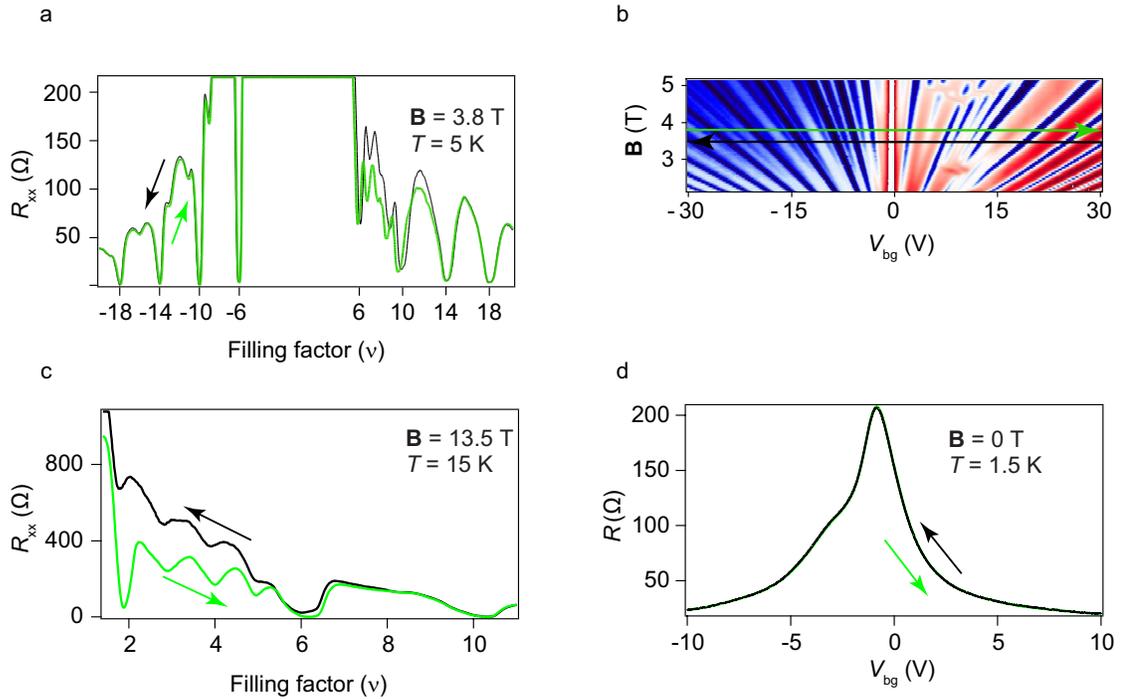

Figure 6.9: **More hysteresis measurements.** a) Hysteresis due to $N_M = 0$ LL at 3.8 T
magnetic field. Green curve shows forward sweep and black curve shows reverse sweep in all
the data. b) Parameter space of the hysteresis study at 3.8 T. c) Hysteresis study at 13.5 T
and at 15 K temperature. d) $V_{bg}$ sweep in forward and reverse direction without magnetic
field, showing no hysteresis.

## 6.7 Theoretical calculations

The Landau levels of trilayer graphene Hamiltonian get broadened in presence of scattering and the density of states of each Landau level (with energy $E$) can be approximated by a Lorentzian

$$\text{DOS}(N, E) = \frac{1}{2\pi l_B^2} \frac{1}{\pi} \left( \frac{\Gamma}{(E - E_N)^2 + \Gamma^2} \right),\qquad(6.3)$$

where $l_B$ is the magnetic length and the effect of disorder strength ($\Gamma$=1.5 meV) is estimated from the Dingle plot, see section **??**.

We calculate the density of states at Fermi energy ($\text{DOS}(E_\text{F})$) as a function of total density $n$ and $B$ to match the experimental data. Density (n) is calculated integrating total $\text{DOS}(E)$ up to Fermi energy $E_\text{F}$):

$$\text{DOS}(E) = \sum_N \text{DOS}(N, E) \quad \text{and} \quad n = \int_{E_0}^{E_F} \text{DOS}(E)\, dE,\qquad(6.4)$$

where $E_0$ is the overall charge neutrality point where experimental Hall conductance is zero.

Fig. 6.10b shows the calculated non-interacting density of states (DOS) in the same parameter range which matches very well with the measured resistance shown in Fig. 6.10a. We find that the low **B** data can be well understood in terms of non-interacting picture and it allows determination of the band parameters.

Now we discuss theoretical calculations to show that electronic interactions are crucial in obtaining a quantitative understanding of the experimental data. The theoretical calculations focus on the $N_\text{M} = 0$ and $N_\text{B} = 2$ LLs, which form the most prominent LL crossing pattern in our data. The effect of disorder is incorporated within a self-consistent Born approximation (SCBA) [98, 99], while electronic interactions are included by considering the exchange corrections to the LL spectrum due to a statically screened Coulomb interaction [100, 101] in a self-consistent way.

The Hamiltonian of the ABA trilayer graphene which is symmetric under the exchange of the two outermost layers (i.e. no electric field between the layers), decouples into a monolayer graphene-like block consisting of co-ordinates antisymmetric under exchange of the outer layers (i.e. $\left( \frac{A_1 - A_3}{\sqrt{2}}, \frac{B_1 - B_3}{\sqrt{2}} \right)$) and a bilayer graphene-like block with the low energy co-ordinates $\left(\frac{A_1+A_3}{\sqrt{2}}, B_2\right)$ [14, 102]. Here $A(B)$ denotes the sublattice and $1, 2, 3$ denote the layer indices. In a perpendicular magnetic field, the energy of the monolayer graphene-like and bilayer graphene-like LLs are given by

$$\epsilon_{N\tau}^{\mathrm{m}} = a + \tau\sqrt{b^2 + \frac{2\hbar^2 v^2}{l_B^2}N} \tag{6.5}$$

$$\epsilon_{N\tau}^{\mathrm{b}} = c + \frac{\hbar^2 v^2}{2\gamma_1^2 l_B^2}(2Nd - d - \tau f) + \sqrt{\left[\left(c + \frac{\hbar^2 v^2}{2\gamma_1^2 l_B^2}(2Nf - \tau d - f)\right)^2 + \left(\frac{\hbar^2}{ml_B^2}\sqrt{N(N-1)}\right)^2\right]}, \tag{6.6}$$

where $a = \left(\frac{\delta}{2} - \frac{\gamma_2}{4} - \frac{\gamma_5}{4}\right)$, $b = \left(-\frac{\delta}{2} - \frac{\gamma_2}{4} + \frac{\gamma_5}{4}\right)$, $c = \frac{\gamma_2}{4}$, $d = (\delta + \frac{\gamma_5}{4})$, and $f = -\frac{\gamma_5}{4}$, with $l_B = \sqrt{\frac{\hbar}{eB}}$, $\frac{1}{m} = \frac{\sqrt{2}v^2}{\gamma_1}$. Here $v$ is the graphene Fermi velocity and $\gamma_i, \delta$ are the band parameters described before. For the $N = 0$ monolayer graphene-like LL, the energies are given by $\epsilon_{0\pm}^{\mathrm{m}} = a \pm b$, leading to a valley splitting of these levels $\sim 2b = 2 meV$ with the chosen band parameters. In terms of the non-relativistic landau level orbitals $|N\rangle$ at a given guiding center position, the eigenfunctions for the $N = 0$ LL are given by [14]

$$\Phi_{0+}^{\mathrm{m}} = \begin{pmatrix} 0 \\ |0\rangle \end{pmatrix} \quad \text{and} \quad \Phi_{0-}^{\mathrm{m}} = \begin{pmatrix} |0\rangle \\ 0 \end{pmatrix} \tag{6.7}$$

and the eigenfunctions for the bilayer graphene-like LLs are given by

$$\Phi_{N+}^{\mathrm{b}} = \begin{pmatrix} u_{N+}|N-2\rangle \\ v_{N+}|N\rangle \end{pmatrix} \quad \text{and} \quad \Phi_{N-}^{\mathrm{b}} = \begin{pmatrix} u_{N-}|N\rangle \\ v_{N-}|N-2\rangle \end{pmatrix}, \tag{6.8}$$

where $u_{N\tau}$ and $v_{N\tau}$ are given by $u_{N\tau}^2 = \frac{1}{2}\left(1 + \frac{c + \frac{\hbar^2 v^2}{2\gamma_1^2 l_B^2}(2Nf - \tau d - f)}{\sqrt{\left(c + \frac{\hbar^2 v^2}{2\gamma_1^2 l_B^2}(2Nf - \tau d - f)\right)^2 + \left(\frac{\hbar^2}{ml_B^2}\sqrt{N(N-1)}\right)^2}}\right)$ and $v_{N\tau}^2 = 1 - u_{N\tau}^2$.

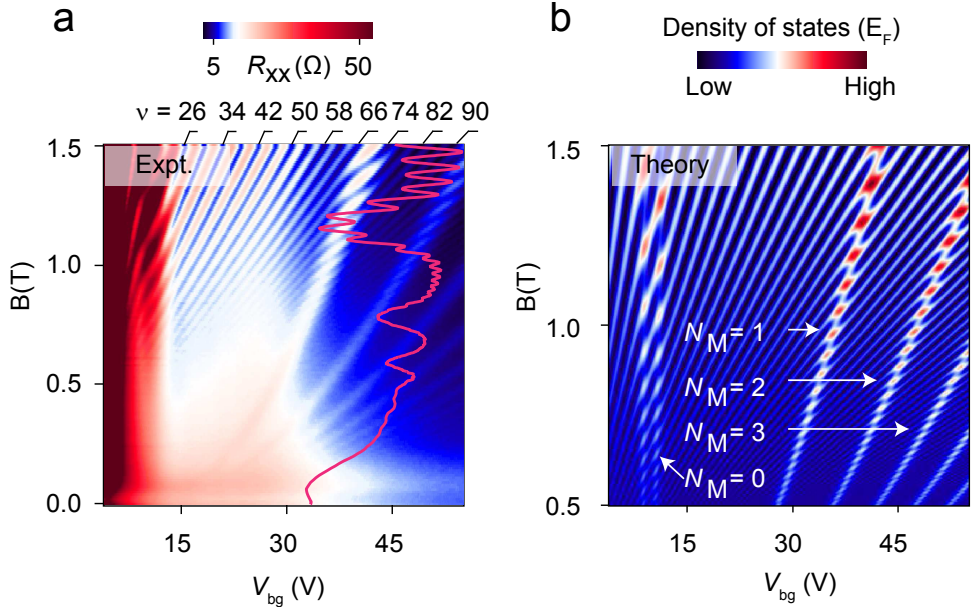

Figure 6.10: **Low magnetic field fan diagram.** a) Colour plot of $R_{xx}$ as a function of $V_{bg}$ and **B** up to 1.5 T. The LL crossings arising from MLG-like band and BLG-like bands are clearly seen. Each parabola is formed by the repetitive crossings of a particular MLG-like LL with other BLG-like LLs. Crossing between any two LLs shows up as $R_{xx}$ maxima in transport measurement due to high density of states at the crossing points. The overlaid magenta line shows a line slice at $V_{bg} = 50$ V. b) DOS corresponding to panel (a). $N_M = 1$ labelled parabola refer to all the crossing points arising from the crossings of $N_M = 1$ LL with other BLG-like LLs. Other labels have a similar meaning. $N_M = 0$ LL does not disperse with **B**, hence the crossings form a straight line parallel to **B** axis. Minimum **B** is taken as 0.5 T to keep finite number of LLs in the calculation. The horizontal axis is converted to an equivalent $V_{bg}$ after normalizing it by the capacitance per unit area ($C_{bg}$) for the ease of comparison with experimental fan diagram. $C_{bg}$ is determined from the high **B** quantum Hall data which matches well with the geometrical capacitance per unit area of 30 nm hBN and 300 nm SiO$_2$: $C_{bg} \sim 105$ μFm$^{-2}$.

We consider the effects of disorder and electronic interactions in the following way. First we assume that the exchange correction leads to modified energy levels $E_{N\sigma\tau}^{\mathrm{m(b)}} = \epsilon_{N\tau}^{\mathrm{m(b)}} + g_0\mu_B B\sigma + \Delta_{N\sigma\tau}^{\mathrm{m(b)}}$, where $\Delta$ is the exchange self-energy. We note that we use the same 2D bare Coulomb potential $V(q) = 2\pi e^2/\kappa q$ for interaction between electrons in
different layers, where we use $\kappa = 4$, corresponding to the dielectric constant of hBN,
since the sample has layers of hBN on either side of the TLG. This neglects variation of
the potential on a scale of layer separation $\sim 3-4$ Å, and is justified when the screening
length is much larger than this scale. This choice, together with the layer exchange
symmetry of the trilayer, keeps the bilayer graphene-like and monolayer graphene-like
levels decoupled even in the presence of electronic interactions. We have checked that
there is no signature of spontaneous breaking of this symmetry due to interactions.
In reality, keeping the $z$ dependence of Coulomb interactions would lead to coupling
of the bilayer graphene-like and monolayer graphene-like LLs, and the effects would
be strongest when the LLs cross. However, precisely in this regime, the large density
of states lead to strong screening and neglecting the $z$ dependence of the Coulomb
interaction is justified. We then use self-consistent Born approximation (SCBA) to
construct the disorder broadened single particle Green's functions [100]

$$G_{N\sigma\tau}^{\mathrm{m(b)}}(\omega) = \frac{2}{\omega - E_{N\sigma\tau}^{\mathrm{m(b)}} + i\sqrt{\Gamma^2 - (\omega - E_{N\sigma\tau}^{\mathrm{m(b)}})^2}} \qquad (6.9)$$

with the corresponding parabolic density of states, $\rho_{N\sigma\tau}(\omega) = \frac{1}{2\pi l_B^2} \frac{2}{\pi\Gamma} \sqrt{1 - (\omega - E_{N\sigma\tau}^{\mathrm{m(b)}})^2/\Gamma^2}$,
when $|\omega - E_{N\sigma\tau}^{\mathrm{m(b)}}| < \Gamma$. The disorder broadening $\Gamma$ is used as an input in our theory.
We use $\Gamma = 1$ meV roughly similar with experimentally estimated $\Gamma \sim 1.5$ meV at $V_{\mathrm{bg}}$
= 23 V. The exchange energy is then given by

$$\Delta_{N\sigma\tau}^{\mathrm{m(b)}} = - \int \frac{d^2 q V(q)}{\varepsilon(q)} A_{N\tau}^{\mathrm{m(b)}}(q) \rho_{N\sigma\tau} \,, \qquad (6.10)$$

where $\rho_{N\sigma\tau}$ is the electronic density in the corresponding state and the matrix el-
ements $A_{N\tau}^{\mathrm{m(b)}}(q) = \left| \int d^2 r \Phi_{N\tau}^{\mathrm{m(b)}\dagger}(r) . (e^{i\mathbf{q}.\mathbf{r}}) \Phi_{N\tau}^{\mathrm{m(b)}}(r) \right|^2$. The dielectric function $\varepsilon(q) =$
$1 + V(q)\Pi(q)$, where the bare polarizability function $\Pi$ is given by

$$\Pi(q) = -\frac{1}{\beta} \sum_{i\omega} \sum_{\substack{N\sigma\tau \\ \{\mathrm{m,b}\}}} G_{N\sigma\tau}^{\mathrm{m(b)}}(i\omega) G_{N\sigma\tau}^{\mathrm{m(b)}}(i\omega) A_{N\tau}^{\mathrm{m(b)}}(q). \qquad (6.11)$$

Here we have neglected inter-LL couplings in calculating the static polarizability func-
tion. With these approximations, the density in each level is calculated self-consistently.

Fig. 6.11a shows the DOS at $E_{\mathrm{F}}$ as a function of $V_{\mathrm{bg}}$ and $\mathbf{B}$ where $V_{\mathrm{bg}}$ is computed by dividing the density by the capacitance of the device. This matches with the experimental results on the $G_{xx}$. Our calculations also provide insight about the polarization of the states inside the ring-like structures in Fig. 6.4f. We find that although the filling factor of region $\triangle$ is the same as that of regions $\nu = 6$ above and below, electronic configurations of these states are different. Fig. 6.11b shows the spin-resolved DOS at $E_F$ as a function of $V_{bg}$ and $\mathbf{B}$. Fig. 6.11b inset shows the calculated exchange enhanced spin $g$-factors. This shows a significant increase over the free electron value of $g$ in the spin-polarized states – in agreement with the large gap observed at $\nu = 3$ and 5 in the experiment. Fig. 6.11c shows the total spin polarization (integrated spin DOS) up to the Fermi energy. We find total spin polarization in region $\triangle$ is non-zero (see Fig. 6.11c), but it vanishes in the $\nu = 6$ regions above and below the ring structure.

The interacting part of the theoretical calculation was done in collaboration with Prof. Rajdeep Sensarma, his students Santanu Dey and Abhisek Samanta.

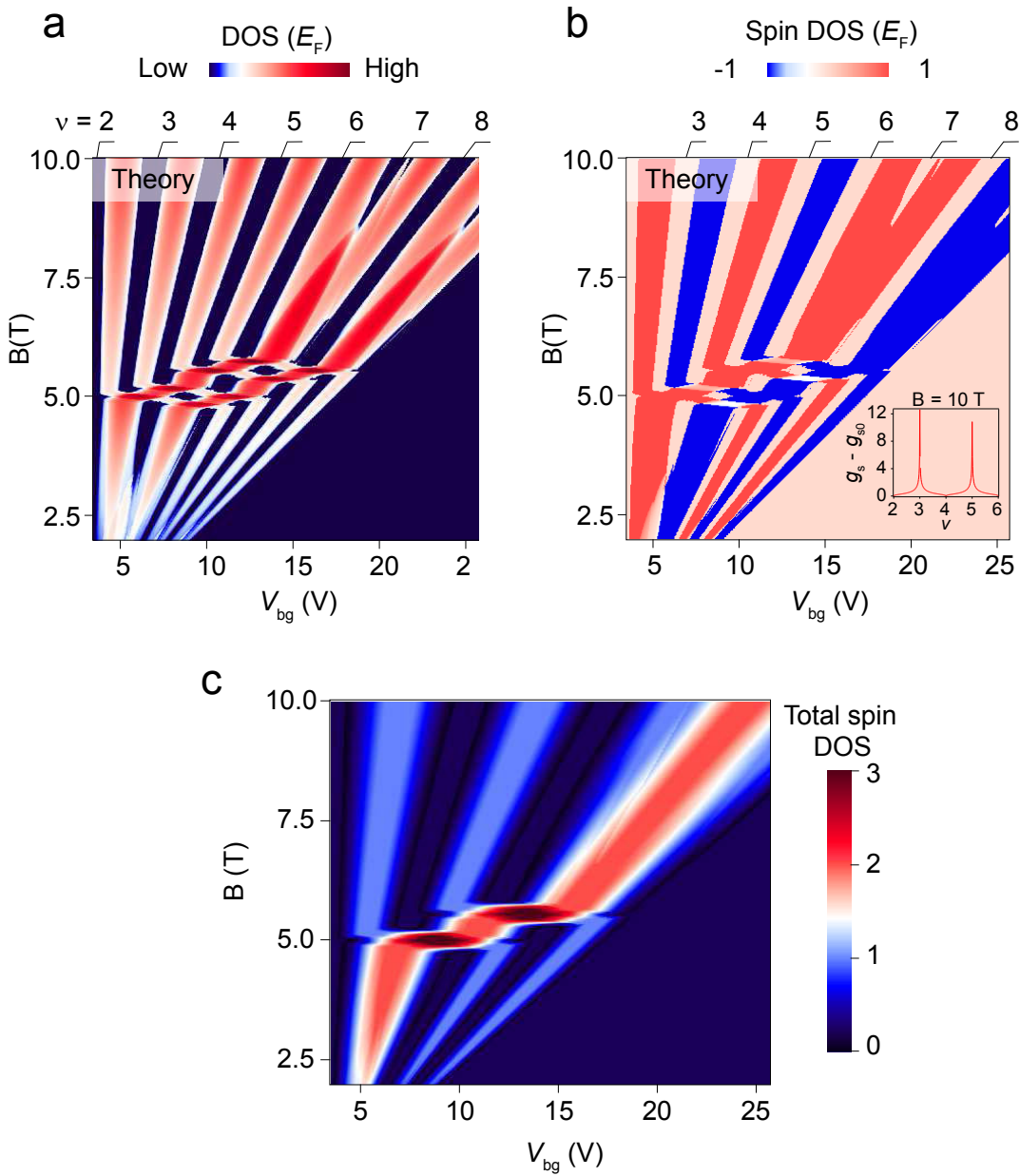

Figure 6.11: **Theoretical calculation of DOS and spin-polarization taking electronic interactions into account.** a) DOS at the Fermi level as a function of $V_{bg}$ and **B**. This matches the fan diagram seen in the experiment. b) The magnetization in the system as a function of $V_{bg}$ and **B**, where density is converted to an equivalent $V_{bg}$. The inset shows calculated enhanced spin $g$-factor above the free electron value 2 for $N_M = 0$ spin and valley split LLs. c) Total spin polarization up to the Fermi energy as a function of $V_{bg}$ and **B**.

## 6.8   Summary

In summary, we have studied the Landau level crossing physics of ABA trilayer graphene
in detail. We see that the Landau levels from the monolayer graphene-like band and
the bilayer graphene-like band cross each other without getting hybridized – a con-
sequence of mirror symmetry at zero electric field. We also see that the interaction
plays an important role to enhance the $g$-factor and favors the formation of QHF states
at low **B** and at a relatively higher temperature. ABA-TLG is the simplest system
that has both massless and massive Dirac fermions, giving rise to an intricate and rich
pattern of LLs that, through their crossings, can allow a detailed study of the effect of
interaction at sufficiently low temperature.

# Landau level sequence and the continuous rotational symmetry breaking in trilayer graphene

In this chapter, we study the Landau level (LL) crossings in mirror symmetry broken Bernal (ABA)-stacked trilayer graphene (TLG). Studying LL crossings as a function of a magnetic field, electric field, and density is a useful way to extract many quantitative details of the system. In particular, we focus on the following two different problems. First, we study the effect of the trigonal warping of the Fermi surface. Specifically, we study how the Landau levels interact with themselves in the presence of a trigonal warping term in the Hamiltonian. Theoretically, it is shown that it leads to a selection rule for the coupling between different Landau levels differing by multiple of 3 in LL index. Experimentally, we observe an anticrossing between two LLs differing by 6 in LL index which is a direct consequence of the broken continuous rotational symmetry to $C_3$ symmetry by the trigonal warping. Second, we study the effect of non-uniform charge distribution in the ABA-stacked TLG. It is theoretically predicted that the charge distribution on the three layers of the ABA-stacked TLG can be non-uniform and this has an important effect on the substructure of the lowest Landau level. In particular, the sequence of the zeroth Landau levels between filling factors -6 to 6 in ABA-stacked trilayer graphene is unknown because it depends sensitively on the non-uniform charge distribution. Using the sensitivity of quantum Hall (QH) data on the electric field and magnetic field, in an ultraclean ABA-stacked TLG sample, we quantitatively estimate the non-uniformity of the electric field and determine the sequence of the zeroth LLs.

## 7.1  Introduction

There is an increasing interest in the electronic properties of few-layer graphene as the dispersion of the bands can be tuned with the number and stacking of the layers in combination with an external electric field. The two variants, namely Bernal (ABA) [12, 16, 70, 72, 74, 76, 85, 103, 104, 105, 106, 107] and rhombohedral (ABC) [108, 109, 110, 111, 112, 113, 114, 115] stackings have been studied for their tunable symmetries. ABA-stacked TLG has a rich low energy band-structure consisting of a monolayer graphene (MLG)-like linear and a bilayer graphene (BLG)-like quadratic bands [14, 27]. Further, the mirror symmetry of ABA-stacked TLG inhibits any coupling between the linear and quadratic bands [14, 73]. In the presence of a transverse electric field ($E^\perp$), however, the mirror symmetry is broken, resulting in the mixing of wavefunctions from two subbands, and thus providing a tunable knob to probe the interesting physics resulting from the mixing of wavefunctions from the linear and quadratic bands [14, 27].

Fig. 7.1a shows an optical image of the dual-gated hexagonal boron nitride sandwiched ABA-stacked TLG device. The fabrication process is discussed in Chapter 4 and is similar to the earlier published methods [41, 42, 116]. The potential energy of each layer of the biased TLG [27] is shown in the schematic Fig. 7.1b. The energy difference between the top and the bottom layer is $2\Delta_1$. Theoretically, an average $E^\perp$ inside the TLG can be defined as $E_{\mathrm{av}}^\perp = \frac{2\Delta_1}{(e)d}$ where d=0.67 nm is the separation between the top and the bottom layer of the TLG. It turns out that the charge distribution in three layers of ABA-stacked TLG is not uniform which gives rise to a difference in the middle layer potential from the average of the top and bottom layers; this difference is defined to be proportional to a parameter $\Delta_2$. Non-zero $\Delta_2$ also generates a non-uniform electric field perpendicular to the layers. In ABA-stacked TLG low energy electronic states are centered on $A_1$-$A_3$, $B_1$-$B_3$ atomic sites (for MLG-like bands) and on $A_1$+$A_3$, $B_2$ atomic sites (for BLG-like bands) [14]. So, electrons primarily prefer to lie on the outer layer atoms to decrease the Coulomb energy by maximizing their spatial separation. This can lead to a positive potential (hence a positive $\Delta_2$) in the middle layer.

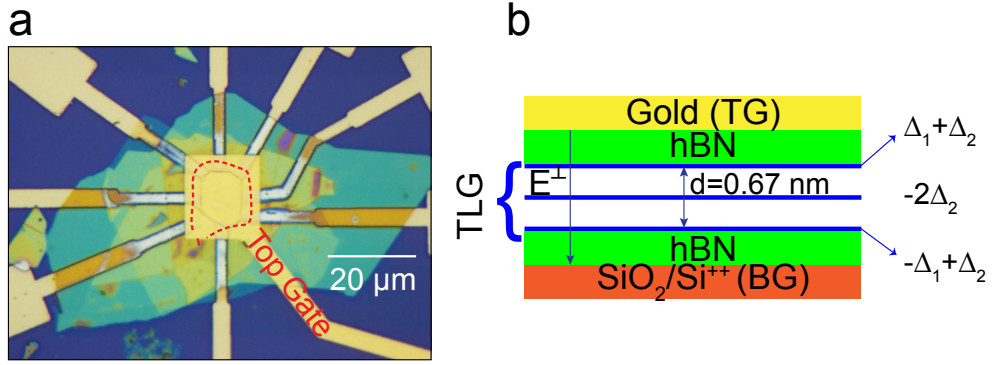

Figure 7.1: **Device image and schematic.** (a) Optical image of the device on 300 nm SiO$_2$/Si$^{++}$ substrate after the top gate fabrication. Multiple hBN (greenish in color) transferred during the fabrication are visible. The red dotted line shows the boundary of the TLG. (b) A schematic of the device showing the potential energy distribution across the three layers of the TLG in the presence of an external $E^\perp$. Non-uniform charge distribution leads to a positive potential in the middle layer which decreases its electrostatic energy.

## 7.2 Field effect mobility and quantum mobility

While studying LLs it is important to understand the difference between the field effect mobility and the quantum mobility. Field effect mobility is mostly insensitive to small angle scattering and mainly depends on the large angle scattering/ back scattering. On the other hand LL quantum mobility which determines the LL broadening is also sensitive to small angle scattering. Field effect mobility and quantum mobility can be an order of magnitude different and they have different physical significance. Here we estimate both the mobilities of our device. Fig. 7.2a shows the colour plot of four-probe resistance as a function of top-gate and back-gate voltages. The yellow and green arrows designate the direction of electric field and density respectively. $E^\perp$ and $n$ can be estimated from the experiment by $E^\perp = E_{bg}^\perp + E_{tg}^\perp = \frac{C_{bg} V_{bg}}{2\kappa_{bg}\epsilon_0} - \frac{C_{tg} V_{tg}}{2\kappa_{tg}\epsilon_0}$ and $n = \frac{1}{e}[C_{bg} V_{bg} + C_{tg} V_{tg}]$ where $V_{bg}$ ($V_{tg}$) is the back (top)-gate voltage, $C_{bg}$ ($C_{tg}$) is the capacitance per unit area of the back (top)-gate, $\kappa_{bg}$ ($\kappa_{tg}$) is the dielectric constant of back (top)-gate dielectric and $\epsilon_0$ is the vacuum permittivity. $C_{bg}$ and $C_{tg}$ are extracted from the QH data: $C_{bg}$=104 µFm$^{-2}$ and $\frac{C_{tg}}{C_{bg}}$=4. The sharp gating curve (Fig. 7.2b) is a typical for high mobility devices. The Drude mobility ($\mu_d = \sigma/ne$) and the field effect mobility ($\mu_f = \frac{1}{C} \frac{L}{W} \frac{dG}{dV_G}$) of our device are very high ($\sim$800,000 cm$^2$V$^{-1}$s$^{-1}$) which can be calculated from the field effect gating measurement data (Fig. 7.2); here C is the capacitance per unit area between the graphene and the gate electrode, L & W are the length and breadth of the device, G is the conductance, $dG/dV_G$ is the slope of the linear part of the gating curve. All the measurements have been done in a He-3 cryostat at 300 mK unless stated otherwise.

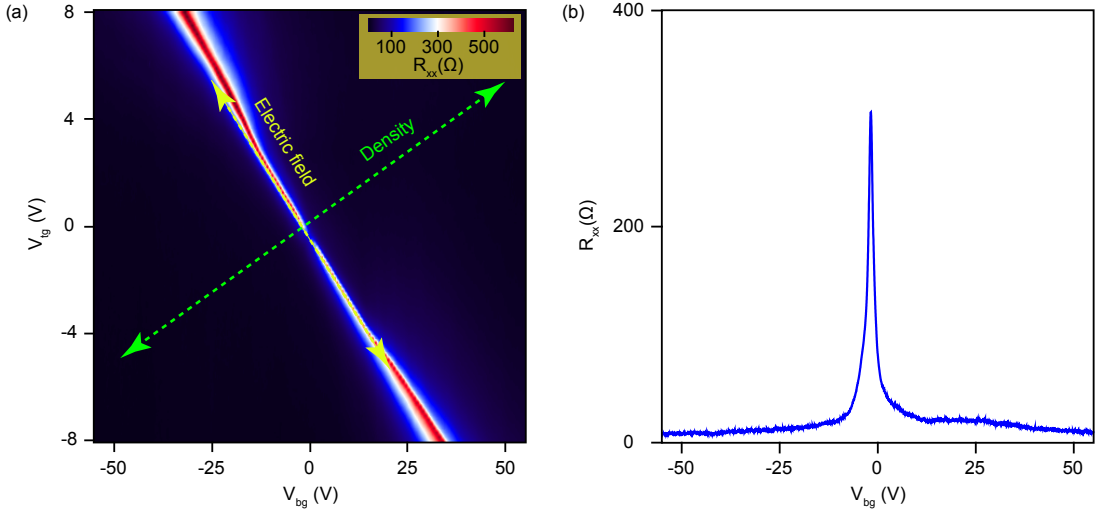

Figure 7.2: **Drude/field effect Mobility from four-probe gating.** (a) Colour plot of four-probe resistance at 0.3 K. (b) Line slice of the colour plot at $V_{tg}$=0 V.

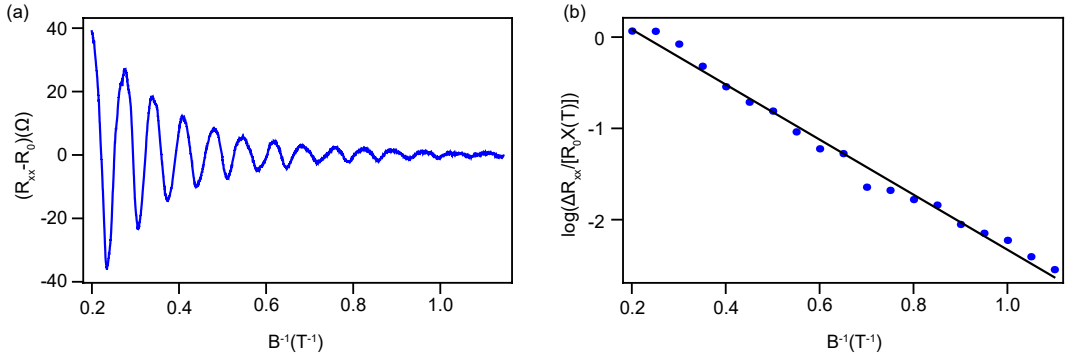

Figure 7.3: **Determination of quantum mobility.** (a) SdH oscillations after subtracting the non-oscillatory background. (b) Straight-line fit in the Dingle plot.

We estimate LL broadening ($\Gamma$) and quantum mobility ($\mu_q$) from the magnetic field dependence of the SdH oscillations [87, 88, 89]. Single particle quantum broadening ($\Gamma$) of the LLs is related to the quantum scattering time ($\tau_q$) by $\Gamma = \frac{\hbar}{2\tau_q}$. Amplitude of the SdH oscillations is given by [87, 88] $\frac{\Delta R}{R_0} = 4X(T)\exp(-\pi/\omega_c\tau_q)$ where $\omega_c = \frac{eB}{m^*}$ is the cyclotron frequency. $\Delta R$ and $R_0$ are the oscillatory and non-oscillatory part of the resistance respectively. $X(T)$ is the temperature dependent amplitude $X(T) = \frac{2\pi^2 k_B T/\hbar\omega_c}{sinh(2\pi^2 k_B T/\hbar\omega_c)}$. Fig. 7.3a shows SdH oscillations at density $1.5 \times 10^{12}$ cm$^{-2}$ after subtracting the non-oscillatory background resistance. Fig. 7.3b shows the so called Dingle plot. Slope of the fitted line is given by $-\pi m^*/e\tau_q$ which gives the quantum scattering time $\tau_q \approx 287$ fs. This corresponds to a disorder energy (LL broadening) scale $\Gamma \sim 1.1$ meV. The quantum scattering time can be related to the quantum mobility $\mu_q = \frac{e\tau_q}{m^*} \approx 10,000$ cm$^2$V$^{-1}$s$^{-1}$. SdH oscillations should be visible at a magnetic field $B_c$ for which $\mu_q B_c \approx 1$; this gives $B_c \approx 1$ T. This is consistent with our observation and affirms that appearance of SdH oscillation depends on quantum mobility instead of the transport mobility.

## 7.3 Electric field tunable hybridization between the linear and quadratic bands

One of the most interesting aspects of few-layer graphene is the tunability of different symmetries. In ABA-TLG a perpendicular electric field can controllably break the mirror symmetry which in turn controls the hybridization between the MLG-like and the BLG-like bands. Fig. 7.4 shows the calculated band structure evolution of ABA-stacked TLG along $k_x$ direction with electric field. It is clear that increasing electric field hybridizes the bands. Hybridization also changes the gap between the tip of the two cones. We notice that for small electric fields (up to $\Delta_1$=5 meV in Fig. 7.4) the band gap between the cones decreases with increasing electric field. Further increment of the electric field leads to the enhancement of the band gap. In QH regime $0_M^+$ and $0_M^-$ LLs originate from the bottom of the MLG-like conduction band and the top of the MLG-like valence bands respectively, see Fig. 7.5. So, the valley splitting of $0_M$ LL is also the band gap of MLG-like bands at a zero magnetic field and hence the valley splitting follows the same electric field dependence as the band gap.

In order to understand our experiments, we calculate LL spectrum of the system using the tight binding model. At zero electric field LLs of the ABA-stacked TLG can be classified into MLG-like and BLG-like LLs [14], see Fig. 7.5b. Each type of LLs can be labelled by a composite quantum number $N_S^{v\sigma}$ consisting of an integer index $N$, a subband index S (M,B) for MLG-like and BLG-like bands, a valley index v (+,-) for $K_+$ and $K_-$ valleys and a spin index $\sigma$ ($\uparrow,\downarrow$) for up and down spins respectively. LLs without spin or valley index implies that spin or valley is degenerate. LLs from the MLG-like bands disperse with the magnetic field as $\sim \sqrt{B}$ whereas LLs from BLG-like bands disperse as $\sim B$.

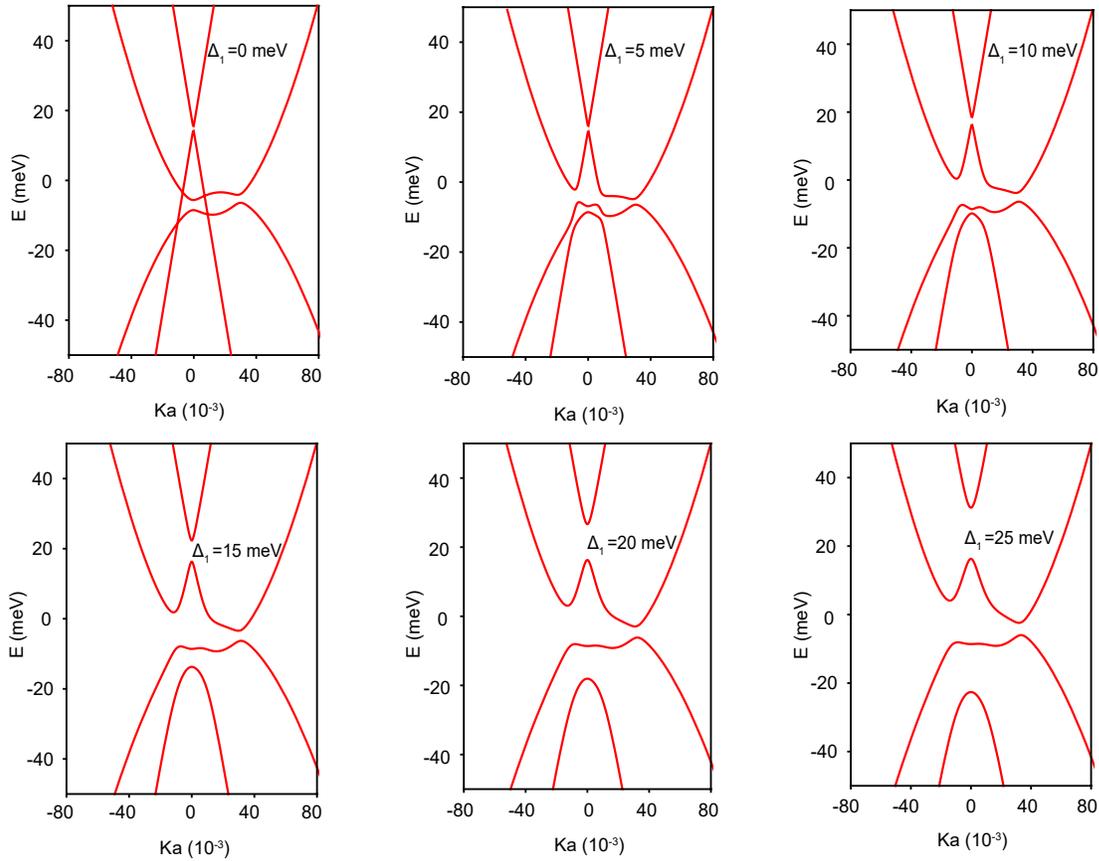

Figure 7.4:  Band structure evolution with electric field of ABA-stacked trilayer graphene along $k_x$ direction around $K_+$ valley. We use the following band parameters: $\gamma_0$=3.1 eV, $\gamma_1$=390 meV, $\gamma_2$=-20 meV, $\gamma_3$=315 meV, $\gamma_4$=120 meV, $\gamma_5$=18 meV, $\delta$=20 meV and $\Delta_2$=4.3 meV. Smaller band parameters are determined by matching the experimental LL crossings with theory at 1.5 T.

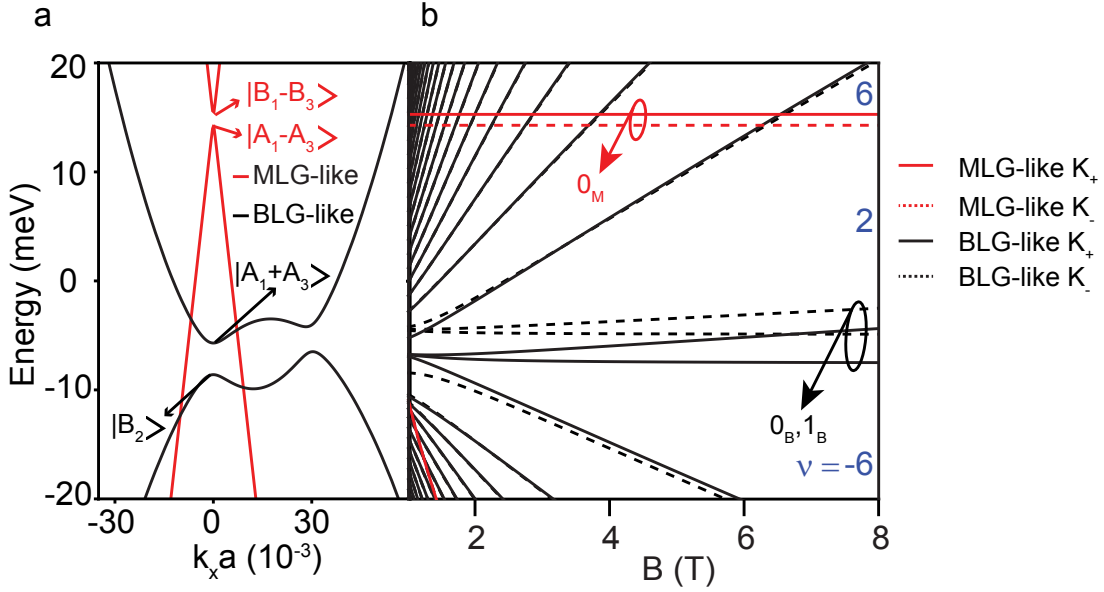

**Figure 7.5: Band structure and Landau level diagram as a function of the magnetic field.** (a) Band structure of ABA-stacked TLG at zero electric field around $K_+$ point in the Brillouin zone. Red and black curves show MLG-like and BLG-like bands respectively. MLG-like bands are spanned by $|A_1 - A_3\rangle$ and $|B_1 - B_3\rangle$ bases. BLG-like bands are spanned by $|A_1 + A_3\rangle$ and $|B_2\rangle$ bases. Wavefunctions of each band at $(k_x, k_y) = (0,0)$ are polarized on different mirror symmetric and antisymmetric basis as labeled in the Fig. (b) LL diagram as a function of the magnetic field. Spin is kept degenerate in this diagram to avoid clutter. Red and black lines denote the MLG-like and BLG-like LLs respectively. Solid and dashed lines denote LLs from $K_+$ and $K_-$ valleys respectively.

We simulate the LLs also as a function of the electric field which can be used to understand our experimental data of LL crossings at a non-zero electric field. Fig. 7.6 shows that the LLs disperse with an electric field. The sensitivity of the LLs on the electric field depends on the symmetry of the LLs and how the electric field affects the symmetry. For instance, electric field breaks the mirror symmetry of the lattice, so finite electric field hybridizes mirror-symmetric states (BLG-like states) with mirror anti-symmetric states (MLG-like states). We know that for a system with two non-degenerate states the hybridization gap between the states depends on the coupling term ($\Delta$) and energy separation of the states before hybridization ($\delta E$) as $\Delta^2/\delta E$. So, closer the two states in energy, better the coupling between two LLs. Moreover, since the electric field term does not appear with any creation or annihilation operator in the Hamiltonian, it couples two states (within the same valley) with same Landau level index.

We can take a few specific examples to understand the nature of coupling due to the electric field. Numerically calculated LL energy diagram (see Fig. 7.6) shows that
$0_B^-$ and $0_M^-$ LLs are very sensitive to the electric field which can be understood in
the following way. We note that wave function of $0_B^-$ and $0_M^-$ LLs are localized at
$|A_1 + A_3\rangle$ and $|A_1 - A_3\rangle$ bases respectively. Since $|A_1 + A_3\rangle$ and $|A_1 - A_3\rangle$ are mirror
symmetric and mirror anti-symmetric respectively, they can be coupled by the electric
field. Moreover, $0_B^-$ and $0_M^-$ LLs have same LL index and these states are close in
energy. These lead to a strong hybridization between $0_B^-$ and $0_M^-$ LLs for which the
hybridization gap steadily increases with increasing $\Delta$. On the other hand, Fig. 7.6b
shows that $0_M^+$ LL is almost insensitive to the electric field. To explain this, we note
that $0_M^+$ LL originates from the MLG-like conduction band edge and its wavefunction
is localized at $|B_1 - B_3\rangle$. Mirror symmetric (BLG-like) partner of $|B_1 - B_3\rangle$ basis state
is $|B_1 + B_3\rangle$. However, we recall that $|B_1 + B_3\rangle$ hosts high energy states – the high
energy BLG-like bands are spanned by $|B_1 + B_3\rangle$ and $|A_2\rangle$ bases. Due to very large
energy separation ($\sim 0.5~eV$) between the MLG-like conduction band edge and the
high energy BLG-like band, the coupling between $|B_1 - B_3\rangle$ and $|B_1 + B_3\rangle$ due to the
electric field is minimal which explains the insensitivity of the $0_M^+$ LL on the electric
field. It is also important to note that the $0_M^+$ LL does not get hybridized to the $0_B^+$
LL (unlike the $0_M^-$ LL gets hybridized with the $0_B^-$ LL) because $0_M^+$ LL is spanned
by $|B_1 - B_3\rangle$ whereas $0_B^+$ LL is spanned by $|B_2\rangle$ – which is not the mirror symmetric
partner of $|B_1 - B_3\rangle$.

With a detailed understanding of the role of the electric field we now focus on the
experimental data. $0_M^-$ and the $0_B^-$ LLs play an important role in our experimental
data since we study low energy physics. We measure the longitudinal conductance
($G_{xx}$) as a function of $E^\perp$ & density for different values of the magnetic field and
encounter LL crossings or anticrossings depending on the symmetries of the underlying
LLs. Fig. 7.7a shows experimentally measured $G_{xx}$ plot as a function of back gate and
top gate voltages. At such low $B$, LLs are not fully resolved and hence locus of the LL
crossing points form curves in $n$-$E^\perp$ space. Fig. 7.7b shows the zoomed-in data shown
in Fig. 7.7a.

Fig. 7.8a shows the same data (shown in Fig. 7.7) plotted as a function of $E^\perp$ and
density which can be directly compared with theoretically calculated density of states
(DOS) shown in Fig. 7.8b. On the electron side at 0.5 T, Fig. 7.8a shows two thick
curves (labelled $0_M^+$ and $0_M^-$) whose separation increases in density axis with increasing
$E^\perp$. Comparing the experimental data (Fig. 7.8a) and the simulation (Fig. 7.6b and

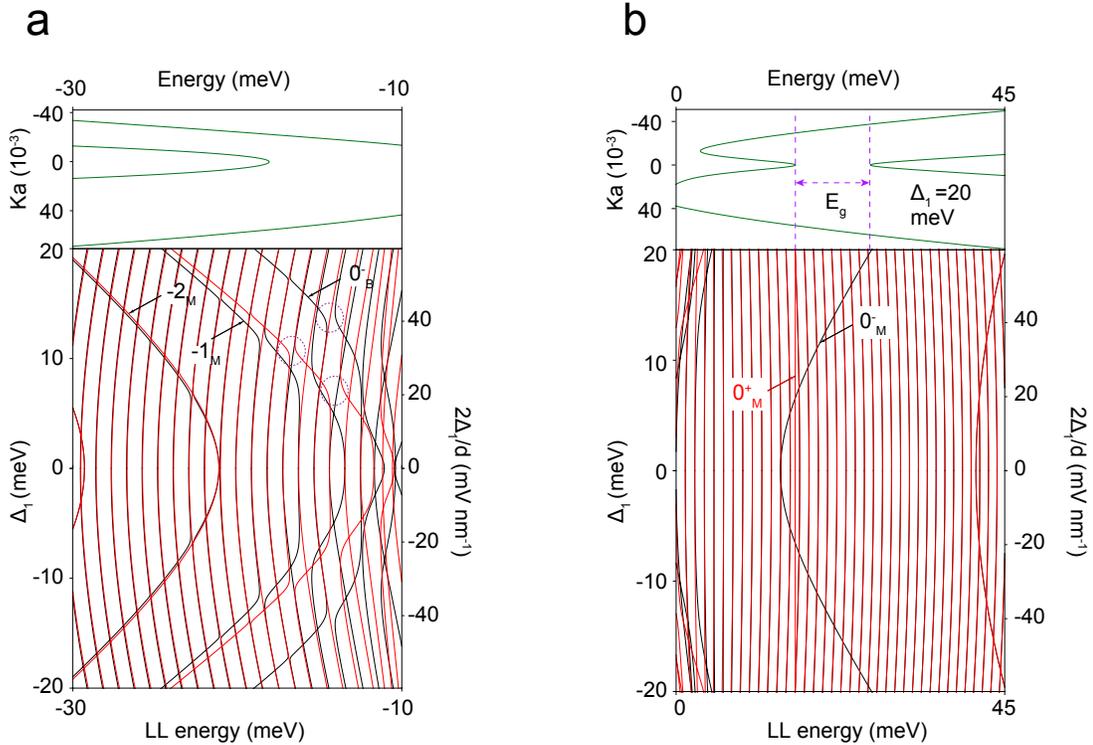

Figure 7.6: **Landau level diagram as a function of electric field.** LL energy as a function of electric field at 0.5 T (a) on the hole side and (b) on the electron side. Spin degeneracy is maintained. Red and black colour denote the $K_+$ valley and $K_-$ valley respectively. Many red lines are on top of the black lines as the spectrum is mostly valley degenerate. The corresponding band diagrams for $\Delta_1$=20 meV at zero magnetic field are also shown in the same energy range. The LL diagram on the electron side shows that the valley splitting of the N=0 MLG-like LL is same as the hybridization gap between the cones.

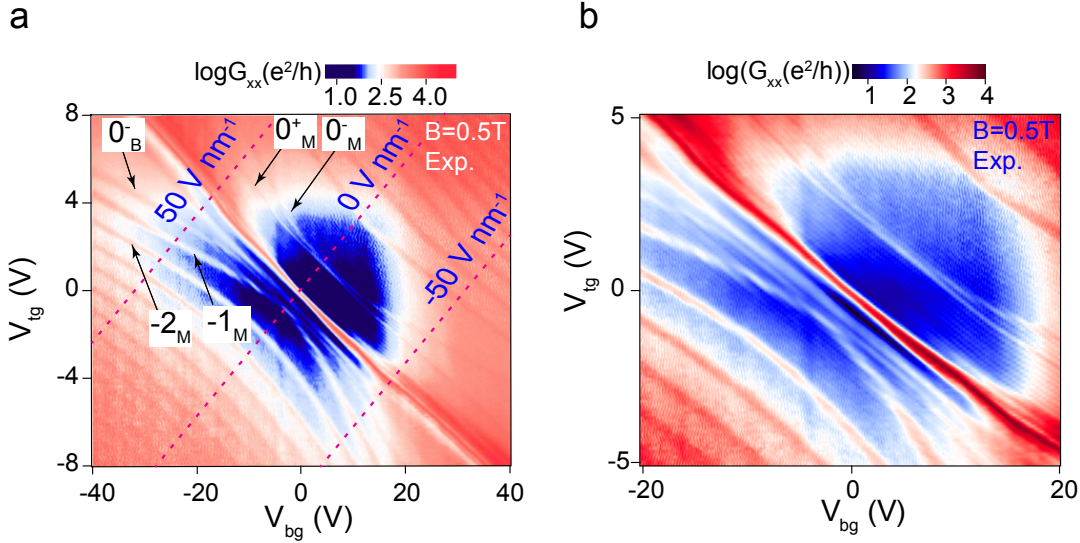

Figure 7.7:  (a) Experimental colour plot of $G_{xx}$ as a function of $V_{bg}$ and $V_{tg}$ at 0.5 T. (b) Zoomed in plot of panel - a.

Fig. 7.8b) these two thick lines can be identified as the crossing points of rapidly dispersing $0_M^+$ and $0_M^-$ LLs with other less dispersive LLs. On the hole side of Fig. 7.8a we see four thick curves dispersing with $E^\perp$. The first thick curve from the right (near zero density) comes due to a group of closely spaced LL crossing with each other. It does not have a particular LL index and hence it is not labeled. Comparing with the simulation (Fig. 7.6a and Fig. 7.8b) the remaining three curves (from the right) can be identified as the crossing points of rapidly dispersing $0_B^-$, -$1_M$ and -$2_M$ LLs with other less dispersive LLs.

Now, using the experimental data on the electron side, as shown in Fig. 7.8a, we can estimate the band gap between the MLG-like bands. We note that the $0_M^+$ and $0_M^-$ marked LLs in Fig. 7.8a originate from the conduction and valence band edges of MLG-like bands [14] (as the bands are split in ABA-stacked TLG). So, the valley splitting of $0_M$ LL is identical to the band gap of MLG-like bands, see Fig. 7.6b. From our experiment, we see that the valley splitting of the $0_M$ LL cannot be resolved at zero electric field (Fig. 7.8a) meaning the valley splitting is ∼1 meV for a LL broadening of ∼1 meV; this means that the band gap of MLG-like bands is also ∼1 meV. Calculation (Fig. 7.6b) shows that the valley splitting of $0_M$ LL changes non-monotonically. At zero electric field $0_M^-$ LL lies below $0_M^+$ LL, however, with increasing electric field $0_M^-$ LL floats up in energy. As a result, the valley splitting of $0_M$ LL decreases with increasing electric field till the two LLs cross (becomes degenerate) at an electric field $E_\perp^d$. After the crossing, the energy of the $0_M^-$ LL continues to increase and hence valley splitting now increases with the increasing electric field for $E_\perp > E_\perp^d$. The larger the initial band gap the larger the $E_\perp^d$ is and hence more evident is the non-monotonic band gap variation. In our experiment, we see (from Fig. 7.8a) that the band gap steadily increases with the electric field i.e. $E_\perp^d$ is quite small ($< 8$ mV nm$^{-1}$). This means that the initial band gap is similar to the LL broadening ($\sim$1 meV) which forbids the observation of the further decrease of the band gap (see Fig. 7.6b and Fig. 7.8b for LL diagram and the calculated DOS respectively). In Fig. 7.9 we show the comparison between the experimental data and calculated DOS for several values of Dirac band gap. It shows that if the Dirac band gap is more than $\sim$1 meV, the valley splitting of $0_M$ LL at zero electric field and the non-monotonicity of the splitting (with the electric field) becomes immediately resolvable. The comparison clearly shows that the Dirac band gap has to be $\sim$1 meV. This band gap was assumed to be much larger in many previous studies [70, 71, 76]. Here we show a comparison table (Table 7.1) for the MLG-like band gap used in literature. Theoretically, the band gap between MLG-like bands at zero electric field can be calculated from the band parameters as $E_g = \delta + \frac{\gamma_2}{2} - \frac{\gamma_5}{2}$. The fitting process to determine the individual band parameters are described in section 7.5. Our work provides a direct way to estimate this band gap by simply keeping track of the LLs originated from the band edges.

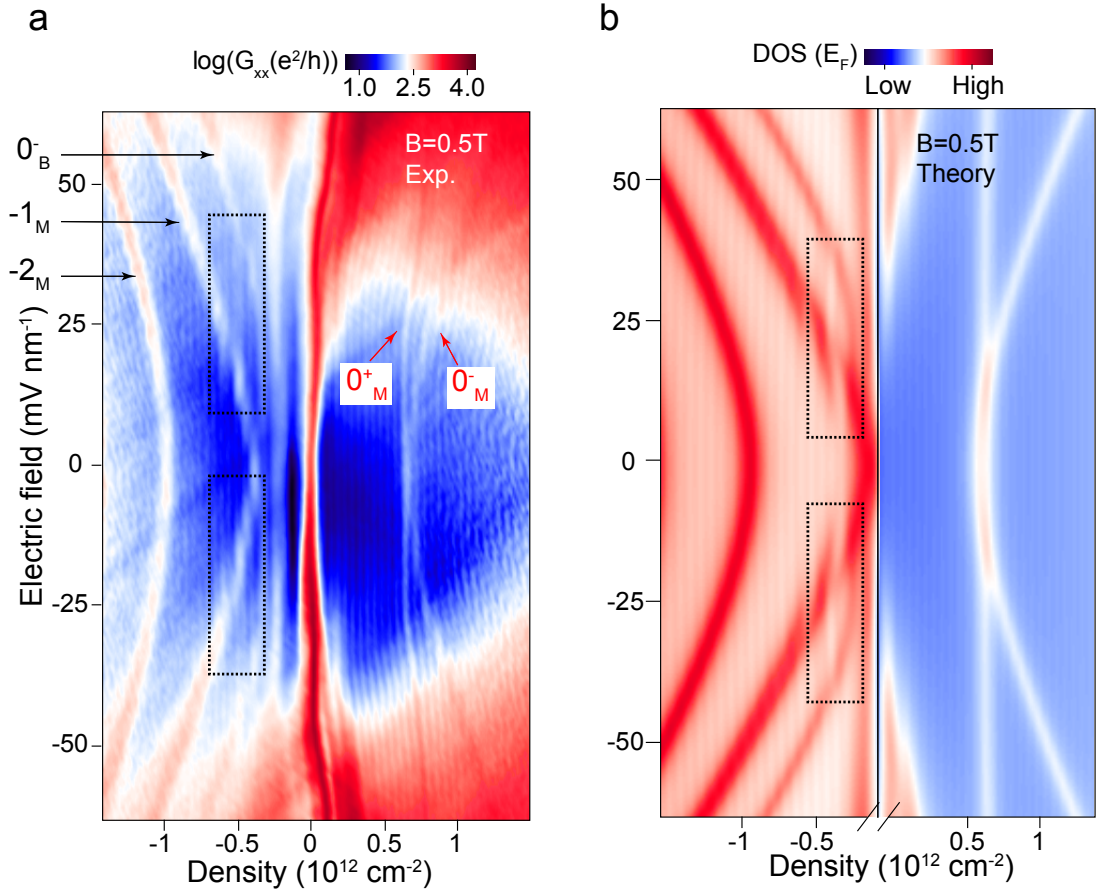

Figure 7.8: (a) $G_{xx}$ plotted as a function of electric field and density at 0.5 T. (b) Theoretically calculated $DOS(E_F)$ in the same electric field and density range.

Table 7.1: Comparison of band gap between MLG-like bands from literature

| Reference | $\delta$ (meV) | $\gamma_2$ (meV) | $\gamma_5$ (meV) | $E_g = \delta + \frac{\gamma_2}{2} - \frac{\gamma_5}{2}$ (meV) |
|---|---|---|---|---|
| This work | 20 | -20 | 18 | 1 |
| Campos, Leonardo C., et al. PRL (2016) [74] | 15 | -18 | 10 | 1 |
| Stepanov, Petr, et al. PRL (2016) [76] | 27 | -32 | 10 | 6 |
| Shimazaki, Yuya, et al. [71] | 46 | -28 | 50 | 7 |
|  | 34 | -28 | 50 | -5 |
| Taychatanapat, Thiti, et al. Nat. Physics (2011) [70] | 46 | -28 | 50 | 7 |
| Asakawa, Yuta, et al. PRL (2017) [72] | 14.3 | -23.7 | 6 | -0.55 |

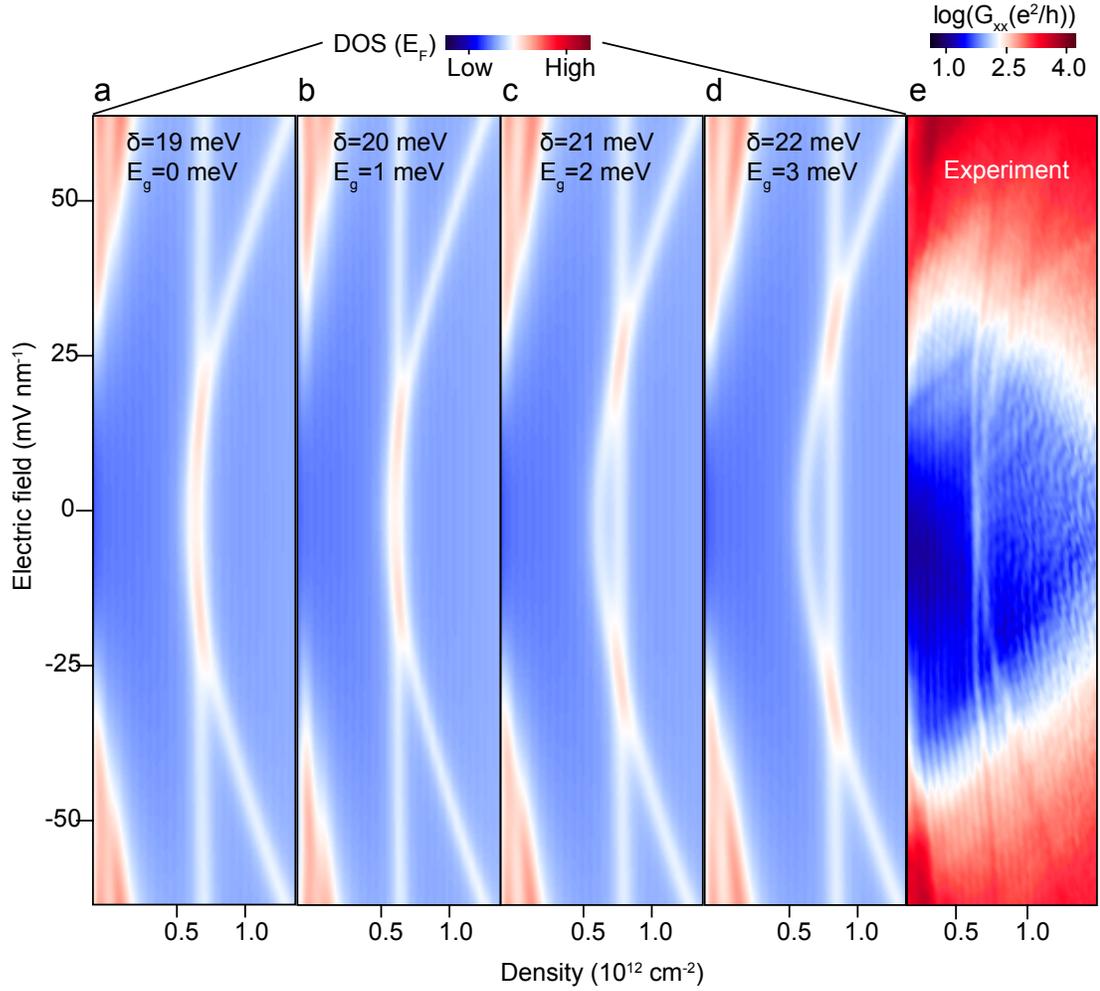

Figure 7.9: **Determination of the Dirac band gap.** (a)-(d) Numerically calculated DOS at 0.5 T for the Dirac band gap 0-3 meV. The band gap is changed by varying the tight binding band parameter $\delta$ from 19 meV to 22 meV. (e) Experimentally measured $G_{xx}$ at 0.5 T.

## 7.4 Consequence of continuous rotational symmetry breaking in transport

Trigonal warping, despite being small in magnitude, has important effects on the low energy physics of few-layer graphene. It stretches the Fermi circle along three directions in the Brillouin zone enforcing discrete three-fold rotational symmetry on the Fermi surface [117]. The consequence of trigonal warping in bilayer graphene and in ABC-stacked TLG is the appearance of Lifshitz transition at low energy and the associated three-fold degenerate LLs which shows up as multiple of 3 in filling factors [117, 118]. In ABA-stacked TLG, however, the effect of trigonal warping is more interesting as it introduces additional selection rules for the coupling between some LLs. Trigonal warping at low energy in few-layer graphene is controlled by the band parameter $\gamma_3$. Fig. 7.10 shows that finite $\gamma_3$ trigonally warps the BLG-like Fermi surface. The effect of trigonal warping in the zero magnetic field transport is not observable. To see the effect of trigonal warping, we focus on the data at a higher magnetic field of 1.5 T where the Landau levels are better resolved. Fig. 7.11a shows LL crossings as a function of the electric field at 1.5 T which can be compared with the theoretical simulations, see Fig. 7.11b and Fig. 7.11c. The first curve from the right labeled $0_B^-$ shows an interesting splitting pattern at $E^\perp \sim 65$ mVnm$^{-1}$ between $\nu$=-22 and -26 marked by a dashed rectangle. By comparing with the theoretical simulation we identify the feature within the rectangle as a LL anticrossing. This anticrossing between two LLs appears due to the trigonal warping ($\gamma_3$) – it couples a BLG-like LL with another BLG-like LL with LL index differing by multiple of 3 [14, 73]. This can be understood physically in the following manner. We note that the low energy BLG-like LLs lie predominantly on the $A_1+A_3$ and $B_2$ lattice sites which get coupled in the presence of $\gamma_3$, as $\gamma_3$ is the hopping parameter between $A_{1,3} \leftrightarrow B_2$ sites. Calculated LL spectra at 1.5 T (Fig. 7.11c) shows that the feature at $E^\perp \sim 65$ mVnm$^{-1}$ in Fig. 7.11a corresponds to the intra-subband anticrossing between $0_B^-$ and -$6_B^-$ LLs. We observe an interesting line-like feature within the anticrossing unlike in traditional GaAs/AlGaAs quantum well systems [119]. This is due to the existence of two nearly degenerate valleys at low $B$. When both -$6_B^+$ and -$6_B^-$ nearly degenerate LLs cross $0_B^-$, only -$6_B^-$ shows anticrossing with $0_B^-$ (Fig. 7.11b). As a result, -$6_B^+$ goes through the anticrossing gap producing a line-like feature within the anticrossing gap.

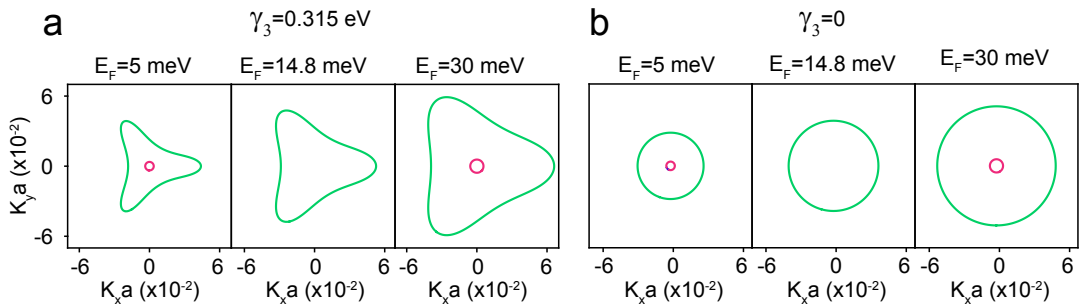

Figure 7.10: **Dependence of trigonal warping on $\gamma_3$.** Fermi surface at three different energies for (a) $\gamma_3$=0.315 meV and (b) $\gamma_3$=0. The red and the green contours are the MLG-like and the BLG-like Fermi surfaces respectively. There is no MLG-like Fermi surface for $E_F$=14.8 meV as it corresponds to the MLG-like band gap.

A closer look at the 0.5 T data at $n \sim$-0.5$\times 10^{12}$ cm$^{-2}$ (Fig. 7.8a) shows a vertical line like feature joining second (marked $0_B^-$) and third (marked -$1_M$) curves on the hole side which is also a signature of anticrossing (see Fig. 7.6a and Fig. 7.8b for the LL diagram and DOS). We infer from theory (Fig. 7.6a) that the anticrossing at $E^\perp \sim 20$ mVnm$^{-1}$ in Fig. 7.8a corresponds to the inter-subband anticrossings of -$1_M$ with other BLG-like LLs. In contrast, the feature at $E^\perp \sim 38$ mVnm$^{-1}$ in Fig. 7.8a corresponds to the intra-subband anticrossing between $0_B^-$ with another BLG-like LL.

The selection rule for the coupling (mediated by $\gamma_3$) between two BLG-like LLs can be understood by considering the approximated $2 \times 2$ BLG-like Hamiltonian and treating the trigonal warping term perturbatively. At not too low energy ($|\epsilon| > \frac{\sqrt{2}\gamma_1}{4} \left( \frac{v_3}{v} \right)^2$) the dominating term of the BLG-like Hamiltonian is quadratic in momentum. Near $K_-$ valley this can be written as

$$H_B = -\frac{1}{2m} \begin{pmatrix} 0 & (a^\dagger)^2 \\ a^2 & 0 \end{pmatrix}.$$

(7.1)

The eigenstates of this Hamiltonian (equation 7.1) are

$$|\psi_N\rangle = \begin{pmatrix} \pm |N\rangle \\ |N-2\rangle \end{pmatrix}.$$

(7.2)

For Fermi energy $|\epsilon| > \frac{\sqrt{2}\gamma_1}{4} \left( \frac{v_3}{v} \right)^2$, the linear trigonal warping term can be treated as a perturbation

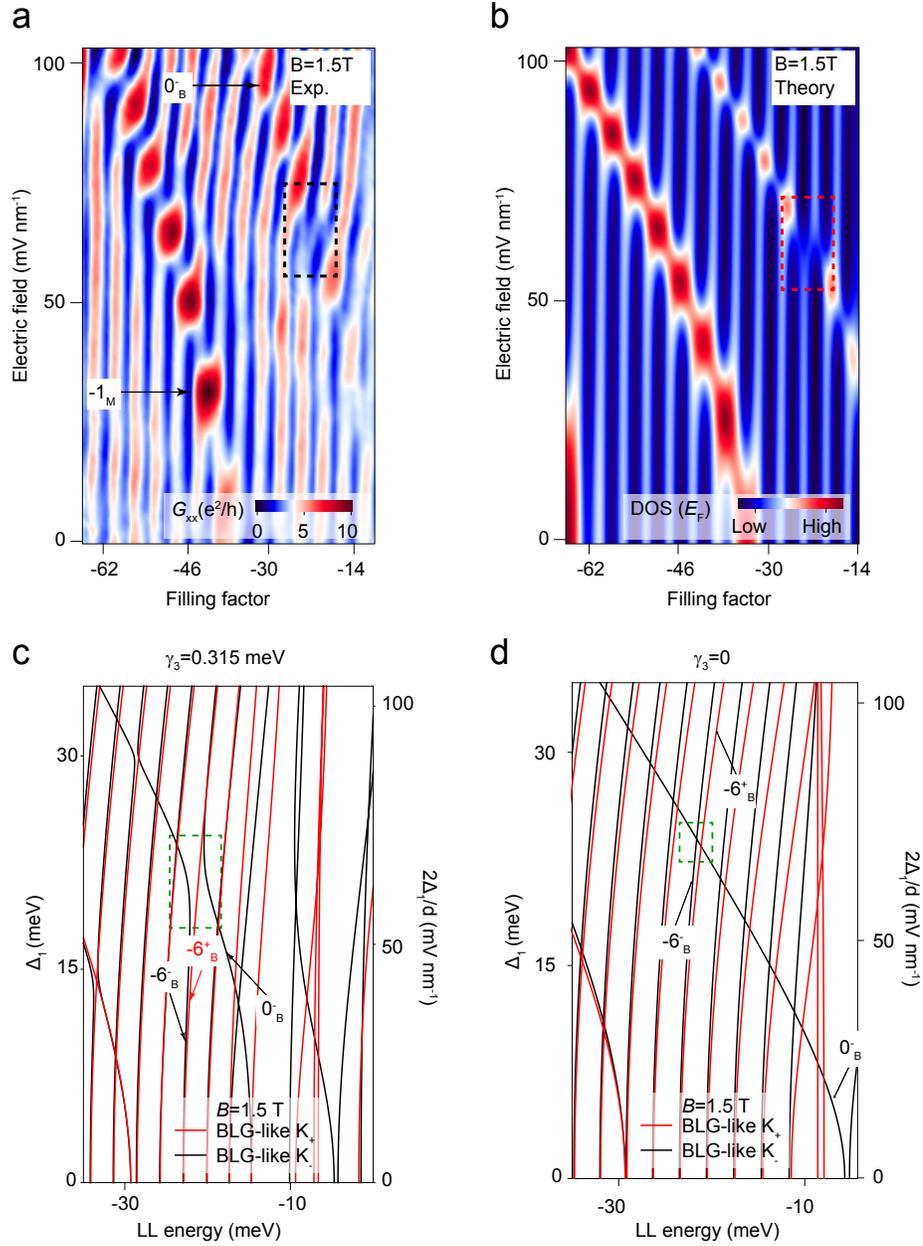

Figure 7.11: **Trigonal warping mediated Landau level anticrossing.** (a) Colour plot of experimental $G_{xx}$ at 1.5 T. The thick curved lines are the locus of the LL crossings. Here, $0_B^-$ refers to the locus of all the crossings between $0_B^-$ and other BLG-like LLs. Other labels have a similar meaning. The dashed rectangle marks the observed LL anti-crossing. (b) Calculated DOS($E_F$) from numerically calculated LL spectrum which can be directly compared to the experimental data in the panel - a. The experimentally observed anti-crossing is also reproduced theoretically at the correct electric field and filling factor. (c) LL energy diagram at 1.5 T which shows $0_B^-$ and $-1_M$ LLs rapidly dispersing with electric field crossing several BLG-like LLs. The green dashed rectangle shows the anticrossing between $0_B^-$ and $-6_B^-$ LLs observed in the experiment. (d) LL energy as a function of the electric field at 1.5 T for $\gamma_3=0$ showing no anticrossings. $0_B^-$ and $-6_B^-$ LLs are marked which show anticrossing in presence of $\gamma_3$.

$$H_{v_3} = \sqrt{2}v_3 \begin{pmatrix} 0 & a \\ a^\dagger & 0 \end{pmatrix}. \tag{7.3}$$

It is easy to see that the effect of $H_{v_3}$ is to couple $|\psi_N\rangle$ and $|\psi_{N\pm3}\rangle$:

$$\langle\psi_{N-3}| H_{v_3} |\psi_N\rangle = \pm\sqrt{2}v_3\sqrt{N-2} \quad \text{and} \quad \langle\psi_{N+3}| H_{v_3} |\psi_N\rangle = \pm\sqrt{2}v_3\sqrt{N+1}. \tag{7.4}$$

Numerical calculation using the full Hamiltonian shows that finite $\gamma_3$ couples $|\psi_N\rangle$ also to $|\psi_{N\pm6}\rangle$, $|\psi_{N\pm9}\rangle$.. with progressively smaller overlap coefficient.

## 7.5   Determination of band parameters

We determine the band parameters by matching the experimental LL crossings with theory at 1.5 T. We find $\gamma_0$=3.1 eV, $\gamma_1$=390 meV, $\gamma_2$=-20 meV, $\gamma_3$=315 meV, $\gamma_4$=120 meV, $\gamma_5$=18 meV, $\delta$=20 meV best describe our data. Precise determination of $\Delta_2$ requires input from high $B$ data and is discussed in more detail in section 7.6. We vary small band parameters which might change from the bulk graphite values. Also, large band parameters affect the high LL index levels whereas low LL index levels are mostly controlled by smaller band parameters (especially $\gamma_2$, $\gamma_5$, $\delta$ and $\Delta_2$). Constraining large band parameters using experimental low LL index crossings are not effective which is evident in the large range of $\gamma_4$ (40-140 meV) in the previous study [74]. In this study, we focus on the zeroth LLs which depend directly on $\gamma_2$, $\gamma_5$, $\delta$ and $\Delta_2$ as mentioned below. For these reasons we determine $\gamma_2$, $\gamma_5$, $\delta$ and $\Delta_2$ from our experiments and use previously used values for the large band parameters $\gamma_0$ =3.1 eV, $\gamma_1$=0.39 eV, $\gamma_3$ =315 meV and $\gamma_4$ =120 meV. Our fitting procedure consists of the following steps.

### 7.5.1   Determination of $\gamma_2$, $\gamma_5$ and $\delta$

We use all the 12 LL crossing points (due to $0^-_{\text{B}}$ and $-1_{\text{M}}$ LLs) visible in Fig. 7.11a (at 1.5 T) to determine the band parameters $\gamma_2$, $\gamma_5$ and $\delta$. The filling factor range of the LL crossings used in the fitting is -62 to -14. We start with the bulk graphite values $\gamma_2$=-20 meV, $\gamma_5$=38 meV, $\delta$=8 meV and vary each of the band parameters around them till we minimize the fitting error in electric field at the correct filling factors. We define the fitting error $= \frac{1}{12}\sqrt{\sum_{i=0}^{12}(E_i^{Exp} - E_i^{Th})^2}$ where $E_i^{Exp}$ and $E_i^{Th}$ are the experimental and theoretical electric fields at the i'th LL crossing point respectively. Fig. 7.12 shows the fitting error in $\delta$, $\gamma_2$ and $\Delta_2$ varying one band parameter at a time (keeping other band parameters fixed to their final values mentioned in the text). Additionally, we find the valley gap of $0_M$ Landau level is $E_g = \delta + \frac{\gamma_2}{2} - \frac{\gamma_5}{2} \sim 1$ meV as discussed in section 7.3, see Fig. 7.9. This additional experimental constraint allows us to estimate $\gamma_2$, $\gamma_5$ and $\delta$ more accurately. The least fitting error in $\Delta_1$ using the band parameters is $\sim 0.8$ meV.

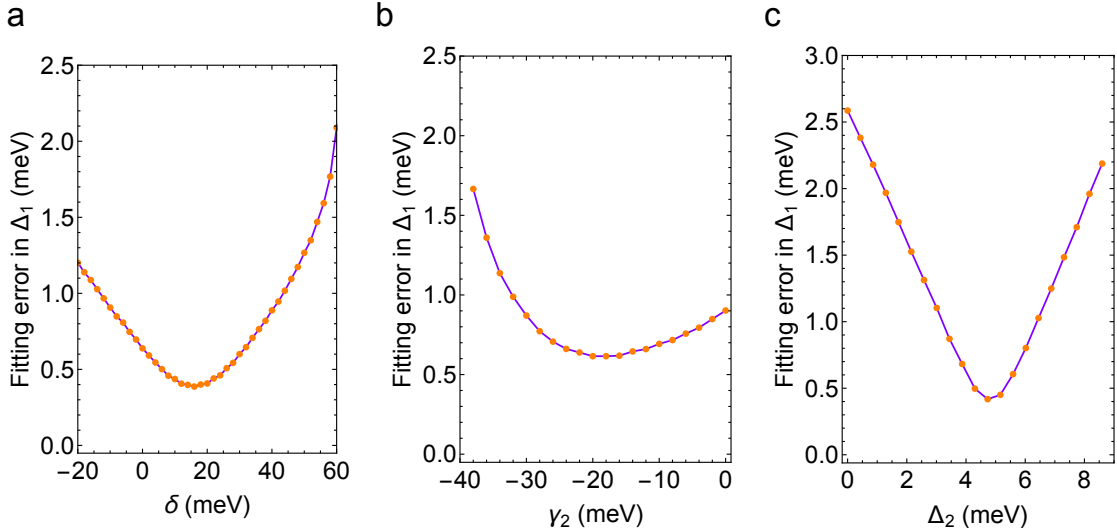

Figure 7.12: **Fitting error of determining the band parameters.** (a), (b) and (c) show the fitting error in $\delta$, $\gamma_2$ and $\Delta_2$ respectively varying one band parameter at a time. Other band parameters are kept fixed to their final values during this variation.

Varying $\Delta_2$ as a free parameter and fitting the low magnetic field data gives us a range for $\Delta_2 \sim 3\text{-}5.5$ meV. Since $\Delta_2$ is small compared to the energy of the LLs in this high filling factor regime, it is not possible to further constrain it with better accuracy. So, we determine the final value of $\Delta_2$ from the crossings within zeroth LLs at high magnetic fields described in section 7.6.

## 7.6 Electric field dependence of the $\nu = 0$ state

An independent verification of the calculated LL spectra using our band parameters can be seen from the fragility of the $\nu = 0$ state in presence of a small $E^\perp$. Fig. 7.13a shows that $\nu = 0$ at 13.5 T is best resolved at zero $E^\perp$ and disappears fast with increasing $E^\perp$. The red dashed line in Fig. 7.13b shows a well-developed minima in $G_{xx}$ at zero density and its disappearance with increasing $E^\perp$. Fig.7.13c shows that the $\nu=0$ quantum Hall plateau is also electric field dependent – the plateau disappears with increasing electric field. We show in the next section that $\nu=0$ state at 14 T appears between $1_B^{+\downarrow}$ and $1_B^{-\uparrow}$ LLs. Calculation shows that $E^\perp$ reduces the energy of $1_B^{+\downarrow}$ shrinking the $\nu=0$ gap which explains the disappearance of $\nu=0$ with increasing $E^\perp$. This is consistent with the electric field dependence of observed $G_{xx}$ in experiment.

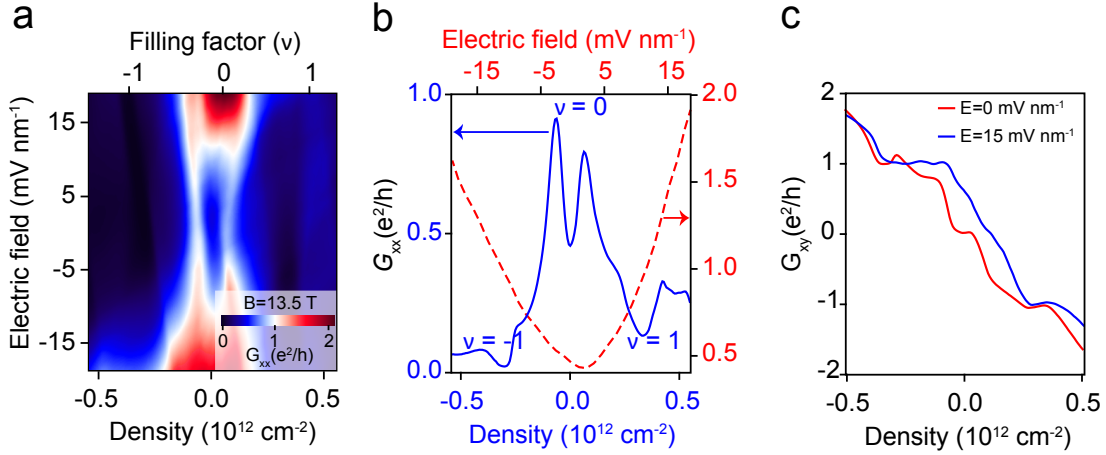

Figure 7.13: **Dependence of $\nu=0$ state on electric field.** (a) Colour plot of $G_{xx}$ showing the $E^\perp$ dependence of the $\nu=0$ state. (b) Blue solid line is the line-cut at zero $E^\perp$ showing a prominent dip at zero density which indicates the formation of the $\nu=0$ state. The red dashed line is the line-cut at zero density showing the disappearance of the minima with increasing $E^\perp$. (c) Electric field dependence of the Hall conductance ($G_{xy}$) corresponding to panel -b.

## 7.7 Role of Non-uniform charge density to dictate the LL sequence

We now show that the charge distribution on the three layers of the ABA-stacked TLG can be non-uniform and this has an important effect on the substructure of the lowest Landau level. Here we show a comparison table (Table 7.2) for the value of $\Delta_2$ used in the literature showing a large distribution of values. We emphasize that $\Delta_2$ plays a very important role and can not be neglected in the following two cases. First, for small filling factors where the order of LLs depends very sensitively on the exact value of $\Delta_2$. Second, for small $\Delta_1$ where the assumption of uniform electric field breaks down dramatically as $\Delta_2$ becomes comparable to $\Delta_1$. For example, the low filling factor LL crossings occur at small $\Delta_1$ ($\leq 10$ meV), see Fig. 7.14. In this regime, even for the maximum $\Delta_1$=10 meV, we show that the perpendicular electric field is very non-uniform. From the potential energy distribution in the three layers of ABA-stacked TLG shown in Fig. 7.1b, the electric field between the top and middle layers is $E_{\text{top}}^{\perp} = \frac{(\Delta_1 + \Delta_2) - (-2\Delta_2)}{d/2} = $ 68.3 mVnm$^{-1}$ where d=0.67 nm and $\Delta_2$=4.3 meV, determined later. In contrast, the electric field between the middle and bottom layers is $E_{\text{bottom}}^{\perp} = \frac{(-2\Delta_2) - (-\Delta_1 + \Delta_2)}{d/2} = $ -8.6 mVnm$^{-1}$. We note that not only the values of electric fields are extremely different, their directions are also opposite. This regime can not be described in terms of an average electric field $E_{\text{av}}^{\perp} = \frac{2\Delta_1}{d}$=29.8 mVnm$^{-1}$. This shows the importance of the $\Delta_2$ and our work shows a systematic way to determine this important parameter.

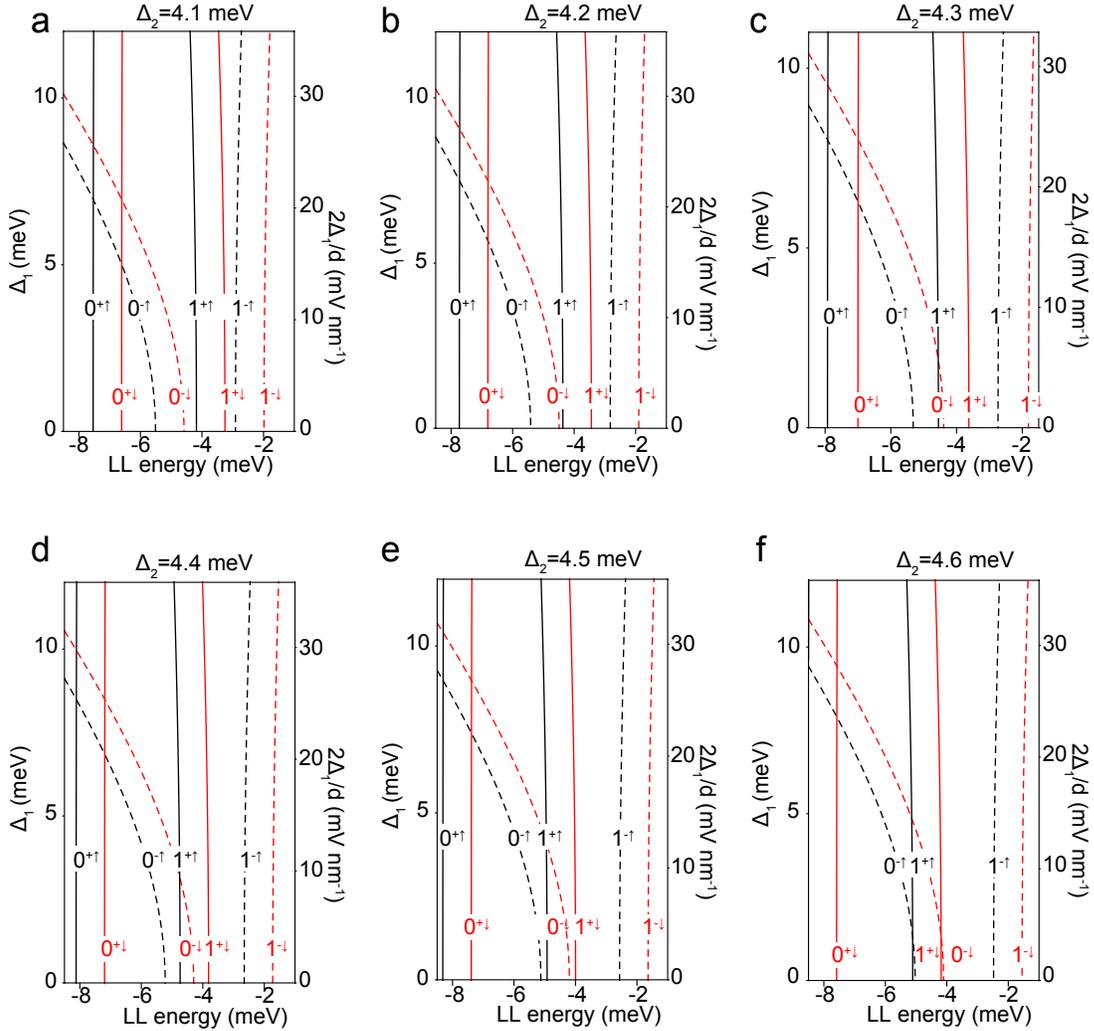

Figure 7.14: **Dependence of LL crossings on $\Delta_2$ at 8 T.** (a)-(f) Electric field dependence of zeroth LLs at 8 T for different values of $\Delta_2$=4.1-4.6 meV. We see that with increasing $\Delta_2$ the valley splitting of $0_B$ LL increases. As a result increasing $\Delta_2$ leads to new crossings between $0_B^-$ and $1_B^+$ LLs. This changes the number of LL crossings. Comparing the number of crossings and their locations with experimental data we can estimate the value of $\Delta_2$.

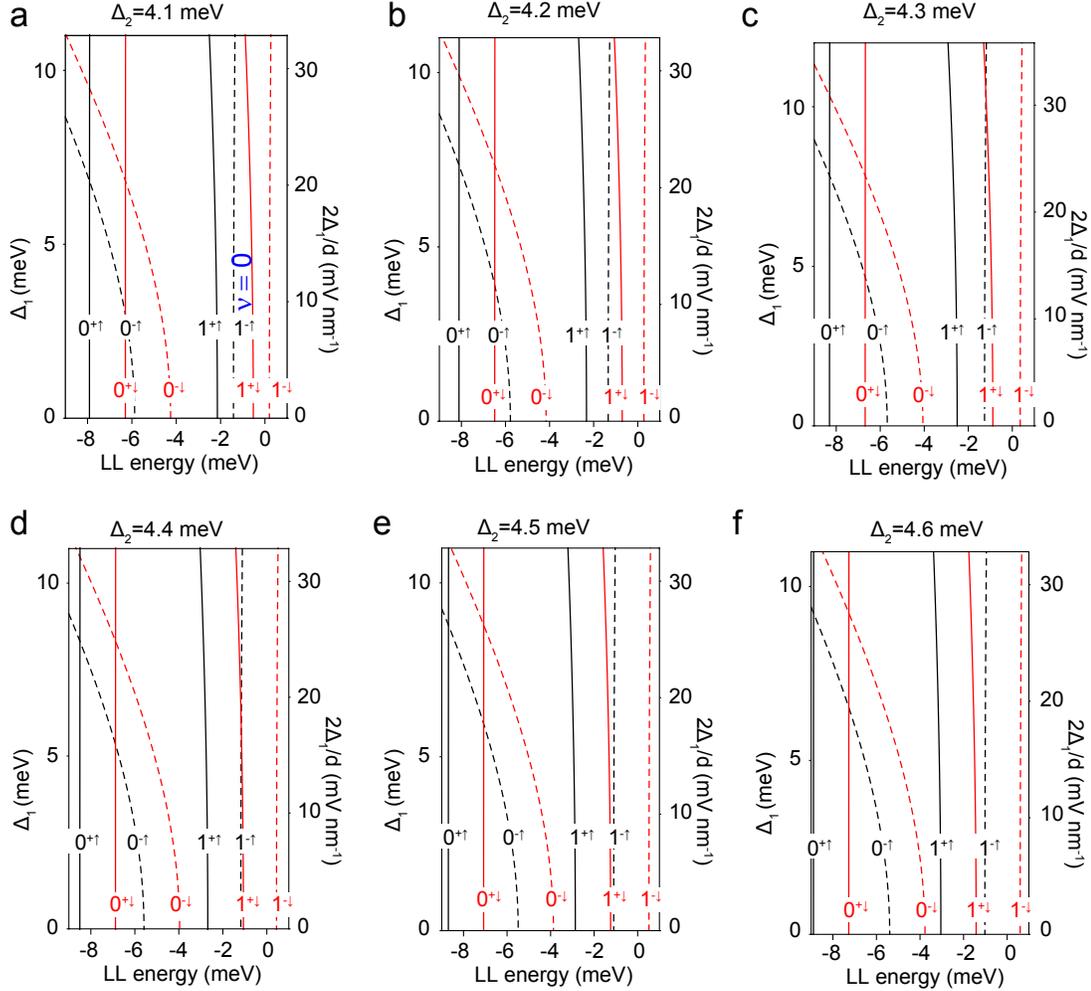

Figure 7.15: **Dependence of LL crossings on $\Delta_2$ at 14 T.** (a)-(f) Electric field dependence of zeroth LLs at 14 T for different values of $\Delta_2$=4.1-4.6 meV. We note that the $\nu$=0 gap is in between $1_B^{+\downarrow}$ and $1_B^{-\uparrow}$ LLs. For $\Delta_2$=4.3-4.4 meV these two LLs cross with each other at a small electric field ($\Delta_1 < 10$ meV) which destroys the $\nu = 0$ gap. Experimentally we see that the $\nu = 0$ gap closes with increasing electric field which allows us to constrain $\Delta_2$.

Table 7.2: Comparison of the $\Delta_2$ value used in literature

| Reference | $\Delta_2$ (meV) |
|---|---|
| This work | 4.3 |
| Campos, Leonardo C., *et al.* PRL (2016) [74] | 1.8 and 5.7 |
| Stepanov, Petr, *et al.* PRL (2016) [76] | 1.8 |
| Shimazaki, Yuya, *et al.* [71] | 0 |
| Taychatanapat, Thiti, *et al.* Nat. Physics (2011) [70] | 0 |
| Asakawa, Yuta, *et al.* PRL (2017) [72] | 0 |

Since $\Delta_2$ is especially important at low energies (for zeroth LLs) and at small $\Delta_1$, we determine $\Delta_2$ from the crossing within zeroth LLs. We focus at the high magnetic field data where the 12 fold approximately degenerate zeroth LL is complete symmetry broken. From our calculated LL spectra at high magnetic fields (Fig. 7.14 and Fig. 7.15), we find that the crossing pattern of the low energy LLs is very sensitive to the value of $\Delta_2$ [14]. This can be easily appreciated if we write down the approximate analytic energy expression of zeroth LLs at the zero electric field:

$$E(0_{\text{B}}^{+}) = -2\Delta_2 \tag{7.5}$$

$$E(0_{\text{B}}^{-}) = -\frac{|\gamma_2|}{2} + \Delta_2 \tag{7.6}$$

$$E(1_{\text{B}}^{+}) = -2\Delta_2 + \xi(\frac{\gamma_5}{2} + \delta + \Delta_2) \tag{7.7}$$

$$E(1_B^-) = -\frac{|\gamma_2|}{2} + \Delta_2 + \xi(\delta - 2\Delta_2). \qquad (7.8)$$

Here $\xi$ is a dimensionless parameter given by $\xi = \frac{\hbar\omega_c}{\sqrt{2}\gamma_1} = \frac{\hbar eBv^2}{\gamma_1^2} = 4.35 \times 10^{-3}B$. We clearly see that the energy of these LLs depends directly on $\Delta_2$ other than a few band parameters. We note that the valley gap of both $0_B$ and $1_B$ LLs (at small electric and magnetic field) is $E_B^{0+}$-$E_B^{0-} = \frac{|\gamma_2|}{2} - 3\Delta_2$ which is very sensitive on $\Delta_2$ as it comes with a factor of 3. So, the valley gap can be negative or positive depending on the value of $\Delta_2$; this determines the sequence of the LLs and hence has a bearing on the LL crossing pattern. Fig. 7.14 and Fig. 7.15 show the calculated LL diagrams for $\Delta_2$ values 4.1-4.6 meV at 8 T and 14 T respectively. This clearly shows that the crossing pattern is very sensitive to $\Delta_2$. Controlling the top and back gate voltages independently we study the LL crossings as a function of $E^\perp$ and density, see Fig. 7.16a and see Fig. 7.16c. Experimental data provides the following constraints on the crossing pattern:

1. At 8 T, there is one LL crossing at each $\nu$=-2,-3 and -5, but at $\nu$=-4 there are two LL crossings, see Fig. 7.16a.

2. At 14 T, $\nu$=-2 crossing disappears, see Fig. 7.16c.

3. For both 8 T and 14 T LL crossings at $\nu$=-3 and $\nu$=-5 are almost at the same electric field, see Fig. 7.16a and Fig. 7.16c.

4. $\nu$=0 gap at 14 T closes with the electric field, see Fig. 7.13.

We note that increasing $\Delta_2$ from zero decreases the valley splitting and their order eventually flips for $\Delta_2 > \frac{|\gamma_2|}{6} = 3.3$ meV, resulting in the $0_B^-$ LL lying above $0_B^+$ LL. This is a crucial fact because depending on the order of the LLs the crossing pattern can change completely. Matching the experimental crossing pattern with theory we conclude that the value of $\Delta_2$ has to be such that $0_B^{-\downarrow}$ lies between spin split $1_B^{+\uparrow}$ and $1_B^{+\downarrow}$ at 8 T for $\Delta_1$=0, see Fig. 7.14. We find a narrow range of positive $\Delta_2$ ∼4.3-4.4 meV which can explain all our experimental data. We consider $\Delta_2$=+4.3 meV in all the calculations; which suggests that the weight of the electronic wavefunction is concentrated more on the outer two layers. The existence of a positive potential in the middle layer can also be understood in quantum Hall regime [14] by noticing that $\nu$=-5 to -2 range on the hole side corresponds to emptying out either $1_B^+$ or both $1_B^+$ and $0_B^+$ LLs (Fig. 7.16b and Fig. 7.16d). Both of these states are polarized in the middle layer, emptying out these states can create a positive potential in the middle layer [14].

Analyzing the LL crossing patterns at 8 T and 14 T we can uniquely determine the ordering of spin and valley resolved LLs. Hence, we can label the LLs responsible for the observed crossings which were not possible in the previous studies [71, 74]. Fig. 7.16b and Fig. 7.16d show the calculated LL spectra on the hole side at 8 T and 14 T respectively. We note that the key change in the LL diagram in going from 8 T to 14 T is flipping the order of $0_B^{-\downarrow}$ and $1_B^{+\uparrow}$ LLs which is shown in Fig. 7.16e. The switching happens because orbital energy of $1_B^{+\uparrow}$ LL increases linearly with $B$ whereas the energy of $0_B^{-\downarrow}$ LL is independent of $B$ other than a small change due to Zeeman interaction. Calculations suggest that the energy of the $0_B^-$ LL decreases rapidly with increasing $\Delta_1$ leading to the crossings with other LLs below it. So, the crossing pattern depends on the sequence of the LLs below $0_B^-$ LL. In our experimental range of electric field, $0_B^-$, $0_B^+$ and $1_B^+$ LLs are solely responsible for all the crossings between $\nu =-5$ and -2. LL diagrams shown in Fig. 7.16b and Fig. 7.16d explain the LL crossings observed in our experiment (Fig. 7.16a and Fig. 7.16c).

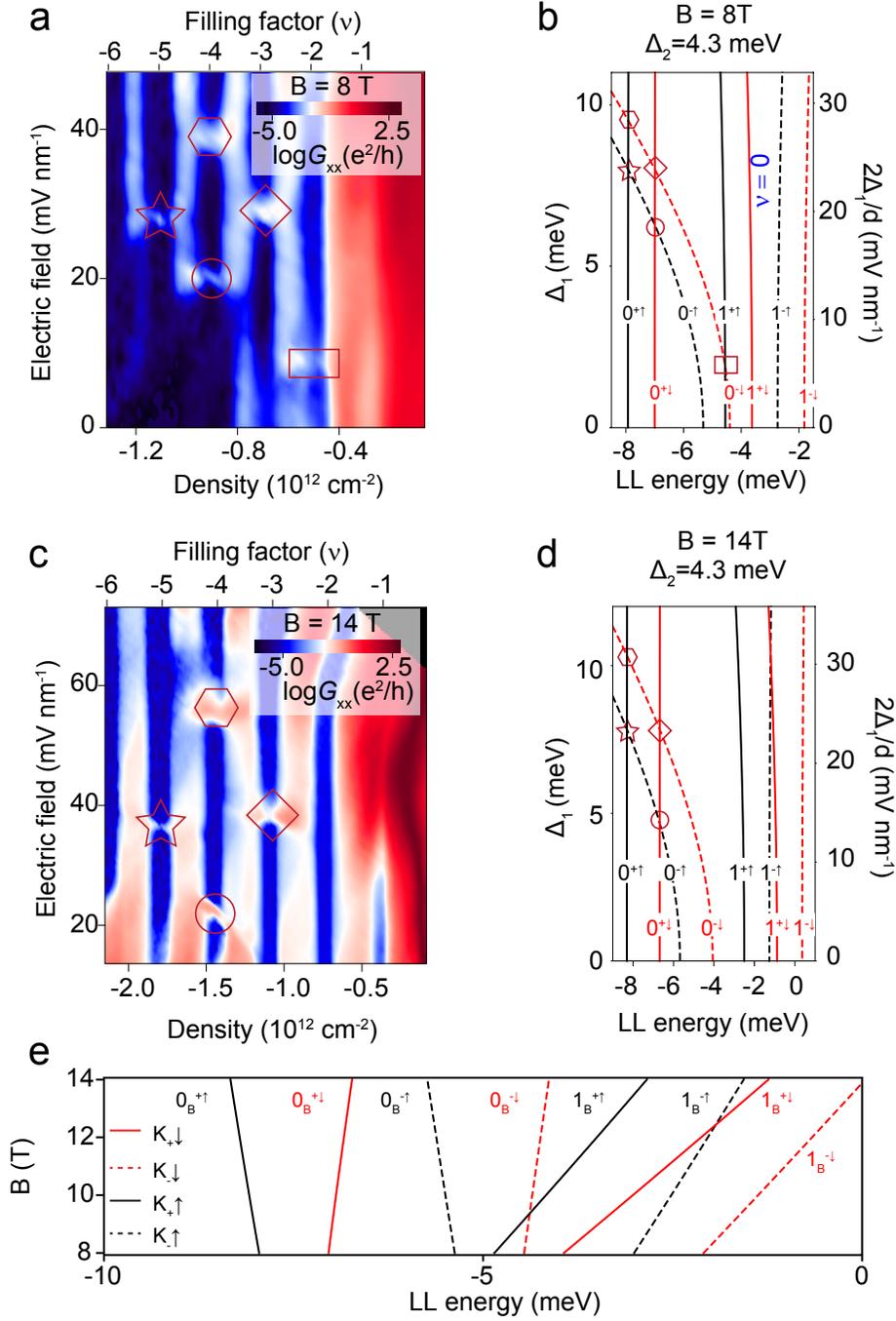

Figure 7.16: **Determination of $\Delta_2$ and the LL sequence.** (a) and (c) Experimental $G_{xx}$ showing the transitions between different QH states as a function of $E^{\perp}$ at 8 T and 14 T respectively. (b) and (d) Calculated LL spectra at 8 T and 14 T showing the BLG-like LLs corresponding to the experimental data. Red and black lines denote the LLs with $\downarrow$ and $\uparrow$ spins respectively. Solid and dashed lines denote the LLs from $K_+$ and $K_-$ valleys respectively. $\square$, $\diamond$, $\star$ symbols designate LL crossings at $\nu$=-2,-3 and -5. $\bigcirc$ and $\bigcirc$ symbols denote the lower and higher electric field crossings respectively at $\nu$=-4. (e) Dispersion of the LLs with magnetic field which shows the change in $0_B^{-\downarrow}$ and $1_B^{+\uparrow}$ LL ordering in going from 8 T to 14 T.

## 7.8    Details of the theoretical calculation

We calculate the Landau level spectra of ABA-stacked trilayer graphene using a tight binding model. The LLs are broadened by a phenomenological parameter to obtain a Lorentzian DOS peaked around the LL energy. We then calculate electron/hole density integrating this DOS up to the chemical potential ($E_F$). We plot the DOS at the Fermi level as a function of density and electric field. The DOS at the Fermi level shows all the features observed in the behavior of longitudinal conductance ($G_{xx}$) in the experiment.

In the presence of the electric field, the reflection symmetry between the top and bottom layer is broken in trilayer graphene and therefore the monolayer-like and bilayer-like bands of trilayer graphene get hybridized. In the rotated basis $\left\{ \frac{A_1-A_3}{\sqrt{2}}, \frac{B_1-B_3}{\sqrt{2}}, \frac{A_1+A_3}{\sqrt{2}}, B_2, A_2, \frac{B_1+B_3}{\sqrt{2}} \right\}$ the trilayer graphene Hamiltonian can be written as [14]

$$H_{TLG} = \begin{pmatrix} \Delta_2 - \frac{\gamma_2}{2} & v_0\pi^\dagger & \Delta_1 & 0 & 0 & 0 \\ v_0\pi & \Delta_2 + \delta - \frac{\gamma_5}{2} & 0 & 0 & 0 & \Delta_1 \\ \Delta_1 & 0 & \Delta_2 + \frac{\gamma_2}{2} & \sqrt{2}v_3\pi & -\sqrt{2}v_4\pi^\dagger & v_0\pi^\dagger \\ 0 & 0 & \sqrt{2}v_3\pi^\dagger & -2\Delta_2 & v_0\pi & -\sqrt{2}v_4\pi \\ 0 & 0 & -\sqrt{2}v_4\pi & v_0\pi^\dagger & \delta - 2\Delta_2 & \sqrt{2}\gamma_1 \\ 0 & \Delta_1 & v_0\pi & -\sqrt{2}v_4\pi^\dagger & \sqrt{2}\gamma_1 & \Delta_2 + \delta + \frac{\gamma_5}{2} \end{pmatrix}, \quad (7.9)$$

where $\Delta_1 = -e.\frac{U_1-U_3}{2}$ and $\Delta_2 = -e.\frac{U_1-2U_2+U_3}{6}$, with the potential of the layer $i$ is given by $U_i$. $\Delta_1$ is related to the average $E^\perp$ by $E^\perp_{av} = \frac{2\Delta_1}{(e)d}$, while $\Delta_2$ controls the asymmetry of the electric field between the layers. The band velocity $v_i (i = 0, 3, 4)$ is related to the tight binding parameter ($\gamma_i$) by $v_i\hbar = \frac{\sqrt{3}}{2}a\gamma_i$, where $a$ is the lattice constant. $\pi$ is defined in different valleys as $\pi = \zeta k_x + ik_y$ with $\zeta = \pm 1$ for $K_+$ and $K_-$ valley respectively. The action of $\pi$ and $\pi^\dagger$ on harmonic oscillator states in different valleys

are given by [14]

$$
\begin{aligned}
K_+ : \pi|n\rangle &= \frac{i\hbar}{l_B}\sqrt{2(n+1)}|n+1\rangle \\
\pi^\dagger|n\rangle &= -\frac{i\hbar}{l_B}\sqrt{2n}|n-1\rangle \\
K_- : \pi|n\rangle &= \frac{i\hbar}{l_B}\sqrt{2n}|n-1\rangle \\
\pi^\dagger|n\rangle &= -\frac{i\hbar}{l_B}\sqrt{2(n+1)}|n+1\rangle ,
\end{aligned}
\tag{7.10}
$$

where magnetic length $l_B = \sqrt{\hbar/eB}$. They satisfy the commutation relation $[\pi, \pi^\dagger] = 1$ in each valley.

The Landau levels for monolayer-like and bilayer-like blocks are hybridized by the off-diagonal matrix elements ($\Delta_1$). In absence of $\Delta_1$, the trilayer graphene Hamiltonian can be block diagonalized into monolayer-like and bilayer-like sectors i.e. $H_{\mathrm{TLG}} = H_{\mathrm{MLG}} \oplus H_{\mathrm{BLG}}$. Let us look at $K_+$ valley first. $H_{\mathrm{MLG}}$ can be diagonalized in an eigenbasis $\{|n-1\rangle, |n\rangle\}$, where $n$ runs from 0 to $\infty$. Similarly in absence of $\gamma_3$, $H_{\mathrm{BLG}}$ can be diagonalized in an eigenbasis $\{|m-2\rangle, |m\rangle, |m-1\rangle, |m-1\rangle\}$, where $m$ runs from 0 to $\infty$. It is evident that in presence of $\Delta_1$, $(n+1)$-th monolayer-like Landau level gets coupled with $m$-th bilayer-like Landau level (where both $n$ and $m$ runs from 0 to $\infty$). In $K_+$ valley $H_{\mathrm{TLG}}$ can be block diagonalized in $6 \times 6$ sectors (one for each $n$) if we choose $m = n + 1$ as following.

$$
\begin{pmatrix}
\Delta_2 - \frac{\gamma_2}{2} & v_0\pi^\dagger & \Delta_1 & 0 & 0 & 0 \\
v_0\pi & \Delta_2 + \delta - \frac{\gamma_5}{2} & 0 & 0 & 0 & \Delta_1 \\
\Delta_1 & 0 & \Delta_2 + \frac{\gamma_2}{2} & 0 & -\sqrt{2}v_4\pi^\dagger & v_0\pi^\dagger \\
0 & 0 & 0 & -2\Delta_2 & v_0\pi & -\sqrt{2}v_4\pi \\
0 & 0 & -\sqrt{2}v_4\pi & v_0\pi^\dagger & \delta - 2\Delta_2 & \sqrt{2}\gamma_1 \\
0 & \Delta_1 & v_0\pi & -\sqrt{2}v_4\pi^\dagger & \sqrt{2}\gamma_1 & \Delta_2 + \delta + \frac{\gamma_5}{2}
\end{pmatrix}
\begin{pmatrix}
c_1|n-1\rangle \\
c_2|n\rangle \\
c_3|n-1\rangle \\
c_4|n+1\rangle \\
c_5|n\rangle \\
c_6|n\rangle
\end{pmatrix}
=
\begin{pmatrix}
c'_1|n-1\rangle \\
c'_2|n\rangle \\
c'_3|n-1\rangle \\
c'_4|n+1\rangle \\
c'_5|n\rangle \\
c'_6|n\rangle
\end{pmatrix}
\tag{7.11}
$$

Similarly, in $K_-$ valley $H_{\mathrm{MLG}}$ can be diagonalized in an eigenbasis $\{|n\rangle, |n-1\rangle\}$ and $H_{\mathrm{BLG}}$ can be diagonalized in an eigenbasis $\{|m\rangle, |m-2\rangle, |m-1\rangle, |m-1\rangle\}$ (in absence of $\gamma_3$), where $n$ and $m$ both run from 0 to $\infty$. In presence of $\Delta_1$, $n$-th monolayer-like Landau level gets coupled with $m$-th bilayer-like Landau level (where both $n$ and $m$ runs from 0 to $\infty$). In $K_-$ valley $H_{\text{TLG}}$ can be block diagonalized in $6 \times 6$ sectors if we choose $m = n$ as following.

$$
\begin{pmatrix}
\Delta_2 - \frac{\gamma_2}{2} & v_0\pi^\dagger & \Delta_1 & 0 & 0 & 0 \\
v_0\pi & \Delta_2 + \delta - \frac{\gamma_5}{2} & 0 & 0 & 0 & \Delta_1 \\
\Delta_1 & 0 & \Delta_2 + \frac{\gamma_2}{2} & 0 & -\sqrt{2}v_4\pi^\dagger & v_0\pi^\dagger \\
0 & 0 & 0 & -2\Delta_2 & v_0\pi & -\sqrt{2}v_4\pi \\
0 & 0 & -\sqrt{2}v_4\pi & v_0\pi^\dagger & \delta - 2\Delta_2 & \sqrt{2}\gamma_1 \\
0 & \Delta_1 & v_0\pi & -\sqrt{2}v_4\pi^\dagger & \sqrt{2}\gamma_1 & \Delta_2 + \delta + \frac{\gamma_5}{2}
\end{pmatrix}
\begin{pmatrix}
c_1|n\rangle \\
c_2|n-1\rangle \\
c_3|n\rangle \\
c_4|n-2\rangle \\
c_5|n-1\rangle \\
c_6|n-1\rangle
\end{pmatrix}
=
\begin{pmatrix}
c_1'|n\rangle \\
c_2'|n-1\rangle \\
c_3'|n\rangle \\
c_4'|n-2\rangle \\
c_5'|n-1\rangle \\
c_6'|n-1\rangle
\end{pmatrix}
$$
$$(7.12)$$

In presence of $\gamma_3$, the rotational symmetry gets broken down to $C_3$ and hence states with quantum numbers differing by 3 couple to each other. This, in principle, makes the diagonalization problem infinite dimensional. However, we put a cut off on the matrix size and diagonalize large but finite matrices to obtain the spectra. We find that $N_{\max} \sim 100$ is sufficient to obtain the dispersion of low lying Landau levels to our desired accuracy.

The Landau levels of trilayer graphene Hamiltonian get broadened in presence of scattering and the density of states of each Landau level (with energy $E$) can be approximated by a Lorentzian

$$\text{DOS}(N, E) = \frac{1}{2\pi l_B^2} \frac{1}{\pi} \left( \frac{\Gamma}{(E - E_N)^2 + \Gamma^2} \right), \qquad (7.13)$$

where the effect of disorder has been incorporated in $\Gamma$, which is related to the relaxation time ($\tau$) of scattering as $\Gamma \sim (1/\tau)^{1/2}$ [100].

We calculate the density of states at Fermi energy ($\text{DOS}(E_{\text{F}})$) as a function of total density $n$ (integrating $\text{DOS}(E)$ up to Fermi energy $E_{\text{F}}$) and $\Delta_1$ to match the experimental data. We have found a good theoretical match with the experimental data with a choice of $\Gamma$=1.2 meV on the electron-side and $\Gamma$=0.85 meV on the hole-side of the dispersion. The choice of different $\Gamma$ is supported by the fact that mobility on the hole-side is more than on the electron-side.

The theoretical calculation was done in collaboration with Prof. Rajdeep Sensarma and his student Abhisek Samanta.

## 7.9    Summary

In summary, we have studied the role of the electric field in the LL crossing physics in ABA-stacked TLG. At a low magnetic field, we have shown that the electric field allows us to observe from simple LL crossings to anticrossings mediated by trigonal warping which we find in the experiment as an additional selection rule for the coupling between two LLs. Matching the LL crossing pattern at multiple magnetic fields allows us to pin down $\Delta_2 \sim 4.3$ meV which in turn helps us to determine the order of the low energy LLs uniquely. This value is surprisingly close to the self-consistently calculated value of $\Delta_2 \sim 4$ meV [14]. Recently it is predicted [120] that bilayer and trilayer graphene can host electric field tunable non-Abelian fractional quantum Hall states. Studying these fractional quantum Hall states in few-layer graphene [121, 122, 123, 124] requires the single particle Landau level diagram in which our work provides an important direction.

**Chapter 8**

# Publications

## Publications related to this thesis

Chapter 5, 6 and 7 have been published in the following papers

1. **Biswajit Datta**, Pratap Chandra Adak, Li-kun Shi, Kenji Watanabe, Takashi Taniguchi, and Mandar M Deshmukh. Non-trivial quantum oscillation geometric phase shift in a trivial band. Science Advances **5**, 10 (2019).

2. **Biswajit Datta**, Santanu Dey, Abhisek Samanta, Hitesh Agarwal, Abhinandan Borah, Kenji Watanabe, Takashi Taniguchi, Rajdeep Sensarma, and Mandar M. Deshmukh. Strong electronic interaction and multiple quantum Hall ferromagnetic phases in trilayer graphene. Nature Communications **8**, 14518 (2017).

3. **Biswajit Datta**, Hitesh Agarwal, Abhisek Samanta, Amulya Ratnakar, Kenji Watanabe, Takashi Taniguchi, Rajdeep Sensarma, and Mandar M. Deshmukh. Landau level diagram and the continuous rotational symmetry breaking in trilayer graphene. Physical Review Letters **121**, 056801 (2018).

# Chapter 9

# Future directions

In a nutshell, in this thesis, we have studied the electronic properties of ABA-trilayer graphene in the presence of a magnetic field. We have seen that, in general, few-layer graphene system has a lot more to offer than a mere 2DEG in semiconductor quantum wells. We have played with the spin, valley, pseudospin, layer degrees of freedom and the dispersion of the bands. We have studied the effect of electronic interaction, the effect of broken symmetries, the effect of the shape of Fermi surface in transport, the effect of different band fillings in a multiband system – these are just tip of the iceberg. There are plenty of other opportunities – For example, ABA-trilayer graphene is predicted to host ferromagnetic states at zero magnetic field when Coulomb interaction is taken into account [125]. Even though we have extensively used the perpendicular electric field and magnetic field to the graphene layers to tune its properties, we did not exploit another experimental knob – the parallel magnetic field in our studies. A parallel magnetic field lifts only the spin degeneracy by Zeeman interaction without coupling the orbital motion because graphene is essentially a zero-thickness system. In ABA-trilayer graphene the coupling between the monolayer graphene-like and the bilayer-layer graphene-like subbands can also be controlled with a parallel magnetic field [72].

Now in the era of "twistronics" one can stack two pieces of a material with an angle in between them and can engineer new band structure [126]. For a specific set of discrete angles – termed magic angles [127] the velocity becomes zero which means the bands become very flat. In twisted bilayer graphene, these flat bands are shown to host several electronic interaction driven states including Mott-like insulator [128], superconductor [10], and ferromagnet [129]. A new study on twisted double bilayer graphene – a system consisting of two bilayer graphene flakes with a relative angle $\approx 1.2°$ also reports correlated insulating states and superconductivity [130]. All these examples show that there are probably many "twisted systems" where the flat bands can host exotic states of matter.

ABC-trilayer graphene – the other stacking variant of trilayer graphene is interesting as well. More recently, a new report shows signatures of tunable superconductivity in the moiré superlattice of ABC-trilayer graphene and hexagonal boron nitride [131]. More works to understand the nature/mechanism of the superconductivity in the moiré superlattices are required. Moreover, coupling these "twisted systems" with supercon-duction electrodes will also be interesting.